\documentclass[aps,prd,reprint,eqsecnum]{revtex4-2}

\usepackage{dcolumn}
\usepackage{bm}
\usepackage{amsfonts}
\usepackage{latexsym}
\usepackage{amssymb}
\usepackage{verbatim}
\usepackage{amsthm}
\usepackage{amsmath}
\usepackage{mathrsfs}
\usepackage{wrapft}
\usepackage{graphicx}
\usepackage{xcolor}

\newcommand{\NF}{{{\mathbb N}}}

\newcommand{\RF}{{{\mathbb R}}}

\newcommand{\calA}{{{\mathcal A}}}
\newcommand{\calB}{{{\mathcal B}}}

\newcommand{\calF}{{{\mathcal F}}}

\newcommand{\calN}{{{\mathcal N}}}

\newcommand{\calP}{{{\mathcal P}}}

\newcommand{\scrD}{{{\mathscr D}}}

\newcommand{\scrF}{{{\mathscr F}}}

\newcommand{\FrakA}{{{\mathfrak A}}}
\newcommand{\FrakB}{{{\mathfrak B}}}

\newcommand{\FrakD}{{{\mathfrak D}}}

\newcommand{\FrakR}{{{\mathfrak R}}}

\begin{document}


\title{Theoretical Detailed Analyses \\ for DC readout and a
  Fabry-P\'erot gravitational-wave detector}


\author{
  Kouji NAKAMURA
}
\email[]{E-mail:dr.kouji.nakamura@gmail.com}
\affiliation{
  Gravitational-Wave Science Project, National Astronomical Observatory,
  Mitaka, Tokyo 181-8588, Japan
}


\date{\today}

\begin{abstract}
  The quantum expectation value and the stationary noise spectral
  density for a Fabry-P\'erot gravitational-wave detector with a DC
  readout scheme are discussed in detail only through the quantum
  electrodynamics of lasers and the Heisenberg equations of mirrors'
  motion.
  We demonstrate that the initial conditions of the mirrors' motion
  concentrate around the fundamental frequency of the pendulum and are
  not related to the frequency range of our interest.
  Although, in the ideal case, there is consensus that the shot-noise
  contribution from the laser to the high-frequency range of the
  signal-referred noise spectral density decreases as the injected
  laser power increases, our derived noise spectral density shows that
  the shot-noise contribution does not decrease.
  This is due to leakage of classical radiation pressure forces from
  the carrier field to the output port, and the carrier field is used
  as the reference in the DC readout scheme.
  Since classical radiation pressure acts as a constant force, it
  shifts the pendulum's equilibrium point of the mirrors' motion.
  To recover the ideal case, we must consider adjusting the
  interferometer's tuning point to place the mirrors at their
  equilibrium positions.
  We investigate the case where the equilibrium tuning is incomplete
  and show that the behavior of the above shot noise is due to this
  incompleteness.
  We also discuss the maximum deviation of the mirror displacements
  from the equilibrium point during incomplete tuning to recover a
  near-ideal case.
\end{abstract}


\maketitle

\section{Introduction}
\label{sec:Introduction}


A decade ago, the first detection of the gravitational-wave signal
GW150914 from a black hole-black hole binary was achieved.
This event was the beginning of gravitational-wave astronomy and
multi-messenger astronomy including gravitational-wave
detection~\cite{B.P.Abbott-et-al-2016a,LIGO-homepage}.
We are now at a stage where we can directly measure gravitational
waves and conduct scientific research using these events.
Many events, mainly from black hole-black hole binaries,
have already been detected.
We can also expect that one of the future directions of
gravitational-wave astronomy is to develop as a ``precise science'',
as we can easily extrapolate from the current achievements~\cite{LIGO-Virgo-KAGRA-publications-homepage}.
These involve detailed studies of source science from the
astrophysical point of view, the tests of general relativity, and the
developments of the global network of gravitational-wave
detectors~\cite{LIGO-homepage,Virgo-home-page,KAGRA-home-page,LIGO-INDIA-home-page}.
In addition to the current network of ground-based detectors, as
future ground-based gravitational-wave detectors, the projects of
Einstein Telescope~\cite{ET-home-page} and Cosmic
Explorer~\cite{CosmicExplorer-home-page} are in progress, aiming to
achieve better sensitivity.


To advance the field of gravitational-wave detections as a more
precise science, it is essential to improve detector sensitivity.
We note that theoretical developments in detector science are also
crucial for enhancing our basic understanding of these
gravitational-wave detectors.
In the high-frequency part of the noise spectral density, current
gravitational-wave detectors are limited by shot noise, one of the
quantum noises in the interferometer.
In interferometric gravitational detectors, it is considered that
there are two kinds of quantum noises, although many plots in the
gravitational-wave community do not distinguish these two quantum
noises, e.g. see
Ref.~\cite{S.L.Danilishin-F.A.Khalili-H.Miao-LRR-2019}.
One is the shot noise of the laser, which limits the sensitivity of
gravitational-wave detectors at high frequencies.
On the other hand, quantum radiation pressure forces are considered a
constraint on sensitivity in the low-frequency region.
Note that, in the current experimental situation, this quantum
radiation-pressure noise remains subdominant in this low-frequency
region.
In the common understanding in the gravitational-wave detection
community, if the injected power is increased, the shot-noise
contribution to the signal-referred noise spectral density is
decreased, while the contribution of the quantum radiation pressure
forces is
increased~\cite{H.J.Kimble-Y.Levin-A.B.Matsko-K.S.Thorne-S.P.Vyatchanin-2001}.
The envelop of this trade-off relation in the signal-referred noise
spectral density is regarded as the ``standard quantum limit.''
It is also a common understanding in the gravitational-wave detection
community that this standard quantum limit is estimated from the
Heisenberg uncertainty relation arising from the noncommutativity of
the quantum positions and momenta of free test
masses~\cite{H.Miao-PhDthesis-2010}.


Furthermore, in the current research on gravitational-wave detection,
there are some reports which state that the current LIGO
gravitational-wave detectors already surpassed the above ``standard
quantum limit''~\cite{H.Yu-et-al-2020,W.Jia-et-al-2024}.
Even from a theoretical perspective, this presents an opportune moment
to revisit and refine the theoretical arguments, aiming for greater
accuracy than those presented in previous works in
Refs.~\cite{H.J.Kimble-Y.Levin-A.B.Matsko-K.S.Thorne-S.P.Vyatchanin-2001,H.Miao-PhDthesis-2010}.


On the other hand, a mathematically rigorous quantum measurement
theory in quantum mechanics has also been developed
(Ref.~\cite{M.Ozawa-2004} and the references therein), introducing new
error-disturbance relations distinct from the Heisenberg uncertainty
relation.
These new error-disturbance relations have already been
confirmed through
experiments~\cite{J.Erhart-et-al-Nature-Phys-2012,L.A.Rozema-et-al-2012,F.Kaneda-S.Baek-M.Ozawa-K.Edamatsu-2014}.
One of the motivations of this development was the detection of
gravitational waves~\cite {M.Ozawa-1988}.
However, the actual application of this mathematical theory to the
gravitational-wave detectors requires its extension to the quantum
field theories, because the quantum noise in gravitational-wave
detectors is analyzed through the quantum field theories of
lasers~\cite{H.J.Kimble-Y.Levin-A.B.Matsko-K.S.Thorne-S.P.Vyatchanin-2001}.
Moreover, in the quantum measurement theory, it is essential to
specify the final measured operator in the quantum measurement process
due to the von Neumann chain problem~\cite{J.vonNeumann-2018}.
In interferometric gravitational-wave detectors, the directly measured
quantum operator is identified within the detectors' ``readout
scheme.''
While current gravitational-wave detectors utilize a feedback control
system, and the final measured data of gravitational-wave detectors
consist of the electric currents from this feedback control, the
readout scheme in gravitational-wave detectors is the optical system
that determines the final measured quantum operator in the optical
fields detected at the photodetectors, which receive signals from the
main interferometer.
Therefore, research into this readout scheme is crucial for advancing
mathematical quantum measurement theory and its application to
gravitational-wave detections.


Due to the von Neumann chain issue, this paper adopts the perspective
that the quantum properties of the photons in the laser
interferometers are preserved until they are detected.
At the photodetectors, the quantum information of photons is
transformed into an electric current.
The processes that occur within photodetectors are complex and vary
depending on the device type.
As a result, we speculate that quantum decoherence occurs during this
photodetection process, and the resulting electric current, derived
from the statistical results of the detection, can be regarded as a
classical current.
If this perspective is incorrect and the quantum nature is retained
even in the feedback electric current, we would need to explore the
concept of a quantum feedback control
system~\cite{H.M.Wiseman-G.J.Milburn-2010}.
This consideration is beyond the current scope of this paper.


Current gravitational-wave detectors utilize a ``DC readout scheme,''
which is explained in
Sec.~\ref{sec:General_arguments_for_the_DC-readout_scheme} of this
paper.
Our first step is to investigate the DC-readout scheme from a quantum
theoretical perspective.
We previously discussed the mathematically rigorous quantum
description of the balanced homodyne detection as a readout scheme in
gravitational-wave detectors~\cite{K.Nakamura-2021}, since the
balanced homodyne detection is planned for installation in future
gravitational-wave detectors~\cite{GWADW2021-home-page}.
In this paper, we will consider the DC readout scheme for
gravitational-wave detectors.
Our investigation in this paper is a natural extension of the quantum
theoretical arguments of a balanced homodyne
detection~\cite{K.Nakamura-2021}.
The key aspect of our argument lies in the specification of the
quantum operator that is finally observed.
We assume that the final observed quantum operator in the
photodetectors is Glauber's photon number~\cite{R.J.Glauber-1963}, as
discussed in
Sec.~\ref{sec:General_arguments_for_the_DC-readout_scheme}.
For the monochromatic case, we consider the number operator
$\hat{a}^{\dagger}(\omega)\hat{a}(\omega)$ for the quantum field
associated with the annihilation operator $\hat{a}(\omega)$.
Glauber's photon number is a natural extension of this single-mode
number operator to the multi-mode electric field.
It is often stated that the final observed quantum operator is the
power operator of the laser.
However, even when we utilize the laser's power operator, the
conclusions we arrive at for the DC readout scheme are consistent with
those related to Glauber's photon number.
Generally speaking, if we employ a different quantum operator, such as
the mode-by-mode number operator, we may draw different
conclusions~\cite{K.Nakamura-M.-K.Fujimoto-2017}.
This variation depends on the photodetection device, as mentioned
above.
Therefore, the specification of our final measured quantum operator is
essential to our arguments.


After discussing the general arguments related to the DC readout
scheme, we will examine the input-output relation for a
Fabry-P\'erot gravitational-wave detector.
A specific input-output relation is essential for a detailed analysis
of the DC readout
scheme~\cite{K.Nakamura-2025-footnote1}.
We aim to derive this input-output relation of a
Fabry-P\'erot gravitational-wave detector, starting without specifying
the small motions of the mirrors.
These small mirror motions encapsulate information about external
forces, including gravitational-wave signals and radiation pressure
from the laser.
In this paper, we describe the small motions of mirrors using the
Heisenberg equation in quantum mechanics for a forced harmonic
oscillator~\cite{K.Nakamura-2025-footnote2}.
While the previous
literature~\cite{H.J.Kimble-Y.Levin-A.B.Matsko-K.S.Thorne-S.P.Vyatchanin-2001,H.Miao-PhDthesis-2010}
has often considered the mirrors' small motions as free except for the
gravitational-wave signal and the radiation pressure forces, we have
introduced the fundamental frequency $\omega_{p}$ of the vibration
isolation system of the pendulum supporting the mirrors.


After solving the Heisenberg equations for the motion of the mirrors,
we find that the initial conditions for the forced harmonic oscillator
concentrate at the frequency $\omega_{p}$, which is outside the
frequency range of our interest.
This indicates that the mirrors' initial conditions, which have
information about the noncommutativity of the position and the
momentum in quantum mechanics, do not relate to the frequency range
relevant to gravitational-wave detectors.


We find that the classical radiation pressure force also acts on the
mirror, which is finite due to the introduced fundamental frequency
$\omega_{p}$ of the pendulums.
This classical radiation pressure manifests as a classical carrier
field that leaks to the output port of the interferometer.
This leakage influences the noise estimation in the DC readout scheme.
We discuss the effects of this leakage of classical radiation pressure
forces in the stationary noise spectral densities.
Notably, when we consider the effects of this leakage, there is a
situation where the shot-noise contribution to the signal-referred
noise spectral density in the high-frequency range does not decrease,
even when the incident  laser power is increased.
To avoid the effects of the leakage of the classical radiation
pressure forces, we have to adjust the tuning point of the
Fabry-P\'erot interferometer very carefully.
We also estimate the maximum deviation from the ideal mirror position,
which realizes the near ideal case.


\begin{figure*}
  \begin{center}
    \includegraphics[width=0.8\textwidth]{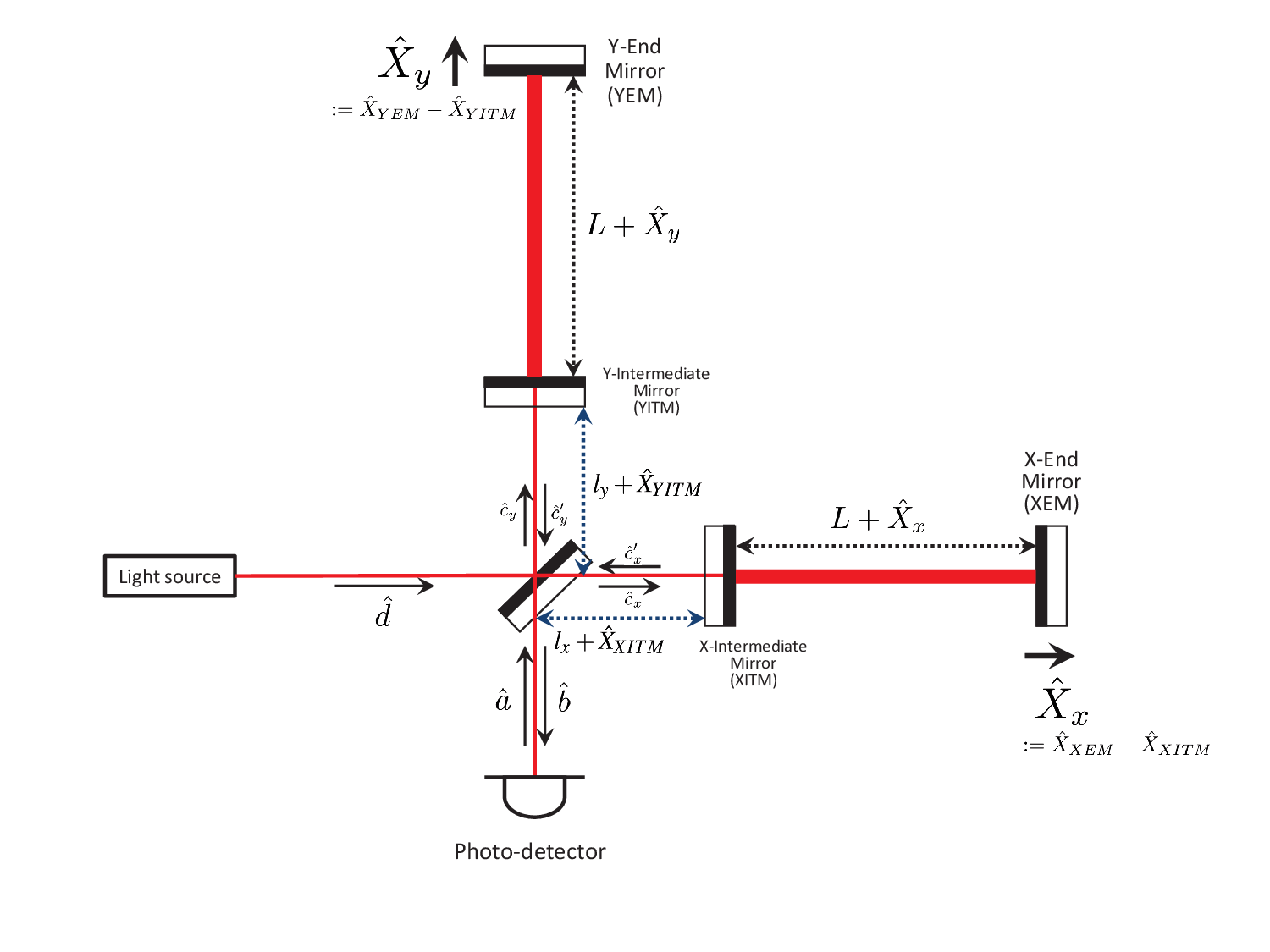}
  \end{center}
  \caption{
    The setup for the Fabry-P\'erot interferometer includes notations
    for photoelectric quadratures.
    We define the classical distances $L$, $l_{x}$, and $l_{y}$, along
    with the small quantum displacements represented by
    $\hat{X}_{XEM}$, $\hat{X}_{XITM}$, $\hat{X}_{YEM}$,
    $\hat{X}_{YITM}$, $\hat{X}_{x}$, and $\hat{X}_{y}$ for the end
    test masses (EMs) and the input test masses (ITMs).
    Additionally, the quadratures for the laser are denoted by
    $\hat{d}$ and $\hat{a}$ as the input quadratures to the
    interferometer, while $\hat{b}$ represents the output quadrature
    from the interferometer.
    The quadratures $\hat{c}_{x}$ and $\hat{c}_{y}$ are for the laser,
    which is separated by the beam splitter (BS), and $\hat{c}_{x}'$
    and $\hat{c}_{y}'$ denote the quadratures that returned from the
    $x$- and $y$-cavities to the BS.
    The notation of the quadratures of the laser between the ITMs
    and EMs is illustrated in
    Fig.~\ref{fig:arm-propagation-Fabry-Perot-setup-notation}.
  }
  \label{fig:kouchan-Fabry-Perot-setup-notation}
\end{figure*}


We need to emphasize that we do not discuss the power recycling, the
signal recycling, or the squeezed state input
techniques~\cite{H.J.Kimble-Y.Levin-A.B.Matsko-K.S.Thorne-S.P.Vyatchanin-2001,H.Miao-PhDthesis-2010,A.Buonanno-Y.Chen-2001,A.Buonanno-Y.Chen-2002,A.Buonanno-Y.Chen-2003}
in this paper.
Our focus is solely on re-evaluating a simple Fabry-P\'erot
gravitational-wave detector.
Therefore, within this paper, we cannot discuss the recent findings
regarding the ``violation of the standard quantum limit'' in LIGO
gravitational-wave detectors~\cite{H.Yu-et-al-2020,W.Jia-et-al-2024}.
However, the main purpose of this paper is to point out two facts.
One is the fact that the quantum uncertainty due to the motion
of test masses does not affect the noise spectral density.
The other is the fact the additional classical-carrier-field leakage
due to the imperfect tuning of the interferometer at the equilibrium
point of the pendulum affect the noise spectral density as an
imperfection of the gravitational-wave detectors.
The realization of the currently operating gravitational-wave
detectors is beyond the current scope of this paper.
Furthermore, we expect that the arguments presented in this paper
can be extended to include techniques such as power recycling, signal
recycling, and squeezed-state input.
In this regard, we are confident that the arguments in this paper are
meaningful.


The organization of this paper is as follows.
In Sec.~\ref{sec:Preliminary}, we summarize the basic notation that is
used in this paper.
In Sec.~\ref{sec:General_arguments_for_the_DC-readout_scheme}, we
develop the general arguments of the DC readout scheme.
In
Sec.~\ref{sec:Input-Output-relation-for-Fabry-Perot_interferometer},
we derive the input-output relation without any specification of the
small mirror displacements.
In Sec.~\ref{sec:Eq_for_mirrors'_motions_and_their_solutions}, we
consider the Heisenberg equations of motion to determine the small
mirror displacements and their solution.
In
Sec.~\ref{sec:Final_input-output-relation_for_Fabry-Perot_GW_Detector},
we derive the final input-output relation by the specifications of the
small mirror displacements through the Heisenberg equation of motion.
We also discuss the comparison with the noise spectral densities in
Ref.~\cite{H.J.Kimble-Y.Levin-A.B.Matsko-K.S.Thorne-S.P.Vyatchanin-2001}
also given in this section.
In Sec.~\ref{sec:DC_readout_scheme_for_FP_GW_Detector}, we discuss the
expectation value and the stationary noise spectral density in the DC
readout from the results in
Secs.~\ref{sec:General_arguments_for_the_DC-readout_scheme}
and~\ref{sec:Final_input-output-relation_for_Fabry-Perot_GW_Detector}.
In this section, the resulting noise spectral density shows that
the shot-noise contribution to the high-frequency region of the
signal-referred noise spectral density does not decrease even as the
injected laser power is increases.
In Sec.~\ref{sec:Changing_Tuning-Point}, we consider changing the
tuning point of the interferometer to recover the ideal noise spectral
density in
Ref.~\cite{H.J.Kimble-Y.Levin-A.B.Matsko-K.S.Thorne-S.P.Vyatchanin-2001}.


Some notations which we use within this paper are illustrated in
Figs.~\ref{fig:kouchan-Fabry-Perot-setup-notation}
and~\ref{fig:arm-propagation-Fabry-Perot-setup-notation}.
In this paper, some numerical values appear in the main text to
provide an estimation of the order of magnitude for certain variables,
but these values are merely estimation measures and have nothing to do
with a specific gravitational-wave detector.


\section{Preliminary}
\label{sec:Preliminary}


\subsection{Electric field notation}
\label{sec:Electric_field_notation}


As in the usual quantum electrodynamics, the one-dimensional electric
field operator $\hat{E}_{a}(t-z)$ at time $t$ and the length $z$ to
the propagation direction in interferometers is described by
\begin{eqnarray}
  \label{eq:K.Nakamura-M.-K.Fujimoto-2018-12}
  && \hat{E}_{a}(t-z) = \hat{E}_{a}^{(+)}(t-z) + \hat{E}_{a}^{(-)}(t-z), \\
  \label{eq:K.Nakamura-M.-K.Fujimoto-2018-13}
  && \hat{E}_{a}^{(-)}(t-z) = \left[\hat{E}_{a}^{(+)}(t-z)\right]^{\dagger}, \\
  \label{eq:K.Nakamura-M.-K.Fujimoto-2018-14}
  && \hat{E}_{a}^{(+)}(t-z) = \int_{0}^{\infty} \frac{d\omega}{2\pi}
     \sqrt{\frac{2\pi\hbar|\omega|}{{\cal A}c}} \hat{a}(\omega) e^{-i\omega(t-z)},
\end{eqnarray}
where $\hat{a}(\omega)$ is the photon annihilation operator associated
with the electric field $\hat{E}_{a}(t-z)$, which satisfies the
commutation relations
\begin{eqnarray}
  \label{eq:K.Nakamura-M.-K.Fujimoto-2018-15}
  &&
     \left[\hat{a}(\omega),\hat{a}^{\dagger}(\omega')\right] = 2 \pi \delta(\omega-\omega'),
  \\
  \label{eq:K.Nakamura-M.-K.Fujimoto-2018-16}
  &&
     \left[\hat{a}(\omega),\hat{a}(\omega')\right] =
     \left[\hat{a}^{\dagger}(\omega),\hat{a}^{\dagger}(\omega')\right]
     = 0.
\end{eqnarray}
${\cal A}$ is the cross-sectional area of the optical beam.
To discuss the input-output relation of the interferometer, based on
one-photon formulation.
Although the two-photon
formulation~\cite{B.L.Schumaker-C.M.Caves-1985a,B.L.Schumaker-C.M.Caves-1985b}
is widely used in the gravitational-wave community, this formulation
includes some approximations.
Since we also want to examine the validity of the two-photon
formulation itself in this paper, we derive the input-output
relation through the one-photon formulation starting from the point
without any approximations.
In the one-photon formulation, it is convenient to introduce the
operator $\hat{A}(\omega)$ defined by
\begin{eqnarray}
  \label{eq:K.Nakamura-M.-K.Fujimoto-2018-17}
  \hat{A}(\omega) := \hat{a}(\omega)\Theta(\omega) + \hat{a}^{\dagger}(-\omega)\Theta(-\omega)
\end{eqnarray}
so that the electric field (\ref{eq:K.Nakamura-M.-K.Fujimoto-2018-12})
is represented as
\begin{eqnarray}
  \label{eq:K.Nakamura-M.-K.Fujimoto-2018-18}
  \hat{E}_{a}(t) = \int_{-\infty}^{+\infty} \frac{d\omega}{2\pi}
  \sqrt{\frac{2\pi\hbar|\omega|}{{\cal A}c}} \hat{A}(\omega)
  e^{-i\omega t}
  ,
\end{eqnarray}
where $\Theta(\omega)$ is the Heaviside step function
\begin{eqnarray}
  \label{eq:K.Nakamura-M.-K.Fujimoto-2018-19}
  \Theta(\omega)
  =
  \left\{
  \begin{array}{ccl}
    1 & \quad & (\omega\leq 0), \\
    0 & \quad & (\omega< 0). \\
  \end{array}
  \right.
\end{eqnarray}
Due to the property of the Dirac $\delta$-function
$\int_{-\infty}^{+\infty} dt
e^{+i(\omega'-\omega)t}=2\pi\delta(\omega'-\omega)$, the inverse
relation of Eq.~(\ref{eq:K.Nakamura-M.-K.Fujimoto-2018-18}) is given
by
\begin{eqnarray}
  \label{eq:K.Nakamura-M.-K.Fujimoto-2018-20}
  \hat{A}(\omega) = \sqrt{\frac{{\cal A}c}{2\pi\hbar|\omega|}}
  \int_{-\infty}^{+\infty} dt e^{+i\omega t} \hat{E}_{a}(t).
\end{eqnarray}
Therefore, the operator $\hat{A}(\omega)$ includes complete
information of the electric field operator $\hat{E}_{a}(t)$ and is
convenient to derive the input-output relation of simple
interferometers.


From the commutation relations of the quadrature operator
$\hat{a}(\omega)$ and $\hat{a}^{\dagger}(\omega)$ defined in
Eq.~(\ref{eq:K.Nakamura-M.-K.Fujimoto-2018-15}) and
(\ref{eq:K.Nakamura-M.-K.Fujimoto-2018-16}), the commutation relations
of electric fields $\hat{E}^{(\pm)}(t)$ are given by
\begin{eqnarray}
  \left[
  \hat{E}^{(+)}_{a}(t), \hat{E}^{(+)}_{a}(t')
  \right]
  &=&
  \left[
  \hat{E}^{(-)}_{a}(t), \hat{E}^{(-)}_{a}(t')
  \right]
  = 0,
  \\
  \left[
  \hat{E}^{(+)}_{a}(t), \hat{E}^{(-)}_{a}(t')
  \right]
  &=&
      \frac{2\pi\hbar}{{\cal A}c} \int_{0}^{+\infty}
      \frac{d\omega}{2\pi} \omega e^{-i\omega(t-t')}
  \nonumber\\
  &=:&
       \frac{2\pi\hbar}{{\cal A}c} \Delta_{a}(t-t')
       .
       \label{eq:Deltaa-def}
\end{eqnarray}
The subscription ``$a$'' of the function $\Delta_{a}(t-t')$ indicates
that this is the vacuum fluctuation originating from the electric
field $\hat{E}_{a}$ with the quadrature $\hat{a}(\omega)$.


We note that the function $\Delta_{a}(t-t')$ has an ultraviolet
divergence in its integration when $\omega\rightarrow\infty$.
However, in the actual measurements of the time sequence of the
variables, the time in a measurement is discrete with a finite time
bin.
This time bin gives the maximum frequency $\omega_{\max}$, which
becomes the natural ultraviolet cut-off of the frequency in the
obtained data.
If the upper frequency of the signal of our interest is 20kHz,
$\omega_{\max}$ must be larger than $2\pi\times 20\times 10^{3}$Hz.
Incidentally, in the actual measurements of the time sequence of the
variables, the whole measurement time is also finite.
It gives the minimum frequency $\omega_{\min}$ which corresponds to a
natural infrared cut-off in frequency.
If the lower frequency of the signal of our interest is 20Hz,
$\omega_{\min}$ must be smaller than $2\pi\times 20$Hz.
Therefore, we may regard that the integration range over $\omega$ in
the definition of the function $\Delta_{a}(t-t')$ in
Eq.~(\ref{eq:Deltaa-def}) is $[\omega_{\min},\omega_{\max}]$ instead
of $[0,+\infty]$.
For this reason, throughout this paper, we do not regard the
divergence in the definition of the function $\Delta_{a}(t-t')$ as a
serious one.
We apply similar arguments when we evaluate the averaged laser power
$I_{0}$ in Sec.~\ref{sec:Coherent_state_ of_the_optical_fields}.


To discuss the quantum properties of the laser, we have to specify the
quantum state of the electric field, which is expressed by
Eqs.~(\ref{eq:K.Nakamura-M.-K.Fujimoto-2018-12})--(\ref{eq:K.Nakamura-M.-K.Fujimoto-2018-14}).
One of the quantum states of the electric fields which is
considered within this paper is the vacuum state $|0\rangle_{a}$
associated with the quadrature $\hat{a}(\omega)$, which is
defined by
\begin{eqnarray}
  \label{eq:vacuum-state-def}
  \hat{a}(\omega)|0\rangle_{a} := 0 \quad {}^{\forall}\omega>0.
\end{eqnarray}
On the other hand, we also consider the coherent state
$|\alpha\rangle_{a}$ associated with the quadrature $\hat{a}(\omega)$,
which is defined by
\begin{eqnarray}
  \label{eq:coherent-state-def}
  \hat{a}(\omega)|\alpha\rangle_{a} := \alpha(\omega)|\alpha\rangle_{a},
\end{eqnarray}
where $\alpha(\omega)$ is a complex function of $\omega$ which has the
dimension $[\mbox{Hz}]^{-1/2}$.
As well-known, the coherent state $|\alpha\rangle_{a}$ and the vacuum
state $|0\rangle_{a}$ are related through the displacement operator
$D_{a}[\alpha]$ as~\cite{H.J.Kimble-Y.Levin-A.B.Matsko-K.S.Thorne-S.P.Vyatchanin-2001}
\begin{eqnarray}
  |\alpha\rangle_{a}
  &=&
       \FrakD_{a}[\alpha]|0\rangle_{a}
       ,
       \label{eq:vacuum-coherent-state-rel}
  \\
  \FrakD_{a}[\alpha]
  &=&
       \exp\left[\int\frac{d\omega}{2\pi}\left(
       \alpha(\omega) \hat{a}^{\dagger}(\omega)
       -
       \alpha^{*}(\omega) \hat{a}(\omega)
       \right)\right]
       .
      \nonumber\\
       \label{eq:displacement-operator-express}
\end{eqnarray}
Here, we note that the subscription ``$a$'' in the state
$|0\rangle_{a}$ and $|\alpha\rangle_{a}$ indicate that these states
are associated with the electric field operator $\hat{E}_{a}(t)$ with
the quadrature $\hat{a}(\omega)$.


\subsection{Multi-mode number and power operators}
\label{sec:Multi-mode_number_and_power_operators}


In this paper, we examine the models of photodetection in which
the photocurrent is proportional to Glauber's multi-mode photon
number $\hat{N}_{b}(t)$ defined by
\begin{eqnarray}
  &&
  \hat{N}_{b}(t)
     \nonumber\\
  &:=&
       \frac{\kappa_{n}c}{2\pi\hbar} {\cal A}
       \hat{E}^{(-)}_{b}(t)
       \hat{E}^{(+)}_{b}(t)
       \nonumber\\
  &=&
      \int_{0}^{\infty}
      \frac{d\omega_{1}}{2\pi}
      \int_{0}^{\infty}
      \frac{d\omega_{2}}{2\pi}
      \sqrt{|\omega_{1}\omega_{2}|}
      \hat{b}^{\dagger}(\omega_{1})
      \hat{b}(\omega_{2})
      e^{+i(\omega_{1}-\omega_{2})t}
      ,
      \nonumber\\
  \label{eq:Glauber-photon-number}
\end{eqnarray}
where $\hat{E^{(\pm)}}_{b}(t)$ are the positive and negative frequency
parts of the output optical electric field $\hat{E}_{b}(t)$ of the
laser and $\kappa_{n}$ is the phenomenological coefficients of
Glauber's photon number $\hat{N}_{b}(t)$ and the photocurrent which
includes the quantum efficiency.
We note that $\kappa_{n}$ has the dimension of [time].


The reasons why we regard the Glauber photon
number~(\ref{eq:Glauber-photon-number}) as the direct observable in
the photodetection were extensively discussed in
Ref.~\cite{K.Nakamura-2021}.
Although the number operator for the single-mode photon is defined by
$\hat{n}(\omega):=\hat{a}^{\dagger}(\omega)\hat{a}(\omega)$,
the superposition of the electric field operator is possible within
the field equations.
In contrast, the superposition of $\hat{n}(\omega)$ is not possible
within the field equations.
A natural extension of the number operator to the multi-mode electric
field is the above Glauber photon number
(\ref{eq:Glauber-photon-number}).


On the other hand, in the gravitational-wave detection community, it
is commonly regarded as the probability of the excitation of the
photocurrent is proportional to the power
$\hat{P}_{b}(t)\propto(\hat{E}_{b}(t))^{2}/(4\pi)$ of the output
optical field $\hat{E}_{b}(t)$.
However, even if we consider the model of photodetection in which the
photocurrent is proportional to the power operator $\hat{P}_{b}$, we
can reach the same conclusion as in
Sec.~\ref{sec:General_arguments_for_the_DC-readout_scheme} in the case
of the DC readout.
For this reason, within this paper, we mainly examine the models of
photodetection in which the photocurrents are proportional to
Glauber's multi-mode photon number (\ref{eq:Glauber-photon-number}),
for simplicity.
We also check the model in which the photocurrent is proportional to
the power $\hat{P}_{b}(t)$ of the optical field $\hat{E}_{b}$ in
Appendix~\ref{sec:Power_counting_photodetection}.
These models yield the same results within this paper.


\section{General arguments for the DC-readout scheme}
\label{sec:General_arguments_for_the_DC-readout_scheme}


In this paper, we consider the situation where the photodetector
measures the output field $\hat{E}_{b}(t)$ whose
quadrature $\hat{b}(\omega)$.
This quadrature $\hat{b}(\omega)$ is
\begin{eqnarray}
  \label{eq:output-quadrature-general}
  \hat{b}(\omega)
  =:
  \langle\hat{b}(\omega)\rangle + \hat{b}_{n}(\omega)
  ,
\end{eqnarray}
where $\langle\hat{b}(\omega)\rangle$ is the expectation value of the
quadrature $\hat{b}(\omega)$ associated with the quantum state
$|\mbox{in}\rangle$ injected to the interferometer and
$\hat{b}_{n}(\omega)$ is the quantum operator which expresses the
quantum noise of the quadrature $\hat{b}(\omega)$ which satisfies
$\langle\hat{b}_{n}(\omega)\rangle=0$.
Here, we express the $\langle\hat{b}(\omega)\rangle$ as
\begin{eqnarray}
  \label{eq:exp-valu-hatb-def}
  \langle\hat{b}(\omega)\rangle
  =
  \FrakA(\omega) + \FrakB 2\pi\delta(\omega-\omega_{0})
  ,
\end{eqnarray}
where $\FrakA(\omega)$ is a classical complex function which includes
gravitational-wave signals as shown in
Sec.~\ref{sec:Final_input-output-relation_for_Fabry-Perot_GW_Detector}.
$\FrakB$ is a complex number which corresponds to the amplitude of the
classical carrier field with the central frequency $\omega_{0}$.


Since we assume that the photodetector measures the quantum operator
$\hat{N}_{b}(t)$ defined by Eq.~(\ref{eq:Glauber-photon-number}), the
Fourier transformation of the observed data is given by
\begin{eqnarray}
  \label{eq:Fourier-Galuber-photon-number}
  \hat{\calN}_{b}(\omega)
  &:=&
       \int_{-\infty}^{+\infty} dt \frac{\kappa_{n}c}{2\pi\hbar} {\cal A}
       \hat{E}^{(-)}_{b}(t) \hat{E}^{(+)}_{b}(t) e^{+i\omega t}
       \\
  &=&
      \kappa_{n} \int_{0}^{\infty} \frac{d\omega_{1}}{2\pi}
      \sqrt{|\omega_{1}(\omega_{1}+\omega)|}
      \hat{b}^{\dagger}(\omega_{1}) \hat{b}(\omega_{1}+\omega)
      .
      \nonumber\\
  \label{eq:Fourier-Galuber-photon-number-expression}
\end{eqnarray}
Substituting Eq.~(\ref{eq:output-quadrature-general}) and
(\ref{eq:exp-valu-hatb-def}) into
Eq.~(\ref{eq:Fourier-Galuber-photon-number-expression}), we obtain
\begin{eqnarray}
  \hat{\calN}_{b}(\omega)
  &=&
      \kappa_{n} \omega_{0}
      |\FrakB|^{2} 2\pi \delta(\omega)
      \nonumber\\
  &&
      +
      \kappa_{n}
      \left(
      \FrakB \sqrt{|\omega_{0}(\omega_{0}-\omega)|}
      \FrakA^{*}(\omega_{0}-\omega)
     \right.
      \nonumber\\
  && \quad\quad\quad
     \left.
      +
      \FrakB^{*} \sqrt{|\omega_{0}(\omega_{0}+\omega)|}
      \FrakA(\omega_{0}+\omega)
      \right)
      \nonumber\\
  &&
      +
      \kappa_{n}
      \left(
      \FrakB \sqrt{|\omega_{0}(\omega_{0}-\omega)|}
      \hat{b}_{n}^{\dagger}(\omega_{0}-\omega)
     \right.
      \nonumber\\
  && \quad\quad\quad
     \left.
      +
      \FrakB^{*} \sqrt{|\omega_{0}(\omega_{0}+\omega)|}
      \hat{b}_{n}(\omega_{0}+\omega)
      \right)
      \nonumber\\
  &&
      +
      O\left(|\FrakB|^{0}\right)
      .
     \label{eq:calN-Fourier-FrakA-FrakB}
\end{eqnarray}
From this, the expectation value of the operator
$\hat{\calN}_{b}(\omega)$ is given by
\begin{eqnarray}
  \langle
  \hat{\calN}_{b}(\omega)
  \rangle
  &:=&
      \langle\mbox{in}|
      \hat{\calN}_{b}(\omega)
      |\mbox{in}\rangle
      \nonumber\\
  &=&
      \kappa_{n} \omega_{0}
      |\FrakB|^{2} 2\pi \delta(\omega)
      \nonumber\\
  &&
      +
      \kappa_{n}
      \left(
      \FrakB \sqrt{|\omega_{0}(\omega_{0}-\omega)|}
      \FrakA^{*}(\omega_{0}-\omega)
     \right.
      \nonumber\\
  && \quad\quad\quad
     \left.
      +
      \FrakB^{*} \sqrt{|\omega_{0}(\omega_{0}+\omega)|}
      \FrakA(\omega_{0}+\omega)
      \right)
      \nonumber\\
  &&
      +
      O\left(|\FrakB|^{0}\right)
      .
  \label{eq:calN-Fourier-FrakA-FrakB-exp}
\end{eqnarray}
As commented above, it will be shown in
Sec.~\ref{sec:Final_input-output-relation_for_Fabry-Perot_GW_Detector}
that $\FrakA(\omega)$ includes gravitational-wave signals.
On the other hand, in the expectation value
(\ref{eq:calN-Fourier-FrakA-FrakB-exp}), the term of order
$|\FrakB|^{2}$ is the amplitude of the classical carrier field of the
laser, which is predictable.
If the amplitude $|\FrakB|$ of the classical carrier field is
sufficiently large, the leading term in
Eq.~(\ref{eq:calN-Fourier-FrakA-FrakB-exp}) is of order
$|\FrakB|^{2}$, and the second leading order is the term of
$|\FrakB|^{1}$ which includes gravitational-wave signals
$\FrakA(\omega)$.
The remaining terms are not interesting in the DC readout scheme.
Since the leading term of order $|\FrakB|^{2}$ is classical,
predictable, and measurable, we can subtract the term of order
$|\FrakB|^{2}$ to extract the sub-leading term of $|\FrakB|^{1}$ which
includes the gravitational-wave signal $\FrakA(\omega)$.
Thus, we can measure the signal term of order $|\FrakB|^{1}$ by the
subtraction of the leading term of order $|\FrakB|^{2}$ from the
expectation value (\ref{eq:calN-Fourier-FrakA-FrakB-exp}).
From this consideration, we may define the signal operator
$\hat{s}_{\calN_{b}}(\omega)$ for the signal $\FrakA(\omega)$ as
\begin{eqnarray}
  \label{eq:signal-operator-Fourier-def}
  \hat{s}_{\calN_{b}}(\omega)
  :=
  \frac{1}{\kappa_{n}} \hat{\calN}_{b}(\omega)
  -
  \omega_{0} |\FrakB|^{2} 2\pi \delta(\omega)
\end{eqnarray}
so that its expectation value is given by
\begin{eqnarray}
  \label{eq:signal-operator-Fourier-exp}
  \langle\hat{s}_{\calN_{b}}(\omega)\rangle
  &=&
      \FrakB \sqrt{|\omega_{0}(\omega_{0}-\omega)|}
      \FrakA^{*}(\omega_{0}-\omega)
      \nonumber\\
  &&
     +
     \FrakB^{*} \sqrt{|\omega_{0}(\omega_{0}+\omega)|}
     \FrakA(\omega_{0}+\omega)
      \nonumber\\
  &&
  +
  O\left(\left|\FrakB\right|^{0}\right)
  .
\end{eqnarray}
Since we consider the situation where $|\FrakB|$ is so large that we
can distinguish the effects due to the different order of $|\FrakB|$,
we ignore the remaining term $O\left(\left|\FrakB\right|^{0}\right)$
because we want to measure the $O(|\FrakB|^{1})$ effect in the
expectation value (\ref{eq:signal-operator-Fourier-exp}).
The time-domain version of this signal operator is given by
\begin{eqnarray}
  \label{eq:signal-operator-time-domain}
  \hat{s}_{N_{b}}(t)
  :=
  \frac{1}{\kappa_{n}} \hat{N}_{b}(t)
  -
  \omega_{0} |\FrakB|^{2}
  .
\end{eqnarray}


From the signal operator $\hat{s}_{N_{b}}(t)$ defined by
Eq.~(\ref{eq:signal-operator-time-domain}), we can define the noise
operator for this measurement scheme as
\begin{eqnarray}
  \label{eq:time-domain-noise-operator-def}
  \hat{s}_{Nn}(t) := \hat{s}_{N_{b}}(t) - \langle\hat{s}_{N_{b}}(t)\rangle.
\end{eqnarray}
Through Eqs.~(\ref{eq:calN-Fourier-FrakA-FrakB}) and
(\ref{eq:signal-operator-time-domain}), the noise operator
$\hat{s}_{Nn}(t)$ defined by
Eq.~(\ref{eq:signal-operator-Fourier-def}) is given by
\begin{eqnarray}
  &&
  \hat{s}_{Nn}(t)
     \nonumber\\
  &=&
      \int_{-\infty}^{+\infty} \frac{d\omega}{2\pi} \left\{
      +
      \FrakB \sqrt{|(\omega_{0}-\omega)\omega_{0}|}
      \hat{b}_{n}^{\dagger}(\omega_{0}-\omega)
      \right.
      \nonumber\\
  && \quad\quad\quad\quad\quad
     \left.
      +
      \FrakB^{*} \sqrt{|\omega_{0}(\omega_{0}+\omega)|}
      \hat{b}_{n}(\omega_{0}+\omega)
      \right\} e^{-i\omega t}
     \nonumber\\
  &&
      +
      O\left(|\FrakB|^{0}\right)
      .
      \label{eq:time-domain-noise-operator-expression}
\end{eqnarray}


As the noise estimation, we consider the time-averaged (stationary)
noise correlation function $C_{(av)s_{Nn}}(\tau)$ defined by
\begin{eqnarray}
  &&
  C_{(av)s_{Nn}}(\tau)
     \nonumber\\
  &:=&
       \lim_{T\rightarrow} \frac{1}{T} \int_{-T/2}^{T/2} dt
       \frac{1}{2} \langle\mbox{in}|
       \hat{s}_{Nn}(t+\tau) \hat{s}_{Nn}(t)
       \nonumber\\
  && \quad\quad\quad\quad\quad\quad\quad\quad\quad
     +
     \hat{s}_{Nn}(t) \hat{s}_{Nn}(t+\tau)
     |\mbox{in}\rangle
     .
     \nonumber\\
  \label{eq:stationary-noise-correlation-function-def}
\end{eqnarray}
From this definition of the time-averaged noise correlation function
(\ref{eq:stationary-noise-correlation-function-def}), we define the
noise spectral density $S_{s_{Nn}}(\omega)$ as the Fourier
transformation of $C_{(av)s_{Nn}}(\tau)$ as
\begin{eqnarray}
  \label{eq:stationary-noise-spectral-density-def}
  S_{s_{Nn}}(\omega)
  :=
  \int_{-\infty}^{+\infty} d\tau
  C_{(av)s_{Nn}}(\tau)
  e^{+i\omega\tau}
  .
\end{eqnarray}
Substituting the expression
(\ref{eq:time-domain-noise-operator-expression}) of the noise operator
$\hat{s}_{Nn}(t)$ into
Eq.~(\ref{eq:stationary-noise-correlation-function-def}), we reach the
expression of the stationary noise-spectral density
$S_{s_{Nn}}(\omega)$ as
\begin{widetext}
\begin{eqnarray}
  \label{eq:stationary-noise-spectral-density-exp}
  S_{s_{Nn}}(\omega)
  &=&
      \frac{1}{2} \omega_{0}
      \int_{0}^{\infty} \frac{d\omega_{1}}{2\pi} \left[
      \FrakB^{2} \sqrt{|(\omega_{0}-\omega)\omega_{1}|}
      f(\omega_{0}-\omega_{1}+\omega)
      \left\langle
      \hat{b}_{n}^{\dagger}(\omega_{0}-\omega)
      \hat{b}_{n}^{\dagger}(\omega_{1})
      +
      \hat{b}_{n}^{\dagger}(\omega_{1})
      \hat{b}_{n}^{\dagger}(\omega_{0}-\omega)
      \right\rangle
      \right.
      \nonumber\\
  && \quad\quad\quad\quad\quad\quad
     \left.
     +
     |\FrakB|^{2} \sqrt{|(\omega_{0}+\omega)\omega_{1}|}
     f(\omega_{0}-\omega_{1}+\omega)
     \left\langle
     \hat{b}_{n}(\omega_{0}+\omega)
     \hat{b}_{n}^{\dagger}(\omega_{1})
     +
     \hat{b}_{n}^{\dagger}(\omega_{1})
     \hat{b}_{n}(\omega_{0}+\omega)
     \right\rangle
     \right.
     \nonumber\\
  && \quad\quad\quad\quad\quad\quad
     \left.
     +
     |\FrakB|^{2} \sqrt{|(\omega_{0}-\omega)\omega_{1}|}
     f(\omega_{0}-\omega_{1}-\omega)
     \left\langle
     \hat{b}_{n}^{\dagger}(\omega_{0}-\omega)
     \hat{b}_{n}(\omega_{1})
     +
     \hat{b}_{n}(\omega_{1})
     \hat{b}_{n}^{\dagger}(\omega_{0}-\omega)
     \right\rangle
     \right.
     \nonumber\\
  && \quad\quad\quad\quad\quad\quad
     \left.
     +
     (\FrakB^{*})^{2} \sqrt{|(\omega_{0}+\omega)\omega_{1}|}
     f(\omega_{0}-\omega_{1}-\omega)
     \left\langle
     \hat{b}_{n}(\omega_{0}+\omega)
     \hat{b}_{n}(\omega_{1})
     +
     \hat{b}_{n}(\omega_{1})
     \hat{b}_{n}(\omega_{0}+\omega)
     \right\rangle
     \right]
     .
\end{eqnarray}
\end{widetext}
Here, we defined the one-point support function $f(a)$ of $a\in\RF$ by
\begin{eqnarray}
  \label{eq:one-point-support-func-def-1}
  f(a)
  &:=&
       \lim_{T\rightarrow+\infty} \frac{1}{T} \int_{-T/2}^{T/2} dt e^{-iat}
       \\
  &=&
      \left\{
      \begin{array}{ccccc}
        1 & \quad & \mbox{for} & \quad & a= 0, \\
        0 & \quad & \mbox{for} & \quad & a\neq 0,
      \end{array}
      \right.
  \label{eq:one-point-support-func-def-2}
\end{eqnarray}
as discussed in Ref.~\cite{K.Nakamura-2021}.
We also note that for a finite function $g(a)$,
\begin{eqnarray}
  \label{eq:one-point-support-func-def-3}
  \int_{-\infty}^{+\infty} \delta(a) g(a) f(a) da = g(0),
\end{eqnarray}
\begin{eqnarray}
  \int_{-\infty}^{+\infty} \delta(b) g(b) f(a) db = f(a)g(0).
  \label{eq:one-point-support-func-def-4}
\end{eqnarray}


Although the final input-output relation will be derived in
Sec.~\ref{sec:Final_input-output-relation_for_Fabry-Perot_GW_Detector},
we compare the final input-output relation
(\ref{eq:input-output-rel-explicit-quad-I0-kappabetahSQL}) and
Eqs.~(\ref{eq:output-quadrature-general}) and
(\ref{eq:exp-valu-hatb-def}) above.
Then, we obtain the expression of the operator
$\hat{b}_{n}(\omega)$, more precisely
$\FrakD_{d}^{\dagger}\hat{b}_{n}(\omega_{0}\pm\Omega)\FrakD_{d}$.
Through this obtained expression of $\hat{b}_{n}(\omega)$, we can
confirm the following expectation values:
\begin{eqnarray}
  \langle
  \hat{b}_{n}^{\dagger}(\omega_{0}-\omega)
  \hat{b}_{n}^{\dagger}(\omega_{0}-\omega_{1})
  \rangle
  &\propto&
  2 \pi \delta(\omega+\omega_{1})
  ,
  \label{eq:bndaggeromega0-omegabndaggeromega3-exp-delta-contribution}
  \\
  \langle
  \hat{b}_{n}(\omega_{0}+\omega)
  \hat{b}_{n}^{\dagger}(\omega_{0}-\omega_{1})
  \rangle
  &\propto&
  2 \pi \delta(\omega+\omega_{1})
  ,
  \label{eq:bnomega0+omegabndaggeromega3-exp-delta-contribution}
     \\
  \langle
  \hat{b}_{n}^{\dagger}(\omega_{0}-\omega)
  \hat{b}_{n}(\omega_{0}+\omega_{1})
  \rangle
  &\propto&
  2 \pi \delta(\omega+\omega_{1})
  ,
  \label{eq:bndaggeromega0-omegabnomega3-exp-delta-contribution}
      \\
  \langle
  \hat{b}_{n}(\omega_{0}+\omega)
  \hat{b}_{n}(\omega_{0}+\omega_{1})
  \rangle
  &\propto&
  2 \pi \delta(\omega+\omega_{1})
  .
  \label{eq:bnomega0+omegabnomega3-exp-delta-contribution}
\end{eqnarray}
\begin{widetext}
Through these expectation values
(\ref{eq:bndaggeromega0-omegabndaggeromega3-exp-delta-contribution})--(\ref{eq:bnomega0+omegabnomega3-exp-delta-contribution})
and the properties of the one-point support function $f(a)$ summarized
in
Eqs.~(\ref{eq:one-point-support-func-def-1})--(\ref{eq:one-point-support-func-def-4}),
we may write
\begin{eqnarray}
  2\pi\delta(\omega-\omega') S_{Nn}(\omega)
  &=&
      \omega_{0}
      |\FrakB|^{2}
      \left[
      e^{+2i\Theta}
      \sqrt{|(\omega_{0}-\omega)(\omega_{0}+\omega')|}
      \langle
      \hat{b}_{n}^{\dagger}(\omega_{0}-\omega)
      \hat{b}_{n}^{\dagger}(\omega_{0}+\omega')
      \rangle
      \right.
      \nonumber\\
  && \quad\quad\quad\quad
     \left.
     +
     \frac{1}{2}
     \sqrt{|(\omega_{0}+\omega)(\omega_{0}+\omega')|}
     \langle
     \hat{b}_{n}(\omega_{0}+\omega)
     \hat{b}_{n}^{\dagger}(\omega_{0}+\omega')
     +
     \hat{b}_{n}^{\dagger}(\omega_{0}+\omega')
     \hat{b}_{n}(\omega_{0}+\omega)
     \rangle
     \right.
     \nonumber\\
  && \quad\quad\quad\quad
     \left.
     +
     \frac{1}{2}
     \sqrt{|(\omega_{0}-\omega)(\omega_{0}-\omega')|}
     \langle
     \hat{b}_{n}(\omega_{0}-\omega')
     \hat{b}_{n}^{\dagger}(\omega_{0}-\omega)
     +
     \hat{b}_{n}^{\dagger}(\omega_{0}-\omega)
     \hat{b}_{n}(\omega_{0}-\omega')
     \rangle
     \right.
     \nonumber\\
  && \quad\quad\quad\quad
     \left.
     +
      e^{-2i\Theta}
     \sqrt{|(\omega_{0}+\omega)(\omega_{0}-\omega')|}
     \langle
     \hat{b}_{n}(\omega_{0}+\omega)
     \hat{b}_{n}(\omega_{0}-\omega')
     \rangle
     \right]
     \nonumber\\
  &&
     + O\left(|\FrakB|^{1},|\FrakB|^{0}\right)
     ,
     \label{eq:Measured-noise-spectral-density-DCreadout}
\end{eqnarray}
\end{widetext}
where we defined $\FrakB=:|\FrakB|e^{i\Theta}$.
Here, we note that the expression
(\ref{eq:Measured-noise-spectral-density-DCreadout}) is described by
the quadrature with the frequency $\omega_{0}\pm\omega$.
This is the motivation of the sideband picture, which describes the
quantum fluctuations around the central frequency $\omega_{0}$ of the
incident laser.
We follow the historical notation in which $\Omega$ denotes the
sideband frequency.
Then, we define the upper- and lower-sideband quadrature
$\hat{b}_{n\pm}(\Omega)$ by
\begin{eqnarray}
  \label{eq:upper-lower-sideband-quadrature-def}
  \hat{b}_{n\pm}(\Omega) := \hat{b}_{n}(\omega_{0}\pm\Omega).
\end{eqnarray}
Furthermore, the amplitude quadrature $\hat{b}_{n1}(\Omega)$ and the
phase quadrature $\hat{b}_{n2}(\Omega)$ are often used, which are
defined by
\begin{eqnarray}
  \label{eq:amplitude-phase-quadrature-def-1}
  \hat{b}_{n+}(\Omega)
  =:
  \frac{1}{\sqrt{2}}\left(
  \hat{b}_{n1}(\Omega)
  +
  i
  \hat{b}_{n2}(\Omega)
  \right)
  ,
  \\
  \label{eq:amplitude-phase-quadrature-def-2}
  \hat{b}_{n-}(\Omega)
  =:
  \frac{1}{\sqrt{2}}\left(
  \hat{b}_{n1}^{\dagger}(\Omega)
  +
  i
  \hat{b}_{n2}^{\dagger}(\Omega)
  \right)
  .
\end{eqnarray}
Moreover, we also introduce the operator $\hat{b}_{n\Theta}(\Omega)$
by
\begin{eqnarray}
  \label{eq:bTheta-noise-quadrature-def}
  \hat{b}_{n\Theta}(\Omega)
  =
  \cos\Theta \hat{b}_{n1}(\Omega)
  +
  \sin\Theta \hat{b}_{n2}(\Omega)
  .
\end{eqnarray}
Through the operator $\hat{b}_{n\Theta}(\Omega)$, the noise spectral
density $S_{s_{Nn}}(\Omega)$ defined by
Eq.~(\ref{eq:Measured-noise-spectral-density-DCreadout}) is given by
\begin{eqnarray}
  &&
     2\pi\delta(\Omega-\Omega') S_{s_{Nn}}(\Omega)
     \nonumber\\
  &=&
      \omega_{0}
      |\FrakB|^{2}
      \left\langle
      \hat{b}_{n\Theta}(\Omega)
      \hat{b}_{n\Theta}^{\dagger}(\Omega')
      +
      \hat{b}_{n\Theta}^{\dagger}(\Omega')
      \hat{b}_{n\Theta}(\Omega)
      \right\rangle
     \nonumber\\
  &&
     + O\left(|\FrakB|^{1},|\FrakB|^{0}\right)
     .
     \label{eq:Measured-noise-spectral-density-DCreadout-bTheta}
\end{eqnarray}
Here, we note that
\begin{eqnarray}
  \label{eq:commutation-of--hat-bTheta}
  \left[
  \hat{b}_{n\Theta}(\Omega), \hat{b}_{n\Theta}^{\dagger}(\Omega)
  \right]
  =
  0
  .
\end{eqnarray}
The formula for the stationary noise-spectral density
$S_{s_{Nn}}(\Omega)$ is the general result under the premise of the
DC-readout scheme.
To derive the result
(\ref{eq:Measured-noise-spectral-density-DCreadout-bTheta}), the
sufficiently large amplitude $|\FrakB|$ of the classical carrier is
essential.
Due to this large $|\FrakB|$, we can neglect the residual term
$O(|\FrakB|^{1},|\FrakB|^{0})$.
This is the main difference of the balanced homodyne detection
discussed in Ref.~\cite{K.Nakamura-2021}.


\section{Input-Output relation of the Fabry-P\'erot interferometer}
\label{sec:Input-Output-relation-for-Fabry-Perot_interferometer}


In this section, we derive the input-output relation of the
Fabry-P\'erot interferometer, which is used in the setup of
gravitational-wave detectors.
The Fabry-P\'erot interferometer consists of the
Beam Splitter (BS), Input Test Masses (ITMs), and the End test Masses
(EMs).
The injected laser to the Fabry-P\'erot interferometer is separated
into the directions of the $x$- and the $y$-arms.
The laser is amplified between the ITM and EM.
In gravitational-wave detectors, BS, ITMs, and EMs are suspended
through the vibration-isolation system to measure the relative
positions of these mirrors precisely.
In these relative positions, gravitational-wave signals are included
through the equation of mirrors' motion as extensively discussed in
Sec.~\ref{sec:Eq_for_mirrors'_motions_and_their_solutions}.


\subsection{Mirror Displacements}
\label{sec:Mirror_Displacements}


To describe the relative positions of BS, ITMs, and EMs, we introduce
the proper reference
frame~\cite{C.W.Misner-T.S.Thorne-J.A.Wheeler-1973} whose origin is
BS.
As depicted in Fig.~\ref{fig:kouchan-Fabry-Perot-setup-notation}, we
denote the coordinate values in the proper reference frame of ITM
(XITM) and EM (XEM) along $x$-arm by $l_{x}+\hat{X}_{XITM}$ and
$l_{x}+L+\hat{X}_{XEM}$, respectively.
Here, $l_{x}$ is the distance between BS and XITM, and $L$ is the
distance between XITM and XEM in a situation where there are no
external forces, including the radiation pressure
forces from the laser and gravitational-wave signals.
We regard $l_{x}$ and $L$ as classical distances.
In addition to the $l_{x}$ and $L$, we introduce the quantum
displacement $\hat{X}_{XITM}$ and $\hat{X}_{XEM}$, which are induced
by the injected laser and other forces, including the gravitational-wave
signal.
In this paper, we regard $\hat{X}_{XITM}$ and $\hat{X}_{XEM}$ as
quantum operators that describe the quantum mirror positions.
In Sec.~\ref{sec:Eq_for_mirrors'_motions_and_their_solutions}, the
operators $\hat{X}_{XITM}$ and $\hat{X}_{XEM}$ are determined as the
solution to the Heisenberg equation of motion.


Similarly, we also introduce the coordinate values in the proper
reference frame of ITM (YITM) and EM (YEM) along $y$-arm
$l_{y}+\hat{X}_{YITM}$ and $l_{y}+L+\hat{X}_{YEM}$, respectively.
We also regard $\hat{X}_{YITM}$ and $\hat{X}_{YEM}$ as describing
quantum displacements, which are induced by the injected laser and
other forces, including the gravitational-wave signal.
As in the case of the operators $\hat{X}_{XITM}$ and $\hat{X}_{XEM}$,
in Sec.~\ref{sec:Eq_for_mirrors'_motions_and_their_solutions}, the
operators $\hat{X}_{YITM}$ and $\hat{X}_{YEM}$ are determined as the
solution to the Heisenberg equation of motion.


Based on the above setup, we calculate the laser's electric field
using quantum electrodynamics.


\subsection{Beam Splitter Junction}
\label{sec:Beam_Splitter_Junction}


First, we consider the junction conditions for optical quadratures at
BS.
Following the notation depicted in
Fig.~\ref{fig:kouchan-Fabry-Perot-setup-notation}, the final output
electric field operator $\hat{E}_{b}(t)$ is given by
\begin{eqnarray}
  \label{eq:output-electric-field-junction}
  \hat{E}_{b}(t)
  =
  \frac{1}{\sqrt{2}} \left[
    \hat{E}_{c_{y}'}(t) - \hat{E}_{c_{x}'}(t)
  \right]
  ,
\end{eqnarray}
where $\hat{E}_{c_{y}'}(t)$ and $\hat{E}_{c_{x}'}(t)$ are electric
field operators injected from the $y$-arm and $x$-arm to BS,
respectively.
Here, we defined
\begin{eqnarray}
  \hat{B}(\omega)
  &:=&
  \hat{b}(\omega) \Theta(\omega)
  +
  \hat{b}^{\dagger}(-\omega) \Theta(-\omega)
  \label{eq:hatB-def}
  , \\
  \hat{C}_{x}'(\omega)
  &:=&
  \hat{c}_{x}'(\omega) \Theta(\omega)
  +
  \hat{c}_{x}^{'\dagger}(-\omega) \Theta(-\omega)
  \label{eq:hatCx'-def}
  , \\
  \hat{C}_{y}'(\omega)
  &:=&
  \hat{c}_{y}'(\omega) \Theta(\omega)
  +
  \hat{c}_{y}^{'\dagger}(-\omega) \Theta(-\omega)
  \label{eq:hatCy'-def}
\end{eqnarray}
as in Eq.~(\ref{eq:K.Nakamura-M.-K.Fujimoto-2018-17}).
In terms of the operators $\hat{B}(\omega)$, $\hat{C}_{x}'(\omega)$,
and $\hat{C}_{y}'(\omega)$, the relation
(\ref{eq:output-electric-field-junction}) is given by
\begin{eqnarray}
  \label{eq:output-electric-field-junction-hatB}
  \hat{B}(\omega)
  =
  \frac{1}{\sqrt{2}} \left(
    \hat{C}_{y}'(\omega) - \hat{C}_{x}'(\omega)
  \right).
\end{eqnarray}
Similarly, the electric-field operators $\hat{E}_{c_{y}}(t)$ and
$\hat{E}_{c_{x}}(t)$, that propagated from BS to each arm, are
also given by the input field operators $\hat{E}_{d}(t)$ and
$\hat{E}_{a}(t)$, which are injected from the light source and
photodetectors to BS, respectively, as follows:
\begin{eqnarray}
  \label{eq:xarm-input-field-junction}
  \hat{E}_{c_{x}}(t)
  &=&
  \frac{1}{\sqrt{2}} \left(
    \hat{E}_{d}(t) - \hat{E}_{a}(t)
  \right)
  , \\
  \label{eq:yarm-input-field-junction}
  \hat{E}_{c_{y}}(t)
  &=&
  \frac{1}{\sqrt{2}} \left(
    \hat{E}_{d}(t) + \hat{E}_{a}(t)
  \right)
  .
\end{eqnarray}
In terms of the quadrature as in Eq.~(\ref{eq:K.Nakamura-M.-K.Fujimoto-2018-17}),
these relations yield
\begin{eqnarray}
  \label{eq:xarm-input-field-junction-hatCx}
  \hat{C}_{x}(\omega)
  &=&
  \frac{1}{\sqrt{2}} \left(
    \hat{D}(\omega) - \hat{A}(\omega)
  \right)
  , \\
  \label{eq:yarm-input-field-junction-hatCy}
  \hat{C}_{y}(\omega)
  &=&
  \frac{1}{\sqrt{2}} \left(
    \hat{D}(\omega) + \hat{A}(\omega)
  \right)
  ,
\end{eqnarray}
where we defined the operators
\begin{eqnarray}
  \hat{C}_{x}(\omega)
  &:=&
  \hat{c}_{x}(\omega) \Theta(\omega)
  +
  \hat{c}_{x}^{\dagger}(-\omega) \Theta(-\omega)
  \label{eq:hatCx-def}
  , \\
  \hat{C}_{y}(\omega)
  &:=&
  \hat{c}_{y}(\omega) \Theta(\omega)
  +
  \hat{c}_{y}^{\dagger}(-\omega) \Theta(-\omega)
  \label{eq:hatCy-def}
  , \\
  \hat{D}(\omega)
  &:=&
  \hat{d}(\omega) \Theta(\omega)
  +
  \hat{d}^{\dagger}(-\omega) \Theta(-\omega)
  \label{eq:hatD-def}
\end{eqnarray}
as in Eq.~(\ref{eq:K.Nakamura-M.-K.Fujimoto-2018-17}).
The notations of the quadratures are also depicted in
Fig.~\ref{fig:kouchan-Fabry-Perot-setup-notation}.


\subsection{Arm Propagation}
\label{sec:Arm_Propagation}


Now, we consider the propagation effects along the $x$- and $y$-arms,
respectively.


\subsubsection{Propagation between BS and ITMs}
\label{sec:Arm_Propagation_BS_ITMs}


From the propagation effect from XITM (YITM) to BS, in the notation
depicted in Fig.~\ref{fig:kouchan-Fabry-Perot-setup-notation}, we have
\begin{eqnarray}
  \hat{E}_{c'_{x}}(t)
  &=&
  \hat{E}_{f'_{x}}\left[t - \frac{l_{x}+\hat{X}_{XITM}(t-l_{x}/c)}{c} \right]
      \nonumber\\
  &=&
  \hat{E}_{f'_{x}}\left[t - \tau_{x}' - \frac{\hat{X}_{XITM}(t-\tau_{x}')}{c} \right]
  \label{eq:xarm-retarded-effect-f'-to-c'}
  , \\
  \hat{E}_{c'_{y}}(t)
  &=&
  \hat{E}_{f'_{y}}\left[t - \tau_{y}' - \frac{\hat{X}_{YITM}(t-\tau_{y}')}{c} \right]
  \label{eq:yarm-retarded-effect-f'-to-c'}
\end{eqnarray}
as the output electric field operators $\hat{E}_{c'_{x}}(t)$ and
$\hat{E}_{c'_{y}}(t)$ from each arm.
Here, we defined $l_{x}=:c\tau_{x}'$ and $l_{y}=:c\tau_{y}'$.
In the Fourier space, Eq.~(\ref{eq:xarm-retarded-effect-f'-to-c'})
is given by
\begin{eqnarray}
  \hat{E}_{c'_{x}}(t)
  &=&
      \int_{-\infty}^{+\infty} \frac{d\omega_{1}}{2\pi}
      \sqrt{\frac{2\pi\hbar|\omega_{1}|}{{\cal A}c}} \hat{C}_{x}'(\omega_{1})
      e^{-i\omega_{1} t}
      \nonumber\\
  &=&
      \int_{-\infty}^{+\infty} \frac{d\omega_{2}}{2\pi}
      \sqrt{\frac{2\pi\hbar|\omega_{2}|}{{\cal A}c}} \hat{F}_{x}'(\omega_{2})
      \nonumber\\
  && \quad
     \times
      \exp\left[ - i \omega_{2} \left(
      t - \tau_{x}' - \frac{\hat{X}_{XITM}(t-\tau_{x}')}{c}
      \right) \right]
      \nonumber\\
  &=&
      \int_{-\infty}^{+\infty} \frac{d\omega_{2}}{2\pi}
      \sqrt{\frac{2\pi\hbar|\omega_{2}|}{{\cal A}c}} \hat{F}_{x}'(\omega_{2})
      e^{ - i \omega_{2} t  }
      e^{ + i \omega_{2} \tau_{x}' }
      \nonumber\\
  && \quad
     \times
      \left(
      1
      +
      i \frac{\omega_{2}}{c}
      \int_{-\infty}^{+\infty}
      \frac{d\omega_{3}}{2\pi}
      \hat{Z}_{XITM}(\omega_{3})
      e^{ - i \omega_{3} (t-\tau_{x}')}
      \right)
     \nonumber\\
  &&
      +
      O\left(\left(\hat{X}_{XITM}\right)^{2}\right)
      ,
      \label{eq:xarm-retarded-effect-f'-to-c'-mod-Fourier-1}
\end{eqnarray}
where we used
\begin{eqnarray}
  \hat{X}_{XITM}
  &=&
  \int_{-\infty}^{+\infty} \frac{d\omega_{3}}{2\pi}
  \hat{Z}_{XITM}(\omega_{3})
  e^{- i \omega_{3} t}
  ,
  \label{eq:XITM-Fourier-def}
  \\
  \hat{X}_{YITM}
  &=&
  \int_{-\infty}^{+\infty} \frac{d\omega_{3}}{2\pi}
  \hat{Z}_{YITM}(\omega_{3})
  e^{- i \omega_{3} t}
  .
  \label{eq:YITM-Fourier-def}
\end{eqnarray}
Operating $\int_{-\infty}^{+\infty}dt e^{+i\omega t}$ to
Eq.~(\ref{eq:xarm-retarded-effect-f'-to-c'-mod-Fourier-1}), we obtain
\begin{eqnarray}
  &&
     \hat{C}_{x}'(\omega)
      \nonumber\\
  &=&
      e^{ + i \omega \tau_{x}' }
      \hat{F}_{x}'(\omega)
      \nonumber\\
  &&
      +
     i
      e^{ + i \omega \tau_{x}' }
      \int_{-\infty}^{+\infty} \frac{d\omega_{2}}{2\pi}
      \sqrt{\frac{|\omega_{2}|}{|\omega|}}
      \frac{\omega_{2}}{c}
      \hat{F}_{x}'(\omega_{2})
      \hat{Z}_{XITM}(\omega - \omega_{2})
      \nonumber\\
  &&
      +
      O\left(\left(\hat{X}_{XITM}\right)^{2}\right)
      ,
      \label{eq:xarm-retarded-effect-f'-to-c'-mod-Fourier}
\end{eqnarray}
for $x$-arm.
Similarly, for $y$-arm, we obtain
\begin{eqnarray}
  &&
  \hat{C}_{y}'(\omega)
     \nonumber\\
  &=&
      e^{ + i \omega \tau_{y}' }
      \hat{F}_{y}'(\omega)
      \nonumber\\
  &&
      +
     i
      e^{ + i \omega \tau_{y}' }
      \int_{-\infty}^{+\infty} \frac{d\omega_{2}}{2\pi}
      \sqrt{\frac{|\omega_{2}|}{|\omega|}}
      \frac{\omega_{2}}{c}
      \hat{F}_{y}'(\omega_{2})
      \hat{Z}_{YITM}(\omega - \omega_{2})
      \nonumber\\
  &&
      +
      O\left(\left(\hat{X}_{YITM}\right)^{2}\right)
      .
      \label{eq:yarm-retarded-effect-f'-to-c'-mod-Fourier}
\end{eqnarray}


Furthermore, the propagation effects of the laser yield the electric
field in each arm by
\begin{eqnarray}
  \label{eq:xarm-retarded-effect-c-to-f-mod}
  \hat{E}_{f_{x}}(t)
  &=&
  \hat{E}_{c_{x}}\left[t - \tau_{x}' - \frac{\hat{X}_{XITM}(t)}{c} \right]
  , \\
  \label{eq:yarm-retarded-effect-c-to-f-mod}
  \hat{E}_{f_{y}}(t)
  &=&
  \hat{E}_{c_{y}}\left[t - \tau_{y}' - \frac{\hat{X}_{YITM}(t)}{c} \right].
\end{eqnarray}
We consider the Fourier transformation of
Eqs.~(\ref{eq:xarm-retarded-effect-c-to-f-mod}) and
(\ref{eq:yarm-retarded-effect-c-to-f-mod}).
As in the case of
Eqs.~(\ref{eq:xarm-retarded-effect-f'-to-c'-mod-Fourier}) and
(\ref{eq:yarm-retarded-effect-f'-to-c'-mod-Fourier}), we consider the
Fourier expansion as Eq.~(\ref{eq:K.Nakamura-M.-K.Fujimoto-2018-18}).
Then, we obtain
\begin{eqnarray}
  &&
  \hat{F}_{x}(\omega)
     \nonumber\\
  &=&
      e^{ + i \omega \tau_{x}' }
      \hat{C}_{x}(\omega)
      \nonumber\\
  &&
     +
     i
     \int_{-\infty}^{+\infty} \frac{d\omega_{1}}{2\pi}
     \sqrt{\frac{|\omega_{1}|}{|\omega|}}
     \frac{\omega_{1}}{c}
     e^{ + i \omega_{1} \tau_{x}' }
     \hat{C}_{x}(\omega_{1})
     \hat{Z}_{XITM}(\omega - \omega_{1})
     \nonumber\\
  &&
     +
     O\left(\hat{X}_{XITM}^{2}\right)
     .
     \label{eq:xarm-retarded-effect-f'-to-c'-mod-Fourier-result}
\end{eqnarray}
Similarly, we obtain
\begin{eqnarray}
  &&
  \hat{F}_{y}(\omega)
     \nonumber\\
  &=&
      e^{ + i \omega \tau_{y}' }
      \hat{C}_{y}(\omega)
      \nonumber\\
  &&
     +
     i
     \int_{-\infty}^{+\infty} \frac{d\omega_{1}}{2\pi}
     \sqrt{\frac{|\omega_{1}|}{|\omega|}}
     \frac{\omega_{1}}{c}
     e^{ + i \omega_{1} \tau_{y}' }
     \hat{C}_{y}(\omega_{1})
     \hat{Z}_{YITM}(\omega - \omega_{1})
     \nonumber\\
  &&
     +
     O\left(\hat{X}_{YITM}^{2}\right)
     .
     \label{eq:yarm-retarded-effect-f'-to-c'-mod-Fourier-result}
\end{eqnarray}


\subsubsection{Junction at ITMs and cavity propagation}
\label{sec:Junction_at_ITMs_Cavity_propagaion}

\begin{figure}
  \begin{center}
    \includegraphics[width=0.48\textwidth]{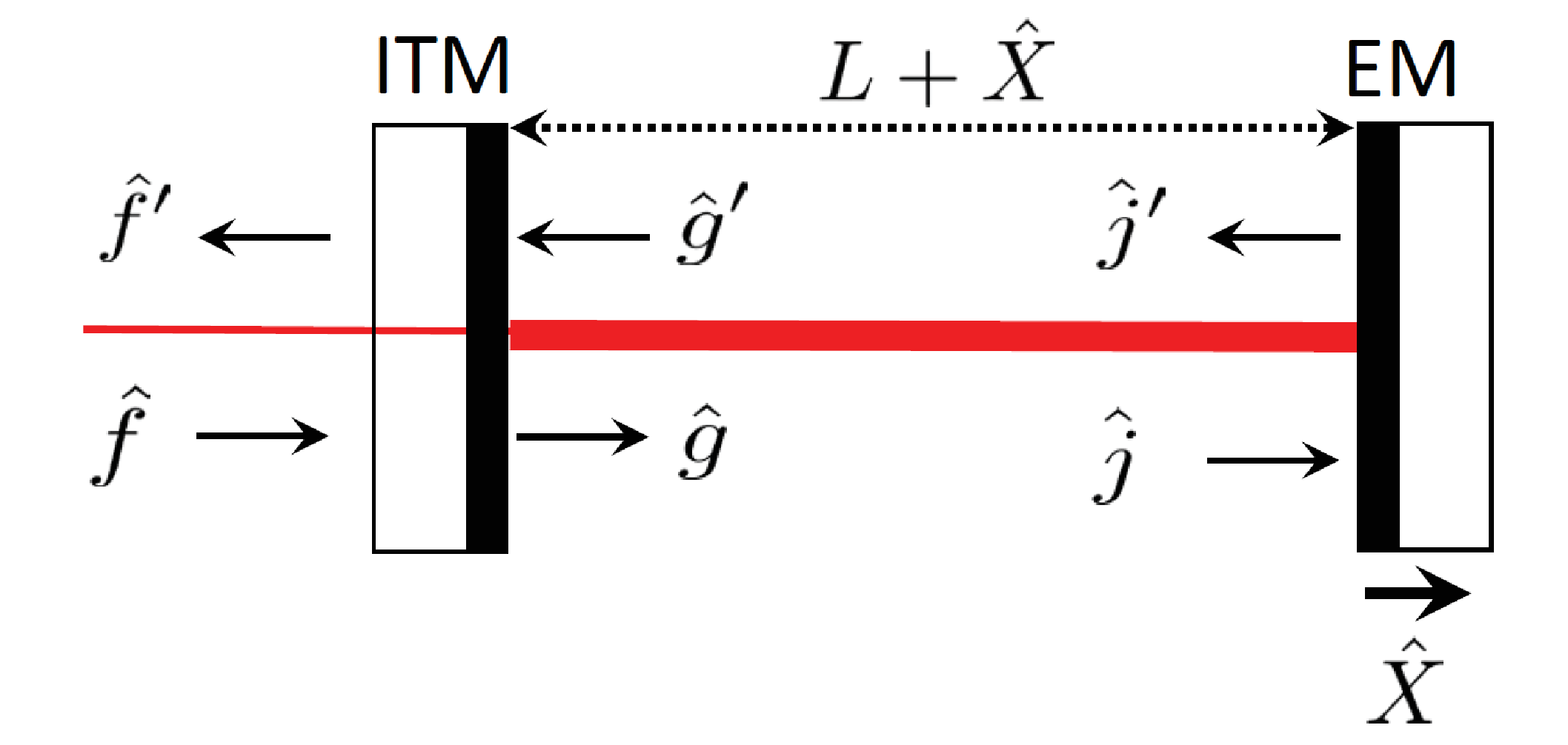}
  \end{center}
  \caption{
    Arm propagation in the Fabry-P\'erot interferometer and
    photoelectric quadratures notations.
    The quadratures $\hat{f}$ and $\hat{f}'$ are those of the laser
    incident from the BS to ITM and the laser from ITM to BS,
    respectively.
    The $\hat{g}$ and $\hat{g}'$ are the quadratures for the lasers
    from ITM to the Fabry-P\'erot cavity and from the Fabry-P\'erot
    cavity to ITM, respectively.
    The quadratures $\hat{j}$ and $\hat{j}'$ are quadratures for
    lasers that reach the EM and are reflected by the EM,
    respectively.
  }
  \label{fig:arm-propagation-Fabry-Perot-setup-notation}
\end{figure}

Next, we consider the laser's propagation along each arm.
We denote the notation of the electric field of the electric field as
depicted in Fig.~\ref{fig:arm-propagation-Fabry-Perot-setup-notation}.


We also denote the power reflection and transmission coefficients by
$R$ and $T$, respectively.
The amplitude reflection and transmission coefficients are chosen to
be real, with signs $\{-\sqrt{R},+\sqrt{T}\}$,
$\{-\sqrt{\tilde{R}},+\sqrt{\tilde{T}}\}$ for the light that impinges
on a mirror from the outside of the cavity at ITMs and EMs,
respectively, and $\{+\sqrt{R},+\sqrt{T}\}$,
$\{+\sqrt{\tilde{R}},+\sqrt{\tilde{T}}\}$ for light that impinges from
the inside of the cavity at ITMs and EMs, respectively.
These satisfy the condition
\begin{eqnarray}
  \label{eq:Mirror-Unitarity-cond}
  R + T = \tilde{R} + \tilde{T} = 1.
\end{eqnarray}
Then, the junction conditions for the electric field operators
$\hat{E}_{f_{x,y}}$, $\hat{E}_{f'_{x,y}}$, $\hat{E}_{g_{x,y}}$, and
$\hat{E}_{g'_{x,y}}$, are given by
\begin{eqnarray}
  \label{eq:Kimble-B8-correspoind-1}
  &&
  \hat{E}_{g_{x,y}}(t) = \sqrt{T} \hat{E}_{f_{x,y}}(t) + \sqrt{R} \hat{E}_{g'_{x,y}}(t)
  ,
  \\
  &&
  \label{eq:Kimble-B8-correspoind-2}
  \hat{E}_{f'_{x,y}}(t) = - \sqrt{R} \hat{E}_{f_{x,y}}(t) + \sqrt{T} \hat{E}_{g'_{x,y}}(t).
\end{eqnarray}


Next, we consider the propagation of the laser within the ITMs and EMs.
In this paper, we assume that the EMs have perfect reflection, i.e.,
\begin{eqnarray}
  \label{eq:End-Mirror-Perfect-reflection}
  \tilde{R} = 1, \quad \tilde{T} = 0.
\end{eqnarray}
and we obtain
\begin{eqnarray}
  \label{eq:End-Mirror-Perfect-reflection-field}
  \hat{E}_{j'_{x,y}}(t) = \hat{E}_{j_{x,y}}(t).
\end{eqnarray}
If we have to consider the loss model of the imperfection of EM as in
Ref.~\cite{H.J.Kimble-Y.Levin-A.B.Matsko-K.S.Thorne-S.P.Vyatchanin-2001},
we have to change Eq.~(\ref{eq:End-Mirror-Perfect-reflection-field})
as described in
Ref.~\cite{H.J.Kimble-Y.Levin-A.B.Matsko-K.S.Thorne-S.P.Vyatchanin-2001}.


Furthermore, the derivations of the electric field relations between
$\hat{E}_{g_{x,y}}$ , $\hat{E}_{j_{x,y}}$, $\hat{E}_{j'_{x,y}}$, and
$\hat{E}_{g'_{x,y}}$ are delicate.
However, the physical effects are just the free propagation of the
laser.
This is because the time-dependence of the relative positions
$\hat{X}_{x}(t):=\hat{X}_{XEM}-\hat{X}_{XITM}$ and
$\hat{X}_{y}(t):=\hat{X}_{YEM}-\hat{X}_{YITM}$.
We have to be careful to treat this time dependence correctly to
estimate the retarded effects on the electric fields in the
interferometer.
Since we apply the perfect reflection of EMs, the relation of
$\hat{E}_{g_{x,y}}$ and $\hat{E}_{g'_{x,y}}$ is identical to the case
of the Michelson interferometer from the setup depicted in Fig.~\ref{fig:arm-propagation-Fabry-Perot-setup-notation}.
Then, we obtain
\begin{eqnarray}
  \label{eq:xarm-retarded-effect-g-to-gprime}
  \hat{E}_{g'_{x}}(t)
  &=&
      \hat{E}_{g_{x}}\left[
      t - 2 \left(\tau + \frac{1}{c}\hat{X}_{x}(t-\tau) \right)
      \right]
  , \\
  \label{eq:yarm-retarded-effect-g-to-gprime}
  \hat{E}_{g'_{y}}(t)
  &=&
      \hat{E}_{g_{y}}\left[
      t - 2 \left(\tau + \frac{1}{c}\hat{X}_{y}(t-\tau) \right)
      \right]
      ,
\end{eqnarray}
where $\tau = L/c$.
Moreover, through Eqs.~(\ref{eq:Kimble-B8-correspoind-1}),
(\ref{eq:Kimble-B8-correspoind-2}),
(\ref{eq:xarm-retarded-effect-g-to-gprime}),
and (\ref{eq:yarm-retarded-effect-g-to-gprime}), we obtain the
relations
\begin{eqnarray}
     \hat{E}_{f'_{x}}(t)
  &=&
      - \sqrt{1-T} \hat{E}_{f_{x}}(t)
      \nonumber\\
  &&
      + \sqrt{T} \hat{E}_{g_{x}}\left[
      t - 2 \left(\tau + \frac{1}{c}\hat{X}_{x}(t-\tau) \right)
      \right]
      ,
  \label{eq:xyarm-prop-fprime-and-f-g}
  \\
     \hat{E}_{g_{x}}(t)
  &=&
      \sqrt{T} \hat{E}_{f_{x}}(t)
      \nonumber\\
  &&
      + \sqrt{1-T} \hat{E}_{g_{x}}\left[
      t - 2 \left(\tau + \frac{1}{c}\hat{X}_{x}(t-\tau) \right)
      \right]
      ,
     \nonumber\\
  \label{eq:xarm-prop-g-and-f-g}
\end{eqnarray}
and
\begin{eqnarray}
     \hat{E}_{f'_{y}}(t)
  &=&
      - \sqrt{1-T} \hat{E}_{f_{y}}(t)
      \nonumber\\
  &&
      + \sqrt{T} \hat{E}_{g_{y}}\left[
      t - 2 \left(\tau + \frac{1}{c}\hat{X}_{y}(t-\tau) \right)
      \right]
      ,
  \label{eq:yarm-prop-fprime-and-f-g}
  \\
     \hat{E}_{g_{y}}(t)
  &=&
      \sqrt{T} \hat{E}_{f_{y}}(t)
      \nonumber\\
  &&
      + \sqrt{1-T} \hat{E}_{g_{y}}\left[
      t - 2 \left(\tau + \frac{1}{c}\hat{X}_{y}(t-\tau) \right)
      \right]
      .
     \nonumber\\
  \label{eq:yarm-prop-g-and-f-g}
\end{eqnarray}
Here, we note that Eq.~(\ref{eq:xarm-prop-g-and-f-g})
(resp. (\ref{eq:yarm-prop-g-and-f-g})) gives the relation between
the electric fields $\hat{E}_{f_{x}}$ and $\hat{E}_{g_{x}}$
(resp. $\hat{E}_{f_{y}}$ and $\hat{E}_{g_{y}}$).
If we can obtain the inverse relation of
Eq.~(\ref{eq:xarm-prop-g-and-f-g})
(resp.~(\ref{eq:yarm-prop-g-and-f-g})), we can obtain the relation
between the electric fields $\hat{E}_{f'_{x}}$ and $\hat{E}_{f_{x}}$
(resp. $\hat{E}_{f'_{y}}$ and $\hat{E}_{f_{y}}$) through the
substitution of this inverse relation of
(\ref{eq:xarm-prop-g-and-f-g})
(resp.~(\ref{eq:yarm-prop-g-and-f-g})) into
Eq.~(\ref{eq:xyarm-prop-fprime-and-f-g})
(resp.~(\ref{eq:yarm-prop-fprime-and-f-g})).


To obtain the expressions of $\hat{E}_{g_{x,y}}$ in terms of
$\hat{E}_{f_{x,y}}$, the treatment of the Fourier transformation is
convenient, since we only consider the linear perturbation level with
respect to $\hat{X}_{x,y}$ within this paper.
First, we consider the Fourier transformation of
Eq.~(\ref{eq:xyarm-prop-fprime-and-f-g}).
Introducing the Fourier transformation of the displacement variable
$\hat{X}_{x,y}(t)$ as
\begin{eqnarray}
  \label{eq:hatXx-Fouriler-def}
  \hat{X}_{x,y}(t)
  &=:&
  \int_{-\infty}^{+\infty} \frac{d\omega_{2}}{2\pi}
  \hat{Z}_{x,y}(\omega_{2})
  e^{-i\omega_{2} t},
\end{eqnarray}
we only consider the effects of the linear level of $\hat{X}_{x}$ in
Eq.~(\ref{eq:xyarm-prop-fprime-and-f-g}), since we regard that the
higher-order of $\hat{X}_{x}$ are negligible.
Furthermore, through the inverse Fourier transformation, we obtain
\begin{eqnarray}
  \hat{F}'_{x}(\omega)
  &=&
      -
      \sqrt{1-T}
      \hat{F}_{x}(\omega)
      +
      \sqrt{T}
      \hat{G}_{x}(\omega)
      e^{ + 2 i \omega \tau }
      \nonumber\\
  &&
     +
     i
     \frac{2}{c}
     \sqrt{T}
     \int_{-\infty}^{+\infty} \frac{d\omega_{1}}{2\pi}
     e^{ + i (\omega_{1}+\omega) \tau }
     \omega_{1}
     \sqrt{\frac{|\omega_{1}|}{|\omega|}}
     \nonumber\\
  && \quad\quad
     \times
     \hat{G}_{x}(\omega_{1})
     \hat{Z}_{x}(\omega-\omega_{1})
      \nonumber\\
  &&
     +
     O\left(\left(\hat{X}_{x}\right)^{2}\right)
     .
     \label{eq:xyarm-prop-fprime-and-f-g-Fourier-Fourier}
\end{eqnarray}


Similarly, we consider the Fourier transformation of
Eq.~(\ref{eq:xarm-prop-g-and-f-g}):
\begin{eqnarray}
     \hat{G}_{x}(\omega)
  &=&
      \sqrt{T}
      \hat{F}_{x}(\omega)
      +
      \sqrt{1-T}
      \hat{G}_{x}(\omega)
      e^{+ 2 i \omega \tau }
      \nonumber\\
  &&
      +
      i
      \frac{2}{c}
      \sqrt{1-T}
      \int_{-\infty}^{+\infty} \frac{d\omega_{1}}{2\pi}
      e^{+ i ( \omega_{1} + \omega )\tau }
      \omega_{1}
      \sqrt{\frac{|\omega_{1}|}{|\omega|}}
      \nonumber\\
  && \quad\quad
     \times
      \hat{G}_{x}(\omega_{1})
      \hat{Z}_{x}(\omega-\omega_{1})
      \nonumber\\
  &&
      +
      O\left(\left(\hat{X}_{x}\right)^{2}\right)
      ,
     \label{eq:xarm-prop-g-and-f-g-Fourier-temp1}
\end{eqnarray}
where we used (\ref{eq:hatXx-Fouriler-def}).
Then, we obtain
\begin{widetext}
\begin{eqnarray}
     \hat{G}_{x}(\omega)
  &=&
      \sqrt{T}
      \left[ 1 - \sqrt{1-T} e^{+ 2 i \omega \tau } \right]^{-1}
      \hat{F}_{x}(\omega)
      \nonumber\\
  &&
      +
      i
      \frac{2}{c}
      \sqrt{1-T}
      \left[ 1 - \sqrt{1-T} e^{+ 2 i \omega \tau } \right]^{-1}
      \int_{-\infty}^{+\infty} \frac{d\omega_{1}}{2\pi}
      e^{+ i ( \omega_{1} + \omega )\tau }
      \omega_{1}
      \sqrt{\frac{|\omega_{1}|}{|\omega|}}
      \hat{G}_{x}(\omega_{1})
      \hat{Z}_{x}(\omega-\omega_{1})
      \nonumber\\
  &&
      +
      O\left(\left(\hat{X}_{x}\right)^{2}\right)
     .
     \label{eq:xarm-prop-g-and-f-g-Fourier-temp2}
\end{eqnarray}
The substitution of $\hat{G}_{x}(\omega_{1})$ in the left-hand
side of Eq.~(\ref{eq:xarm-prop-g-and-f-g-Fourier-temp2}) into
$\hat{G}_{x}(\omega_{1})$ in the right-hand side of
Eq.~(\ref{eq:xarm-prop-g-and-f-g-Fourier-temp2}) yields
\begin{eqnarray}
  &&
     \hat{G}_{x}(\omega)
     \nonumber\\
  &=&
      \sqrt{T}
      \left[ 1 - \sqrt{1-T} e^{+ 2 i \omega \tau } \right]^{-1}
      \hat{F}_{x}(\omega)
      \nonumber\\
  &&
      +
      i
      \frac{2}{c}
      \sqrt{T(1-T)}
      \left[ 1 - \sqrt{1-T} e^{+ 2 i \omega \tau } \right]^{-1}
      \int_{-\infty}^{+\infty} \frac{d\omega_{1}}{2\pi}
      e^{+ i ( \omega_{1} + \omega )\tau }
      \omega_{1}
      \sqrt{\frac{|\omega_{1}|}{|\omega|}}
      \left[ 1 - \sqrt{1-T} e^{+ 2 i \omega_{1} \tau } \right]^{-1}
      \hat{F}_{x}(\omega_{1})
      \hat{Z}_{x}(\omega-\omega_{1})
      \nonumber\\
  &&
      +
      O\left(\left(\hat{X}_{x}\right)^{2}\right)
      .
      \label{eq:xarm-prop-gx-is-f-z-Fourier}
\end{eqnarray}
Substituting Eq.~(\ref{eq:xarm-prop-gx-is-f-z-Fourier}) into
Eq.~(\ref{eq:xyarm-prop-fprime-and-f-g-Fourier-Fourier}), we obtain
\begin{eqnarray}
  &&
     \hat{F}'_{x}(\omega)
     \nonumber\\
  &=&
      \left[ 1 - \sqrt{1-T} e^{ - 2 i \omega \tau } \right]
      \left[ 1 - \sqrt{1-T} e^{+ 2 i \omega \tau } \right]^{-1}
      e^{ + 2 i \omega \tau }
      \hat{F}_{x}(\omega)
      \nonumber\\
  &&
     +
     i
     \frac{2}{c}
     T
     e^{ + 2 i \omega \tau }
     \left[ 1 - \sqrt{1-T} e^{+ 2 i \omega \tau } \right]^{-1}
     \int_{-\infty}^{+\infty} \frac{d\omega_{1}}{2\pi}
     \omega_{1}
     \sqrt{\frac{|\omega_{1}|}{|\omega|}}
     e^{ -  i (\omega-\omega_{1}) \tau }
     \left[ 1 - \sqrt{1-T} e^{+ 2 i \omega_{1} \tau } \right]^{-1}
     \hat{F}_{x}(\omega_{1})
     \hat{Z}_{x}(\omega-\omega_{1})
      \nonumber\\
  &&
     +
     O\left(\left(\hat{X}_{x}\right)^{2}\right)
     .
     \label{eq:xarm-prop-fprime-and-f-Fourier-final}
\end{eqnarray}
Furthermore, the substitution of
Eq.~(\ref{eq:xarm-prop-fprime-and-f-Fourier-final}) into
Eq.~(\ref{eq:xarm-retarded-effect-f'-to-c'-mod-Fourier}) yields
\begin{eqnarray}
     \hat{C}_{x}'(\omega)
  &=&
      \left[ 1 - \sqrt{1-T} e^{ - 2 i \omega \tau } \right]
      \left[ 1 - \sqrt{1-T} e^{+ 2 i \omega \tau } \right]^{-1}
      e^{ + 2 i \omega \tau }
      e^{ + i \omega \tau_{x}' }
      \hat{F}_{x}(\omega)
      \nonumber\\
  &&
     +
     i
     \frac{2}{c}
     T
     e^{ + 2 i \omega \tau }
      e^{ + i \omega \tau_{x}' }
     \left[ 1 - \sqrt{1-T} e^{+ 2 i \omega \tau } \right]^{-1}
     \nonumber\\
  && \quad\quad\quad
     \times
     \int_{-\infty}^{+\infty} \frac{d\omega_{1}}{2\pi}
     \omega_{1}
     \sqrt{\frac{|\omega_{1}|}{|\omega|}}
     e^{ -  i (\omega-\omega_{1}) \tau }
     \left[ 1 - \sqrt{1-T} e^{+ 2 i \omega_{1} \tau } \right]^{-1}
     \hat{F}_{x}(\omega_{1})
     \hat{Z}_{x}(\omega-\omega_{1})
     \nonumber\\
  &&
     +
     e^{ + i \omega \tau_{x}' }
     \int_{-\infty}^{+\infty} \frac{d\omega_{1}}{2\pi}
     i \frac{\omega_{1}}{c}
     \sqrt{\frac{|\omega_{2}|}{|\omega|}}
     \left[ 1 - \sqrt{1-T} e^{ - 2 i \omega_{1} \tau } \right]
     \left[ 1 - \sqrt{1-T} e^{+ 2 i \omega_{1} \tau } \right]^{-1}
     e^{ + 2 i \omega_{1} \tau }
     \hat{F}_{x}(\omega_{1})
     \hat{Z}_{XITM}(\omega - \omega_{1})
     \nonumber\\
  &&
     +
     O\left(\left(\hat{X}_{XITM}\right)^{2},\left(\hat{X}_{x}\right)^{2},\hat{X}_{x}\hat{X}_{XITM}\right)
     .
     \label{eq:xarm-prop-cprime-and-f-Fourier-final}
\end{eqnarray}


For $y$-arm quadrature relation we replace $x\rightarrow y$ and
$XITM\rightarrow YITM$ as
\begin{eqnarray}
     \hat{C}_{y}'(\omega)
  &=&
      \left[ 1 - \sqrt{1-T} e^{ - 2 i \omega \tau } \right]
      \left[ 1 - \sqrt{1-T} e^{+ 2 i \omega \tau } \right]^{-1}
      e^{ + 2 i \omega \tau }
      e^{ + i \omega \tau_{y}' }
      \hat{F}_{y}(\omega)
      \nonumber\\
  &&
     +
     i
     \frac{2}{c}
     T
     e^{ + 2 i \omega \tau }
     e^{ + i \omega \tau_{y}' }
     \left[ 1 - \sqrt{1-T} e^{+ 2 i \omega \tau } \right]^{-1}
     \nonumber\\
  && \quad\quad\quad
     \times
     \int_{-\infty}^{+\infty} \frac{d\omega_{1}}{2\pi}
     \omega_{1}
     \sqrt{\frac{|\omega_{1}|}{|\omega|}}
     e^{ -  i (\omega-\omega_{1}) \tau }
     \left[ 1 - \sqrt{1-T} e^{+ 2 i \omega_{1} \tau } \right]^{-1}
     \hat{F}_{y}(\omega_{1})
     \hat{Z}_{y}(\omega-\omega_{1})
     \nonumber\\
  &&
     +
     e^{ + i \omega \tau_{y}' }
     \int_{-\infty}^{+\infty} \frac{d\omega_{1}}{2\pi}
     i \frac{\omega_{1}}{c}
     \sqrt{\frac{|\omega_{1}|}{|\omega|}}
     \left[ 1 - \sqrt{1-T} e^{ - 2 i \omega_{1} \tau } \right]
     \left[ 1 - \sqrt{1-T} e^{+ 2 i \omega_{1} \tau } \right]^{-1}
     e^{ + 2 i \omega_{1} \tau }
     \hat{F}_{y}(\omega_{1})
     \hat{Z}_{YITM}(\omega - \omega_{1})
     \nonumber\\
  &&
     +
     O\left(\left(\hat{X}_{YITM}\right)^{2},\left(\hat{X}_{y}\right)^{2},\hat{X}_{y}\hat{X}_{YITM}\right)
     .
     \label{eq:yarm-prop-cprime-and-f-Fourier-final}
\end{eqnarray}


Moreover, the substitution
Eq.~(\ref{eq:xarm-retarded-effect-f'-to-c'-mod-Fourier-result}) into
Eq.~(\ref{eq:xarm-prop-cprime-and-f-Fourier-final}) yields
\begin{eqnarray}
  \hat{C}_{x}'(\omega)
  &=&
      \left[ 1 - \sqrt{1-T} e^{ - 2 i \omega \tau } \right]
      \left[ 1 - \sqrt{1-T} e^{+ 2 i \omega \tau } \right]^{-1}
      e^{ + 2 i \omega \tau }
      e^{ + 2 i \omega \tau_{x}' }
      \hat{C}_{x}(\omega)
      \nonumber\\
  &&
     +
     i
     e^{ + 2 i \omega \tau_{x}' }
     e^{ + 2 i \omega \tau }
     \left[ 1 - \sqrt{1-T} e^{ - 2 i \omega \tau } \right]
     \left[ 1 - \sqrt{1-T} e^{+ 2 i \omega \tau } \right]^{-1}
     \int_{-\infty}^{+\infty} \frac{d\omega_{1}}{2\pi}
     \sqrt{\frac{|\omega_{1}|}{|\omega|}}
     \frac{\omega_{1}}{c}
     \hat{C}_{x}(\omega_{1})
     \hat{Z}_{XITM}(\omega - \omega_{1})
     \nonumber\\
  &&
     +
     i
     \frac{2}{c}
     T
     e^{ + i \omega \tau }
     e^{ + i \omega \tau_{x}' }
     \left[ 1 - \sqrt{1-T} e^{+ 2 i \omega \tau } \right]^{-1}
     \nonumber\\
  && \quad\quad\quad
     \times
     \int_{-\infty}^{+\infty} \frac{d\omega_{1}}{2\pi}
     \omega_{1}
     \sqrt{\frac{|\omega_{1}|}{|\omega|}}
     e^{ +  i \omega_{1}) \tau }
     e^{ +  i \omega_{1} \tau_{x}' }
     \left[ 1 - \sqrt{1-T} e^{+ 2 i \omega_{1} \tau } \right]^{-1}
     \hat{C}_{x}(\omega_{1})
     \hat{Z}_{x}(\omega-\omega_{1})
     \nonumber\\
  &&
     +
     e^{ + i \omega \tau_{x}' }
     \int_{-\infty}^{+\infty} \frac{d\omega_{1}}{2\pi}
     i \frac{\omega_{1}}{c}
     \sqrt{\frac{|\omega_{1}|}{|\omega|}}
     \left[ 1 - \sqrt{1-T} e^{ - 2 i \omega_{1} \tau } \right]
     \left[ 1 - \sqrt{1-T} e^{+ 2 i \omega_{1} \tau } \right]^{-1}
      \nonumber\\
  && \quad\quad
     \times
     e^{ + 2 i \omega_{1} \tau }
     e^{ + i \omega_{1} \tau_{x}' }
     \hat{C}_{x}(\omega_{1})
     \hat{Z}_{XITM}(\omega - \omega_{1})
     \nonumber\\
  &&
     +
     O\left(\left(\hat{X}_{XITM}\right)^{2},\left(\hat{X}_{x}\right)^{2},\hat{X}_{x}\hat{X}_{XITM}\right)
     .
     \label{eq:xarm-prop-cprime-and-c-Fourier-final}
\end{eqnarray}
Similarly, for $C_{y}'(\omega)$, we have
\begin{eqnarray}
  \hat{C}_{y}'(\omega)
  &=&
      \left[ 1 - \sqrt{1-T} e^{ - 2 i \omega \tau } \right]
      \left[ 1 - \sqrt{1-T} e^{+ 2 i \omega \tau } \right]^{-1}
      e^{ + 2 i \omega \tau }
      e^{ + 2 i \omega \tau_{y}' }
      \hat{C}_{y}(\omega)
      \nonumber\\
  &&
     +
     i
     e^{ + 2 i \omega \tau_{y}' }
     e^{ + 2 i \omega \tau }
     \left[ 1 - \sqrt{1-T} e^{ - 2 i \omega \tau } \right]
     \left[ 1 - \sqrt{1-T} e^{+ 2 i \omega \tau } \right]^{-1}
     \int_{-\infty}^{+\infty} \frac{d\omega_{1}}{2\pi}
     \sqrt{\frac{|\omega_{1}|}{|\omega|}}
     \frac{\omega_{1}}{c}
     \hat{C}_{y}(\omega_{1})
     \hat{Z}_{YITM}(\omega - \omega_{1})
     \nonumber\\
  &&
     +
     i
     \frac{2}{c}
     T
     e^{ +  i \omega \tau }
     e^{ + i \omega \tau_{y}' }
     \left[ 1 - \sqrt{1-T} e^{+ 2 i \omega \tau } \right]^{-1}
     \nonumber\\
  && \quad\quad
     \times
     \int_{-\infty}^{+\infty} \frac{d\omega_{1}}{2\pi}
     \omega_{1}
     \sqrt{\frac{|\omega_{1}|}{|\omega|}}
     e^{ +  i \omega_{1} \tau }
     e^{ +  i \omega_{1} \tau_{y}' }
     \left[ 1 - \sqrt{1-T} e^{+ 2 i \omega_{1} \tau } \right]^{-1}
     \hat{C}_{y}(\omega_{1})
     \hat{Z}_{y}(\omega-\omega_{1})
     \nonumber\\
  &&
     +
     e^{ + i \omega \tau_{y}' }
     \int_{-\infty}^{+\infty} \frac{d\omega_{1}}{2\pi}
     i \frac{\omega_{1}}{c}
     \sqrt{\frac{|\omega_{1}|}{|\omega|}}
     \left[ 1 - \sqrt{1-T} e^{ - 2 i \omega_{1} \tau } \right]
     \left[ 1 - \sqrt{1-T} e^{+ 2 i \omega_{1} \tau } \right]^{-1}
     \nonumber\\
  && \quad\quad
     \times
     e^{ + 2 i \omega_{1} \tau }
     e^{ + i \omega_{1} \tau_{y}' }
     \hat{C}_{y}(\omega_{1})
     \hat{Z}_{YITM}(\omega - \omega_{1})
     \nonumber\\
  &&
     +
     O\left(\left(\hat{X}_{YITM}\right)^{2},\left(\hat{X}_{y}\right)^{2},\hat{X}_{y}\hat{X}_{YITM}\right)
     .
     \label{eq:yarm-prop-cprime-and-c-Fourier-final}
\end{eqnarray}


\subsection{Input-output relation with mirrors' motion}
\label{sec:Input-output_relation_with_mirrors_motion}


From Eqs.~(\ref{eq:output-electric-field-junction-hatB}),
(\ref{eq:xarm-input-field-junction-hatCx}),
(\ref{eq:yarm-input-field-junction-hatCy}), and the arm propagation
relations (\ref{eq:xarm-prop-cprime-and-c-Fourier-final}) and
(\ref{eq:yarm-prop-cprime-and-c-Fourier-final}), we obtain
\begin{eqnarray}
  &&
     \!\!\!\!\!\!\!
     \hat{B}(\omega)
     \nonumber\\
  &=&
      \frac{1}{\sqrt{2}}
      \left[ 1 - \sqrt{1-T} e^{ - 2 i \omega \tau } \right]
      \left[ 1 - \sqrt{1-T} e^{+ 2 i \omega \tau } \right]^{-1}
      e^{ + 2 i \omega \tau }
      e^{ + i \omega (\tau_{y}' + \tau_{x}' ) }
      \left[
      e^{ + i \omega (\tau_{y}' - \tau_{x}' ) }
      \hat{C}_{y}(\omega)
      -
      e^{ -  i \omega (\tau_{y}' - \tau_{x}' ) }
      \hat{C}_{x}(\omega)
      \right]
      \nonumber\\
  &&
     +
     i
     \frac{1}{\sqrt{2}}
     e^{ + 2 i \omega \tau }
     e^{ + i \omega ( \tau_{y}' + \tau_{x}' ) }
     \left[ 1 - \sqrt{1-T} e^{ - 2 i \omega \tau } \right]
     \left[ 1 - \sqrt{1-T} e^{+ 2 i \omega \tau } \right]^{-1}
     \int_{-\infty}^{+\infty} \frac{d\omega_{1}}{2\pi}
     \sqrt{\frac{|\omega_{1}|}{|\omega|}}
     \frac{\omega_{1}}{c}
      \nonumber\\
  && \quad\quad\quad
     \times
     \left[
     e^{ + i \omega ( \tau_{y}' - \tau_{x}' ) }
     \hat{C}_{y}(\omega_{1})
     \hat{Z}_{YITM}(\omega - \omega_{1})
     -
     e^{ - i \omega ( \tau_{y}' - \tau_{x}' ) }
     \hat{C}_{x}(\omega_{1})
     \hat{Z}_{XITM}(\omega - \omega_{1})
     \right]
     \nonumber\\
  &&
     +
     i
     \frac{1}{\sqrt{2}}
     \frac{2}{c}
     T
     \left[ 1 - \sqrt{1-T} e^{+ 2 i \omega \tau } \right]^{-1}
     \int_{-\infty}^{+\infty} \frac{d\omega_{1}}{2\pi}
     \omega_{1}
     \sqrt{\frac{|\omega_{1}|}{|\omega|}}
     \left[ 1 - \sqrt{1-T} e^{+ 2 i \omega_{1} \tau } \right]^{-1}
     e^{ + i (\omega+\omega_{1}) \tau }
     e^{ + i (\omega+\omega_{1}) \frac{\tau_{y}' + \tau_{x}'}{2}}
      \nonumber\\
  && \quad\quad\quad
     \times
     \left[
     e^{ + i (\omega+\omega_{1}) \frac{\tau_{y}' - \tau_{x}'}{2}}
     \hat{C}_{y}(\omega_{1})
     \hat{Z}_{y}(\omega-\omega_{1})
      -
     e^{ - i (\omega+\omega_{1}) \frac{\tau_{y}' - \tau_{x}'}{2}}
     \hat{C}_{x}(\omega_{1})
     \hat{Z}_{x}(\omega-\omega_{1})
     \right]
     \nonumber\\
  &&
     +
     \frac{1}{\sqrt{2}}
     \int_{-\infty}^{+\infty} \frac{d\omega_{1}}{2\pi}
     i \frac{\omega_{1}}{c}
     \sqrt{\frac{|\omega_{1}|}{|\omega|}}
     \left[ 1 - \sqrt{1-T} e^{ - 2 i \omega_{1} \tau } \right]
     \left[ 1 - \sqrt{1-T} e^{+ 2 i \omega_{1} \tau } \right]^{-1}
     e^{ + 2 i \omega_{1} \tau }
     e^{ + i (\omega+\omega_{1}) \frac{\tau_{y}' + \tau_{x}'}{2} }
      \nonumber\\
  && \quad\quad\quad
     \times
     \left[
     e^{ + i (\omega+\omega_{1}) \frac{\tau_{y}' - \tau_{x}'}{2} }
     \hat{C}_{y}(\omega_{1})
     \hat{Z}_{YITM}(\omega - \omega_{1})
     -
     e^{ - i (\omega+\omega_{1}) \frac{\tau_{y}' - \tau_{x}'}{2} }
     \hat{C}_{x}(\omega_{1})
     \hat{Z}_{XITM}(\omega - \omega_{1})
     \right]
     \nonumber\\
  &&
     +
     O\left(\left(\hat{X}\right)^{2}\right)
     .
     \label{eq:output-electric-field-junction-hatB-4}
\end{eqnarray}
\end{widetext}
The first line is the direct shot noise of this interferometer.
The integral in the second and third lines in
Eq.~(\ref{eq:output-electric-field-junction-hatB-4}) is the retarded
effect during the propagation from the beam splitter to the input test
masses due to the modification of the proper distance between the beam
splitter and the input test masses $l_{x}+\hat{X}_{XITM}$ (
$l_{y}+\hat{X}_{YITM}$ ), respectively.
We note that these second and third lines do not include the
propagation of the Fabry-Perot cavities between the ITMs and EMs,
respectively.
The integral in the fourth and fifth line in
Eq.~(\ref{eq:output-electric-field-junction-hatB-4}), which includes
$\hat{X}_{x}:=\hat{X}_{XEM}-\hat{X}_{XITM}$ and
$\hat{X}_{y}:=\hat{X}_{XEM}-\hat{X}_{XITM}$, is also the retarded
effect during the propagation of the laser in the Fabry-P\'erot
cavities.
The integral in the sixth and seventh line of
Eq.~(\ref{eq:output-electric-field-junction-hatB-4}), which includes
$\hat{Z}_{YITM}$ and $\hat{Z}_{XITM}$, are the retarded effects during
the propagation of the laser from the input test masses to the beam
splitter due to the modification of the proper distance between the
beam splitter and the input test masses $l_{x}+\hat{X}_{XITM}$ (
$l_{y}+\hat{X}_{YITM}$ ), respectively.
The difference of these terms from the second and third lines in
Eq.~(\ref{eq:output-electric-field-junction-hatB-4}) is the fact that
the sixth and seventh lines of
Eq.~(\ref{eq:output-electric-field-junction-hatB-4}) include the
information of the propagation in the Fabry-P\'erot cavities.
The last line in Eq.~(\ref{eq:output-electric-field-junction-hatB-4})
symbolically represents the terms of the order of $O(\hat{X}_{x}^{2}$,
$\hat{X}_{y}^{2}$, $\hat{X}_{XITM}^{2}$, $\hat{X}_{YITM}^{2})$.


Substituting Eqs.~(\ref{eq:xarm-input-field-junction-hatCx}) and
(\ref{eq:yarm-input-field-junction-hatCy}) into
Eq.~(\ref{eq:output-electric-field-junction-hatB-4}), we obtain the
output field quadrature $\hat{B}(\omega)$.
Furthermore, for our convention, we define
$\hat{Z}_{com,diff}(\omega)$ by
\begin{eqnarray}
  \label{eq:x-com+diff-motion-Fourier}
  \hat{Z}_{x}(\omega)
  &=:&
      \hat{Z}_{com}(\omega)
      +
      \hat{Z}_{diff}(\omega)
       ,
  \\
  \label{eq:y-com-diff-motion-Fourier}
  \hat{Z}_{y}(\omega)
  &=:&
      \hat{Z}_{com}(\omega)
      -
      \hat{Z}_{diff}(\omega)
      ,
\end{eqnarray}
where $\hat{Z}_{com}(\omega)$ represents the $x$-arm and $y$-arm
common motion of the relative motion between the end test mass and the
input test mass and $\hat{Z}_{diff}(\omega)$ represents the $x$-arm and
$y$-arm differential motion of the relative motion the end test mass
and input test mass.
Furthermore, we also define $\hat{Z}_{comITM,diffITM}(\omega)$ by
\begin{eqnarray}
  \label{eq:XITM-com+diffITM-motion-Fourier}
  \hat{Z}_{XITM}(\omega)
  &=:&
      \hat{Z}_{comITM}(\omega)
      +
      \hat{Z}_{diffITM}(\omega)
       ,
  \\
  \label{eq:YITM-com-diffITM-motion-Fourier}
  \hat{Z}_{YITM}(\omega)
  &=:&
      \hat{Z}_{comITM}(\omega)
      -
      \hat{Z}_{diffITM}(\omega)
      .
\end{eqnarray}
Substituting Eqs.~(\ref{eq:x-com+diff-motion-Fourier}),
(\ref{eq:y-com-diff-motion-Fourier}),
(\ref{eq:XITM-com+diffITM-motion-Fourier}), and
(\ref{eq:YITM-com-diffITM-motion-Fourier}) into
Eq.~(\ref{eq:output-electric-field-junction-hatB-4}), we obtain
\begin{widetext}
\begin{eqnarray}
  &&
     \!\!\!\!\!\!\!
     \hat{B}(\omega)
     \nonumber\\
  &=&
      \left[ 1 - \sqrt{1-T} e^{ - 2 i \omega \tau } \right]
      \left[ 1 - \sqrt{1-T} e^{+ 2 i \omega \tau } \right]^{-1}
      e^{ + 2 i \omega \tau }
      e^{ + i \omega (\tau_{y}' + \tau_{x}' ) }
      \nonumber\\
  && \quad\quad
     \times
      \left[
      i \sin( \omega (\tau_{y}' - \tau_{x}' ) )
      \hat{D}(\omega)
      +
      \cos( \omega (\tau_{y}' - \tau_{x}' ) )
      \hat{A}(\omega)
      \right]
      \nonumber\\
  &&
     +
     i
     e^{ + 2 i \omega \tau }
     e^{ + i \omega ( \tau_{y}' + \tau_{x}' ) }
     \left[ 1 - \sqrt{1-T} e^{ - 2 i \omega \tau } \right]
     \left[ 1 - \sqrt{1-T} e^{+ 2 i \omega \tau } \right]^{-1}
     \int_{-\infty}^{+\infty} \frac{d\omega_{1}}{2\pi}
     \sqrt{\frac{|\omega_{1}|}{|\omega|}}
     \frac{\omega_{1}}{c}
      \nonumber\\
  && \quad\quad\quad
     \times
     \left[
     +
     \left(
     i \sin( \omega ( \tau_{y}' - \tau_{x}' ) )
     \hat{D}(\omega_{1})
     +
     \cos( \omega ( \tau_{y}' - \tau_{x}' ) )
     \hat{A}(\omega_{1})
     \right)
     \hat{Z}_{comITM}(\omega - \omega_{1})
     \right.
      \nonumber\\
  && \quad\quad\quad\quad\quad
     \left.
     -
     \left(
     \cos( \omega ( \tau_{y}' - \tau_{x}' ) )
     \hat{D}(\omega_{1})
     +
     i \sin( \omega ( \tau_{y}' - \tau_{x}' ) )
     \hat{A}(\omega_{1})
     \right)
     \hat{Z}_{diffITM}(\omega - \omega_{1})
     \right]
     \nonumber\\
  &&
     +
     i
     \frac{2}{c}
     T
     \left[ 1 - \sqrt{1-T} e^{+ 2 i \omega \tau } \right]^{-1}
     \int_{-\infty}^{+\infty} \frac{d\omega_{1}}{2\pi}
     \omega_{1}
     \sqrt{\frac{|\omega_{1}|}{|\omega|}}
     \left[ 1 - \sqrt{1-T} e^{+ 2 i \omega_{1} \tau } \right]^{-1}
     e^{ + i (\omega+\omega_{1}) \tau }
     e^{ + i (\omega+\omega_{1}) \frac{\tau_{y}' + \tau_{x}'}{2}}
      \nonumber\\
  && \quad\quad\quad
     \times
     \left[
     +
     \left(
     i \sin\left( (\omega+\omega_{1}) \frac{\tau_{y}' - \tau_{x}'}{2} \right)
     \hat{D}(\omega_{1})
     +
     \cos\left( (\omega+\omega_{1}) \frac{\tau_{y}' - \tau_{x}'}{2} \right)
     \hat{A}(\omega_{1})
     \right)
     \hat{Z}_{com}(\omega-\omega_{1})
     \right.
      \nonumber\\
  && \quad\quad\quad\quad\quad
     \left.
     -
     \left(
     \cos\left( (\omega+\omega_{1}) \frac{\tau_{y}' - \tau_{x}'}{2} \right)
     \hat{D}(\omega_{1})
     +
     i \sin\left( (\omega+\omega_{1}) \frac{\tau_{y}' - \tau_{x}'}{2} \right)
     \hat{A}(\omega_{1})
     \right)
     \hat{Z}_{diff}(\omega-\omega_{1})
     \right]
     \nonumber\\
  &&
     +
     \int_{-\infty}^{+\infty} \frac{d\omega_{1}}{2\pi}
     i \frac{\omega_{1}}{c}
     \sqrt{\frac{|\omega_{1}|}{|\omega|}}
     \left[ 1 - \sqrt{1-T} e^{ - 2 i \omega_{1} \tau } \right]
     \left[ 1 - \sqrt{1-T} e^{+ 2 i \omega_{1} \tau } \right]^{-1}
     e^{ + 2 i \omega_{1} \tau }
     e^{ + i (\omega+\omega_{1}) \frac{\tau_{y}' + \tau_{x}'}{2} }
      \nonumber\\
  && \quad\quad\quad
     \times
     \left[
     \left(
     i \sin\left( (\omega+\omega_{1}) \frac{\tau_{y}' - \tau_{x}'}{2} \right)
     \hat{D}(\omega_{1})
     +
     \cos\left( (\omega+\omega_{1}) \frac{\tau_{y}' - \tau_{x}'}{2} \right)
     \hat{A}(\omega_{1})
     \right)
     \hat{Z}_{comITM}(\omega - \omega_{1})
     \right.
      \nonumber\\
  && \quad\quad\quad\quad\quad
     \left.
     -
     \left(
     \cos\left( (\omega+\omega_{1}) \frac{\tau_{y}' - \tau_{x}'}{2} \right)
     \hat{D}(\omega_{1})
     +
     i \sin\left( (\omega+\omega_{1}) \frac{\tau_{y}' - \tau_{x}'}{2} \right)
     \hat{A}(\omega_{1})
     \right)
     \hat{Z}_{diffITM}(\omega - \omega_{1})
     \right]
     \nonumber\\
  &&
     +
     O\left(\left(\hat{X}\right)^{2}\right)
     .
     \label{eq:input-output-relation-general}
\end{eqnarray}
\end{widetext}


\subsection{Coherent state of the optical fields}
\label{sec:Coherent_state_ of_the_optical_fields}


Here, Eq.~(\ref{eq:input-output-relation-general}) implies that the
output operator $\hat{B}$ is given by the operators $\hat{A}$,
$\hat{D}$, $\hat{Z}_{diff}$, $\hat{Z}_{com}$, $\hat{Z}_{diffITM}$, and
$\hat{Z}_{comITM}$.
Later, we see that the displacements $\hat{Z}_{diff}$,
$\hat{Z}_{com}$, $\hat{Z}_{diffITM}$, and $\hat{Z}_{comITM}$ are given
by $\hat{A}$ and $\hat{D}$ together with the gravitational-wave signal
through the equations of motions for end test masses and the input
test masses.
Therefore, to discuss the information from the output operator
$\hat{B}$, we have to specify the quantum states associated with the
operators $\hat{A}$ and $\hat{D}$.
The state associated with the operator $\hat{A}$ is the state of the
electric field that is injected from the anti-symmetric port.
At this anti-symmetric port, the photodetector is located as in
Fig.~\ref{fig:kouchan-Fabry-Perot-setup-notation}.
On the other hand, the state associated with the operator $\hat{D}$ is
the state of the electric field that is injected from the symmetric
port.
At this symmetric port, the light source exists as depicted in
Fig.~\ref{fig:kouchan-Fabry-Perot-setup-notation}.
The total state of photon in the output port $\hat{B}$, i.e.,
$\hat{b}$, is determined by the specification of the states associated
with the operators $\hat{D}$ and $\hat{A}$, i.e., the annihilation
and creation operators $(\hat{d},\hat{d}^{\dagger})$ and
$(\hat{a},\hat{a}^{\dagger})$, respectively.


Within this paper, we assume that there is no entanglement in the
states associated with the operators $\hat{d}$ and $\hat{a}$.
Furthermore, we assume that the state associated with the operator
$\hat{d}$ is a coherent state with the complex amplitude
$\alpha(\omega)$ and the state associated with the operator $\hat{a}$
is the vacuum state.
Then, the total state $|\mbox{in}\rangle$ of the photon is given by
the direct product of the photon states of each frequency as
\begin{eqnarray}
  |\mbox{in}\rangle
  &=&
  \prod_{\omega} |\alpha(\omega)\rangle_{d}\otimes|0\rangle_{a}
  =
  \prod_{\omega} \FrakD_{d}[\alpha(\omega)]|0\rangle_{d}\otimes|0\rangle_{a}
  \nonumber\\
  &=:&
  \FrakD_{d}|0\rangle_{d}\otimes|0\rangle_{a}
  ,
  \label{eq:total-state-def}
       \\
  \FrakD_{d}
  &:=&
       \prod_{\omega} \FrakD_{d}(\alpha(\omega))
       \nonumber\\
  &=&
      \exp\left[
      \int \frac{d\omega}{2\pi}\left(
      \alpha(\omega) d^{\dagger}(\omega)
      -
      \alpha^{*}(\omega) d(\omega)
      \right)
      \right]
      .
  \label{eq:displacement-operator-for-hatd}
\end{eqnarray}
These states are the reasonable realization of electric field
operators that represent the injection of the coherent state laser
from the light source, and no input sources except for vacuum
fluctuations from the output port.
We also note that the unitarity of the operation of the beam splitter
on the electric field is guaranteed by the injection of the vacuum
state associated with the operator $\hat{a}$ from the output port.
Therefore, these setups are essential to the interferometer's
arguments.


In the Heisenberg picture, the operator $\hat{d}$ is replaced as
\begin{eqnarray}
  \label{eq:coherent-state-hatd-Heisenberg}
  \FrakD_{d}^{\dagger}\hat{d}(\omega)\FrakD_{d}
  =
  \hat{d}(\omega) + \alpha(\omega)
  ,
  \\
  \label{eq:coherent-state-hatddagger-Heisenberg}
  \FrakD_{d}^{\dagger}\hat{d}^{\dagger}(\omega)\FrakD_{d}
  =
  \hat{d}^{\dagger}(\omega) + \alpha^{*}(\omega),
\end{eqnarray}
through the displacement operator $\FrakD_{d}$ defined by Eq.~(\ref{eq:displacement-operator-for-hatd}).
Then, the operator $\hat{D}(\omega)$ defined by
Eq.~(\ref{eq:hatD-def}) is transformed as
\begin{eqnarray}
  \label{eq:coherent-state-hatD-Heisenberg}
  \FrakD_{d}^{\dagger}\hat{D}(\omega)\FrakD_{d}
  =
  \hat{D}_{c}(\omega) + \hat{D}_{v}(\omega),
\end{eqnarray}
where
\begin{eqnarray}
  \label{eq:hatDc-def}
  \hat{D}_{c}(\omega)
  :=
  \alpha(\omega)\Theta(\omega)
  +
  \alpha^{*}(-\omega)\Theta(-\omega)
  ,
  \\
  \label{eq:hatDv-def}
  \hat{D}_{v}(\omega)
  :=
  \hat{d}(\omega)\Theta(\omega)
  +
  \hat{d}^{\dagger}(-\omega)\Theta(-\omega)
  .
\end{eqnarray}


To evaluate the output signal expectation value and its fluctuations
from the input-output relation (\ref{eq:input-output-relation-general})
under the coherent state of the quadrature $\hat{D}(\omega)$, the
expression of $\FrakD_{d}^{\dagger}\hat{B}(\omega)\FrakD_{d}$ is
useful instead of $\hat{B}(\omega)$, because we treat operators in the
Heisenberg picture.
Furthermore, we regard that
$\FrakD_{d}^{\dagger}\hat{Z}_{com}(\Omega)\FrakD_{d}$,
$\FrakD_{d}^{\dagger}\hat{Z}_{diff}(\Omega)\FrakD_{d}$,
$\FrakD_{d}^{\dagger}\hat{Z}_{comITM}(\Omega)\FrakD_{d}$,
and $\FrakD_{d}^{\dagger}\hat{Z}_{diffITM}(\Omega)\FrakD_{d}$ are
small correction due to the radiation-pressure noise and
gravitational-wave signals.
Since $\hat{D}_{v}(\omega)$ and $\hat{A}(\omega)$ also describe small
fluctuations, we neglect the quadratic terms
$\hat{D}_{v}(\omega)\FrakD_{d}^{\dagger}\hat{Z}_{*}(\Omega)\FrakD_{d}$
and $\hat{A}(\omega)\FrakD_{d}^{\dagger}\hat{Z}_{*}(\Omega)\FrakD_{d}$ in the
expression of $\FrakD_{d}^{\dagger}\hat{B}(\omega)\FrakD_{d}$.


Operating $\FrakD_{d}^{\dagger}$ and $\FrakD_{d}$ to Eq.~(\ref{eq:input-output-relation-general}), substituting
Eqs.~(\ref{eq:coherent-state-hatD-Heisenberg}) into this
$\FrakD_{d}^{\dagger}$-$\FrakD_{d}$-operated version of
Eq.~(\ref{eq:input-output-relation-general}), and applying the above
approximation, then, we obtain the input-output relation as
\begin{widetext}
\begin{eqnarray}
  &&
     \!\!\!\!\!\!
     \FrakD_{d}^{\dagger}\hat{B}(\omega)\FrakD_{d}
     \nonumber\\
  &=&
      i
      \left[ 1 - \sqrt{1-T} e^{ - 2 i \omega \tau } \right]
      \left[ 1 - \sqrt{1-T} e^{+ 2 i \omega \tau } \right]^{-1}
      e^{ + 2 i \omega \tau }
      e^{ + i \omega (\tau_{y}' + \tau_{x}' ) }
      \sin( \omega (\tau_{y}' - \tau_{x}' ) )
      \hat{D}_{c}(\omega)
      \nonumber\\
  &&
      +
      \left[ 1 - \sqrt{1-T} e^{ - 2 i \omega \tau } \right]
      \left[ 1 - \sqrt{1-T} e^{+ 2 i \omega \tau } \right]^{-1}
      e^{ + 2 i \omega \tau }
      e^{ + i \omega (\tau_{y}' + \tau_{x}' ) }
      \nonumber\\
  && \quad\quad
     \times
      \left[
      i \sin( \omega (\tau_{y}' - \tau_{x}' ) )
      \hat{D}_{v}(\omega)
      +
      \cos( \omega (\tau_{y}' - \tau_{x}' ) )
      \hat{A}(\omega)
      \right]
      \nonumber\\
  &&
     +
     \frac{i}{c}
     e^{ + 2 i \omega \tau }
     e^{ + i \omega ( \tau_{y}' + \tau_{x}' ) }
     \left[ 1 - \sqrt{1-T} e^{ - 2 i \omega \tau } \right]
     \left[ 1 - \sqrt{1-T} e^{+ 2 i \omega \tau } \right]^{-1}
     \int_{-\infty}^{+\infty} \frac{d\omega_{1}}{2\pi}
     \sqrt{\frac{|\omega_{1}|}{|\omega|}}
     \omega_{1}
     \hat{D}_{c}(\omega_{1})
      \nonumber\\
  && \quad\quad
     \times
     \left[
     i \sin( \omega ( \tau_{y}' - \tau_{x}' ) )
     \FrakD_{d}^{\dagger}\hat{Z}_{comITM}(\omega - \omega_{1})\FrakD_{d}
     -
     \cos( \omega ( \tau_{y}' - \tau_{x}' ) )
     \FrakD_{d}^{\dagger}\hat{Z}_{diffITM}(\omega - \omega_{1})\FrakD_{d}
     \right]
     \nonumber\\
  &&
     +
     i
     \frac{2}{c}
     T
     \left[ 1 - \sqrt{1-T} e^{+ 2 i \omega \tau } \right]^{-1}
     \int_{-\infty}^{+\infty} \frac{d\omega_{1}}{2\pi}
     \omega_{1}
     \sqrt{\frac{|\omega_{1}|}{|\omega|}}
     \left[ 1 - \sqrt{1-T} e^{+ 2 i \omega_{1} \tau } \right]^{-1}
     e^{ + i (\omega+\omega_{1}) \tau }
     e^{ + i (\omega+\omega_{1}) \frac{\tau_{y}' + \tau_{x}'}{2}}
     \hat{D}_{c}(\omega_{1})
      \nonumber\\
  && \quad\quad
     \times
     \left[
     i \sin\left( (\omega+\omega_{1}) \frac{\tau_{y}' - \tau_{x}'}{2} \right)
     \FrakD_{d}^{\dagger}\hat{Z}_{com}(\omega-\omega_{1})\FrakD_{d}
     -
     \cos\left( (\omega+\omega_{1}) \frac{\tau_{y}' - \tau_{x}'}{2} \right)
     \FrakD_{d}^{\dagger}\hat{Z}_{diff}(\omega-\omega_{1})\FrakD_{d}
     \right]
     \nonumber\\
  &&
     +
     \frac{i}{c}
     \int_{-\infty}^{+\infty} \frac{d\omega_{1}}{2\pi}
     \omega_{1}
     \sqrt{\frac{|\omega_{1}|}{|\omega|}}
     \left[ 1 - \sqrt{1-T} e^{ - 2 i \omega_{1} \tau } \right]
     \left[ 1 - \sqrt{1-T} e^{+ 2 i \omega_{1} \tau } \right]^{-1}
     e^{ + 2 i \omega_{1} \tau }
     e^{ + i (\omega+\omega_{1}) \frac{\tau_{y}' + \tau_{x}'}{2} }
     \hat{D}_{c}(\omega_{1})
      \nonumber\\
  && \quad\quad
     \times
     \left[
     i \sin\left( (\omega+\omega_{1}) \frac{\tau_{y}' - \tau_{x}'}{2} \right)
     \FrakD_{d}^{\dagger}\hat{Z}_{comITM}(\omega - \omega_{1})\FrakD_{d}
     -
     \cos\left( (\omega+\omega_{1}) \frac{\tau_{y}' - \tau_{x}'}{2} \right)
     \FrakD_{d}^{\dagger}\hat{Z}_{diffITM}(\omega - \omega_{1})\FrakD_{d}
     \right]
     \nonumber\\
  &&
     +
     O\left(\left(\hat{X}\right)^{2}, \hat{D}_{v}\hat{X}, \hat{A}\hat{X}\right)
     .
     \label{eq:input-output-relation-general-pert}
\end{eqnarray}
\end{widetext}


The input-output relation
(\ref{eq:input-output-relation-general-pert}) is the most general
input-output relation within our consideration.
The first term in the first line of
Eq.~(\ref{eq:input-output-relation-general-pert}) is the leakage of
the classical carrier field due to the phase offset
$\omega(\tau_{y}'-\tau_{x}')$.
The terms in the second line are the vacuum fluctuations that
correspond to the shot noise in the conventional input-output
relations in
Refs.~\cite{H.J.Kimble-Y.Levin-A.B.Matsko-K.S.Thorne-S.P.Vyatchanin-2001}.
The terms in the third and fourth lines are the response of the
input-test-mass motion, which includes the radiation pressure noise
through the motions of the input test masses
$\FrakD_{d}^{\dagger}\hat{Z}_{comITM}(\Omega)\FrakD_{d}$ and
$\FrakD_{d}^{\dagger}\hat{Z}_{diffITM}(\Omega)\FrakD_{d}$.
The terms in the fifth and sixth lines are the response of the
relative motion of the end test masses and the input test masses
$\FrakD_{d}^{\dagger}\hat{Z}_{com}(\Omega)\FrakD_{d}$ and
$\FrakD_{d}^{\dagger}\hat{Z}_{diff}(\Omega)\FrakD_{d}$.
Finally, the terms in the seventh and eighth lines are responses of
the input-test-mass motion, which includes the radiation pressure
noise through the motion of the input test masses
$\FrakD_{d}^{\dagger}\hat{Z}_{comITM}(\Omega)\FrakD_{d}$ and
$\FrakD_{d}^{\dagger}\hat{Z}_{diffITM}(\Omega)\FrakD_{d}$.
These come from the retarded effects which were explained after
Eq.~(\ref{eq:output-electric-field-junction-hatB-4}).
The last line in Eq.~(\ref{eq:input-output-relation-general-pert})
symbolically represents the terms of the order of $O(\hat{X}_{x}^{2}$,
$\hat{X}_{y}^{2}$, $\hat{X}_{XITM}^{2}$, $\hat{X}_{YITM}^{2}$,
$\hat{D}_{v}\hat{X}_{x}$, $\hat{D}_{v}\hat{X}_{y}$,
$\hat{D}_{v}\hat{X}_{XITM}$, $\hat{D}_{v}\hat{X}_{YITM}$,
$\hat{A}\hat{X}_{x}$, $\hat{A}\hat{X}_{y}$, $\hat{A}\hat{X}_{XITM}$,
$\hat{A}\hat{X}_{YITM})$.
The input-output relation
(\ref{eq:input-output-relation-general-pert}) is one of the main
results of this paper.


To conduct the actual evaluation of the input-output relation
(\ref{eq:input-output-relation-general-pert}), we have to evaluate
$\FrakD_{d}^{\dagger}\hat{Z}_{comITM}(\Omega)\FrakD_{d}$,
$\FrakD_{d}^{\dagger}\hat{Z}_{diffITM}(\Omega)\FrakD_{d}$,
$\FrakD_{d}^{\dagger}\hat{Z}_{com}(\Omega)\FrakD_{d}$ and
$\FrakD_{d}^{\dagger}\hat{Z}_{diff}(\Omega)\FrakD_{d}$ in some way.
In the case of gravitational-wave detectors,
$\FrakD_{d}^{\dagger}\hat{Z}_{comITM}(\Omega)\FrakD_{d}$,
$\FrakD_{d}^{\dagger}\hat{Z}_{diffITM}(\Omega)\FrakD_{d}$,
$\FrakD_{d}^{\dagger}\hat{Z}_{com}(\Omega)\FrakD_{d}$ and
$\FrakD_{d}^{\dagger}\hat{Z}_{diff}(\Omega)\FrakD_{d}$ are evaluated
through the equations of motions for the end test masses as discussed
in the next section.


\subsubsection{Monochromatic coherent amplitude}
\label{sec:Monochromatic_coherent_amplitude}


Before evaluating the mirror displacements, we consider the
coherent state of the incident laser.
In the conventional gravitational-wave detectors, the state of the
optical beam from the light source is in the single-mode coherent
state with the complex amplitude
\begin{eqnarray}
  \label{eq:mohochromatic-alphaomega}
  \alpha(\omega) = 2\pi N \delta(\omega-\omega_{0}).
\end{eqnarray}
Here, $N$ is given by
\begin{eqnarray}
  N
  =
  \sqrt{\frac{I_{0}}{\hbar\omega_{0}}}
  ,
  \label{eq:N-classical-power-relation}
\end{eqnarray}
where $I_{0}$ is the averaged classical power of the laser defined
by Eq.~(\ref{eq:classical-power}) below.
We note that $\alpha(\omega)$ is real.
The corresponding electric field with the amplitude
(\ref{eq:mohochromatic-alphaomega}) of $\alpha(\omega)$ is the
continuous monochromatic carrier field with the frequency
$\omega_{0}$.
The electric field $\hat{E}_{d}$ is given through
Eq.~(\ref{eq:K.Nakamura-M.-K.Fujimoto-2018-18}) as
\begin{eqnarray}
  \label{eq:K.Nakamura-M.-K.Fujimoto-2018-18-d}
  \hat{E}_{d}(t) = \int_{-\infty}^{+\infty} \frac{d\omega}{2\pi}
  \sqrt{\frac{2\pi\hbar|\omega|}{{\cal A}c}} \hat{D}(\omega)
  e^{-i\omega t}
  .
\end{eqnarray}
Then, we have obtain
\begin{eqnarray}
  &&
     \FrakD_{d}^{\dagger}\hat{E}_{d}(t)\FrakD_{d}
     \nonumber\\
  &=&
  \int_{-\infty}^{+\infty} \frac{d\omega}{2\pi}
  \sqrt{\frac{2\pi\hbar|\omega|}{{\cal A}c}} \left(
  \hat{D}_{c}(\omega)
  +
  \hat{D}_{v}(\omega)
  \right)
  e^{-i\omega t}
  .
  \nonumber\\
  \label{eq:K.Nakamura-M.-K.Fujimoto-2018-18-d-replace}
\end{eqnarray}
Here, we note that Eq.~(\ref{eq:hatDc-def}) yields that
\begin{eqnarray}
  \label{eq:hatDc-def-monochromatic}
  \hat{D}_{c}(\omega)
  &:=&
  \alpha(\omega)\Theta(\omega)
  +
  \alpha^{*}(-\omega)\Theta(-\omega)
       \nonumber\\
  &=&
  2 \pi N \left\{
  \delta(\omega-\omega_{0}) \Theta(\omega)
  +
  \delta(\omega+\omega_{0}) \Theta(-\omega)
  \right\}
  .
  \nonumber\\
\end{eqnarray}
The quantum expectation value $I$ of the power of the electric field
$\FrakD_{d}^{\dagger}\hat{E}_{d}(t)\FrakD_{d}$ is given by
\begin{eqnarray}
  I
  =
  N^{2}
  \hbar\omega_{0}
  (1 + \cos(2\omega_{0}t))
  +
  \frac{\hbar}{2}
  \int_{0}^{+\infty} \frac{d\omega}{2\pi}
  \omega
  .
  \label{eq:power-expectation-value}
\end{eqnarray}
The first term in Eq.~(\ref{eq:power-expectation-value}) is the contribution
from the classical carrier field $\alpha(\omega)$, and the second term
comes from the vacuum fluctuations of the electric field operator
$\FrakD_{d}^{\dagger}\hat{E}_{d}(t)\FrakD_{d}$.
The second term diverges as well-known.
We neglect it when we estimate the classical power.
However, if we take into account the cut-off in the frequency range
due to the time bin and maximal observation time as noted in
Sec.~\ref{sec:Electric_field_notation}, the second term is estimated
as a negligible term.
To evaluate the averaged classical power $I_{0}$ by neglecting the
second term and by the time-average of
Eq.~(\ref{eq:power-expectation-value}) as
\begin{eqnarray}
  I_{0}
  &=&
      \lim_{T\rightarrow\infty}
      \frac{1}{T}
      \int_{-T}^{T}dt
      N^{2}
      \hbar\omega_{0}
      (1 + \cos(2\omega_{0}t))
      \nonumber\\
  &=&
      N^{2}
      \hbar\omega_{0}
      .
      \label{eq:classical-power}
\end{eqnarray}
Thus, we denote the factor $N$ in the function $\alpha(\omega)$ is
given by Eq.~(\ref{eq:N-classical-power-relation}).


\subsubsection{Final Input-output relation with mirrors' motion}
\label{sec:Final_Input-output_relation_with_mirrors_motion}


Now, we consider the input-output relation
(\ref{eq:input-output-relation-general-pert}) with the monochromatic
condition (\ref{eq:hatDc-def-monochromatic}).
Furthermore, we consider the sideband picture
$\omega=\omega_{0}\pm\Omega$ of the input-output relation under this
monochromatic condition (\ref{eq:hatDc-def-monochromatic}).
In this sideband picture, the terms include
$\delta(2\omega_{0}\pm\Omega)$,
$\FrakD_{d}^{\dagger}\hat{Z}_{diff}(2\omega_{0}\pm\Omega)\FrakD_{d}$, or
$\FrakD_{d}^{\dagger}\hat{Z}_{com}(2\omega_{0}\pm\Omega)\FrakD_{d}$
appears in the expression of the input-output relation.
Here, we neglect these terms because they behave as
\begin{eqnarray}
  &&
     \int_{-\infty}^{+\infty} \frac{d\Omega}{2\pi}
     f(2\omega_{0}+\Omega) e^{ - i (\Omega t}
     \nonumber\\
  &=&
     \int_{-\infty}^{+\infty} \frac{d\Omega}{2\pi}
     f(2\omega_{0}+\Omega) e^{ - i (2\omega_{0}+\Omega-2\omega_{0}) t}
     \nonumber\\
  &=&
     e^{ + 2 i \omega_{0} t}
     \int_{-\infty}^{+\infty} \frac{d\omega'}{2\pi}
     f(\omega') e^{ - i \omega' t}
     ,
  \\
  &&
     \int_{-\infty}^{+\infty} \frac{d\Omega}{2\pi}
     f(2\omega_{0}-\Omega) e^{ - i (\Omega t}
     \nonumber\\
  &=&
     \int_{-\infty}^{+\infty} \frac{d\Omega}{2\pi}
     f(2\omega_{0}-\Omega) e^{ + i ( 2\omega_{0}-\Omega+2\omega_{0}) t}
     \nonumber\\
  &=&
     e^{ + 2 i \omega_{0} t}
     \int_{-\infty}^{+\infty} \frac{d\omega'}{2\pi}
     f(\omega') e^{ + i \omega' t}
     .
\end{eqnarray}
These show that the terms of
$\FrakD_{d}^{\dagger}\hat{Z}_{diff}(2\omega_{0}\pm\Omega)\FrakD_{d}$ and
$\FrakD_{d}^{\dagger}\hat{Z}_{com}(2\omega_{0}\pm\Omega)\FrakD_{d}$
have the behavior of the rapid oscillation with the frequency
$2\omega_{0}$ in the time domain.
Since we concentrate only on the behavior of the sideband frequency
$\Omega$, these rapid oscillation terms are outside of the
frequency range of interest.
For these reasons, we ignore the terms $\delta(2\omega_{0}\pm\Omega)$,
$\FrakD_{d}^{\dagger}\hat{Z}_{diff}(2\omega_{0}\pm\Omega)\FrakD_{d}$, and
$\FrakD_{d}^{\dagger}\hat{Z}_{com}(2\omega_{0}\pm\Omega)\FrakD_{d}$ as
an approximation~\cite{K.Nakamura-2025-footnote3}.
Applying this approximation, the input-output relation
(\ref{eq:input-output-relation-general-pert}) is given by
\begin{widetext}
\begin{eqnarray}
  &&
     \FrakD_{d}^{\dagger}\hat{B}(\omega_{0}\pm\Omega)\FrakD_{d}
     \nonumber\\
  &=&
      N
      \left[ 1 - \sqrt{1-T} e^{ - 2 i \omega_{0} \tau } \right]
      \left[ 1 - \sqrt{1-T} e^{+ 2 i \omega_{0} \tau } \right]^{-1}
      e^{ + 2 i \omega_{0} \tau }
      e^{ + i \omega_{0} (\tau_{y}' + \tau_{x}' ) }
      i \sin( \omega_{0} (\tau_{y}' - \tau_{x}' ) )
      2 \pi \delta(\Omega)
      \nonumber\\
  &&
      +
      \left[ 1 - \sqrt{1-T} e^{ - 2 i (\omega_{0}\pm\Omega) \tau } \right]
      \left[ 1 - \sqrt{1-T} e^{+ 2 i (\omega_{0}\pm\Omega) \tau } \right]^{-1}
      e^{ + 2 i (\omega_{0}\pm\Omega) \tau }
      e^{ + i (\omega_{0}\pm\Omega) (\tau_{y}' + \tau_{x}' ) }
     \nonumber\\
  && \quad\quad
     \times
      \left[
      i \sin( (\omega_{0}\pm\Omega) (\tau_{y}' - \tau_{x}' ) )
      \hat{D}_{v}(\omega_{0}\pm\Omega)
      +
      \cos( (\omega_{0}\pm\Omega) (\tau_{y}' - \tau_{x}' ) )
      \hat{A}(\omega_{0}\pm\Omega)
      \right]
      \nonumber\\
  &&
     +
     N
     i
     e^{ + 2 i (\omega_{0}\pm\Omega) \tau }
     e^{ + i (\omega_{0}\pm\Omega) ( \tau_{y}' + \tau_{x}' ) }
     \left[ 1 - \sqrt{1-T} e^{ - 2 i (\omega_{0}\pm\Omega) \tau } \right]
     \left[ 1 - \sqrt{1-T} e^{+ 2 i (\omega_{0}\pm\Omega) \tau } \right]^{-1}
     \sqrt{\frac{|\omega_{0}|}{|\omega_{0}\pm\Omega|}}
     \frac{\omega_{0}}{c}
      \nonumber\\
  && \quad\quad
     \times
     \left[
     i \sin( (\omega_{0}\pm\Omega) ( \tau_{y}' - \tau_{x}' ) )
     \FrakD_{d}^{\dagger}\hat{Z}_{comITM}(\pm\Omega)\FrakD_{d}
     -
     \cos( (\omega_{0}\pm\Omega) ( \tau_{y}' - \tau_{x}' ) )
     \FrakD_{d}^{\dagger}\hat{Z}_{diffITM}(\pm\Omega)\FrakD_{d}
     \right]
     \nonumber\\
  &&
     +
     N
     i
     \frac{2}{c}
     T
     \left[ 1 - \sqrt{1-T} e^{+ 2 i (\omega_{0}\pm\Omega) \tau } \right]^{-1}
     \omega_{0}
     \sqrt{\frac{|\omega_{0}|}{|\omega_{0}\pm\Omega|}}
     \left[ 1 - \sqrt{1-T} e^{+ 2 i \omega_{0} \tau } \right]^{-1}
     e^{ + i (2\omega_{0}\pm\Omega) \tau }
     e^{ + i (2\omega_{0}\pm\Omega) \frac{\tau_{y}' + \tau_{x}'}{2}}
      \nonumber\\
  && \quad\quad
     \times
     \left[
     i \sin\left( (2\omega_{0}\pm\Omega) \frac{\tau_{y}' - \tau_{x}'}{2} \right)
     \FrakD_{d}^{\dagger}\hat{Z}_{com}(\pm\Omega)\FrakD_{d}
     -
     \cos\left( (2\omega_{0}\pm\Omega) \frac{\tau_{y}' - \tau_{x}'}{2} \right)
     \FrakD_{d}^{\dagger}\hat{Z}_{diff}(\pm\Omega)\FrakD_{d}
     \right]
     \nonumber\\
  &&
     +
     N
     i
     \frac{\omega_{0}}{c}
     \sqrt{\frac{|\omega_{0}|}{|\omega_{0}\pm\Omega|}}
     \left[ 1 - \sqrt{1-T} e^{ - 2 i \omega_{0} \tau } \right]
     \left[ 1 - \sqrt{1-T} e^{+ 2 i \omega_{0} \tau } \right]^{-1}
     e^{ + 2 i \omega_{0} \tau }
     e^{ + i (2\omega_{0}\pm\Omega) \frac{\tau_{y}' + \tau_{x}'}{2} }
      \nonumber\\
  && \quad\quad
     \times
     \left[
     i \sin\left( (2\omega_{0}\pm\Omega) \frac{\tau_{y}' - \tau_{x}'}{2} \right)
     \FrakD_{d}^{\dagger}\hat{Z}_{comITM}(\pm\Omega)\FrakD_{d}
     -
     \cos\left( (2\omega_{0}\pm\Omega) \frac{\tau_{y}' - \tau_{x}'}{2} \right)
     \FrakD_{d}^{\dagger}\hat{Z}_{diffITM}(\pm\Omega)\FrakD_{d}
     \right]
     \nonumber\\
  &&
     +
     O\left(\left(\hat{X}\right)^{2}, \hat{D}_{v}\hat{X}, \hat{A}\hat{X}\right)
     \nonumber\\
  &&
     +
     \mbox{``rapid oscillation terms with the frequency $2\omega_{0}\pm\omega$"}
     .
     \label{eq:input-output-relation-general-pert-sideband-approx}
\end{eqnarray}
\end{widetext}


To complete the evaluation of this input-output relation
(\ref{eq:input-output-relation-general-pert-sideband-approx}), we have
to specify the Fourier transformations
$\FrakD_{d}^{\dagger}\hat{Z}_{com}(\pm\Omega)\FrakD_{d}$,
$\FrakD_{d}^{\dagger}\hat{Z}_{diff}(\pm\Omega)\FrakD_{d}$,
$\FrakD_{d}^{\dagger}\hat{Z}_{comITM}(\pm\Omega)\FrakD_{d}$, and
$\FrakD_{d}^{\dagger}\hat{Z}_{diffITM}(\pm\Omega)\FrakD_{d}$ of the
mirrors' displacement operators.
These are determined through the Heisenberg equation
of motion in the next section.


\section{Equations for mirrors' motions and their solutions}
\label{sec:Eq_for_mirrors'_motions_and_their_solutions}


The purpose of this section is the evaluation of the Fourier
transformations
$\FrakD_{d}^{\dagger}\hat{Z}_{com}(\pm\Omega)\FrakD_{d}$, $\FrakD_{d}^{\dagger}\hat{Z}_{diff}(\pm\Omega)\FrakD_{d}$,
$\FrakD_{d}^{\dagger}\hat{Z}_{comITM}(\pm\Omega)\FrakD_{d}$, and
$\FrakD_{d}^{\dagger}\hat{Z}_{diffITM}(\pm\Omega)\FrakD_{d}$ of the
mirrors' displacement operators.
To discuss this evaluation within quantum mechanics, we first discuss
a quantum mechanical model of a forced harmonic oscillator in
Sec.~\ref{sec:Quantum_Mechanical_model_for_mirror_motions_and_its_sol}.
In Sec.~\ref{sec:Heisenberg_Equations_for_mirrors'_motions}, we
discuss the equations of mirrors' motion in the proper reference
frame~\cite{C.W.Misner-T.S.Thorne-J.A.Wheeler-1973} whose center is BS
of the interferometer.
In Sec.~\ref{sec:Evaluation_of_Radiation_pressure_forces}, we evaluate
the radiation pressure forces that act on each mirror.
In Sec.~\ref{sec:Sol_to_Heisenberg_Eq_for_mirrors}, we derived the
explicit form of the Fourier transformations
$\FrakD_{d}^{\dagger}\hat{Z}_{com}(\pm\Omega)\FrakD_{d}$,
$\FrakD_{d}^{\dagger}\hat{Z}_{diff}(\pm\Omega)\FrakD_{d}$,
$\FrakD_{d}^{\dagger}\hat{Z}_{comITM}(\pm\Omega)\FrakD_{d}$, and
$\FrakD_{d}^{\dagger}\hat{Z}_{diffITM}(\pm\Omega)\FrakD_{d}$ of the
mirrors' displacement operators based on the arguments in Sec.~\ref{sec:Quantum_Mechanical_model_for_mirror_motions_and_its_sol}.
The results of this section are used to evaluate the input-output
relation of the Fabry-P\'erot interferometer in
Sec.~\ref{sec:Final_input-output-relation_for_Fabry-Perot_GW_Detector}.


\subsection{Quantum Mechanical model for mirror motions and its solutions}
\label{sec:Quantum_Mechanical_model_for_mirror_motions_and_its_sol}


In this subsection, we consider the quantum mechanical Hamiltonian
$\hat{H}$ for a forced harmonic oscillator with the position
operator $\hat{X}(t)$ and the momentum operator $\hat{P}(t)$ with
external quantum force $\hat{F}(t)$:
\begin{eqnarray}
  \label{eq:QuanatumHamiltonian-hatX}
  \hat{H} = \frac{\hat{P}^{2}}{2\mu}
  + \frac{1}{2}\mu \omega_{p}^{2}\hat{X}^{2} - \hat{F}(t)\hat{X}.
\end{eqnarray}
The position operator $\hat{X}$ corresponds to the mirror displacement
operators $\hat{X}_{x}$, $\hat{X}_{y}$, $\hat{X}_{XITM}$, and
$\hat{X}_{YITM}$.
$\mu$ and $\omega_{p}$ correspond to the mass and the
pendulum fundamental frequency of mirrors, respectively.
The operators $\hat{X}(t)$ and $\hat{P}(t)$ satisfy the canonical
commutation relation
\begin{eqnarray}
  \label{eq:canonical-commutation}
  \left[\hat{X}(t),\hat{P}(t)\right] = i\hbar.
\end{eqnarray}
We assume that
\begin{eqnarray}
  \label{eq:hatXhatF-commutation-again}
  \left[\hat{X},\hat{F}(t)\right] = \left[\hat{P},\hat{F}(t)\right] = 0.
\end{eqnarray}
This assumption (\ref{eq:hatXhatF-commutation-again}) is valid for the
mirrors' motion if we ignore the optical spring effects in
Appendix~\ref{sec:Explicit_evaluation_of_the_radiation_pressure_forces}
due to the appropriate tuning conditions as discussed
in Sec.~\ref{sec:Changing_Tuning-Point}.
From the Hamiltonian (\ref{eq:QuanatumHamiltonian-hatX}), Heisenberg's
equations of motion are given by
\begin{eqnarray}
  \label{eq:Heisenberg-eqs-mirrors-1}
  \frac{d}{dt}\hat{X} &=& \frac{1}{i\hbar} \left[\hat{X},\hat{H}\right]
  = \frac{1}{\mu}\hat{P}, \\
  \label{eq:Heisenberg-eqs-mirrors-2}
  \frac{d}{dt}\hat{P} &=& \frac{1}{i\hbar} \left[\hat{P},\hat{H}\right]
  = - \mu \omega_{p}^{2} \hat{X} + \hat{F}(t).
\end{eqnarray}


When the external force $\hat{F}(t)=0$,
Eqs.~(\ref{eq:Heisenberg-eqs-mirrors-1}) and
(\ref{eq:Heisenberg-eqs-mirrors-2}) are the Heisenberg equations
of motion for the simple harmonic oscillator.
Then the solution to this equation is given by~\cite{J.J.Sakurai-1985}
\begin{eqnarray}
  \label{eq:solutions-harmonic-oscillator-Heisenberg-eq-hatX-again}
  \hat{X}_{\hat{F}=0}(t)
  &=&
      \hat{X}(0) \cos\omega_{p}t
      +
      \left[\frac{\hat{P}(0)}{\mu\omega_{p}}\right]\sin\omega_{p} t
      ,
  \\
  \label{eq:solutions-harmonic-oscillator-Heisenberg-eq-hatP-again}
  \hat{P}_{\hat{F}=0}(t)
  &=&
      - \mu \omega_{p} \hat{X}(0) \sin\omega_{p}t
      +
      \hat{P}(0) \cos\omega_{p}t
      .
\end{eqnarray}


Substituting Eq.~(\ref{eq:Heisenberg-eqs-mirrors-1}) into
Eq.~(\ref{eq:Heisenberg-eqs-mirrors-2}), the Heisenberg equation is
given by
\begin{eqnarray}
  \label{eq:Heisenberg-eqs-mirrors-3}
  \mu \frac{d^{2}}{dt^{2}}\hat{X} + \mu \omega_{p}^{2} \hat{X}
  - \hat{F}(t) = 0.
\end{eqnarray}
To solve this equation, we use the usual method to solve the
second-order differential equation, i.e., we consider the solution
\begin{eqnarray}
  \label{eq:Heisenberg-eqs-mirrors-undetermined-coef}
  \hat{X}(t) = \hat{\calA}(t) \cos\omega_{p}t + \hat{\calB}(t) \sin\omega_{p}t
\end{eqnarray}
with the condition
\begin{eqnarray}
  \left(\frac{d}{dt}\hat{\calA}(t)\right) \cos\omega_{p}t
  +
  \left(\frac{d}{dt}\hat{\calB}(t)\right) \sin\omega_{p}t = 0.
  \label{eq:Heisenberg-eqs-mirrors-additional-condidtion}
\end{eqnarray}
From Eq.~(\ref{eq:Heisenberg-eqs-mirrors-3}) with
Eqs.~(\ref{eq:Heisenberg-eqs-mirrors-undetermined-coef}) and
(\ref{eq:Heisenberg-eqs-mirrors-additional-condidtion}), we obtain
\begin{eqnarray}
  - \left(\frac{d}{dt}\hat{\calA}(t)\right) \sin\omega_{p}t
  + \left(\frac{d}{dt}\hat{\calB}(t)\right) \cos\omega_{p}t
  =
  \frac{1}{\mu\omega_{p}} \hat{F}(t).
  \nonumber\\
  \label{eq:Heisenberg-eqs-mirrors-4}
\end{eqnarray}
From Eqs.~(\ref{eq:Heisenberg-eqs-mirrors-additional-condidtion}) and
(\ref{eq:Heisenberg-eqs-mirrors-4}), we obtain the solution operators
$\hat{\calA}(t)$ and $\hat{\calB}(t)$.
Then, we reach the solution $\hat{X}(t)$ to Eq.~(\ref{eq:Heisenberg-eqs-mirrors-3})
\begin{eqnarray}
  \hat{X}(t)
  &=&
      \hat{X}(-\infty) \cos\omega_{p}t
      + \frac{1}{\mu\omega_{p}} \hat{P}(-\infty) \sin\omega_{p}t
      \nonumber\\
  &&
     + \frac{1}{\mu\omega_{p}} \int_{-\infty}^{+\infty} dt'
     \Theta(t-t') \hat{F}(t') \sin(\omega_{p}(t-t'))
     .
     \nonumber\\
  \label{eq:Heisenberg-eqs-mirrors-3-sol}
\end{eqnarray}
Here, we introduced the Heaviside function $\Theta(t-t')$ to change the
range of the time $t'$ integration from $[-\infty,t]$ to
$[-\infty,+\infty]$.
Here, $\hat{X}(-\infty)$ and $\hat{P}(-\infty)$ are quantum operators
which correspond to the initial condition of the position
$\hat{X}(t)$ and the momentum $\hat{P}(t)$.
These operators satisfy the usual commutation relation
\begin{eqnarray}
  \label{eq:canonical-commutation-initial}
  \left[\hat{X}(-\infty),\hat{P}(-\infty)\right] = i\hbar.
\end{eqnarray}


Here, we consider the Fourier transformation of the solution
(\ref{eq:Heisenberg-eqs-mirrors-3-sol}).
The Heaviside step function $\Theta(t-t')$ is a distribution.
Therefore, we have to be careful about the order of the limit when we
consider the Fourier transformation of the Heaviside step function,
which is given by
\begin{eqnarray}
  &&
     \int_{-\infty}^{+\infty} dx \Theta(x) \phi(x)
     \nonumber\\
  &=&
      \lim_{\epsilon\rightarrow 0}
      \int_{-\infty}^{+\infty} dx
      \int_{-\infty}^{+\infty} \frac{dk}{2\pi i}
      \frac{e^{+i(k-i\epsilon)x}}{k-i\epsilon} \phi(x)
      \label{eq:Fourier-Heaviside-step-function}
\end{eqnarray}
for any smooth function $\phi(x)$.
Taking care of the order of the limit of the step function, through
the Fourier transformations
\begin{eqnarray}
  \label{eq:calF-def}
  \hat{F}(t)
  &=:&
      \int_{-\infty}^{+\infty} \frac{d\omega_{F}}{2\pi}
      \calF(\omega_{F}) e^{-i\omega_{F}t}
      ,
  \\
  \label{eq:Zomega-def}
  \hat{X}(t)
  &=:&
      \int_{-\infty}^{+\infty} \frac{d\omega}{2\pi}
      \hat{Z}(\omega) e^{-i\omega t}
      ,
\end{eqnarray}
we obtain
\begin{eqnarray}
  \hat{Z}(\omega)
  &=&
      \frac{1}{2}
      \left(
      \hat{X}(-\infty)
      -
      \frac{i\hat{P}(-\infty)}{\mu\omega_{p}}
      \right)
      2 \pi \delta( \omega_{p} + \omega )
      \nonumber\\
  &&
      +
      \frac{1}{2}
      \left(
      \hat{X}(-\infty)
      +
      \frac{i\hat{P}(-\infty)}{\mu\omega_{p}}
      \right)
      2 \pi \delta( \omega_{p} - \omega )
      \nonumber\\
  &&
     -
     \frac{1}{\mu(\omega^{2} - \omega_{p}^{2})}
     \hat{\calF}(\omega)
     .
     \label{eq:J.J.Sakurai-2.3.46-1-again-assumption-sol-t=-infty-final}
\end{eqnarray}
Equation (\ref{eq:J.J.Sakurai-2.3.46-1-again-assumption-sol-t=-infty-final})
indicates that the quantum fluctuations of the mirror motions' initial
conditions concentrate to the frequency $\omega=\omega_{p}$.
Therefore, if $\omega=\omega_{p}$ is outside of the frequency range of
our interest, the initial quantum fluctuations from $\hat{X}(-\infty)$
and $\hat{P}(-\infty)$ does not contribute to the quantum noise of the
operator $\hat{Z}(\omega)$ in the frequency range of our interest.


Through the canonical commutation relations
(\ref{eq:canonical-commutation}) and
(\ref{eq:canonical-commutation-initial}), we can derive the
consistency relation of the solution
(\ref{eq:Heisenberg-eqs-mirrors-3-sol}) as the commutation relation of
$\hat{\calF}(\omega)/\mu$ as follows:
\begin{eqnarray}
  \label{eq:calF-different-omega-commutation-relation-Fourier}
  &&
     \int_{-\infty}^{+\infty} \frac{d\omega_{1}}{2\pi}
     \frac{
     \omega-\omega_{1}
     }{
     \left(\omega_{1}^{2} - \omega_{p}^{2}\right)
     \left((\omega-\omega_{1})^{2} - \omega_{p}^{2}\right)
     }
     \nonumber\\
  && \quad\quad
     \times
     \left[
     \frac{1}{\mu}
     \hat{\calF}(\omega_{1})
     ,
     \frac{1}{\mu}
     \hat{\calF}(\omega-\omega_{1})
     \right]
     =
     0
     .
\end{eqnarray}
This consistency relation is checked in
Appendix~\ref{sec:Consistency_relation_of_calF} through the explicit
form of $\hat{\calF}(\omega)/\mu$ obtained in
Sec.~\ref{sec:Evaluation_of_Radiation_pressure_forces}.


\subsection{Equations of motion for each mirror}
\label{sec:Heisenberg_Equations_for_mirrors'_motions}


Now, we consider the equation of motion for EMs and ITMs to evaluate
the Fourier transformations
$\FrakD_{d}^{\dagger}\hat{Z}_{comITM}(\pm\Omega)\FrakD_{d}$,
$\FrakD_{d}^{\dagger}\hat{Z}_{diffITM}(\pm\Omega)\FrakD_{d}$,
$\FrakD_{d}\hat{Z}_{com}(\Omega)\FrakD_{d}$, and
$\FrakD_{d}^{\dagger}\hat{Z}_{diff}(\Omega)\FrakD_{d}$ in the
input-output relation
(\ref{eq:input-output-relation-general-pert-sideband-approx}).
We assume the masses of ITMs are equal to $m_{ITM}$, and those of EMs
are equal to $m_{EM}$.
Since we apply the proper reference frame whose center is BS, the
displacement of BS vanishes, which means that we have to control the
external forces to BS vanish.
Furthermore, we only consider the motions of mirrors are restricted so
that XITM and XEM move only along the $x$-direction and YITM and YEM
move only along the $y$-direction and $x$-direction and $y$-direction
is orthogonal.
In the proper reference frame, the effects of gravitational waves are
represented by the components $R_{txtx}$, $R_{tyty}$ of the Riemann
curvature tensor~\cite{C.W.Misner-T.S.Thorne-J.A.Wheeler-1973}.
Moreover, we denote the radiation pressure force to mirrors, XITM,
YITM, XEM, and YEM from the laser in the interferometer by
$\hat{F}_{rpXITM}$, $\hat{F}_{rpYITM}$, $\hat{F}_{rpXEM}$,
$\hat{F}_{rpYEM}$, respectively.
Then, the equations of motion for $\hat{X}_{XITM}$, $\hat{X}_{YITM}$,
$\hat{X}_{XEM}$, and $\hat{X}_{YEM}$, we obtain
\begin{eqnarray}
  \label{eq:Eq-of-motion-XITM-3}
     \frac{d^{2}}{dt^{2}}\hat{X}_{XITM}
  &=&
      -
      \omega_{p}^{2} \hat{X}_{XITM}
      +
      \frac{1}{m_{ITM}}
      \hat{F}_{rpXITM}
      \nonumber\\
  &&
      - R_{txtx} l_{x}
      ,
  \\
  \label{eq:Eq-of-motion-YITM-3}
     \frac{d^{2}}{dt^{2}}\hat{X}_{YITM}
  &=&
      -
      \omega_{p}^{2} \hat{X}_{YITM}
      +
      \frac{1}{m_{ITM}}
      \hat{F}_{rpYITM}
      \nonumber\\
  &&
      - R_{tyty} l_{y}
      ,
  \\
  \label{eq:Eq-of-motion-XEM-3}
     \frac{d^{2}}{dt^{2}}\hat{X}_{XEM}
  &=&
      -
      \omega_{p}^{2} \hat{X}_{XEM}
      +
      \frac{1}{m_{EM}}
      \hat{F}_{rpXEM}
     \nonumber\\
  &&
      - R_{txtx} (L+l_{x})
      ,
  \\
  \label{eq:Eq-of-motion-YEM-3}
     \frac{d^{2}}{dt^{2}}\hat{X}_{YEM}
  &=&
      -
      \omega_{p}^{2} \hat{X}_{YEM}
      +
      \frac{1}{m_{EM}}
      \hat{F}_{rpYEM}
     \nonumber\\
  &&
      - R_{tyty} (L+l_{y}).
\end{eqnarray}
Here, we assumed that $\omega_{p}$ is the fundamental frequency of the
mirror pendulums of the vibration isolation system of XITM, YITM, XEM,
and YEM.
We also applied the approximation $L,l_{x},l_{y}$ $\gg$
$\hat{X}_{XITM}$, $\hat{X}_{YITM}$, $\hat{X}_{XEM}$, $\hat{X}_{YEM}$.


As the components of the Riemann curvature tensor, we only consider
the plus modes of gravitational waves associated with the $x$- and
$y$-directions of the interferometer.
In this case, the components of $R_{txtx}$ and $R_{tyty}$ the Riemann
curvature tensor are given by
\begin{eqnarray}
  \label{eq:Riemann-compnennts}
  R_{txtx} = - \frac{1}{2} \frac{\partial^{2}}{\partial t^{2}}h_{+} =
  - R_{tyty}, \quad h := h_{+}.
\end{eqnarray}
Furthermore, we ignore the gravitational-wave term in
Eqs.~(\ref{eq:Eq-of-motion-XITM-3}) and (\ref{eq:Eq-of-motion-YITM-3})
due to the approximation $L\gg l_{x},l_{y}$.
Then, the equations
(\ref{eq:Eq-of-motion-XITM-3})--(\ref{eq:Eq-of-motion-YEM-3}) of
motion are given by
\begin{eqnarray}
  \label{eq:Eq-of-motion-XITM-4}
  \!\!\!\!
  \frac{d^{2}}{dt^{2}}\hat{X}_{XITM}
  &=&
      -
      \omega_{p}^{2} \hat{X}_{XITM}
      +
      \frac{1}{m_{ITM}}
      \hat{F}_{rpXITM}
      ,
\end{eqnarray}
\begin{eqnarray}
  \label{eq:Eq-of-motion-YITM-4}
  \!\!\!\!
  \frac{d^{2}}{dt^{2}}\hat{X}_{YITM}
  &=&
      -
      \omega_{p}^{2} \hat{X}_{YITM}
      +
      \frac{1}{m_{ITM}}
      \hat{F}_{rpYITM}
      ,
\end{eqnarray}
\begin{eqnarray}
  \label{eq:Eq-of-motion-XEM-4}
  \!\!\!\!
  \frac{d^{2}}{dt^{2}}\hat{X}_{XEM}
  &=&
      -
      \omega_{p}^{2} \hat{X}_{XEM}
      +
      \frac{1}{m_{EM}}
      \hat{F}_{rpXEM}
      \nonumber\\
  &&
      +
      \frac{1}{2} L \frac{d^{2}}{dt^{2}}h(t,L)
      ,
\end{eqnarray}
\begin{eqnarray}
  \label{eq:Eq-of-motion-YEM-4}
  \!\!\!\!
     \frac{d^{2}}{dt^{2}}\hat{X}_{YEM}
     &=&
     -
     \omega_{p}^{2} \hat{X}_{YEM}
     +
     \frac{1}{m_{EM}}
     \hat{F}_{rpYEM}
      \nonumber\\
  &&
     -
     \frac{1}{2} L \frac{d^{2}}{dt^{2}}h(t,L)
     .
\end{eqnarray}


From
Eqs.~(\ref{eq:Eq-of-motion-XITM-4})--(\ref{eq:Eq-of-motion-YEM-4}),
the equations for the relative displacements
$\hat{X}_{x}:=\hat{X}_{XEM}-\hat{X}_{XITM}$ and
$\hat{X}_{y}:=\hat{X}_{YEM}-\hat{X}_{YITM}$ are given by
\begin{eqnarray}
     \frac{d^{2}}{dt^{2}}\hat{X}_{x}
     + \omega_{p}^{2} \hat{X}_{x}
  &=&
      \frac{1}{m_{EM}} \hat{F}_{rpXEM}
      -  \frac{1}{m_{ITM}}\hat{F}_{rpXITM}
      \nonumber\\
  &&
     +  \frac{1}{2} L \frac{d^{2}}{dt^{2}}h(t,L)
     ,
     \label{eq:Eq-of-motion-X-mirrors-differential-original}
\end{eqnarray}
\begin{eqnarray}
     \frac{d^{2}}{dt^{2}}\hat{X}_{y}
     + \omega_{p}^{2} \hat{X}_{y}
  &=&
      \frac{1}{m_{EM}} \hat{F}_{rpYEM}
      -  \frac{1}{m_{ITM}} \hat{F}_{rpYITM}
      \nonumber\\
  &&
     -  \frac{1}{2} L \frac{d^{2}}{dt^{2}}h(t,L)
     .
     \label{eq:Eq-of-motion-Y-mirrors-differential-original}
\end{eqnarray}


The equations (\ref{eq:Eq-of-motion-XITM-4}),
(\ref{eq:Eq-of-motion-YITM-4}),
(\ref{eq:Eq-of-motion-X-mirrors-differential-original}), and
(\ref{eq:Eq-of-motion-Y-mirrors-differential-original}) of motion for
$\hat{X}_{XITM}$, $\hat{X}_{YITM}$, and the relative motion
$\hat{X}_{x}$ and $\hat{X}_{y}$ have the same form as the Heisenberg
equation~(\ref{eq:Heisenberg-eqs-mirrors-3}).
Then, the solution to Eqs.~(\ref{eq:Eq-of-motion-XITM-4}),
(\ref{eq:Eq-of-motion-YITM-4}),
(\ref{eq:Eq-of-motion-X-mirrors-differential-original}), and
(\ref{eq:Eq-of-motion-Y-mirrors-differential-original}) should have
the form of Eq.~(\ref{eq:Heisenberg-eqs-mirrors-3-sol}).
Furthermore, their Fourier transformations are given in the form (\ref{eq:J.J.Sakurai-2.3.46-1-again-assumption-sol-t=-infty-final}).
The Fourier transformations of the operators $\hat{X}_{XITM}$,
$\hat{X}_{YITM}$, $\hat{X}_{x}$, and $\hat{X}_{y}$ are expressed by
Eqs.~(\ref{eq:XITM-Fourier-def}), (\ref{eq:YITM-Fourier-def}), and
(\ref{eq:hatXx-Fouriler-def}).
Moreover, we introduce the Fourier transformations of the radiation
pressure forces by the Fourier transformations
$\scrF_{rpXITM}(\omega)$, $\scrF_{rpYITM}(\omega)$,
$\scrF_{rpx}(\omega)$, and $\scrF_{rpy}(\omega)$ of the radiation
pressure forces, which are defined by
\begin{eqnarray}
  \scrF_{rpXITM}(\omega)
  :=
  \int_{-\infty}^{+\infty} dt e^{+i\omega t} \frac{1}{m_{ITM}}
  \hat{F}_{rpXITM}(t)
  ,
  \nonumber\\
  \label{eq:XITM-radi-press-Fourier}
\end{eqnarray}
\begin{eqnarray}
  \scrF_{rpYITM}(\omega)
  :=
  \int_{-\infty}^{+\infty} dt e^{+i\omega t} \frac{1}{m_{ITM}}
  \hat{F}_{rpYITM}(t)
  ,
  \nonumber\\
  \label{eq:YITM-radi-press-Fourier}
\end{eqnarray}
\begin{eqnarray}
  \scrF_{rpx}(\omega)
  &:=&
       \int_{-\infty}^{+\infty} dt e^{+i\omega t}
       \left[
       \frac{1}{m_{EM}} \hat{F}_{rpXEM}(t)
       \right.
       \nonumber\\
  && \quad\quad\quad\quad\quad\quad\quad
     \left.
     -
     \frac{1}{m_{ITM}}\hat{F}_{rpXITM}(t)
     \right]
     ,
     \nonumber\\
  \label{eq:radiation-pressure-Fourier-X-def}
\end{eqnarray}
\begin{eqnarray}
  \scrF_{rpy}(\omega)
  &:=&
  \int_{-\infty}^{+\infty} dt e^{+i\omega t}
  \left[
  \frac{1}{m_{EM}} \hat{F}_{rpYEM}(t)
       \right.
       \nonumber\\
  && \quad\quad\quad\quad\quad\quad\quad
     \left.
  -
  \frac{1}{m_{ITM}}\hat{F}_{rpYITM}(t)
  \right]
  .
  \nonumber\\
  \label{eq:radiation-pressure-Fourier-Y-def}
\end{eqnarray}


Since we express the mirrors' displacement in the input-output
relation
(\ref{eq:input-output-relation-general-pert-sideband-approx}) by their
common modes and differential modes as
$\hat{Z}_{com}(\pm\Omega)$, $\hat{Z}_{diff}(\pm\Omega)$,
$\hat{Z}_{comITM}(\pm\Omega)$, and $\hat{Z}_{diffITM}(\pm\Omega)$,
we use the definitions
(\ref{eq:x-com+diff-motion-Fourier})--(\ref{eq:YITM-com-diffITM-motion-Fourier})
for these variables.


To describe the Fourier transformations of
Eqs.~(\ref{eq:Eq-of-motion-XITM-4}), (\ref{eq:Eq-of-motion-YITM-4}),
(\ref{eq:Eq-of-motion-X-mirrors-differential-original}), and
(\ref{eq:Eq-of-motion-Y-mirrors-differential-original}), we have to
introduce the Fourier transformation $H(\omega)$ of the gravitational
wave signal $h(t)$ as
\begin{eqnarray}
  \label{eq:GW-signal-Fourier}
  h(t,z) =: \int_{-\infty}^{+\infty} \frac{d\omega}{2\pi}
  e^{-i\omega_{2}t} H(\omega,z),
\end{eqnarray}
where $z$ is proper distance to the propagation direction of the laser
from BS in the interferometer.


Through the above preparations, the solutions to
Eqs.~(\ref{eq:Eq-of-motion-XITM-4}), (\ref{eq:Eq-of-motion-YITM-4}),
(\ref{eq:Eq-of-motion-X-mirrors-differential-original}), and
(\ref{eq:Eq-of-motion-Y-mirrors-differential-original}) are given in
the form of
Eq.~(\ref{eq:J.J.Sakurai-2.3.46-1-again-assumption-sol-t=-infty-final}).
Here, we note that in the gravitational-wave detectors, the pendulum
frequency $\omega_{p}$ is usually set so that it is outside of the
frequency range of interest.
Therefore, we ignore the first- and the second-term in
Eq.~(\ref{eq:J.J.Sakurai-2.3.46-1-again-assumption-sol-t=-infty-final}),
which depends on the quantum initial condition $\hat{X}(-\infty)$ and
$\hat{P}(-\infty)$ of the mirrors.
Then, the Fourier transformations of the solutions to
Eqs.~(\ref{eq:Eq-of-motion-XITM-4}), (\ref{eq:Eq-of-motion-YITM-4}),
(\ref{eq:Eq-of-motion-X-mirrors-differential-original}), and
(\ref{eq:Eq-of-motion-Y-mirrors-differential-original}) are given by
\begin{eqnarray}
  \hat{Z}_{comITM}(\omega)
  &=&
  -
  \frac{
  \scrF_{rpYITM}(\omega)
  +
  \scrF_{rpXITM}(\omega)
  }{2(\omega^{2}-\omega_{p}^{2})}
  ,
  \nonumber\\
  \label{eq:Eq-of-motion-XYITM-com-Fourier-approx}
  \\
  \hat{Z}_{diffITM}(\omega)
  &=&
  -
  \frac{
  \scrF_{rpXITM}(\omega)
  -
  \scrF_{rpYITM}(\omega)
  }{2(\omega^{2}-\omega_{p}^{2})}
  ,
  \nonumber\\
  \label{eq:Eq-of-motion-XYITM-dif-Fourier-approx}
  \\
  \hat{Z}_{com}(\omega)
  &=&
  -
  \frac{
  \scrF_{rpy}(\omega)
  +
  \scrF_{rpx}(\omega)
  }{2(\omega^{2}-\omega_{p}^{2})}
  ,
  \label{eq:Eq-of-motion-XY-com-Fourier-approx}
  \\
  \hat{Z}_{diff}(\omega)
  &=&
      -
      \frac{
      \scrF_{rpx}(\omega)
      -
      \scrF_{rpy}(\omega)
      }{2(\omega^{2}-\omega_{p}^{2})}
      \nonumber\\
  &&
     +
     \frac{\omega^{2}}{2(\omega^{2}-\omega_{p}^{2})}
     H(\omega,L) L
     .
     \label{eq:Eq-of-motion-XY-dif-Fourier-approx}
\end{eqnarray}


In the input-output relation
(\ref{eq:input-output-relation-general-pert-sideband-approx}),
$\hat{Z}_{diff}(\omega)$, $\hat{Z}_{com}(\omega)$,
$\hat{Z}_{comITM}(\omega)$, and $\hat{Z}_{diffITM}(\omega)$ do not
appear in this expression, but
$\FrakD_{d}^{\dagger}\hat{Z}_{diff}(\omega)\FrakD_{d}$,
$\FrakD_{d}^{\dagger}\hat{Z}_{com}(\omega)\FrakD_{d}$,
$\FrakD_{d}^{\dagger}\hat{Z}_{comITM}(\omega)\FrakD_{d}$, and
$\FrakD_{d}^{\dagger}\hat{Z}_{diffITM}(\omega)\FrakD_{d}$ appear in Eq.~(\ref{eq:input-output-relation-general-pert-sideband-approx}).
Through Eqs.~(\ref{eq:Eq-of-motion-XY-com-Fourier-approx}) and
(\ref{eq:Eq-of-motion-XY-dif-Fourier-approx}),
$\hat{Z}_{diff}(\omega)$ and $\hat{Z}_{com}(\omega)$ are
linearly related to $\scrF_{rpx}(\omega)$ and $\scrF_{rpy}(\omega)$.
Furthermore, Eqs.~(\ref{eq:Eq-of-motion-XYITM-com-Fourier-approx}) and
(\ref{eq:Eq-of-motion-XYITM-dif-Fourier-approx}) yields that
$\hat{Z}_{comITM}(\omega)$ and $\hat{Z}_{diffITM}(\omega)$ linearly
depends on $\scrF_{rpXITM}(\omega)$ and $\scrF_{rpYITM}(\omega)$.
Therefore, we have to evaluate
$\FrakD_{d}^{\dagger}\scrF_{rpx}(\omega)\FrakD_{d}$,
$\FrakD_{d}^{\dagger}\scrF_{rpy}(\omega)\FrakD_{d}$,
$\FrakD_{d}^{\dagger}\scrF_{rpXITM}(\omega)\FrakD_{d}$, and
$\FrakD_{d}^{\dagger}\scrF_{rpYITM}(\omega)\FrakD_{d}$ to the direct
evaluation of the input-output relation
(\ref{eq:input-output-relation-general-pert-sideband-approx}).
These evaluations are discussed in the next subsection.


\subsection{Evaluation of Radiation pressure forces}
\label{sec:Evaluation_of_Radiation_pressure_forces}


Now, we evaluate the radiation pressure forces
$\FrakD_{d}^{\dagger}\scrF_{rpx}(\omega)\FrakD_{d}$,
$\FrakD_{d}^{\dagger}\scrF_{rpy}(\omega)\FrakD_{d}$,
$\FrakD_{d}^{\dagger}\scrF_{rpXITM}(\omega)\FrakD_{d}$, and
$\FrakD_{d}^{\dagger}\scrF_{rpYITM}(\omega)\FrakD_{d}$.
The details of the evaluation of these radiation pressure forces are
given in
Appendix~\ref{sec:Explicit_evaluation_of_the_radiation_pressure_forces}.
In this section, we only explain the outline of
Appendix~\ref{sec:Explicit_evaluation_of_the_radiation_pressure_forces}.


As described in
Ref.~\cite{H.J.Kimble-Y.Levin-A.B.Matsko-K.S.Thorne-S.P.Vyatchanin-2001},
the radiation pressure forces on the mirrors are determined by the
power operator ${\cal A}\hat{E}^{2}(t)/4\pi$, where $\hat{E}(t)$ is
the electric field operator that touches the mirror.
Since we assume the perfect reflection at EMs, the radiation pressure
force on EMs are given by
\begin{eqnarray}
  \label{eq:radiation-pressure-EM-is-j}
  2 \times \frac{{\cal A}}{4\pi} \hat{E}_{j_{x,y}}^{2}(t)
\end{eqnarray}
as depicted in
Fig.~\ref{fig:arm-propagation-Fabry-Perot-setup-notation}.
From the propagation effect from the ITMs to EMs, the radiation
pressure forces to XEM and YEM are given by
\begin{eqnarray}
  \hat{F}_{rpXEM}(t)
  &=&
  \frac{2{\cal A}}{4\pi}
  \hat{E}_{g_{x}}^{2}\left[t - \tau-\frac{1}{c}\hat{X}_{x}(t-\tau)\right]
  ,
  \label{eq:radiation-pressure-force-XEM}
  \\
  \hat{F}_{rpYEM}(t)
  &=&
  \frac{2{\cal A}}{4\pi}
  \hat{E}_{g_{y}}^{2}\left[t - \tau-\frac{1}{c}\hat{X}_{y}(t-\tau)\right]
  ,
  \label{eq:radiation-pressure-force-YEM}
\end{eqnarray}
respectively.
We evaluate the radiation pressure forces
(\ref{eq:radiation-pressure-force-XEM}) and
(\ref{eq:radiation-pressure-force-YEM}) through the propagation
effects (\ref{eq:xarm-retarded-effect-c-to-f-mod}) and
(\ref{eq:yarm-retarded-effect-c-to-f-mod}) from BS to ITMs for the
electric field operator, the junction condition
(\ref{eq:Kimble-B8-correspoind-1}) for the electric field operator at
the ITMs, the propagation effects
(\ref{eq:xarm-retarded-effect-g-to-gprime}) and
(\ref{eq:yarm-retarded-effect-g-to-gprime}) between ITMs and EMs, and
the junction conditions (\ref{eq:xarm-input-field-junction}) and
(\ref{eq:yarm-input-field-junction}) for the electric field operator
at BS, and their Fourier transformations.
Then, we obtain the expression of the Fourier transformations of the
radiation pressure forces $\hat{F}_{rpXEM}(t)$ and
$\hat{F}_{rpYEM}(t)$ as
Eqs.~(\ref{eq:radiation-pressure-Fourier-X-first-term-in-DAZx-sum})
and (\ref{eq:radiation-pressure-Fourier-Y-first-term-in-DAZy-sum}),
respectively.


On the other hand, the radiation pressure forces on ITMs are more
complicated than those on EMs.
As depicted in
Fig.~\ref{fig:arm-propagation-Fabry-Perot-setup-notation}, four
electric fields contribute to the radiation pressure forces to ITMs,
which are $\hat{E}_{f_{x}}(t)$, $\hat{E}_{f_{x}'}(t)$,
$\hat{E}_{g_{x}}(t)$, and $\hat{E}_{f_{x}'}(t)$.
Taking into account the directions of the forces from these electric
fields to ITMs, the radiation pressure force to XITM is given by
\begin{eqnarray}
     \hat{F}_{rpXITM}(t)
  &=&
      \frac{{\cal A}}{4\pi}\left(\hat{E}_{f_{x}}(t)\right)^{2}
      +
      \frac{{\cal A}}{4\pi}\left(\hat{E}_{f_{x}'}(t)\right)^{2}
      \nonumber\\
  &&
      -
      \frac{{\cal A}}{4\pi}\left(\hat{E}_{g_{x}}(t)\right)^{2}
      -
      \frac{{\cal A}}{4\pi}\left(\hat{E}_{g_{x}'}(t)\right)^{2}
      .
     \nonumber\\
  \label{eq:total-radiation-pressure-force-XITM}
\end{eqnarray}
Similarly, the radiation pressure force on the YITM is given by
\begin{eqnarray}
     \hat{F}_{rpYITM}(t)
  &=&
      \frac{{\cal A}}{4\pi}\left(\hat{E}_{f_{y}}(t)\right)^{2}
      +
      \frac{{\cal A}}{4\pi}\left(\hat{E}_{f_{y}'}(t)\right)^{2}
      \nonumber\\
  &&
      -
      \frac{{\cal A}}{4\pi}\left(\hat{E}_{g_{y}}(t)\right)^{2}
      -
      \frac{{\cal A}}{4\pi}\left(\hat{E}_{g_{y}'}(t)\right)^{2}
      .
     \nonumber\\
  \label{eq:total-radiation-pressure-force-YITM}
\end{eqnarray}
In
Appendix~\ref{sec:Explicit_evaluation_of_the_radiation_pressure_forces},
we evaluate these radiation pressure forces
(\ref{eq:radiation-pressure-force-XEM})--(\ref{eq:total-radiation-pressure-force-YITM})
within the first-order of the displacements $\hat{X}_{x}$,
$\hat{X}_{y}$, $\hat{X}_{XITM}$, and $\hat{X}_{YITM}$ through the
Fourier transformations (\ref{eq:XITM-Fourier-def}),
(\ref{eq:YITM-Fourier-def}), and (\ref{eq:hatXx-Fouriler-def}).
In this evaluation, we use the junction conditions
(\ref{eq:xarm-input-field-junction}) and
(\ref{eq:yarm-input-field-junction}) for the electric field operators
at BS, the propagation effects
(\ref{eq:xarm-retarded-effect-c-to-f-mod}) and
(\ref{eq:yarm-retarded-effect-c-to-f-mod}) between BS and ITMs, the
junction conditions (\ref{eq:Kimble-B8-correspoind-1}) at ITMs, the
propagation effects (\ref{eq:xarm-retarded-effect-g-to-gprime}) and
(\ref{eq:yarm-retarded-effect-g-to-gprime}), and their Fourier
transformations.
Then, we reach the expression of the Fourier transformations of the
radiation pressure forces $\hat{F}_{rpXITM}(t)$ and
$\hat{F}_{rpYITM}(t)$ as
Eqs.~(\ref{eq:total-radiation-pressure-force-to-XITM-DAZx-sum}) and (\ref{eq:total-radiation-pressure-force-to-YITM-DA-sum}),
respectively.


To consider the situation where the state for the incident electric
field from the light source in the coherent state with the complex
amplitude $\alpha(\omega)$, we apply the
displacement operator $\FrakD_{d}^{\dagger}$ from the left and
$\FrakD_{d}$ from the right, which are defined by
Eq.~(\ref{eq:displacement-operator-for-hatd}), to the obtained
radiation pressure forces
(\ref{eq:radiation-pressure-Fourier-X-first-term-in-DAZx-sum}), (\ref{eq:radiation-pressure-Fourier-Y-first-term-in-DAZy-sum}),
(\ref{eq:total-radiation-pressure-force-to-XITM-DAZx-sum}), and
(\ref{eq:total-radiation-pressure-force-to-YITM-DA-sum}).
Using (\ref{eq:coherent-state-hatD-Heisenberg}), we separate the
incident electric field quadrature $\hat{D}(\omega)$ from the light
source into its classical part and the part of quantum fluctuations.
Then, we evaluate the classical part, the linear-order contributions
of the quantum fluctuations from $\hat{D}_{v}(\omega)$ and
$\hat{A}(\omega)$, and the linear-order contributions from the mirror
displacements $\hat{X}_{x}$, $\hat{X}_{y}$, $\hat{X}_{XITM}$, and
$\hat{X}_{YITM}$.
After this evaluation, we consider the situation where the incident
electric field from the light source is in the monochromatic coherent
state with the central frequency $\omega_{0}$ through the choice of
the complex amplitude $\alpha(\omega)$ for the coherent state by
Eq.~(\ref{eq:mohochromatic-alphaomega}).


Thus, we reach the expressions of the Fourier transformation of the
radiation pressure forces
$\FrakD_{d}^{\dagger}\scrF_{rpXITM}(\Omega)\FrakD_{d}$,
$\FrakD_{d}^{\dagger}\scrF_{rpYITM}(\Omega)\FrakD_{d}$,
$\FrakD_{d}^{\dagger}\scrF_{rpx}(\Omega)\FrakD_{d}$, and
$\FrakD_{d}^{\dagger}\scrF_{rpy}(\Omega)\FrakD_{d}$ as
Eqs.~(\ref{eq:DdaggerD-rad-pres-F-XITM-DAZ-Dmono-hineg}),
(\ref{eq:DdaggerD-rad-pres-F-YITM-in-DA-Dmono-hineg}),
(\ref{eq:DdaggerD-rad-pres-Frpx-Dmono-hineg}), and
(\ref{eq:DdaggerD-rad-pres-Frpy-Dmono-hineg}), respectively.


Here, we consider the tuning condition of the Fabry-P\'erot cavity by
\begin{eqnarray}
  \label{eq:omega0-L-tune-cond}
  \omega_{0} \tau = 2n\pi, \quad n\in\NF.
\end{eqnarray}
Due to this tuning condition, the radiation pressure forces
$\FrakD_{d}^{\dagger}\scrF_{rpXITM}(\Omega)\FrakD_{d}$,
$\FrakD_{d}^{\dagger}\scrF_{rpYITM}(\Omega)\FrakD_{d}$,
$\FrakD_{d}^{\dagger}\scrF_{rpx}(\Omega)\FrakD_{d}$, and
$\FrakD_{d}^{\dagger}\scrF_{rpy}(\Omega)\FrakD_{d}$ given by
Eqs.~(\ref{eq:DdaggerD-rad-pres-F-XITM-DAZ-Dmono-hineg}),
(\ref{eq:DdaggerD-rad-pres-F-YITM-in-DA-Dmono-hineg}),
(\ref{eq:DdaggerD-rad-pres-Frpx-Dmono-hineg}), and
(\ref{eq:DdaggerD-rad-pres-Frpy-Dmono-hineg}), respectively, are
expressed as
\begin{widetext}
\begin{eqnarray}
  &&
     \FrakD_{d}^{\dagger}\scrF_{rpXITM}(\Omega)\FrakD_{d}
     \nonumber\\
  &=&
      -
      2 N^{2} \sqrt{1-T}
      \frac{\hbar\omega_{0}}{m_{ITM}c}
      \left[ 1 - \sqrt{1-T} \right]^{-1}
      2 \pi \delta(\Omega)
      \nonumber\\
  &&
     -
     2 N \sqrt{1-T}
     \frac{\hbar}{m_{ITM}c}
     e^{ + i \Omega \tau }
     e^{ + i \Omega \tau_{x}' }
     \left[ 1 - \sqrt{1-T} e^{+ 2 i \Omega \tau } \right]^{-1}
     \cos(\Omega\tau)
     \nonumber\\
  && \quad\quad
     \times
     \left[
     \sqrt{|\omega_{0}(\Omega-\omega_{0})|}
     \left(
     \hat{D}_{v}(\Omega-\omega_{0}) - \hat{A}(\Omega-\omega_{0})
     \right)
     +
     \sqrt{|\omega_{0}(\Omega+\omega_{0})|}
     \left(
     \hat{D}_{v}(\Omega+\omega_{0}) - \hat{A}(\Omega+\omega_{0})
     \right)
     \right]
     ,
     \label{eq:DdaggerD-rad-pres-F-XITM-DAZ-Dmono-hineg-toXITM-tautune}
\end{eqnarray}
\begin{eqnarray}
  &&
     \FrakD_{d}^{\dagger}\scrF_{rpYITM}(\Omega)\FrakD_{d}
     \nonumber\\
  &=&
      -
      2 N^{2} \sqrt{1-T}
      \frac{\hbar\omega_{0}}{m_{ITM}c}
      \left[ 1 - \sqrt{1-T} \right]^{-1}
      2 \pi \delta(\Omega)
      \nonumber\\
  &&
     -
     2 N \sqrt{1-T}
     \frac{\hbar}{m_{ITM}c}
     e^{ + i \Omega \tau_{y}' }
     e^{ + i \Omega \tau }
     \left[ 1 - \sqrt{1-T} e^{+ 2 i \Omega \tau } \right]^{-1}
     \cos(\Omega\tau)
     \nonumber\\
  && \quad\quad
     \times
     \left[
     \sqrt{|\omega_{0}(\Omega-\omega_{0})|}
     \left(
     \hat{D}_{v}(\Omega-\omega_{0}) + \hat{A}(\Omega-\omega_{0})
     \right)
     +
     \sqrt{|\omega_{0}(\Omega+\omega_{0})|}
     \left(
     \hat{D}_{v}(\Omega+\omega_{0}) + \hat{A}(\Omega+\omega_{0})
     \right)
     \right]
     ,
     \label{eq:DdaggerD-rad-pres-F-YITM-in-DA-Dmono-hineg-toYITM-tautune}
\end{eqnarray}
\begin{eqnarray}
  &&
     \FrakD_{d}^{\dagger}\scrF_{rpx}(\Omega)\FrakD_{d}
     \nonumber\\
  &=&
      +
      \frac{N^{2}\hbar\omega_{0}}{c}
      \left[ 1 - \sqrt{1-T} \right]^{-2}
     \left[
     \frac{T}{m_{EM}}
     +
     \frac{2\sqrt{1-T}}{m_{ITM}} \left[ 1 - \sqrt{1-T} \right]
     \right]
     2 \pi \delta(\Omega)
      \nonumber\\
  &&
     +
     \frac{N\hbar}{c}
     e^{ + i \Omega \tau_{x}' }
     e^{ + i \Omega \tau }
     \left[ 1 - \sqrt{1-T} \right]^{-1}
     \left[ 1 - \sqrt{1-T} e^{+ 2 i \Omega \tau } \right]^{-1}
     \left[
     \frac{T}{m_{EM}}
     +
     \frac{2\sqrt{1-T}}{m_{ITM}}
     \cos(\Omega\tau)
     \left[ 1 - \sqrt{1-T} \right]
     \right]
     \nonumber\\
  && \quad\quad
     \times
     \left[
     \sqrt{|\omega_{0}(\Omega-\omega_{0})|}
     \left(
     \hat{D}_{v}(\Omega-\omega_{0}) - \hat{A}(\Omega-\omega_{0})
     \right)
     +
     \sqrt{|\omega_{0}(\Omega+\omega_{0})|}
     \left(
     \hat{D}_{v}(\Omega+\omega_{0}) - \hat{A}(\Omega+\omega_{0})
     \right)
     \right]
     ,
     \label{eq:DdaggerD-rad-pres-Frpx-Dmono-hineg-tautune}
\end{eqnarray}
\begin{eqnarray}
  &&
     \FrakD_{d}^{\dagger}\scrF_{rpy}(\omega)\FrakD_{d}
     \nonumber\\
  &=&
      +
      N^{2}
      \frac{\hbar\omega_{0}}{c}
      \left[ 1 - \sqrt{1-T} \right]^{-2}
      \left[
      \frac{T}{m_{EM}}
      +
      \frac{2 \sqrt{1-T}}{m_{ITM}} \left[ 1 -\sqrt{1-T} \right]
      \right]
      2 \pi \delta(\Omega)
      \nonumber\\
  &&
     +
     N
     \frac{\hbar}{c}
     e^{ + i \Omega \tau }
     e^{ + i \Omega \tau_{y}' }
     \left[ 1 - \sqrt{1-T} \right]^{-1}
     \left[ 1 - \sqrt{1-T} e^{+ 2 i \Omega \tau } \right]^{-1}
     \left[
     \frac{T}{m_{EM}}
     + \frac{2 \sqrt{1-T}}{m_{ITM}}
     \cos( \Omega \tau )
     \left[ 1 - \sqrt{1-T} \right]
     \right]
     \nonumber\\
  && \quad\quad
     \times
     \left[
     \sqrt{|(\Omega-\omega_{0})\omega_{0}|}
     \left(
     \hat{D}_{v}(\Omega-\omega_{0}) + \hat{A}(\Omega-\omega_{0})
     \right)
     +
     \sqrt{|(\Omega+\omega_{0})\omega_{0}|}
     \left(
     \hat{D}_{v}(\Omega+\omega_{0}) + \hat{A}(\Omega+\omega_{0})
     \right)
     \right]
     ,
     \label{eq:DdaggerD-rad-pres-Frpy-Dmono-hineg-tautune}
\end{eqnarray}
respectively.
Here, we note that
Eqs.~(\ref{eq:DdaggerD-rad-pres-F-XITM-DAZ-Dmono-hineg-toXITM-tautune})--(\ref{eq:DdaggerD-rad-pres-Frpy-Dmono-hineg-tautune})
includes the terms which proportional to $\delta(\Omega)$.
These terms represent the classical constant force in time.
Due to the existence of these terms, we have to develop further
discussion about the tuning condition (\ref{eq:omega0-L-tune-cond}) in
Sec.~\ref{sec:Changing_Tuning-Point}.
Before we reach the discussions in
Sec.~\ref{sec:Changing_Tuning-Point}, we keep the tuning condition
(\ref{eq:omega0-L-tune-cond}) to observe the results caused by these
classical constant forces.


\subsection{Solutions to the Heisenberg equations for mirrors' motion}
\label{sec:Sol_to_Heisenberg_Eq_for_mirrors}


Now, we can write down the solution to the Heisenberg equations
(\ref{eq:Eq-of-motion-XYITM-com-Fourier-approx})--(\ref{eq:Eq-of-motion-XY-dif-Fourier-approx}).
Substituting the expression of the radiation pressure forces
(\ref{eq:DdaggerD-rad-pres-F-XITM-DAZ-Dmono-hineg-toXITM-tautune})--(\ref{eq:DdaggerD-rad-pres-Frpy-Dmono-hineg-tautune})
into
Eqs.~(\ref{eq:Eq-of-motion-XYITM-com-Fourier-approx})--(\ref{eq:Eq-of-motion-XY-dif-Fourier-approx}),
we can obtain the explicit solution to the Heisenberg equations as follows:
\begin{eqnarray}
  &&
     \FrakD_{d}^{\dagger}\hat{Z}_{com}(\Omega)\FrakD_{d}
     \nonumber\\
  &=&
      +
      \frac{N^{2}\hbar\omega_{0}}{c\omega_{p}^{2}}
      \left[ 1 - \sqrt{1-T} \right]^{-2}
      \left[ \frac{T}{m_{EM}} + \frac{2\left[\sqrt{1-T}-1+T\right]}{m_{ITM}} \right]
      2 \pi \delta(\Omega)
      \nonumber\\
  &&
     -
     \frac{N\hbar}{c(\Omega^{2}-\omega_{p}^{2})}
     e^{ + i \Omega \tau }
     e^{ + i \Omega \left(\frac{\tau_{y}' + \tau_{x}' }{2}\right)}
     \left[ 1 - \sqrt{1-T} \right]^{-1}
     \left[ 1 - \sqrt{1-T} e^{+ 2 i \Omega \tau } \right]^{-1}
     \left[ \frac{T}{m_{EM}} + \frac{2\cos(\Omega\tau)\left[\sqrt{1-T}-1+T\right]}{m_{ITM}} \right]
     \nonumber\\
  && \quad
     \times
     \left[
     \sqrt{|\omega_{0}(\Omega-\omega_{0})|}
     \left(
     \cos\left( \Omega \left(\frac{\tau_{y}' - \tau_{x}' }{2} \right)\right)
     \hat{D}_{v}(\Omega-\omega_{0})
     +
     i \sin\left( \Omega \left(\frac{\tau_{y}' - \tau_{x}' }{2} \right)\right)
     \hat{A}(\Omega-\omega_{0})
     \right)
     \right.
     \nonumber\\
  && \quad\quad\quad
     \left.
     +
     \sqrt{|\omega_{0}(\Omega+\omega_{0})|}
     \left(
     \cos\left( \Omega \left(\frac{\tau_{y}' - \tau_{x}' }{2} \right)\right)
     \hat{D}_{v}(\Omega+\omega_{0})
     +
     i \sin\left( \Omega \left(\frac{\tau_{y}' - \tau_{x}' }{2} \right)\right)
     \hat{A}(\Omega+\omega_{0})
     \right)
     \right]
     ,
     \label{eq:DdaggerD-rad-pres-com-DA-Dmono-tuned-wobifreq-sol-sum}
\end{eqnarray}
\begin{eqnarray}
  &&
     \FrakD_{d}^{\dagger}\hat{Z}_{diff}(\Omega)\FrakD_{d}
     \nonumber\\
  &=&
      -
      \frac{N\hbar}{c(\Omega^{2}-\omega_{p}^{2})}
      e^{ + i \Omega \tau }
      e^{ + i \Omega \left(\frac{\tau_{x}' + \tau_{y}' }{2}\right)}
      \left[ 1 - \sqrt{1-T} \right]^{-1}
      \left[ 1 - \sqrt{1-T} e^{+ 2 i \Omega \tau } \right]^{-1}
      \left[ \frac{T}{m_{EM}} + \frac{2\cos(\Omega\tau)\left[\sqrt{1-T}-1+T\right]}{m_{ITM}} \right]
      \nonumber\\
  && \quad
     \times
     \left[
     \sqrt{|\omega_{0}(\Omega-\omega_{0})|}
     \left[
     i \sin\left( \Omega \left(\frac{\tau_{x}' - \tau_{y}' }{2} \right) \right)
     \hat{D}_{v}(\Omega-\omega_{0})
     -
     \cos\left( \Omega \left(\frac{\tau_{x}' - \tau_{y}' }{2} \right) \right)
     \hat{A}(\Omega-\omega_{0})
     \right]
     \right.
     \nonumber\\
  && \quad\quad
     \left.
     +
     \sqrt{|\omega_{0}(\Omega+\omega_{0})|}
     \left[
     i \sin\left( \Omega \left(\frac{\tau_{x}' - \tau_{y}' }{2} \right) \right)
     \hat{D}_{v}(\Omega+\omega_{0})
     -
     \cos\left( \Omega \left(\frac{\tau_{x}' - \tau_{y}' }{2} \right) \right)
     \hat{A}(\Omega+\omega_{0})
     \right]
     \right]
     \nonumber\\
  &&
     +
     \frac{\Omega^{2}}{2(\Omega^{2}-\omega_{p}^{2})}
     H(\Omega,L+l) L
     ,
     \label{eq:DdaggerD-rad-pres-dif-DA-Dmono-tuned-wobifreq-sol-sum}
\end{eqnarray}
\begin{eqnarray}
  &&
     \FrakD_{d}^{\dagger}\hat{Z}_{comITM}(\Omega)\FrakD_{d}
     \nonumber\\
  &=&
      -
      \frac{2N^{2}\hbar\omega_{0}}{m_{ITM}c\omega_{p}^{2}}
      \sqrt{1-T}
      \left[ 1 - \sqrt{1-T} \right]^{-1}
      2 \pi \delta(\Omega)
      \nonumber\\
  &&
     +
     \frac{2 N \sqrt{1-T}\hbar}{m_{ITM}c(\Omega^{2}-\omega_{p}^{2})}
     e^{ + i \Omega \tau }
     \cos(\Omega\tau)
     \left[ 1 - \sqrt{1-T} e^{+ 2 i \Omega \tau } \right]^{-1}
     e^{ + i \Omega \left(\frac{\tau_{x}' + \tau_{y}' }{2}\right) }
     \nonumber\\
  && \quad
     \times
     \left[
     +
     \sqrt{|\omega_{0}(\Omega-\omega_{0})|}
     \left[
     \cos\left( \Omega \left(\frac{\tau_{x}' - \tau_{y}' }{2}\right) \right)
     \hat{D}_{v}(\Omega-\omega_{0})
     -
     i \sin\left( \Omega \left(\frac{\tau_{x}' - \tau_{y}' }{2}\right)\right)
     \hat{A}(\Omega-\omega_{0})
     \right]
     \right.
     \nonumber\\
  && \quad\quad\quad
     \left.
     +
     \sqrt{|\omega_{0}(\Omega+\omega_{0})|}
     \left[
     \cos\left( \Omega \left( \frac{\tau_{x}' - \tau_{y}' }{2} \right) \right)
     \hat{D}_{v}(\Omega+\omega_{0})
     -
     i \sin\left( \Omega \left( \frac{\tau_{x}' - \tau_{y}' }{2} \right) \right)
     \hat{A}(\Omega+\omega_{0})
     \right]
     \right]
     ,
     \label{eq:Eq-of-motion-XYITM-com-tuned-wobifreq-sol-sum}
\end{eqnarray}
\begin{eqnarray}
  &&
     \FrakD_{d}^{\dagger}\hat{Z}_{diffITM}(\Omega)\FrakD_{d}
     \nonumber\\
  &=&
      -
      \frac{2N \sqrt{1-T}\hbar}{m_{ITM}c(\Omega^{2}-\omega_{p}^{2})}
      e^{ + i \Omega \tau }
      \left[ 1 - \sqrt{1-T} e^{+ 2 i \Omega \tau } \right]^{-1}
      \cos(\Omega\tau)
      e^{ + i \Omega \left(\frac{\tau_{x}' + \tau_{y}' }{2}\right)}
      \nonumber\\
  && \quad
     \times
     \left[
     \sqrt{|\omega_{0}(\Omega-\omega_{0})|}
     \left[
     - i \sin\left( \Omega \left( \frac{\tau_{x}' - \tau_{y}' }{2} \right) \right)
     \hat{D}_{v}(\Omega-\omega_{0})
     +
     \cos\left( \Omega \left( \frac{\tau_{x}' - \tau_{y}' }{2}\right) \right)
     \hat{A}(\Omega-\omega_{0})
     \right]
     \right.
     \nonumber\\
  && \quad\quad
     \left.
     +
     \sqrt{|\omega_{0}(\Omega+\omega_{0})|}
     \left[
     -
     i \sin\left( \Omega \left( \frac{\tau_{x}' - \tau_{y}' }{2} \right) \right)
     \hat{D}_{v}(\Omega+\omega_{0})
     +
     \cos\left( \Omega \left( \frac{\tau_{x}' - \tau_{y}' }{2} \right) \right)
     \hat{A}(\Omega+\omega_{0})
     \right]
     \right]
     .
     \label{eq:Eq-of-motion-XYITM-dif-tuned-wobifreq-sol-sum}
\end{eqnarray}
\end{widetext}


Here, we evaluate the order of the phase $\Omega(\tau_{y}'-\tau_{x}')/2$.
In
Sec.~\ref{sec:Final_input-output-relation_for_Fabry-Perot_GW_Detector},
we choose the phase offset of the interferometer so that
$\omega_{0}(\tau_{y}' - \tau_{x}')$ $\sim$ $O(1)$.
Then, the order of $\Omega(\tau_{y}'-\tau_{x}')/2$ is given by
\begin{eqnarray}
  \Omega\frac{\tau_{y}'-\tau_{x}'}{2}
  &\sim&
         10^{-10}
         \left(\frac{\Omega}{2\pi \times 10^{4} \mbox{Hz}}\right)
         \left(\frac{2\pi \times10^{14} \mbox{Hz}}{\omega_{0}}\right)
         .
         \nonumber\\
         \label{eq:Omegatauy-taux-half}
\end{eqnarray}
This indicates that we may regard that $\Omega(\tau_{y}'-\tau_{x}')/2$
is zero.
Furthermore, we choose $(\tau_{x}'+\tau_{y}')/2$ so that
\begin{eqnarray}
  \Omega \frac{\tau_{x}'+\tau_{y}'}{2}
  &\sim&
         2 \times 10^{-3}
         \left(\frac{\Omega}{2\pi \times 10^{4} \mbox{Hz}}\right)
         \left(\frac{(l_{x}+l_{y})/2}{10 \mbox{m}}\right)
         .
         \nonumber\\
         \label{eq:Omegatauy+taux-half}
\end{eqnarray}
We may also regard that $\Omega(\tau_{x}'+\tau_{y}')/2$ is zero.


Then, we apply the approximation $\Omega(\tau_{y}'-\tau_{x}')/2\sim 0$
and $\Omega(\tau_{x}'+\tau_{y}')/2\sim 0$.
In addition, we apply the approximation $\Omega\ll\omega_{0}$.
Through these approximations, (\ref{eq:DdaggerD-rad-pres-com-DA-Dmono-tuned-wobifreq-sol-sum})--(\ref{eq:Eq-of-motion-XYITM-dif-tuned-wobifreq-sol-sum})
to the Heisenberg equations for mirrors' motion~\cite{K.Nakamura-2025-footnote4}
\begin{widetext}
\begin{eqnarray}
  \FrakD_{d}^{\dagger}\hat{Z}_{com}(\Omega)\FrakD_{d}
  &=&
      \frac{N^{2}\hbar\omega_{0}}{c\omega_{p}^{2}}
      \left[ 1 - \sqrt{1-T} \right]^{-2}
      \left[ \frac{T}{m_{EM}} + \frac{2\left[\sqrt{1-T}-1+T\right]}{m_{ITM}} \right]
      2 \pi \delta(\Omega)
      \nonumber\\
  &&
     -
     \frac{N\hbar\omega_{0}}{c(\Omega^{2}-\omega_{p}^{2})}
     e^{ + i \Omega \tau }
     \left[ 1 - \sqrt{1-T} \right]^{-1}
     \left[ 1 - \sqrt{1-T} e^{+ 2 i \Omega \tau } \right]^{-1}
     \nonumber\\
  && \quad\quad
     \times
     \left[ \frac{T}{m_{EM}} + \frac{2\cos(\Omega\tau)\left[\sqrt{1-T}-1+T\right]}{m_{ITM}} \right]
     \left[
     \hat{D}_{v}(\Omega-\omega_{0})
     +
     \hat{D}_{v}(\Omega+\omega_{0})
     \right]
     ,
     \label{eq:DdaggerD-rad-pres-com-DA-Dmono-tuned-wobifreq-sol-sum-approx}
  \\
  \FrakD_{d}^{\dagger}\hat{Z}_{diff}(\Omega)\FrakD_{d}
  &=&
      \frac{N\hbar\omega_{0}}{c(\Omega^{2}-\omega_{p}^{2})}
      e^{ + i \Omega \tau }
      \left[ 1 - \sqrt{1-T} \right]^{-1}
      \left[ 1 - \sqrt{1-T} e^{+ 2 i \Omega \tau } \right]^{-1}
      \nonumber\\
  && \quad\quad
     \times
      \left[ \frac{T}{m_{EM}} + \frac{2\cos(\Omega\tau)\left[\sqrt{1-T}-1+T\right]}{m_{ITM}} \right]
     \left[
     \hat{A}(\Omega-\omega_{0})
     +
     \hat{A}(\Omega+\omega_{0})
     \right]
     \nonumber\\
  &&
     +
     \frac{\Omega^{2}}{2(\Omega^{2}-\omega_{p}^{2})}
     H(\Omega,L+l) L
     ,
     \label{eq:DdaggerD-rad-pres-dif-DA-Dmono-tuned-wobifreq-sol-sum-approx}
  \\
  \FrakD_{d}^{\dagger}\hat{Z}_{comITM}(\Omega)\FrakD_{d}
  &=&
      -
      \frac{2N^{2}\hbar\omega_{0}}{m_{ITM}c\omega_{p}^{2}}
      \sqrt{1-T}
      \left[ 1 - \sqrt{1-T} \right]^{-1}
      2 \pi \delta(\Omega)
      \nonumber\\
  &&
     +
     \frac{2 N \sqrt{1-T}\hbar\omega_{0}}{m_{ITM}c(\Omega^{2}-\omega_{p}^{2})}
     e^{ + i \Omega \tau }
     \cos(\Omega\tau)
     \left[ 1 - \sqrt{1-T} e^{+ 2 i \Omega \tau } \right]^{-1}
     \left[
     \hat{D}_{v}(\Omega-\omega_{0})
     +
     \hat{D}_{v}(\Omega+\omega_{0})
     \right]
     ,
     \nonumber\\
     \label{eq:Eq-of-motion-XYITM-com-tuned-wobifreq-sol-sum-approx}
  \\
  \FrakD_{d}^{\dagger}\hat{Z}_{diffITM}(\Omega)\FrakD_{d}
  &=&
      -
      \frac{2N \sqrt{1-T}\hbar\omega_{0}}{m_{ITM}c(\Omega^{2}-\omega_{p}^{2})}
      e^{ + i \Omega \tau }
      \left[ 1 - \sqrt{1-T} e^{+ 2 i \Omega \tau } \right]^{-1}
      \cos(\Omega\tau)
      \left[
      \hat{A}(\Omega-\omega_{0})
      +
      \hat{A}(\Omega+\omega_{0})
      \right]
      .
  \label{eq:Eq-of-motion-XYITM-dif-tuned-wobifreq-sol-sum-approx}
\end{eqnarray}
\end{widetext}


We have to note that the classical constant terms in
Eqs.~(\ref{eq:DdaggerD-rad-pres-com-DA-Dmono-tuned-wobifreq-sol-sum-approx})
and (\ref{eq:Eq-of-motion-XYITM-com-tuned-wobifreq-sol-sum-approx})
diverge if we choose $\omega_{p}=0$, i.e., if we assume the mirrors
are in completely free motion.
This is the reason why we introduced the nonvanishing pendulum
fundamental frequency, $\omega_{p}$, in each equation of motion for
the mirrors.
This introduction of the nonvanishing pendulum fundamental frequency
$\omega_{p}$ is a natural consequence from the viewpoint of the actual
ground-based gravitational-wave detectors.
From a mathematical point of view, this introduction of $\omega_{p}$
is a kind of regularization of the divergence.
However, we can regard the introduction of this regularization as
reasonable for the actual ground-based gravitational-wave detectors.
In this sense, we have to emphasize that this introduction of
$\omega_{p}$ has a physical meaning more than a mere mathematical
regularization.


Although the original motivation for the introduction of the pendulum
fundamental frequency $\omega_{p}$ is due to the regularization of the
divergence of the classical radiation pressure force, we have a
by-product due to the introduction of $\omega_{p}$.
We also derived the general solution
(\ref{eq:J.J.Sakurai-2.3.46-1-again-assumption-sol-t=-infty-final}) of
our quantum forced harmonic oscillator model with the Hamiltonian
(\ref{eq:QuanatumHamiltonian-hatX}).
This solution
(\ref{eq:J.J.Sakurai-2.3.46-1-again-assumption-sol-t=-infty-final})
indicates that the initial conditions $\hat{X}(-\infty)$ and
$\hat{P}(-\infty)$ are concentrate to the frequency
$\omega=\pm\omega_{p}$.
In usual quantum mechanics, the uncertainty relation between the
position $\hat{X}(t)$ and momentum $\hat{P}(t)$ is derived from the
commutation relation (\ref{eq:canonical-commutation}).
Furthermore, through the solution
(\ref{eq:J.J.Sakurai-2.3.46-1-again-assumption-sol-t=-infty-final}),
the noncommutativity (\ref{eq:canonical-commutation}) arise from the
commutation relation (\ref{eq:canonical-commutation-initial}) for the
initial conditions $\hat{X}(-\infty)$ and $\hat{P}(-\infty)$.
However, the solution
(\ref{eq:J.J.Sakurai-2.3.46-1-again-assumption-sol-t=-infty-final})
indicates that the contribution of this initial condition
$\hat{X}(-\infty)$ and $\hat{P}(-\infty)$ concentrates to the
frequency $\omega=\pm\omega_{p}$.
This indicates that the uncertainty which is due to the commutation
relation (\ref{eq:canonical-commutation}) concentrates to the
frequency $\omega=\pm\omega_{p}$.
This is an important implication which is obtained by the introduction
of the pendulum fundamental frequency $\omega_{p}$.


\section{Final input-output relation for a Fabry-P\'erot Graviational-Wave Detector}
\label{sec:Final_input-output-relation_for_Fabry-Perot_GW_Detector}


Now, we return to the evaluation of the input-output relation
(\ref{eq:input-output-relation-general-pert-sideband-approx}).
First, we introduce the offset phase $\theta$ by
\begin{eqnarray}
  \label{eq:offset-def}
  \theta := \omega_{0}(\tau_{y}'-\tau_{x}').
\end{eqnarray}
In the DC readout scheme, we dare to set up the interferometer to leak the
classical carrier field in the dark port to use this classical carrier
field as a reference of the measurement of the small fluctuations in
the dark port.
Within this paper, we control this leakage of the classical carrier
field through the control the parameter $\theta$ defined by
Eq.~(\ref{eq:offset-def}), though there are many different methods
to make a leakage of the classical carrier field at the dark port.
We also apply the approximation $\Omega(\tau_{y}'-\tau_{x}')/2\sim
0$ and $\Omega(\tau_{x}'+\tau_{y}')/2\sim 0$ as estimated in
Eqs.~(\ref{eq:Omegatauy-taux-half}) and
(\ref{eq:Omegatauy+taux-half}).
Furthermore, we consider the tuning condition of the Fabry-P\'erot
cavity by Eq.~(\ref{eq:omega0-L-tune-cond}).
In addition, we employ the tuning condition of the lengths between the
BS and ITMs by
\begin{eqnarray}
  \label{eq:omega0-lxly-tune-cond}
  \omega_{0} \frac{\tau_{x}'+\tau_{y}'}{2} = 2m\pi, \quad m\in\NF.
\end{eqnarray}
Even if we specified the offset angle $\theta$ defined by
Eq.~(\ref{eq:offset-def}), the average $(\tau_{x}'+\tau_{y}')/2$ is
still free.
Therefore, we have to tune this average distance to guarantee the
output port is nearly dark.


Moreover, we apply the approximation $\Omega\ll \omega_{0}$.
\begin{widetext}
\begin{eqnarray}
  &&
     \FrakD_{d}^{\dagger}\hat{B}(\omega_{0}\pm\Omega)\FrakD_{d}
     \nonumber\\
  &=&
      +
      i N
      \sin\theta
      2 \pi \delta(\Omega)
      \nonumber\\
  &&
     +
     \left[ 1 - \sqrt{1-T} e^{ \mp 2 i \Omega \tau } \right]
     \left[ 1 - \sqrt{1-T} e^{ \pm 2 i \Omega \tau } \right]^{-1}
     e^{ \pm 2 i \Omega \tau }
     \left[
     i
     \sin\theta
     \hat{D}_{v}(\omega_{0}\pm\Omega)
     +
     \cos\theta
     \hat{A}(\omega_{0}\pm\Omega)
     \right]
     \nonumber\\
  &&
     +
     i
     \frac{N\omega_{0}}{c}
     e^{ \pm 2 i \Omega \tau }
     \left[ 1 - \sqrt{1-T} e^{ \mp 2 i \Omega \tau } \right]
     \left[ 1 - \sqrt{1-T} e^{ \pm 2 i \Omega \tau } \right]^{-1}
     \left[
     i
     \sin\theta
     \FrakD_{d}^{\dagger}\hat{Z}_{comITM}(\pm\Omega)\FrakD_{d}
     -
     \cos\theta
     \FrakD_{d}^{\dagger}\hat{Z}_{diffITM}(\pm\Omega)\FrakD_{d}
     \right]
     \nonumber\\
  &&
     -
     i
     \frac{2NT\omega_{0}}{c}
     e^{ \pm i \Omega \tau }
     \left[ 1 - \sqrt{1-T} e^{ \pm 2 i \Omega \tau } \right]^{-1}
     \left[ 1 - \sqrt{1-T} \right]^{-1}
     \left[
     i
     \sin\theta
     \FrakD_{d}^{\dagger}\hat{Z}_{com}(\pm\Omega)\FrakD_{d}
     +
     \cos\theta
     \FrakD_{d}^{\dagger}\hat{Z}_{diff}(\pm\Omega)\FrakD_{d}
     \right]
     \nonumber\\
  &&
     +
     \frac{i N\omega_{0}}{c}
     \left[
     i
     \sin\theta
     \FrakD_{d}^{\dagger}\hat{Z}_{comITM}(\pm\Omega )\FrakD_{d}
     -
     \cos\theta
     \FrakD_{d}^{\dagger}\hat{Z}_{diffITM}(\pm\Omega )\FrakD_{d}
     \right]
     .
     \label{eq:input-output-relation-general-pert-sideband-approx-2}
\end{eqnarray}


Substituting the radiation pressure forces
(\ref{eq:DdaggerD-rad-pres-com-DA-Dmono-tuned-wobifreq-sol-sum-approx})--(\ref{eq:Eq-of-motion-XYITM-dif-tuned-wobifreq-sol-sum-approx})
into
Eq.~(\ref{eq:input-output-relation-general-pert-sideband-approx-2}),
and applying the approximation $L\gg l_{x,y}$, we reached the
input-output relation
\begin{eqnarray}
  &&
     \FrakD_{d}^{\dagger}\hat{B}(\omega_{0}\pm\Omega)\FrakD_{d}
     \nonumber\\
  &=&
      +
      i N \sin\theta
      \left[
      1
      +
      i
      \frac{2N^{2}\hbar\omega_{0}^{2}}{c^{2}\omega_{p}^{2}}
      \left[ 1 - \sqrt{1-T} \right]^{-4}
      \left[
      \frac{T^{2}}{m_{EM}}
      +
      \frac{4 (1-T) \left[ 1 - \sqrt{1-T} \right]^{2}}{m_{ITM}}
      \right]
      \right]
      2 \pi \delta(\Omega)
      \nonumber\\
  &&
     +
     \left[ 1 - \sqrt{1-T} e^{ \mp 2 i \Omega \tau } \right]
     \left[ 1 - \sqrt{1-T} e^{ \pm 2 i \Omega \tau } \right]^{-1}
     e^{ \pm 2 i \Omega \tau }
     \left[
     i \sin\theta \hat{D}_{v}(\omega_{0}\pm\Omega)
     +  \cos\theta \hat{A}(\omega_{0}\pm\Omega)
     \right]
     \nonumber\\
  &&
     -
     2 i
     \frac{N^{2}\hbar\omega_{0}^{2}}{c^{2}\Omega^{2}}
     e^{ \pm 2 i \Omega \tau }
     \left[ 1 - \sqrt{1-T} \right]^{-2}
     \left[ 1 - \sqrt{1-T} e^{ \pm 2 i \Omega \tau } \right]^{-2}
     \nonumber\\
  && \quad\quad\quad
     \times
     \left[
     \frac{T^{2}}{m_{EM}}
     +
     2 \sqrt{1-T} \cos(\Omega\tau)
     \left[ 1 - \sqrt{1-T} \right]
     \left[ T - \left[ 1 - \sqrt{1-T} \right]^{2} \cos(\Omega\tau) \right]
     \frac{1}{m_{ITM}}
     \right]
     \nonumber\\
  && \quad\quad\quad
     \times
     \left[
     i \sin\theta \left( \hat{D}_{v}(-(\omega_{0}\mp\Omega)) + \hat{D}_{v}(\omega_{0}\pm\Omega) \right)
     +  \cos\theta \left( \hat{A}(-(\omega_{0}\mp\Omega)) + \hat{A}(\omega_{0}\pm\Omega) \right)
     \right]
     \nonumber\\
  &&
     -
     i
     e^{ \pm i \Omega \tau }
     \cos\theta
     \frac{NTL\omega_{0}}{c}
     \left[ 1 - \sqrt{1-T} e^{ \pm 2 i \Omega \tau } \right]^{-1}
     \left[ 1 - \sqrt{1-T} \right]^{-1}
     H(\pm\Omega,L)
     .
     \label{eq:input-output-rel-approx-tuned-theta-explicit}
\end{eqnarray}
\end{widetext}


Here, from the definition of the quadrature
(\ref{eq:K.Nakamura-M.-K.Fujimoto-2018-17}) and the situation
$\omega_{0}\gg\Omega$, we note that the operator
$\hat{B}(\omega_{0}\pm\Omega)$, $\hat{D}_{v}(\omega_{0}\pm\Omega)$,
$\hat{D}_{v}(-(\omega_{0}\mp\Omega))$, $\hat{A}(\omega_{0}\pm\Omega)$,
and $\hat{A}(-(\omega_{0}\mp\Omega))$ are given by
\begin{eqnarray}
  \label{eq:hatBomega0pmOmega}
  \hat{B}(\omega_{0}\pm\Omega)
  &=&
      \hat{b}(\omega_{0}\pm\Omega) =: \hat{b}_{\pm}(\Omega)
      ,
  \\
  \label{eq:hatDvomega0pmOmega}
  \hat{D}_{v}(\omega_{0}\pm\Omega)
  &=&
      \hat{d}(\omega_{0}\pm\Omega) =: \hat{d}_{\pm}(\Omega)
      ,
  \\
  \label{eq:hatDv-omega0mpOmega}
  \hat{D}_{v}(-(\omega_{0}\mp\Omega))
  &=&
      \hat{d}^{\dagger}(\omega_{0}\mp\Omega) =: \hat{d}_{\mp}^{\dagger}(\Omega)
      ,
  \\
  \label{eq:hatA-omega0pmOmega}
  \hat{A}(\omega_{0}\mp\Omega)
  &=&
      \hat{a}(\omega_{0}\mp\Omega) =: \hat{a}_{\pm}(\Omega)
      ,
  \\
  \label{eq:hatA-omega0mpOmega}
  \hat{A}(-(\omega_{0}\mp\Omega))
  &=&
      \hat{a}^{\dagger}(\omega_{0}\mp\Omega) =: \hat{a}_{\mp}^{\dagger}(\Omega)
      .
\end{eqnarray}
Furthermore, using Eq.~(\ref{eq:N-classical-power-relation}), we
define the variable $\FrakR$, the phase $\beta$, $\kappa$, and
$h_{SQL}$ by
\begin{eqnarray}
  \FrakR
  &:=&
       \frac{2I_{0}\omega_{0}T^{2}}{c^{2}\omega_{p}^{2}}
       \left[ 1 - \sqrt{1-T} \right]^{-4}
       \nonumber\\
  && \quad
     \times
     \left[
     \frac{1}{m_{EM}}
     +
     \frac{1}{m_{ITM}}
     \frac{4 (1-T) \left[ 1 - \sqrt{1-T} \right]^{2}}{T^{2}}
     \right]
     ,
     \nonumber\\
  \label{eq:FrakR-def}
  \\
  e^{\pm 2 i \beta}
  &:=&
       \frac{
       e^{ \pm 2 i \Omega \tau }
       \left[ 1 - \sqrt{1-T} e^{ \mp 2 i \Omega \tau } \right]
       }{
       \left[ 1 - \sqrt{1-T} e^{ \pm 2 i \Omega \tau } \right]
       }
       .
       \label{eq:beta-def}
\end{eqnarray}
\begin{widetext}
\begin{eqnarray}
  \kappa
  &:=&
      \frac{4I_{0}T^{2}\omega_{0}}{c^{2}\Omega^{2}}
      \left[ 1 - \sqrt{1-T} \right]^{-2}
      \left[ 2 - T - 2 \sqrt{1-T} \cos(2\Omega\tau) \right]^{-1}
      \nonumber\\
  && \quad\quad\quad
     \times
     \left[
     \frac{1}{m_{EM}}
     +
     \frac{ 2 \sqrt{1-T} \left[ 1 - \sqrt{1-T} \right] \cos(\Omega\tau) }{T}
     \left[ 1 - \frac{\left[ 1 - \sqrt{1-T} \right]^{2}}{T} \cos(\Omega\tau) \right]
     \frac{1}{m_{ITM}}
     \right]
     ,
     \label{eq:kappa-def}
\end{eqnarray}
\begin{eqnarray}
   h_{SQL}
  &:=&
       \sqrt{\frac{4\hbar}{\Omega^{2}L^{2}}}
       \left[
       \frac{1}{m_{EM}}
       +
       \frac{ 2 \sqrt{1-T}  \left[ 1 - \sqrt{1-T} \right] \cos(\Omega\tau) }{T}
       \left[ 1 - \frac{\left[ 1 - \sqrt{1-T} \right]^{2}}{T} \cos(\Omega\tau) \right]
       \frac{1}{m_{ITM}}
       \right]^{1/2}
       .
       \label{eq:hSQL-def}
\end{eqnarray}
Here, ``SQL'' in $h_{SQL}$ means ``Standard Quantum Limit'' which is
given by
Eq.~(\ref{eq:Kimble-noise-spectral-density-is-greater-than-hSQL2})
below.
We may regard that the definition of $h_{SQL}$ comes from the
inequality
Eq.~(\ref{eq:Kimble-noise-spectral-density-is-greater-than-hSQL2}).
Through the quadratures
(\ref{eq:hatBomega0pmOmega})--(\ref{eq:hatA-omega0mpOmega}), variables
$\FrakR$, $\beta$, $\kappa$, and $h_{SQL}$, the input-output relation
(\ref{eq:input-output-rel-approx-tuned-theta-explicit}) is given by
\begin{eqnarray}
     \FrakD_{d}^{\dagger}\hat{b}_{\pm}(\Omega)\FrakD_{d}
  &=&
      +
      i \sqrt{\frac{I_{0}}{\hbar\omega_{0}}} \sin\theta
      \left[
      1
      +
      i \FrakR
      \right]
      2 \pi \delta(\Omega)
      \nonumber\\
  &&
     +
     e^{\pm 2 i \beta}
     \left[
     i \sin\theta \hat{d}_{\pm}(\Omega)
     +  \cos\theta \hat{a}_{\pm}(\Omega)
     \right]
     \nonumber\\
  &&
     +
     e^{\pm 2 i \beta}
     \frac{\kappa}{2}
     \left[
     \sin\theta \left( \hat{d}^{\dagger}_{\mp}(\Omega) + \hat{d}_{\pm}(\Omega) \right)
     - i \cos\theta \left( \hat{a}^{\dagger}_{\mp}(\Omega) + \hat{a}_{\pm}(\Omega) \right)
     \right]
     \nonumber\\
  &&
     -
     i
     e^{\pm i \beta}
     \sqrt{\kappa}
     \cos\theta
     \frac{H(\pm\Omega,L)}{h_{SQL}}
     .
     \label{eq:input-output-rel-explicit-quad-I0-kappabetahSQL}
\end{eqnarray}
\end{widetext}


Here, we consider the representation of the input-output relation
(\ref{eq:input-output-rel-explicit-quad-I0-kappabetahSQL}) in
terms of the two-photon formulation.
In the two-photon formulation, we introduce the amplitude operators
$\hat{a}_{1}$, $\hat{b}_{1}$, $\hat{d}_{1}$ and the phase operators
$\hat{a}_{2}$, $\hat{b}_{2}$, $\hat{d}_{2}$ as follows:
\begin{eqnarray}
  \hat{a}_{1}
  &:=&
      \frac{1}{\sqrt{2}}
      (\hat{a}_{+}+\hat{a}_{-}^{\dagger}), \quad
      \hat{a}_{2}
      :=
      \frac{1}{\sqrt{2}i}
      (\hat{a}_{+}-\hat{a}_{-}^{\dagger}),
      \label{eq:a1-a2-operator-def-2}
  \\
  \hat{b}_{1}
  &:=&
      \frac{1}{\sqrt{2}}
      (\hat{b}_{+}+\hat{b}_{-}^{\dagger}), \quad
      \hat{b}_{2}
      :=
      \frac{1}{\sqrt{2}i}
      (\hat{b}_{+}-\hat{b}_{-}^{\dagger}),
      \label{eq:b1-b2-operator-def-2}
  \\
  \hat{d}_{1}
  &:=&
      \frac{1}{\sqrt{2}}
      (\hat{d}_{+}+\hat{d}_{-}^{\dagger}), \quad
      \hat{d}_{2}
      :=
      \frac{1}{\sqrt{2}i}
      (\hat{d}_{+}-\hat{d}_{-}^{\dagger}).
      \label{eq:d1-d2-operator-def-2}
\end{eqnarray}
From these definitions, we obtain
\begin{eqnarray}
  \FrakD_{d}^{\dagger}\hat{b}_{1}(\Omega)\FrakD_{d}
  &=&
      -
      \sqrt{\frac{2I_{0}}{\hbar\omega_{0}}}
      \FrakR
      \sin\theta
      2 \pi \delta(\Omega)
      \nonumber\\
  &&
     +
     e^{ + 2 i \beta} \left[ - \sin\theta \hat{d}_{2}(\Omega) + \cos\theta \hat{a}_{1}(\Omega) \right]
     \nonumber\\
  &&
     +
     e^{ + 2 i \beta} \kappa \sin\theta \hat{d}_{1}(\Omega)
     .
     \label{eq:amplitude-quadrature-ITM-M-inc-sum}
\end{eqnarray}
\begin{eqnarray}
  \FrakD_{d}^{\dagger}\hat{b}_{2}(\Omega)\FrakD_{d}
  &=&
      \sqrt{\frac{2I_{0}}{\hbar\omega_{0}}} \sin\theta
      2 \pi \delta(\Omega)
      \nonumber\\
  &&
     +
     e^{ + 2 i \beta} \left[ \sin\theta \hat{d}_{1}(\Omega) + \cos\theta \hat{a}_{2}(\Omega) \right]
     \nonumber\\
  &&
     - e^{ + 2 i \beta} \cos\theta \kappa \hat{a}_{1}(\Omega)
     \nonumber\\
  &&
     - e^{ + i \beta} \sqrt{2\kappa} \cos\theta \frac{H(\Omega,L)}{h_{SQL}}
     .
     \label{eq:phase-quadrature-ITM-M-inc-sum}
\end{eqnarray}


In
Ref.~\cite{H.J.Kimble-Y.Levin-A.B.Matsko-K.S.Thorne-S.P.Vyatchanin-2001},
$\FrakD_{d}^{\dagger}\hat{b}_{2}\FrakD_{d}$ is regarded as the signal
operator in gravitational-wave detectors.
The first line of Eq.~(\ref{eq:phase-quadrature-ITM-M-inc-sum}) is the
contribution from the classical coherent light.
The second line of Eq.~(\ref{eq:phase-quadrature-ITM-M-inc-sum})
corresponds to the shot noise due to the uncertainties of the photon
number estimation in the coherent state of the laser.
The third line of Eq.~(\ref{eq:phase-quadrature-ITM-M-inc-sum}) is the
contribution from the radiation pressure noise.
Then, the final line of Eq.~(\ref{eq:phase-quadrature-ITM-M-inc-sum})
corresponds to the gravitational-wave signal cooperating with the
response function $e^{+i\beta}\sqrt{2\kappa} \cos\theta/h_{SQL}$.


In
Ref.~\cite{H.J.Kimble-Y.Levin-A.B.Matsko-K.S.Thorne-S.P.Vyatchanin-2001},
the first line of the classical coherent light in
Eqs.~(\ref{eq:amplitude-quadrature-ITM-M-inc-sum}) and
(\ref{eq:phase-quadrature-ITM-M-inc-sum}) are regarded as trivial, and
are neglected.
Furthermore, in
Ref.~\cite{H.J.Kimble-Y.Levin-A.B.Matsko-K.S.Thorne-S.P.Vyatchanin-2001},
the quantum noise against the signal $H(\Omega,L)$ of gravitational
waves is determined by the quantum operator $\hat{h}_{n}$, which is
given by
\begin{eqnarray}
  \hat{h}_{n}
  &:=&
       e^{ - i \beta} \frac{h_{SQL}}{\sqrt{2\kappa} \cos\theta}
       \nonumber\\
  && \quad
     \times
     \left[
     \left(
     \FrakD_{d}^{\dagger}\hat{b}_{2}(\Omega)\FrakD_{d}
     -
     \sqrt{\frac{2I_{0}}{\hbar\omega_{0}}} \sin\theta
     2 \pi \delta(\Omega)
     \right)
     \right.
     \nonumber\\
  &&
     \quad\quad\quad
     \left.
     + e^{ + i \beta} \sqrt{2\kappa} \cos\theta \frac{H(\Omega,L)}{h_{SQL}}
     \right]
     \nonumber\\
  &=&
      e^{ - i \beta} \frac{h_{SQL}}{\sqrt{2\kappa} \cos\theta}
      \nonumber\\
  && \quad
     \times
     \left[
     e^{ + 2 i \beta} \left[ \sin\theta \hat{d}_{1}(\Omega) + \cos\theta \hat{a}_{2}(\Omega) \right]
     \right.
     \nonumber\\
  &&
     \quad\quad\quad
     \left.
     - e^{ + 2 i \beta} \cos\theta \kappa \hat{a}_{1}(\Omega)
     \right]
     .
     \label{eq:phase-quadrature-ITM-M-inc-sum-noise-operator-def}
\end{eqnarray}
Moreover, in
Ref.~\cite{H.J.Kimble-Y.Levin-A.B.Matsko-K.S.Thorne-S.P.Vyatchanin-2001},
the quantum noise spectral density is estimated by
\begin{eqnarray}
  &&
     \frac{1}{2} 2\pi \delta(\Omega-\Omega') S_{(K)}(\Omega)
     \nonumber\\
  &:=&
       \frac{1}{2} \langle\mbox{in}|
       \hat{h}_{n}(\Omega)
       \hat{h}_{n}^{\dagger}(\Omega')
       +
       \hat{h}_{n}^{\dagger}(\Omega')
       \hat{h}_{n}(\Omega)
       |\mbox{in}\rangle
       .
       \label{eq:Kimble-noise-spectral-density}
\end{eqnarray}
without any reason except for the citation of
Refs.~\cite{B.L.Schumaker-C.M.Caves-1985a,B.L.Schumaker-C.M.Caves-1985b}.
Incidentally, in
Refs.~\cite{B.L.Schumaker-C.M.Caves-1985a,B.L.Schumaker-C.M.Caves-1985b},
there is no description that the noise spectral density
$S_{(K)}(\Omega)$ defined by
Eq.~(\ref{eq:Kimble-noise-spectral-density}) is the actual quantum
noise in measurements of gravitational waves.
They only suggested that $S_{(K)}(\Omega)$ is an estimation measure of
the amplitude of the quantum noises in the lasers of interferometers.
However, the main purpose of
Ref.~\cite{H.J.Kimble-Y.Levin-A.B.Matsko-K.S.Thorne-S.P.Vyatchanin-2001}
was to show some techniques to surpass the ``standard quantum limit''
using the squeezed states techniques and comparison with them, and it
was not a serious problem of the definition of the noise spectral
density at that time.
After the publication of
Ref.~\cite{H.J.Kimble-Y.Levin-A.B.Matsko-K.S.Thorne-S.P.Vyatchanin-2001},
general arguments of quantum measurement theory were
proposed~\cite{M.Ozawa-2004}.
Then, the definitions and the meanings of ``noises'' began to be
discussed seriously.


Besides the problem whether $S_{(K)}(\Omega)$ defined by
Eq.~(\ref{eq:Kimble-noise-spectral-density}) is a physically
reasonable noise estimation or not, we can calculate $S_{(K)}(\Omega)$
through the definitions (\ref{eq:Kimble-noise-spectral-density}) of
$S_{(K)}(\Omega)$ and the definition
(\ref{eq:phase-quadrature-ITM-M-inc-sum-noise-operator-def}) of the
noise operator $\hat{h}_{n}$.
As a result of this calculation, we obtain the result
\begin{eqnarray}
  \label{eq:Kimble-noise-spectral-density-explicit}
  S_{(K)}(\Omega)
  &=&
       \frac{h_{SQL}^{2}(\Omega)}{2\cos^{2}\theta} \left(
       \frac{1}{\kappa(\Omega)} + \cos^{2}\theta\kappa(\Omega)
       \right)
       .
\end{eqnarray}
It is a famous fact that $1/\kappa$ in the bracket of the right-hand
side of Eq.~(\ref{eq:Kimble-noise-spectral-density-explicit}) is the
contribution from the shot noise of the coherent state of the laser
and $\kappa$ in the bracket of the right-hand side of
Eq.~(\ref{eq:Kimble-noise-spectral-density-explicit}) is the
contribution from the radiation pressure noise.
It is also a famous fact that $\kappa$ depends on the frequency $\Omega$,
which corresponds to the frequency of the gravitational-wave signal,
and the laser power $I_{0}$ as shown in Eq.~(\ref{eq:kappa-def}).
Furthermore, we can also show that $S_{(K)}$ is bounded below as
\begin{eqnarray}
  \label{eq:Kimble-noise-spectral-density-is-greater-than-hSQL2}
  S_{(K)}(\Omega)
  \geq h_{SQL}^{2}.
\end{eqnarray}
This is called ``standard quantum limit'' in the context of the
measurement of
gravitational-waves~\cite{H.J.Kimble-Y.Levin-A.B.Matsko-K.S.Thorne-S.P.Vyatchanin-2001,H.Miao-PhDthesis-2010}.
Here, we have to mention that these variables $\beta$, $\kappa$, and
$h_{SQL}$ defined by Eqs.~(\ref{eq:beta-def}), (\ref{eq:kappa-def}),
and (\ref{eq:hSQL-def}), respectively, are defined so that Kimble's noise
spectral density $S_{(K)}$ is realized in the form of
Eq.~(\ref{eq:Kimble-noise-spectral-density-explicit}).


One of the main problems that we discuss in this paper is whether the
Kimble noise spectral density $S_{(K)}(\Omega)$ is regarded as the
noise in the measurement process of gravitational waves in some sense,
or not.


Before discussing this main problem within this paper, we compare the
variables $\beta$, $\kappa$, and $h_{SQL}$ defined by
Eqs.~(\ref{eq:beta-def}), (\ref{eq:kappa-def}), and
(\ref{eq:hSQL-def}), with those in
Ref.~\cite{H.J.Kimble-Y.Levin-A.B.Matsko-K.S.Thorne-S.P.Vyatchanin-2001}.
First, we evaluate the phase $\beta$ and obtain the result
\begin{eqnarray}
  \label{eq:beta-gamma-tan-inv}
  \beta
  &=&
      - \frac{1}{2} \arctan\left(2\frac{Tc}{4\Omega L}\right)
      \nonumber\\
  &&
     + O(T^{2},(2\Omega\tau)^{3},T(2\Omega\tau))
\end{eqnarray}
and we may regard
\begin{eqnarray}
  \label{eq:beta-gamma-leading}
  \beta = - \frac{Tc}{4\Omega L} + O(T^{2},(2\Omega\tau)^{3},T(2\Omega\tau)).
\end{eqnarray}
This coincides with the Kimble's $\beta$ in
Ref.~\cite{H.J.Kimble-Y.Levin-A.B.Matsko-K.S.Thorne-S.P.Vyatchanin-2001}.
In this paper, we denote this $\beta$ as
\begin{eqnarray}
  \label{eq:betaK-def}
  \beta^{(K)} := - \frac{Tc}{4\Omega L} =: - \frac{\gamma}{\Omega},
\end{eqnarray}
where we defined the resonant frequency $\gamma$ by
\begin{eqnarray}
  \label{eq:gamma-def}
  \gamma := \frac{Tc}{4L}.
\end{eqnarray}

\begin{figure}
  \begin{center}
    \includegraphics[width=0.48\textwidth]{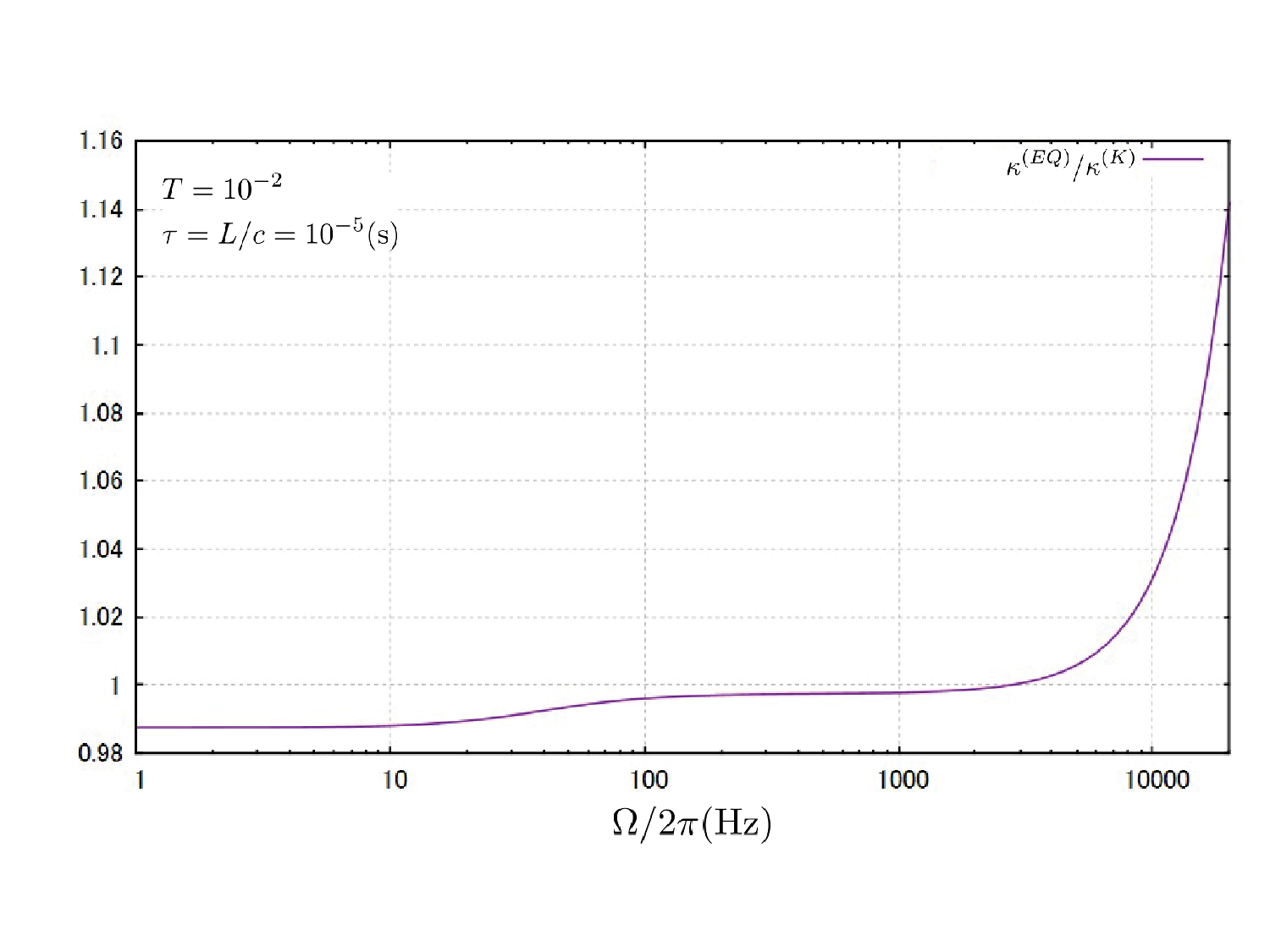}
  \end{center}
  \caption{
    The ratio $\kappa^{(EQ)}/\kappa^{(K)}$ given by
    Eqs.~(\ref{eq:Kimble-kappa-def-3}) and (\ref{eq:kappaEQ-def}).
    We choose the transmissivity $T=10^{-2}$ and $\tau=L/c=10^{-5}$s.
    This figure indicates that $\kappa^{EQ}$ is 10\% greater than
    $\kappa^{(K)}$ at the high frequency region ($\sim 20$kHz) and 1\%
    smaller than $\kappa^{(K)}$ at low frequency region ($\sim 10$Hz).
  }
  \label{fig:KappaEQKappaK-ratio}
\end{figure}

On the other hand, $\kappa$ defined by Eq.~(\ref{eq:kappa-def}) is
evaluated as
\begin{eqnarray}
  \kappa
  &=&
      4 \frac{I_{0}\omega_{0}}{L^{2}\Omega^{2}(\Omega^{2}+\gamma^{2})}
      \left[
      \frac{1}{m_{EM}}
      +
      \frac{1}{m_{ITM}}
      \right]
      \nonumber\\
  &&
     + O(T^{3},T(\Omega\tau)^{2},(\Omega\tau)^{4})
     .
  \label{eq:kappa-def-ITM-inclusion-explicit-approx-3}
\end{eqnarray}
When $m_{EM}=m_{ITM}=:m$, we obtain
\begin{eqnarray}
  \kappa
  &=&
      \frac{2(I_{0}/I_{SQL})\gamma^{4}}{\Omega^{2}(\Omega^{2}+\gamma^{2})}
      + O(T^{3},T(\Omega\tau)^{2},(\Omega\tau)^{4})
      ,
      \label{eq:kappa-def-ITM-inclusion-explicit-approx-eq-mass}
\end{eqnarray}
where
\begin{eqnarray}
  \label{eq:ISQL-def}
  I_{SQL} := \frac{mL^{2}\gamma^{4}}{4\omega_{0}}.
\end{eqnarray}
Eq.~(\ref{eq:kappa-def-ITM-inclusion-explicit-approx-eq-mass}) is a
realization of Kimble's $\kappa$ in
Ref.~\cite{H.J.Kimble-Y.Levin-A.B.Matsko-K.S.Thorne-S.P.Vyatchanin-2001}.
In this paper, we denote Kimble's $\kappa$ as
\begin{eqnarray}
  \kappa^{(K)}
  &:=&
      \frac{2(I_{0}/I_{SQL})\gamma^{4}}{\Omega^{2}(\Omega^{2}+\gamma^{2})}
      \nonumber\\
  &=&
      \frac{4I_{0}\omega_{0}}{mc^{2}\Omega^{2}}
      \frac{32}{\left(16(\Omega\tau)^{2}+T^{2}\right)}
      .
      \label{eq:Kimble-kappa-def-3}
\end{eqnarray}
Furthermore, we also define $\kappa^{(EQ)}$ from $\kappa$ defined by
Eq.~(\ref{eq:kappa-def}) as
\begin{eqnarray}
  \label{eq:kappaEQ-def}
  \kappa^{(EQ)} := \left.\kappa\right|_{m=m_{EM}=m_{ITM}}.
\end{eqnarray}
The ratio $\kappa^{(EQ)}/\kappa^{(K)}$ are depicted in
Fig.~\ref{fig:KappaEQKappaK-ratio}.
In Fig.~\ref{fig:KappaEQKappaK-ratio}, we choose the transmissivity
$T=10^{-2}$ and $\tau=L/c=10^{-5}$s.
This figure indicates that $\kappa^{EQ}$ is 10\% greater than
$\kappa^{(K)}$ at the high frequency region ($\sim 20$kHz) and 1\%
smaller than $\kappa^{(K)}$ at low frequency region ($\sim 10$Hz).


\begin{figure}[htb]
  \begin{center}
    \includegraphics[width=0.48\textwidth]{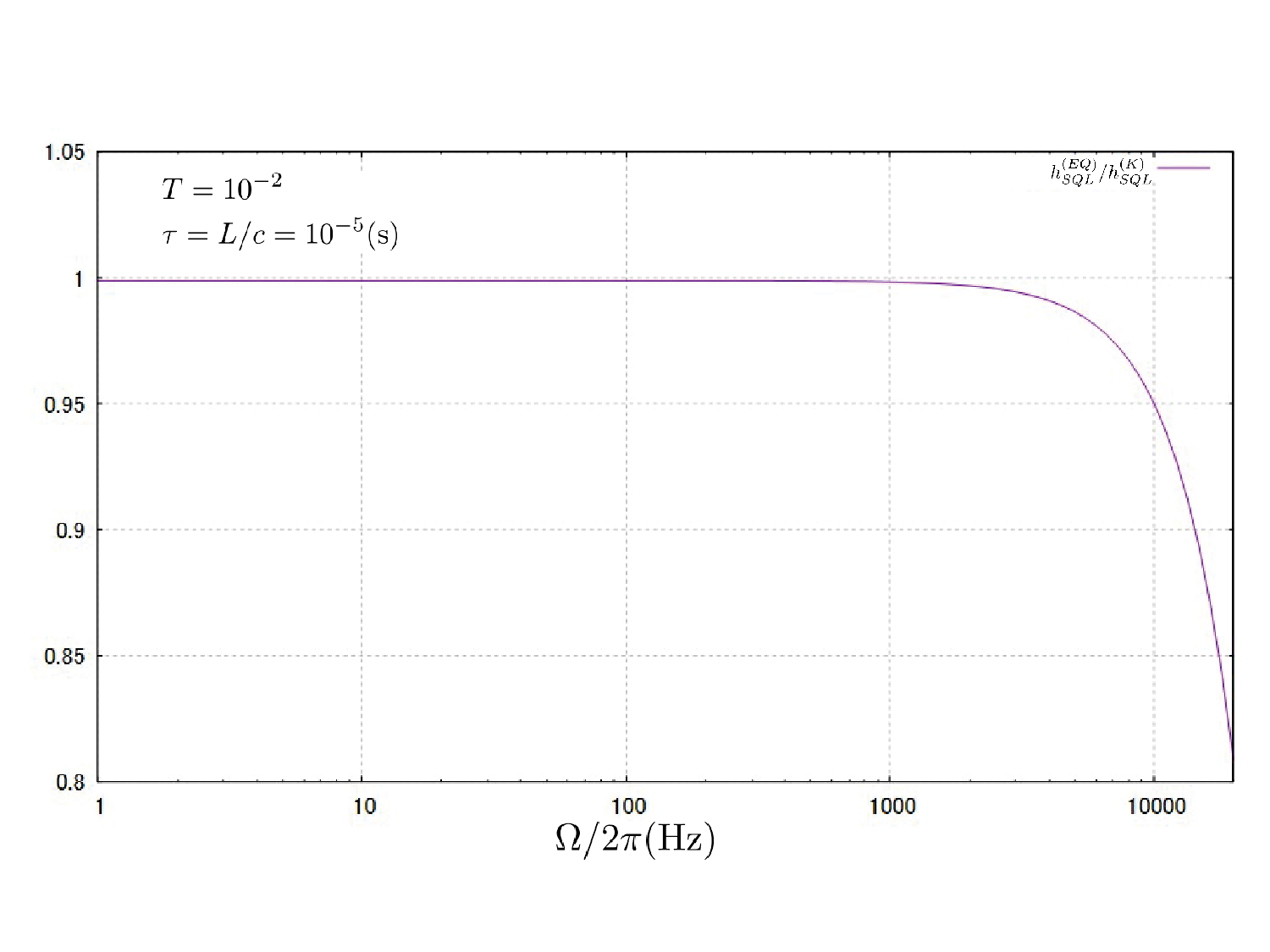}
  \end{center}
  \caption{
    The ratio $h_{SQL}^{(EQ)}/h_{SQL}^{(K)}$ is depicted.
    $h_{SQL}^{(K)}$ and $h_{SQL}^{(EQ)}$ are defined in
    Eqs.~(\ref{eq:hSQLK-def}) and (\ref{eq:hSQLK-def}), respectively.
    The difference between $h_{SQL}^{(EQ)}$ and $h_{SQL}^{(K)}$ is
    $\sim 15\%$ at the high frequency range $\sim 20$kHz.
    This is due to the $\cos(\Omega\tau)$-dependence in
    Eq.~(\ref{eq:hSQL-def}).
  }
  \label{fig:hSQLEQhSQLK-ratio}
\end{figure}

Next, we consider $h_{SQL}$ defined by Eq.~(\ref{eq:hSQL-def}), which
is estimated as
\begin{eqnarray}
  h_{SQL}
  &=&
      \sqrt{\frac{4\hbar}{\Omega^{2}L^{2}}}
      \left(\frac{1}{m_{EM}}+\frac{1}{m_{ITM}}\right)^{1/2}
      \nonumber\\
  &&
     +
     O(T,(\Omega\tau)^{2})
     .
     \label{eq:hSQL-leading}
\end{eqnarray}
Here again, we consider the case $m:=m_{EM}=m_{ITM}$ and we define
\begin{eqnarray}
  \label{eq:hSQLK-def}
  h_{SQL}^{(K)}
  &:=&
       \sqrt{\frac{8\hbar}{m\Omega^{2}L^{2}}}
       ,
  \\
  \label{eq:hSQLEQ-def}
  h_{SQL}^{(EQ)}
  &:=&
       \left.h_{SQL}\right|_{m=m_{EM}=m_{ITM}}
       .
\end{eqnarray}
$h_{SQL}^{(K)}$ is the $\kappa$ in
Ref.~\cite{H.J.Kimble-Y.Levin-A.B.Matsko-K.S.Thorne-S.P.Vyatchanin-2001}.
Then, Eq.~(\ref{eq:hSQL-leading}) yields that the leading term of
$h_{SQL}$ defined by Eq.~(\ref{eq:hSQL-def}) coincides with the
$h_{SQL}$ in
Ref.~\cite{H.J.Kimble-Y.Levin-A.B.Matsko-K.S.Thorne-S.P.Vyatchanin-2001}.
We also show the ratio $h_{SQL}^{(EQ)}/h_{SQL}^{(K)}$ in Fig.~\ref{fig:hSQLEQhSQLK-ratio}.
The difference between $h_{SQL}^{(EQ)}$ and $h_{SQL}^{(K)}$ is $\sim 15\%$ at
the high frequency range $\sim 20$kHz.
This is due to the $\cos(\Omega\tau)$-dependence in
Eq.~(\ref{eq:hSQL-def}) which is introduced by the phase difference
in the radiation pressure forces affecting the EM and ITM.
Essentially, this difference was introduced by the motion of ITMs in
the input-output relation, which are ignored in
Ref.~\cite{H.J.Kimble-Y.Levin-A.B.Matsko-K.S.Thorne-S.P.Vyatchanin-2001}.


Although $\kappa$ and $h_{SQL}$ were defined so that the noise
spectral density $S_{(K)}$ is given by
Eq.~(\ref{eq:Kimble-noise-spectral-density-explicit}) which leads to
the inequality
(\ref{eq:Kimble-noise-spectral-density-is-greater-than-hSQL2}), the
input-output relation (\ref{eq:input-output-rel-approx-tuned-theta-explicit})
is more accurate than that in
Ref.~\cite{H.J.Kimble-Y.Levin-A.B.Matsko-K.S.Thorne-S.P.Vyatchanin-2001}.
The $10\%\sim 20\%$ difference from those in
Ref.~\cite{H.J.Kimble-Y.Levin-A.B.Matsko-K.S.Thorne-S.P.Vyatchanin-2001}
can be seen in the high-frequency region.


Finally, we consider the difference $S_{(K)}(\Omega)$ defined by
Eq.~(\ref{eq:Kimble-noise-spectral-density-explicit}) from the noise
spectral density in
Ref.~\cite{H.J.Kimble-Y.Levin-A.B.Matsko-K.S.Thorne-S.P.Vyatchanin-2001}.
To discuss this difference, we again consider the case
$m:=m_{EM}=m_{ITM}$ and we define $S_{(K)}^{(EQ))}(\Omega)$ by
\begin{eqnarray}
  \label{eq:SKEQ-def}
  S_{(K)}^{(EQ)}(\Omega) := \left.S_{(K)}(\Omega)\right|_{m:=m_{EM}=m_{ITM},\theta=0}.
\end{eqnarray}
We also define the other noise spectral density
$S_{(K)}^{(K)}(\Omega)$ by
\begin{eqnarray}
  \label{eq:SKK-def}
  S_{(K)}^{(K)}(\Omega)
  :=
  \frac{(h_{SQL}^{(K)})^{2}}{2} \left(
  \frac{1}{\kappa^{(K)}} + \kappa^{(K)}
  \right)
  ,
\end{eqnarray}
where $\kappa^{(K)}$ and $h_{SQL}^{(K)}$ are defined
Eqs.~(\ref{eq:Kimble-kappa-def-3}) and (\ref{eq:hSQLK-def}),
respectively.
This $S_{(K)}^{(K)}(\Omega)$ coincides with the noise spectral density
for the Fabry-P\'erot interferometer in
Ref.~\cite{H.J.Kimble-Y.Levin-A.B.Matsko-K.S.Thorne-S.P.Vyatchanin-2001}.

\begin{figure}[htb]
  \begin{center}
    \includegraphics[width=0.5\textwidth]{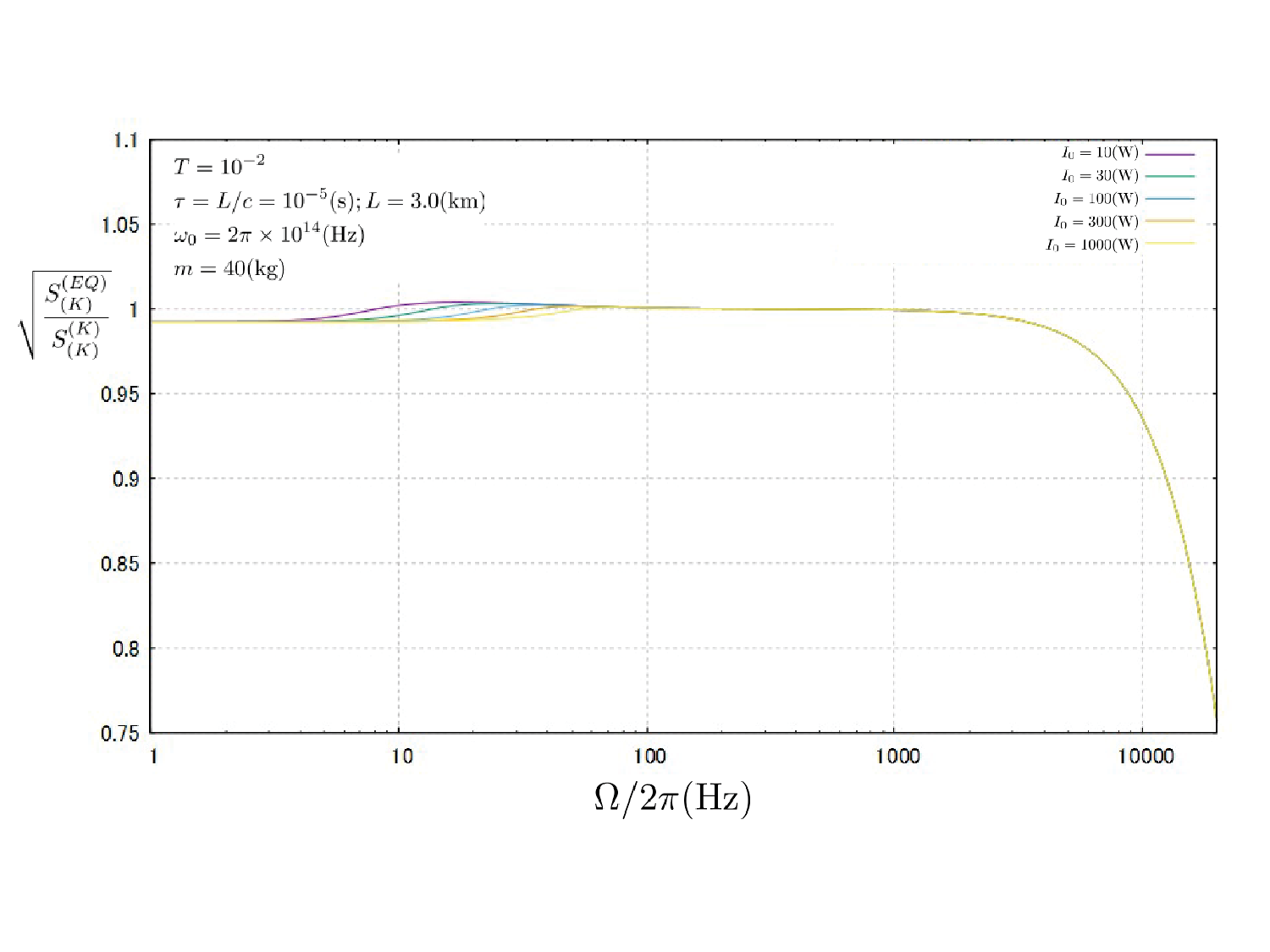}
  \end{center}
  \caption{
    The square root of the ratio $\sqrt{S_{(K)}^{(EQ)}/S_{(K)}^{(K)}}$
    defined by Eqs.~(\ref{eq:SKEQ-def}) and (\ref{eq:SKK-def}) in the
    range 1Hz to 20kHz.
    $\sqrt{S_{(K)}^{(EQ)}}$ is 20\% smaller than
    $\sqrt{S_{(K)}^{(K)}}$ at the high frequency range $\sim 20$kHz.
    We also observe a small power dependence in the range 1Hz to 100
    Hz.
    This difference in the range 1Hz to 100Hz is roughly $1\%$.
  }
  \label{fig:SKEQ-SKK-ratio-1-20k}
\end{figure}

The square root of the ratio
$\sqrt{S_{(K)}^{(EQ)}(\Omega)/S_{(K)}^{(K)}(\Omega)}$ of these noise
spectral densities is depicted in Fig.~\ref{fig:SKEQ-SKK-ratio-1-20k}.
Fig.~\ref{fig:SKEQ-SKK-ratio-1-20k} shows that  the noise spectral
density $\sqrt{S_{(K)}^{(EQ)}}$ is 20\% smaller than the Kimble's
noise spectral density  $\sqrt{S_{(K)}^{(K)}}$ at the high frequency
range $\sim 20$kHz.
We also observe a small power dependence within the range of 1Hz to
100 Hz.
This difference in the range 1Hz to 100Hz is roughly $1\%$.


From Fig.~\ref{fig:SKEQ-SKK-ratio-1-20k}, we may say that our accurate
input-output
relation~\ref{eq:input-output-rel-explicit-quad-I0-kappabetahSQL}
leads small corrections to the Kimble noise spectral density
$S_{(K)}^{(K)}$ in
Ref.~\cite{H.J.Kimble-Y.Levin-A.B.Matsko-K.S.Thorne-S.P.Vyatchanin-2001}
if we apply Eq.~(\ref{eq:Kimble-noise-spectral-density}) as the
definition of the noise spectral density.
We have to mention that we ignore the quantum initial operators
$\hat{X}(-\infty)$ and $\hat{P}(-\infty)$ for mirrors in the solutions
(\ref{eq:Eq-of-motion-XYITM-com-Fourier-approx})--(\ref{eq:Eq-of-motion-XY-dif-Fourier-approx})
to the Heisenberg equations.
As mentioned in
Sec.~\ref{sec:Heisenberg_Equations_for_mirrors'_motions}, these terms
contribute to the frequency $\Omega=\omega_{p}$.
In Fig.~\ref{fig:SKEQ-SKK-ratio-1-20k}, we choose $\omega_{p}=1$Hz.
Therefore, the curves in the neighborhood of 1Hz are meaningless.
However, we do not take this difference seriously in this paper.


In the next section, we derive the quantum noise spectral density from
the quantum fluctuations of photons at the photodetector through a DC
readout scheme, which is discussed in
Sec.~\ref{sec:General_arguments_for_the_DC-readout_scheme}.


\section{DC readout scheme for a Fabry-P\'erot Graviational-Wave Detector}
\label{sec:DC_readout_scheme_for_FP_GW_Detector}


Here, we apply the general arguments on the DC readout scheme for a
Fabry-P\'erot interferometer with the input-output relation
(\ref{eq:input-output-rel-explicit-quad-I0-kappabetahSQL}) or
equivalently Eqs.~(\ref{eq:amplitude-quadrature-ITM-M-inc-sum}) and
(\ref{eq:phase-quadrature-ITM-M-inc-sum}).
Comparing Eqs.~(\ref{eq:output-quadrature-general}) and
(\ref{eq:exp-valu-hatb-def}) and
(\ref{eq:input-output-rel-explicit-quad-I0-kappabetahSQL}), we obtain
\begin{eqnarray}
     \FrakA(\omega_{0}\pm\Omega)
     &=&
     -
     i
     e^{\pm i \beta}
     \sqrt{\kappa}
     \cos\theta
     \frac{H(\pm\Omega,L)}{h_{SQL}}
     ,
     \label{eq:input-output-relation-FrakA}
  \\
     \FrakB
     &=&
     i
     \sin\theta
     \sqrt{\frac{I_{0}}{\hbar\omega_{0}}}
     \left[
     1
     +
     i \FrakR
     \right]
     ,
     \label{eq:input-output-relation-FrakB}
\end{eqnarray}
\begin{eqnarray}
  &&
     \FrakD_{d}^{\dagger} \hat{b}_{n}(\omega_{0}\pm\Omega) \FrakD_{d}
     =
      \FrakD_{d}^{\dagger} \hat{b}_{n\pm}(\Omega) \FrakD_{d}
     \nonumber\\
  &=&
      e^{\pm2i\beta}
      \left[
      i \sin\theta \hat{d}_{\pm}(\Omega)
      +  \cos\theta \hat{a}_{\pm}(\Omega)
      \right]
      \nonumber\\
  &&
     +
     e^{\pm2i\beta} \frac{\kappa}{2}
     \left[
     \sin\theta \left( \hat{d}_{\mp}^{\dagger}(\Omega) + \hat{d}_{\pm}(\Omega) \right)
     \right.
     \nonumber\\
  && \quad\quad\quad\quad\quad
     \left.
     - i \cos\theta \left( \hat{a}_{\mp}^{\dagger}(\Omega) + \hat{a}_{\pm}(\Omega) \right)
     \right]
     .
     \label{eq:input-output-relation-hatbn}
\end{eqnarray}
Here, we note that $\FrakR$ in
Eq.~(\ref{eq:input-output-relation-FrakB}) is an extra-term which is
not included in the previous
works~\cite{H.J.Kimble-Y.Levin-A.B.Matsko-K.S.Thorne-S.P.Vyatchanin-2001}.
Therefore, we estimate the order of magnitude of $\FrakR$ as
\begin{eqnarray}
  \FrakR
  &=&
       \frac{32 I_{0}\omega_{0}}{c^{2}\omega_{p}^{2}T^{2}}
       \left[
       \frac{1}{m_{EM}}
       +
       \frac{1}{m_{ITM}}
       \right.
       \nonumber\\
  && \quad\quad\quad\quad
     \left.
     -
     \left(
     \frac{1}{m_{EM}}
     +
     \frac{3}{2m_{ITM}}
     \right)
     T
     + O(T^{2})
     \right]
     .
     \nonumber\\
     \label{eq:FrakR-defq-2-approx}
\end{eqnarray}
Here, we consider the case where $m:=m_{EM}=m_{ITM}$.
In the case where $m:=m_{EM}=m_{ITM}$,
Eq.~(\ref{eq:FrakR-defq-2-approx}) is given by
\begin{eqnarray}
  \FrakR
  :=
  \frac{64 I_{0}\omega_{0}}{mc^{2}\omega_{p}^{2}T^{2}}
  \left[
  1
  -
  \frac{5}{4}
  T
  + O(T^{2})
  \right]
  .
  \label{eq:FrakR-defq-2-approx-eqmas}
\end{eqnarray}
Then, we have the leading term of $\FrakR$ for the situation $T\ll 1$
as
\begin{eqnarray}
  \FrakR
  &\sim&
         \frac{ 64 I_{0} \omega_{0} }{ m c^{2} \omega_{p}^{2} T^{2}}
         \nonumber\\
  &=&
      3\times 10^{2}
      \times
      \left(
      \frac{I_{0}}{10^{2}W}
      \right)
      \left(
      \frac{\omega_{0}}{2\pi\times 10^{14}\mbox{Hz}}
      \right)
      \left(
      \frac{40 \mbox{kg}}{m}
      \right)
      \nonumber\\
  && \quad\quad\quad\quad
     \times
     \left(
     \frac{2\pi\times 1 \mbox{Hz}}{\omega_{p}}
     \right)^{2}
     \left(
     \frac{10^{-2}}{T}
     \right)^{2}
     .
     \label{eq:FrakR-eval-smallT-leading}
\end{eqnarray}


Now, we consider the signal operator $\hat{s}_{\calN_{b}}(\Omega)$
defined by Eq.~(\ref{eq:signal-operator-Fourier-def}), its
expectation value (\ref{eq:signal-operator-Fourier-exp}) and the
stationary noise spectral density
(\ref{eq:Measured-noise-spectral-density-DCreadout-bTheta}).
First, from Eqs.~(\ref{eq:input-output-relation-FrakA}) and
(\ref{eq:input-output-relation-FrakB}), the expectation value
(\ref{eq:signal-operator-Fourier-exp}) of the signal operator
$\hat{s}_{\calN_{b}}(\Omega)$ for the DC readout is given by
\begin{eqnarray}
  \langle
  \hat{s}_{\calN_{b}}(\Omega)
  \rangle
  &=&
      -
      \omega_{0}\sin(2\theta)
      \sqrt{\frac{I_{0}}{\hbar\omega_{0}}}
      \frac{\sqrt{\kappa}}{h_{SQL}}
      e^{+ i \beta}
      H(\Omega,L)
      \nonumber\\
  &&
     +
     O((I_{0})^{0})
     .
     \label{eq:GW-signal-operator-exp-value}
\end{eqnarray}
We note that the additional classical part $\FrakR$ defined by
Eq.~(\ref{eq:FrakR-def}) in $\FrakB$ of
Eq.~(\ref{eq:input-output-relation-FrakA}) does not contribute to the
expectation value (\ref{eq:GW-signal-operator-exp-value}) due to the
reality condition of the Fourier transformation $H(\Omega,L)$ of the
gravitational-wave signal.
Here, we note that the expectation value
(\ref{eq:GW-signal-operator-exp-value}) is maximized when the offset
$\theta=\pi/4$, while we do not measure $H(\Omega,L)$ when the
complete dark port $\theta=0$.
In the case of $\theta=0$, the gravitational-wave signal $H(\Omega,L)$
is included in the term of $O((I_{0})^{0})$.
It is well-known that the complete dark-port condition $\theta=0$ is
meaningless in the DC-readout scheme.
The factor $\sin(2\theta)$ in
Eq.~(\ref{eq:GW-signal-operator-exp-value}) comes from the fact that
$\FrakA(\omega_{0}\pm\Omega)$ in
Eq.~(\ref{eq:input-output-relation-FrakA}) depends on $\cos\theta$ and
$\FrakB$ in Eq.~(\ref{eq:input-output-relation-FrakB}) is proportional
to $\sin\theta$.
Thus, we conclude that $\theta=\pi/4$ maximize the expectation value
(\ref{eq:GW-signal-operator-exp-value}).


From Eq.~(\ref{eq:GW-signal-operator-exp-value}), we obtain the
gravitational-wave signal $H(\Omega,L)$ through the expectation value
(\ref{eq:GW-signal-operator-exp-value}) of the signal operator
$\hat{s}_{\calN_{b}}(\Omega)$ of the DC-readout by
\begin{eqnarray}
  H(\Omega,L)
  &=&
      -
      e^{- i \beta}
      \frac{1}{\omega_{0}\sqrt{\kappa}\sin(2\theta)}
      \sqrt{\frac{\hbar\omega_{0}}{I_{0}}}
      h_{SQL}
      \langle
      \hat{s}_{\calN_{b}}(\Omega)
      \rangle
      \nonumber\\
  &&
     +
     O((I_{0})^{-1/2})
     .
     \label{eq:GW-signal-operator-exp-value-inverse}
\end{eqnarray}
The factor
\begin{eqnarray}
  \label{eq:response-function}
      -
      \omega_{0}\sin(2\theta)
      \sqrt{\frac{I_{0}}{\hbar\omega_{0}}}
      \frac{\sqrt{\kappa}}{h_{SQL}}
      e^{+ i \beta}
\end{eqnarray}
in Eq.~(\ref{eq:GW-signal-operator-exp-value}) is regarded as the
response function for the gravitational-wave signal.


Next, we consider the stationary noise spectral density
(\ref{eq:Measured-noise-spectral-density-DCreadout-bTheta}).
From the explicit expression (\ref{eq:input-output-relation-hatbn}) of
the output operator
$\FrakD_{d}^{\dagger}\hat{b}_{n\pm}(\Omega)\FrakD_{d}$, the amplitude-
and phase-quadrature
$\FrakD_{d}^{\dagger}\hat{b}_{n1}(\Omega)\FrakD_{d}$ and
$\FrakD_{d}^{\dagger}\hat{b}_{n2}(\Omega)\FrakD_{d}$ defined by Eqs.~(\ref{eq:amplitude-phase-quadrature-def-1}) and
(\ref{eq:amplitude-phase-quadrature-def-2}), respectively, are given
by
\begin{eqnarray}
  &&
     \FrakD_{d}^{\dagger} \hat{b}_{n1}(\Omega) \FrakD_{d}
     \nonumber\\
  &=&
      e^{+ 2 i \beta}
      \left[
      -
      \sin\theta
      \hat{d}_{2}(\Omega)
      +
      \cos\theta
      \hat{a}_{1}(\Omega)
      +
      \kappa
      \sin\theta
      \hat{d}_{1}(\Omega)
      \right]
      ,
      \nonumber\\
      \label{eq:B1-noise-final}
      \\
  &&
     \FrakD_{d}^{\dagger} \hat{b}_{n2}(\Omega) \FrakD_{d}
     \nonumber\\
  &=&
      e^{+ 2 i \beta}
      \left[
      \sin\theta
      \hat{d}_{1}(\Omega)
      +
      \cos\theta
      \hat{a}_{2}(\Omega)
      -
      \kappa
      \cos\theta
      \hat{a}_{1}(\Omega)
      \right]
      .
      \nonumber\\
      \label{eq:B2-noise-final}
\end{eqnarray}
From the expressions (\ref{eq:B1-noise-final}) and
(\ref{eq:B2-noise-final}) of the amplitude- and phase-quadrature, the
operator $\FrakD_{d}^{\dagger}\hat{b}_{n\Theta}\FrakD_{d}$ defined by
Eq.~(\ref{eq:bTheta-noise-quadrature-def}) is given by
\begin{eqnarray}
  &&
     \FrakD_{d}^{\dagger}\hat{b}_{n\Theta}(\Omega)\FrakD_{d}
     \nonumber\\
  &=&
      e^{+ 2 i \beta}
      \left[
      \sin\theta
      \left(
      -
      \cos\Theta
      \hat{d}_{2}(\Omega)
      +
      \sin\Theta
      \hat{d}_{1}(\Omega)
      \right)
      \right.
      \nonumber\\
  && \quad\quad\quad
     \left.
     +
     \cos\theta
     \left(
     \cos\Theta
     \hat{a}_{1}(\Omega)
     +
     \sin\Theta
     \hat{a}_{2}(\Omega)
     \right)
     \right.
     \nonumber\\
  && \quad\quad\quad
     \left.
      +
      \kappa
      \left(
      \cos\Theta
      \sin\theta
      \hat{d}_{1}(\Omega)
      -
      \sin\Theta
      \cos\theta
      \hat{a}_{1}(\Omega)
      \right)
      \right]
      .
     \nonumber\\
      \label{eq:hatbnTheta-explicit}
\end{eqnarray}
We also have
\begin{eqnarray}
  &&
     \FrakD_{d}^{\dagger}\hat{b}_{n\Theta}^{\dagger}(\Omega)\FrakD_{d}
     \nonumber\\
  &=&
      e^{- 2 i \beta}
      \left[
      \sin\theta
      \left(
      -
      \cos\Theta
      \hat{d}_{2}^{\dagger}(\Omega)
      +
      \sin\Theta
      \hat{d}_{1}^{\dagger}(\Omega)
      \right)
      \right.
      \nonumber\\
  && \quad\quad\quad
     \left.
     +
     \cos\theta
     \left(
     \cos\Theta
     \hat{a}_{1}^{\dagger}(\Omega)
     +
     \sin\Theta
     \hat{a}_{2}^{\dagger}(\Omega)
     \right)
     \right.
     \nonumber\\
  && \quad\quad\quad
     \left.
     +
     \kappa
     \left(
     \cos\Theta
     \sin\theta
     \hat{d}_{1}^{\dagger}(\Omega)
     -
     \sin\Theta
     \cos\theta
     \hat{a}_{1}^{\dagger}(\Omega)
     \right)
     \right]
     .
     \nonumber\\
     \label{eq:hatbnThetadagger-explicit}
\end{eqnarray}
From the commutation relations
(\ref{eq:K.Nakamura-M.-K.Fujimoto-2018-15}) and
(\ref{eq:K.Nakamura-M.-K.Fujimoto-2018-16}), the definitions of the
sideband quadrature (\ref{eq:upper-lower-sideband-quadrature-def}),
and the amplitude- and phase-quadratures
(\ref{eq:amplitude-phase-quadrature-def-1}) and
(\ref{eq:amplitude-phase-quadrature-def-2}), or equivalently
Eqs.~(\ref{eq:a1-a2-operator-def-2})--(\ref{eq:d1-d2-operator-def-2}),
non-vanishing commutation relations for $\hat{a}_{n1}$ and
$\hat{a}_{n2}$ are summarized as
\begin{eqnarray}
  &&
     \left[
     \hat{a}_{n1}(\Omega)
     ,
     \hat{a}_{n2}^{\dagger}(\Omega')
     \right]
     =
     2 \pi i \delta(\Omega-\Omega')
     ,
     \label{eq:hatan1-hatan2-commutation-relations-non-vanish-1}
     \\
  &&
     \left[
     \hat{a}_{n2}(\Omega)
     ,
     \hat{a}_{n1}^{\dagger}(\Omega')
     \right]
     =
     - 2 \pi i \delta(\Omega-\Omega')
     ,
     \label{eq:hatan1-hatan2-commutation-relations-non-vanish-2}
     \\
  &&
     \left[
     \hat{a}_{n1}^{\dagger}(\Omega)
     ,
     \hat{a}_{n2}(\Omega')
     \right]
     =
     2 \pi i \delta(\Omega'-\Omega)
     ,
     \label{eq:hatan1-hatan2-commutation-relations-non-vanish-3}
     \\
  &&
     \left[
     \hat{a}_{n2}^{\dagger}(\Omega)
     ,
     \hat{a}_{n1}(\Omega')
     \right]
     =
     -
     2 \pi i \delta(\Omega-\Omega')
     .
     \label{eq:hatan1-hatan2-commutation-relations-non-vanish-4}
\end{eqnarray}
We also obtain the corresponding commutation relations for the
amplitude- and phase-quadratures $\hat{d}_{n1}$ and $\hat{d}_{n2}$.
From these commutation relations
(\ref{eq:hatan1-hatan2-commutation-relations-non-vanish-1})--(\ref{eq:hatan1-hatan2-commutation-relations-non-vanish-4})
for the operators $\hat{a}_{n1}$ and $\hat{a}_{n2}$, and corresponding
commutation relations for the operators $\hat{d}_{n1}$ and
$\hat{d}_{n2}$, we can confirm the commutation relation
(\ref{eq:commutation-of--hat-bTheta}).


Through the operators
$\FrakD_{d}^{\dagger}\hat{b}_{n\Theta}(\Omega)\FrakD_{d}$ above, the
straightforward calculations for the right-hand side of
Eq.~(\ref{eq:Measured-noise-spectral-density-DCreadout-bTheta}) yields
\begin{eqnarray}
  S_{Nn}(\Omega)
  &=&
      \omega_{0}^{2}
      |\FrakB|^{2}
      \left[
      1
      -
      \kappa
      \sin(2\Theta)
      \cos(2\theta)
      \right.
      \nonumber\\
  && \quad\quad\quad\quad
      \left.
      +
      \frac{1}{2}
      \kappa^{2}
      \left( 1 - \cos(2\Theta) \cos(2\theta) \right)
      \right]
      .
     \nonumber\\
      \label{eq:Glaubers-photon-number-noise-spectral-density}
\end{eqnarray}
From this expression
(\ref{eq:Glaubers-photon-number-noise-spectral-density}) of the
stationary noise-spectral density and comparing with
Eq.~(\ref{eq:Kimble-noise-spectral-density-explicit}), we can see that
$\Theta$ $=$ $\pi/2$ gives $\left. S_{Nn}(\Omega)
\right|_{\Theta=\pi/2}$ $\propto$ $S_{(K)}$.


\begin{figure}
  \begin{center}
    \includegraphics[width=0.5\textwidth]{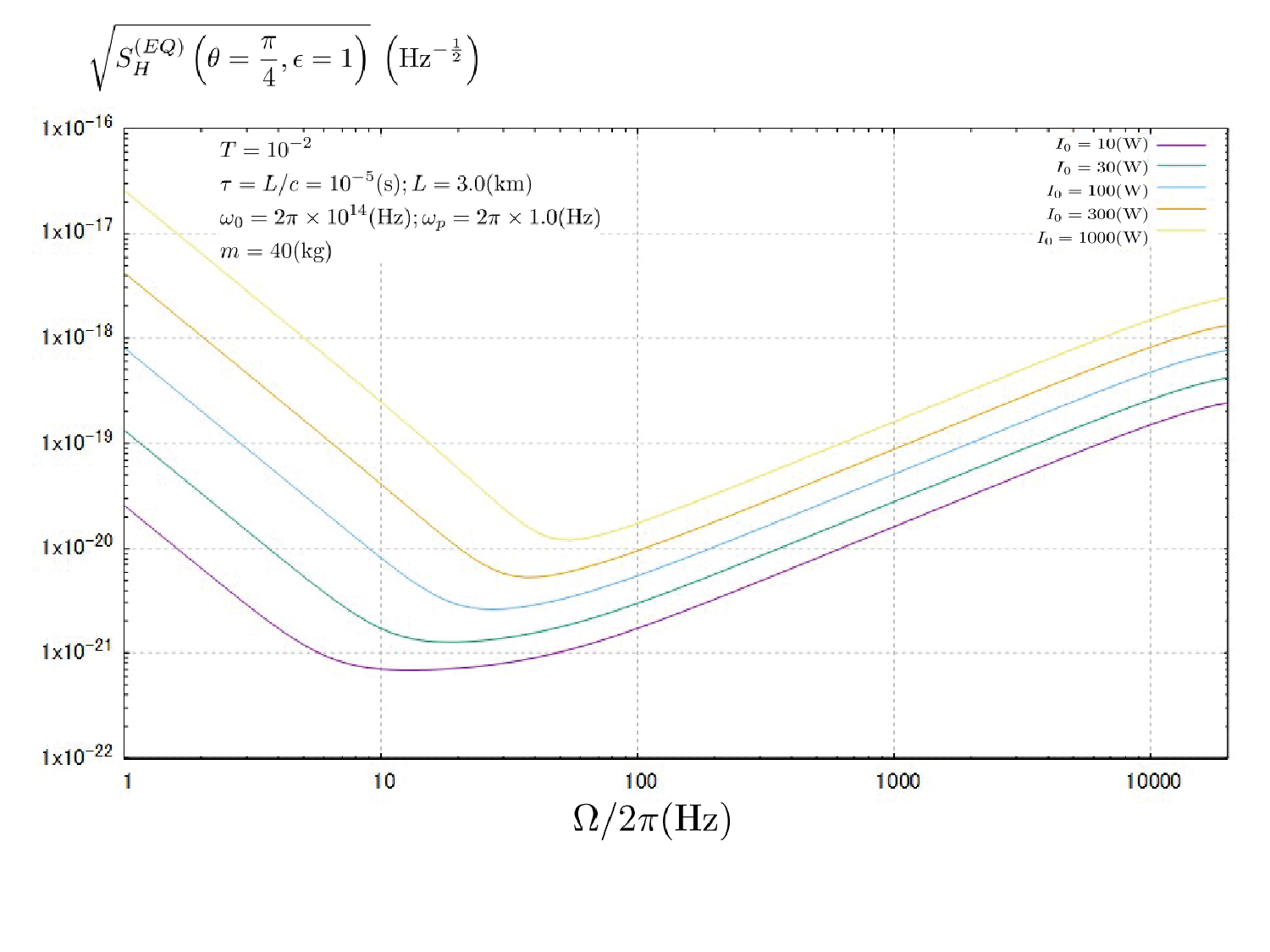}
  \end{center}
  \caption{
    The square root of the noise spectral density $\sqrt{S_{H}}$ for
    the $m_{EM}=m_{ITM}$ and the optimal offset $\theta=\pi/4$ case.
    (EQ) indicates $m_{EM}=m_{ITM}$ case.
    The noise spectral density $S_{H}$ is defined by
    Eq.~(\ref{eq:strain-referred-noise-spectrum-density}).
    $\epsilon$ is the incomplete parameter discussed
    Sec.~\ref{sec:Changing_Tuning-Point}, while $\epsilon=0$
    corresponds to the fact that the tuning point is the equilibrium
    point of the pendulum and $\epsilon=1$ corresponds to the tuning
    without taking into account of the displacement of the equilibrium
    point of the pendulum due to the classical radiation pressure from
    the laser.
    (See Sec.~\ref{sec:Changing_Tuning-Point}.)
  }
  \label{fig:SHEQFULL-1-20k}
\end{figure}


\begin{figure}
  \begin{center}
    \includegraphics[width=0.5\textwidth]{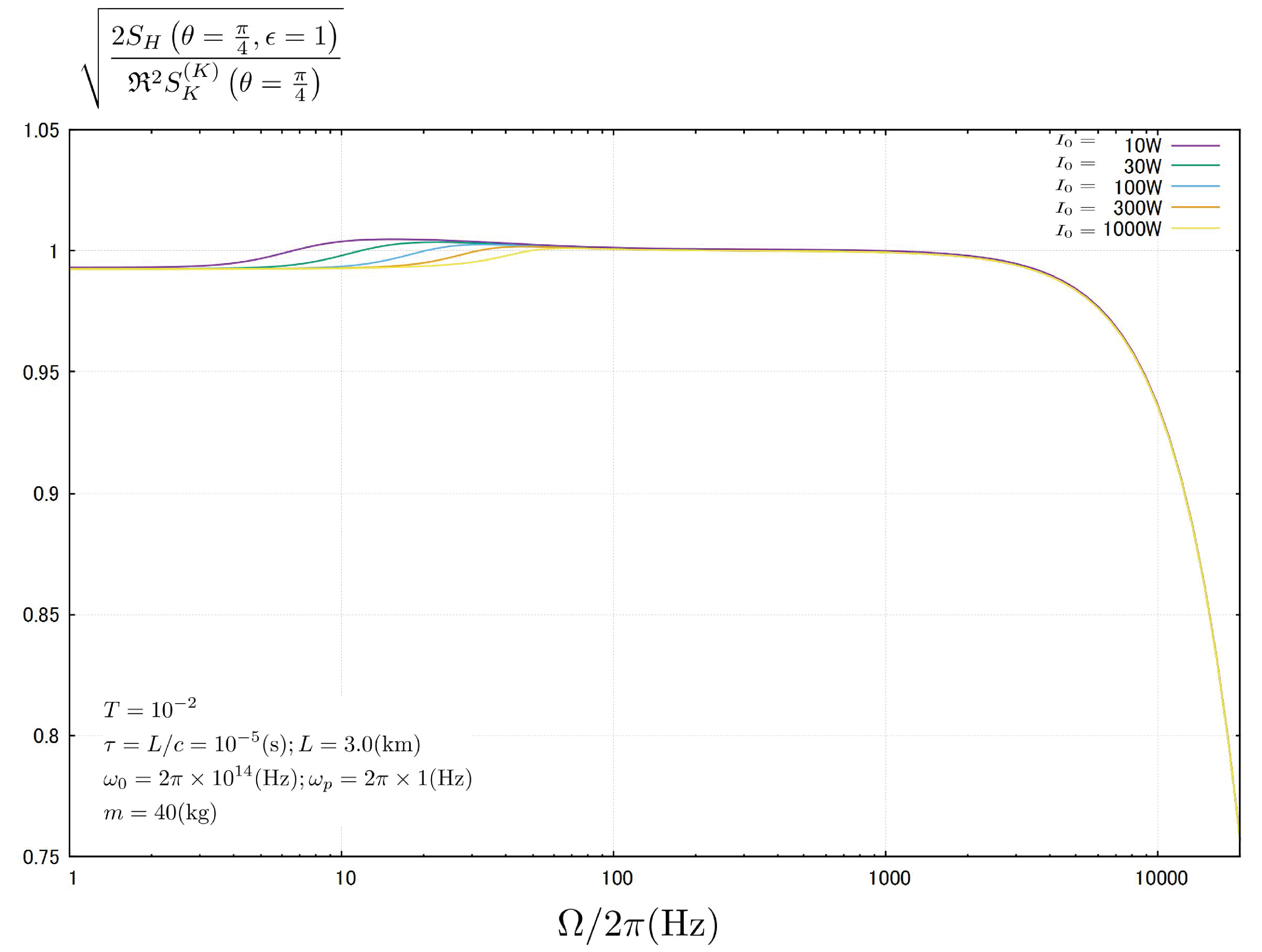}
  \end{center}
  \caption{
    The square root ratio of the noise spectral density
    $S_{H}^{(EQ)}/\FrakR^{2}$ and $S_{(K)}^{(K)}$ where
    $S_{(K)}^{(K)}$ is defined by Eq.~(\ref{eq:SKK-def}).
    This ratio is approximately unity except in the high-frequency
    range $\sim 20$ kHz.
    At the high frequency range $\sim 20$kHz, the noise spectral
    density $\sqrt{2S_{H}^{(EQ)}}/\FrakR$ is 20\% smaller than the
    noise spectral density $S_{(K)}^{(K)}$.
    We also observe a small power dependence around 10 Hz, indicating
    roughly 1\% differences from $S_{(K)}^{(K)}$.
  }
  \label{fig:SQRTSHEQFULLoverFrakRSKK-1-20k}
\end{figure}


From Eq.~(\ref{eq:signal-operator-Fourier-exp}) with
$\Omega\ll\omega_{0}$, Eqs.~(\ref{eq:input-output-relation-FrakA}) and
(\ref{eq:input-output-relation-FrakB}), the square of the absolute
value of Eq.~(\ref{eq:GW-signal-operator-exp-value}) is also given by
\begin{eqnarray}
  |\langle\hat{s}_{\calN_{b}}(\Omega)\rangle|^{2}
  &=&
      \omega_{0}^{2}
      |\FrakB|^{2}
      \frac{1}{1 + \FrakR^{2}}
      \kappa
      \cos^{2}\theta
      \frac{4}{h_{SQL}^{2}}
      \left|H(\Omega,L)\right|^{2}
      \nonumber\\
  &&
      +
      O(|\FrakB|^{1},|\FrakB|^{0})
      .
      \label{eq:exp-val-signal-operator-large-B-small-Omega-abs2-2}
\end{eqnarray}
Furthermore, through the definition of $\Theta$
($\FrakB=:|\FrakB|e^{i\Theta}$), we obtain
\begin{eqnarray}
  \label{eq:sinTheta-cosTheta-FrakR}
  \sin(2\Theta) = \frac{2\FrakR}{\FrakR^{2}+1}, \quad
  \cos(2\Theta) = \frac{\FrakR^{2}-1}{\FrakR^{2}+1}.
\end{eqnarray}
Then, the signal-to-noise ratio at the photodetector is evaluated as
$S_{Nn}(\Omega)/|\langle \hat{s}_{\calN_{b}}(\Omega)\rangle|^{2}$.
Since we regard the Fourier transformation $H(\Omega,L)$ of the 
gravitational waves as the signal of our measurement, the
signal-to-noise ratio at the photodetector is converted to the
signal-referred noise $S_{H}(\Omega)$ by
\begin{eqnarray}
  \label{eq:signal-referred-noise-spectral-density-def}
  \frac{S_{H}(\Omega)}{|H(\Omega,L)|^{2}}
  :=
  \frac{S_{Nn}(\Omega)}{|\langle \hat{s}_{\calN}(\Omega)\rangle|^{2}}
  .
\end{eqnarray}
Through Eqs.~(\ref{eq:Glaubers-photon-number-noise-spectral-density}),
(\ref{eq:exp-val-signal-operator-large-B-small-Omega-abs2-2}), and
(\ref{eq:sinTheta-cosTheta-FrakR}), the signal-referred noise spectral
density $S_{H}(\Omega)$ defined by
Eq.~(\ref{eq:signal-referred-noise-spectral-density-def}) is given by
\begin{eqnarray}
  S_{H}(\Omega)
  &=&
      \frac{h_{SQL}^{2}}{4\cos^{2}\theta}
      \left[
      \left( \frac{1}{\kappa} + \frac{\kappa}{2} \left( 1 + \cos(2\theta) \right) \right)
      - 2 \FrakR \cos(2\theta)
      \right.
      \nonumber\\
  && \quad\quad\quad\quad
      \left.
      +   \left( \frac{1}{\kappa} + \frac{\kappa}{2} \left( 1 - \cos(2\theta) \right) \right) \FrakR^{2}
      \right]
      .
      \label{eq:strain-referred-noise-spectrum-density}
\end{eqnarray}


In the case where $\FrakR\rightarrow 0$, we realize Kimble's noise
spectral density up to the factor $1/2$: $S_{H}(\Omega)$ $=$
$S_{(K)}(\Omega)/2$.
In the situation $\theta=\pi/4$, the signal referred stationary
noise-spectral density $S_{H}(\Omega)$ is given by
\begin{eqnarray}
  S_{H}(\Omega,\theta=\pi/4)
  &=&
      \frac{h_{SQL}^{2}}{2}
      \left( \frac{1}{\kappa} + \frac{\kappa}{2} \right)
      \left(
      1 + \FrakR^{2}
      \right)
      .
      \label{eq:strain-referred-noise-spectrum-density-piover4}
\end{eqnarray}

\begin{figure}
  \begin{center}
    \includegraphics[width=0.5\textwidth]{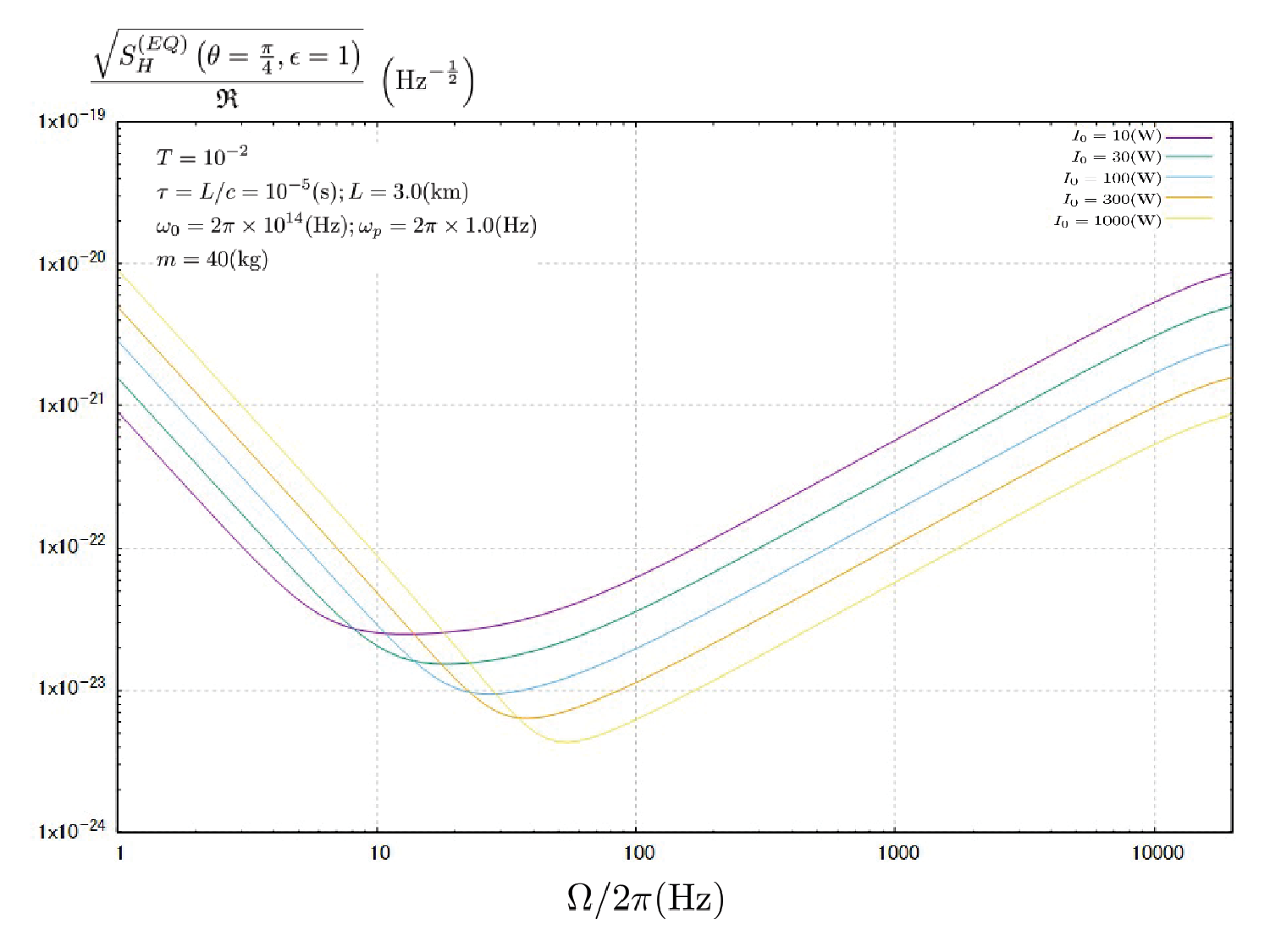}
  \end{center}
  \caption{
    The square root of the noise spectral density
    $\sqrt{S_{H}}/\FrakR$ with equal mass $(EQ)$ case where
    $m_{EM}=m_{ITM}$ and the offset is $\theta=\pi/4$.
    The behavior of this ratio is similar to Kimble's signal-referred
    noise spectral density $\sqrt{S_{K}}$.
  }
  \label{fig:SQRTSHEQFULLoverFrakR-1-20k}
\end{figure}

The square root signal-referred stationary noise spectral
density $S_{H}(\Omega)$ with the offset $\theta=\pi/4$ in the case
$m_{EM}=m_{ITM}$ is depicted in Fig.~\ref{fig:SHEQFULL-1-20k}.
Since the noise spectral density $S_{H}$ coincides with $S_{K}/2$ and
$\FrakR\neq 0$ case, the dominant order of $\FrakR$ in $S_{H}$ is
$O(\FrakR^{2})$.
Therefore, we show the ratio
\begin{eqnarray}
  \label{eq:sqr2sHoverFrakRSKK}
  \sqrt{\frac{2S_{H}}{\FrakR^{2} S_{(K)}^{(K)}}}
\end{eqnarray}
in Fig.~\ref{fig:SQRTSHEQFULLoverFrakRSKK-1-20k} to clarify the
difference between the derived noise-spectral density $S_{H}$ and the
original Kimble's noise spectral density.
This ratio is approximately unity except for the high-frequency range
$\sim 20$kHz.
At the high frequency range $\sim 20$kHz, the noise spectral density
$\sqrt{2S_{H}^{(EQ)}}/\FrakR$ is 20\% smaller than the
noise spectral density $S_{(K)}^{(K)}$ defined by
Eq.~(\ref{eq:SKK-def}).
We also observe a small power dependence around 10 Hz, indicating
roughly 1\% differences from $S_{(K)}^{(K)}$.
Since the expectation value
$|\langle\hat{s}_{\calN_{b}}(\Omega)\rangle|^{2}$ given by
Eq.~(\ref{eq:GW-signal-operator-exp-value}) is the largest when
$\theta=\pi/4$, we show
Figures~\ref{fig:SHEQFULL-1-20k},
\ref{fig:SQRTSHEQFULLoverFrakRSKK-1-20k},
and~\ref{fig:SQRTSHEQFULLoverFrakR-1-20k} with $\theta=\pi/4$.


The factor $\FrakR$ in the signal referred noise spectral density
$S_{H}(\Omega)$ comes from (\ref{eq:input-output-relation-FrakB}).
Equation~(\ref{eq:input-output-relation-FrakB}) indicates that the
classical carrier field is modified by $\FrakR$.
In DC readout schemes, we use the classical carrier field as the
reference to measure gravitational-wave signals.
Due to this modification of the classical carrier field appears to the
signal referred noise spectral density $S_{H}(\Omega)$, although it
does not appear in the expectation value
(\ref{eq:GW-signal-operator-exp-value}) of the output signal due to
the reality condition of the gravitational-wave signals
$H(\Omega)=H^{*}(-\Omega)$.
Furthermore, we also note that the existence of the factor $\FrakR$
leads to the signal-referred noise spectral density in which
the shot-noise contribution is not reduced even if the
injected power $I_{0}$ is increased.
This contradicts the consensus in the community of
gravitational-wave experiments.


Originally, the additional classical carrier field $\FrakR$ comes from
the solutions
(\ref{eq:DdaggerD-rad-pres-com-DA-Dmono-tuned-wobifreq-sol-sum-approx})
and (\ref{eq:Eq-of-motion-XYITM-com-tuned-wobifreq-sol-sum-approx}) to
the Heisenberg equations as the terms proportional to $2\pi\delta(\Omega)$.
Furthermore, these terms come from the radiation pressure forces
(\ref{eq:DdaggerD-rad-pres-F-XITM-DAZ-Dmono-hineg-toXITM-tautune})--(\ref{eq:DdaggerD-rad-pres-Frpy-Dmono-hineg-tautune})
as the classical constant force in the time domain.
This classical constant force in the time domain appears in the
Heisenberg equations of motion (\ref{eq:Eq-of-motion-XITM-4}),
(\ref{eq:Eq-of-motion-YITM-4}),
(\ref{eq:Eq-of-motion-X-mirrors-differential-original}), and
(\ref{eq:Eq-of-motion-Y-mirrors-differential-original}).
In both classical and quantum mechanics of a forced harmonic
oscillator, the constant force in the harmonic oscillator leads to
changes in the equilibrium points of the oscillator.
In actual experiments, the interferometers should be operated at the
mirrors' equilibrium positions.
Therefore, to take into account the effects of the constant forces,
we have to change the tuning condition (\ref{eq:omega0-L-tune-cond})
so that the laser is tuned at the equilibrium points of mirrors.
Therefore, we have to consider the change of the tuning condition
(\ref{eq:omega0-L-tune-cond}) in the next
section~\ref{sec:Changing_Tuning-Point}.
To do this, we have to reconsider the Heisenberg equations for mirrors
in Sec.~\ref{sec:Eq_for_mirrors'_motions_and_their_solutions}.


\section{Changing Tuning Point}
\label{sec:Changing_Tuning-Point}


\subsection{Complete equilibrium tuning}
\label{sec:Complete_equilibrium_tuning}


Now, we consider the modification of the tuning points from the tuning
conditions (\ref{eq:omega0-L-tune-cond}) and
(\ref{eq:omega0-lxly-tune-cond}).
First, we consider the modification of the tuning condition
(\ref{eq:omega0-L-tune-cond}).
To do this, we note that the cavity propagation
conditions~(\ref{eq:xarm-retarded-effect-g-to-gprime}) and
(\ref{eq:yarm-retarded-effect-g-to-gprime}) yield that the retarded
effects in the electric field operators are determined by
\begin{eqnarray}
  \tau + \frac{1}{c}\hat{X}_{x}(t-\tau)
  &=&
      \frac{L + \hat{X}_{XEM}(t-\tau) - \hat{X}_{XITM}(t-\tau)}{c}
      ,
      \nonumber\\
  \label{eq:retarded-effect-cavity-X}
  \\
  \tau + \frac{1}{c}\hat{X}_{y}(t-\tau)
  &=&
      \frac{L + \hat{X}_{YEM}(t-\tau) - \hat{X}_{YITM}(t-\tau)}{c}
      .
      \nonumber\\
  \label{eq:retarded-effect-cavity-Y}
\end{eqnarray}
Even when we evaluate the radiation pressure forces which affect XEM
and YEM, we use Eqs.~(\ref{eq:radiation-pressure-force-XEM}) and
(\ref{eq:radiation-pressure-force-YEM}) in which the retarded effects
in the electric field operator are determined by
Eqs.~(\ref{eq:retarded-effect-cavity-X}) and
(\ref{eq:retarded-effect-cavity-Y}).
From Eqs.~(\ref{eq:retarded-effect-cavity-X}) and
(\ref{eq:retarded-effect-cavity-Y}), we can see that if the operators
$\hat{X}_{XEM} - \hat{X}_{XITM}$ and $\hat{X}_{YEM} - \hat{X}_{YITM}$
have the constant terms in time, we may always include these constant
terms in $L$, i.e., in $\tau=L/c$.


As we see in
Appendix~\ref{sec:Explicit_evaluation_of_the_radiation_pressure_forces},
the Fourier transformations
(\ref{eq:DdaggerD-rad-pres-F-XEM-in-DA-Dmono-hineg}) and
(\ref{eq:DdaggerD-rad-pres-F-YEM-in-DA-Dmono-hineg}) of the radiation
pressure forces to XEM and YEM, respectively, include the term
proportional to $2\pi\delta(\Omega)$.
In the time domain, these terms represent the constant forces in time.
These terms are the same in the radiation pressure forces for XEM and YEM.
We denote these terms, which are proportional to $2\pi\delta(\Omega)$,
as
\begin{eqnarray}
  &&
     \omega_{p}^{2} \scrD_{EM} 2\pi \delta(\Omega)
     \nonumber\\
  &:=&
       N^{2} T^{2} \frac{\hbar\omega_{0}}{m_{EM}c}
       \left[ 1 - \sqrt{1-T} e^{+2i\omega_{0}\tau} \right]^{-1}
       \nonumber\\
  && \quad
     \times
       \left[ 1 - \sqrt{1-T} e^{ - 2i\omega_{0}\tau} \right]^{-1}
       2 \pi \delta(\Omega)
     .
       \label{eq:classical-terms-of-RPF-EM}
\end{eqnarray}
On the other hand, for ITMs, the Fourier transformations
(\ref{eq:DdaggerD-rad-pres-F-XITM-DAZ-Dmono-hineg}) and
(\ref{eq:DdaggerD-rad-pres-F-YITM-in-DA-Dmono-hineg}) of the radiation
pressure forces affect XITM and YITM, respectively, and include the
term proportional to $2\pi\delta(\Omega)$.
In the time domain, these terms represent the constant forces in time.
These terms are the same in the radiation pressure forces for XITM and
YITM.
We denote these terms, which are proportional to $2\pi\delta(\Omega)$,
as
\begin{eqnarray}
  &&
     \omega_{p}^{2} \scrD_{ITM} 2\pi \delta(\Omega)
     \nonumber\\
  &:=&
       - 2 N^{2} \sqrt{1-T} \frac{\hbar\omega_{0}}{m_{ITM}c}
       \left[ 1 - \sqrt{1-T} e^{+2i\omega_{0}\tau} \right]^{-1}
       \nonumber\\
  && \quad
     \times
       \left[ 1 - \sqrt{1-T} e^{ - 2i\omega_{0}\tau} \right]^{-1}
       \nonumber\\
  && \quad
     \times
       \left[ \cos(2\omega_{0}\tau)  - \sqrt{1-T} \right]
       2 \pi \delta(\Omega)
     .
       \label{eq:classical-terms-of-RPF-ITM}
\end{eqnarray}


================ at 17:50 on 13th April in 2026

Keep in our mind the existence of these constant terms $\scrD_{EM}$
and $\scrD_{ITM}$, the Heisenberg equations
(\ref{eq:Eq-of-motion-XITM-4})--(\ref{eq:Eq-of-motion-YEM-4}) of
mirrors' motion are given by
\begin{eqnarray}
  \label{eq:Eq-of-motion-XITM-separation-classical-quantm}
  \!\!\!\!
  \frac{d^{2}}{dt^{2}}\hat{X}_{XITM}
  &=&
      -
      \omega_{p}^{2} \hat{X}_{XITM}
      + \omega_{p}^{2} \scrD_{ITM}
      \nonumber\\
  &&
      +
      \frac{1}{m_{ITM}}
      \hat{F}^{(fluc)}_{rpXITM}
      ,
\end{eqnarray}
\begin{eqnarray}
  \label{eq:Eq-of-motion-YITM-separation-classical-quantm}
  \!\!\!\!
  \frac{d^{2}}{dt^{2}}\hat{X}_{YITM}
  &=&
      -
      \omega_{p}^{2} \hat{X}_{XITM}
      + \omega_{p}^{2} \scrD_{ITM}
      \nonumber\\
  &&
      +
      \frac{1}{m_{ITM}}
      \hat{F}^{(fluc)}_{rpYITM}
      ,
\end{eqnarray}
\begin{eqnarray}
  \label{eq:Eq-of-motion-XEM-separation-classical-quantm}
  \!\!\!\!
  \frac{d^{2}}{dt^{2}}\hat{X}_{XEM}
  &=&
      -
      \omega_{p}^{2} \hat{X}_{XEM}
      + \omega_{p}^{2} \scrD_{EM}
      \nonumber\\
  &&
      +
      \frac{1}{m_{EM}}
      \hat{F}^{(fluc)}_{rpXEM}
      +
      \frac{1}{2} L \frac{d^{2}}{dt^{2}}h(t,L)
      ,
\end{eqnarray}
\begin{eqnarray}
  \label{eq:Eq-of-motion-YEM-separation-classical-quantm}
  \!\!\!\!
  \frac{d^{2}}{dt^{2}}\hat{X}_{YEM}
  &=&
         -
         \omega_{p}^{2} \hat{X}_{YEM}
         + \omega_{p}^{2} \scrD_{EM}
      \nonumber\\
  &&
     +
     \frac{1}{m_{EM}}
     \hat{F}^{(fluc)}_{rpYEM}
     -
     \frac{1}{2} L \frac{d^{2}}{dt^{2}}h(t,L)
     ,
\end{eqnarray}
where
$\hat{F}^{(fluc)}_{rpXITM}$,
$\hat{F}^{(fluc)}_{rpYITM}$,
$\hat{F}^{(fluc)}_{rpXEM}$, and
$\hat{F}^{(fluc)}_{rpYEM}$ the fluctuation parts of the
radiation pressure forces.
These do not include the classical constant part of the forces.
The solutions to
Eqs.~(\ref{eq:Eq-of-motion-XITM-separation-classical-quantm})--(\ref{eq:Eq-of-motion-YEM-separation-classical-quantm})
are given by
\begin{eqnarray}
  \label{eq:EOM-XITM-separation-classical-quantm-sol}
  \hat{X}_{XITM}
  &=:&
      \hat{X}^{(fluc)}_{XITM}
      + \scrD_{ITM}
      ,
\end{eqnarray}
\begin{eqnarray}
  \label{eq:EOM-YITM-separation-classical-quantm-sol}
  \hat{X}_{YITM}
  &=:&
      \hat{X}^{(fluc)}_{XITM}
      + \scrD_{ITM}
      ,
\end{eqnarray}
\begin{eqnarray}
  \label{eq:EOM-XEM-separation-classical-quantm-sol}
  \hat{X}_{XEM}
  &=:&
      \hat{X}^{(fluc)}_{XEM}
      + \scrD_{EM}
      ,
\end{eqnarray}
\begin{eqnarray}
  \label{eq:EOM-YEM-separation-classical-quantm-sol}
  \hat{X}_{YEM}
  &=:&
      \hat{X}^{(fluc)}_{YEM}
      + \scrD_{EM}
     ,
\end{eqnarray}
where $\hat{X}^{(fluc)}_{XITM}$, $\hat{X}^{(fluc)}_{YITM}$,
$\hat{X}^{(fluc)}_{XEM}$, and $\hat{X}^{(fluc)}_{YEM}$ are solutions
to
Eqs.~(\ref{eq:Eq-of-motion-XITM-separation-classical-quantm})--(\ref{eq:Eq-of-motion-YEM-separation-classical-quantm})
with $\scrD_{ITM}=\scrD_{EM}=0$.


Through the expressions
(\ref{eq:EOM-XITM-separation-classical-quantm-sol})--(\ref{eq:EOM-YEM-separation-classical-quantm-sol}),
the retarded effects (\ref{eq:retarded-effect-cavity-X}) and
(\ref{eq:retarded-effect-cavity-Y}) are given by
\begin{eqnarray}
  \tau + \frac{1}{c}\hat{X}_{x}(t-\tau)
  &=&
      \frac{L + \scrD_{EM} - \scrD_{ITM}}{c}
      \nonumber\\
  &&
      +
      \frac{1}{c} \hat{X}_{x}^{(fluc)}(t-\tau)
      \nonumber\\
  &=:&
      \tau^{(o)}
      +
      \frac{1}{c} \hat{X}_{x}^{(fluc)}(t-\tau)
      \nonumber\\
  &=&
      \tau^{(o)}
      +
      \frac{1}{c} \hat{X}_{x}^{(fluc)}(t-\tau^{(o)})
      +
      O(X_{x}^{2})
      ,
      \nonumber\\
  \label{eq:retarded-effect-cavity-X-renorm}
  \\
  \tau + \frac{1}{c}\hat{X}_{y}(t-\tau)
  &=&
      \frac{L + \scrD_{EM} - \scrD_{ITM}}{c}
      \nonumber\\
  &&
      +
      \frac{1}{c} \hat{X}_{y}^{(fluc)}(t-\tau)
      \nonumber\\
  &=:&
      \tau^{(o)}
      +
      \frac{1}{c} \hat{X}_{y}^{(fluc)}(t-\tau)
      \nonumber\\
  &=&
      \tau^{(o)}
      +
      \frac{1}{c} \hat{X}_{y}^{(fluc)}(t-\tau^{(o)})
      +
      O(X_{y}^{2})
      .
      \nonumber\\
  \label{eq:retarded-effect-cavity-Y-renom}
\end{eqnarray}
where
$\hat{X}_{x}^{(fluc)}:=\hat{X}^{(fluc)}_{XEM}-\hat{X}_{XITM}^{(fluc)}$ and
$\hat{X}_{y}^{(fluc)}:=\hat{X}^{(fluc)}_{YEM}-\hat{X}_{YITM}^{(fluc)}$.
Here, $\tau^{(o)}$ is regarded as the operation point of the cavity
length of the Fabry-P\'erot interferometer.
Equations
(\ref{eq:retarded-effect-cavity-X-renorm}) and
(\ref{eq:retarded-effect-cavity-Y-renom}) indicate that we may replace
\begin{eqnarray}
  \label{eq:operation-point-replacements-cavity-tau}
  && \tau \rightarrow \tau^{(o)}, \\
  \label{eq:operation-point-replacements-cavity-Xxy}
  && \hat{X}_{x,y}(t-\tau) \rightarrow \hat{X}_{x,y}^{(fluc)}(t-\tau^{(o)})
\end{eqnarray}
in Eqs.~(\ref{eq:xarm-retarded-effect-g-to-gprime}),
(\ref{eq:yarm-retarded-effect-g-to-gprime}),
(\ref{eq:xyarm-prop-fprime-and-f-g})--(\ref{eq:yarm-prop-g-and-f-g}),
(\ref{eq:radiation-pressure-force-XEM}),
(\ref{eq:radiation-pressure-force-YEM}), and any equations derived
from them within the accuracy up to $O(X_{x,y}^{2})$.


On the other hand, the retarded effects between BS and ITMs, which are
evaluated in Eqs.~(\ref{eq:xarm-retarded-effect-f'-to-c'}),
(\ref{eq:yarm-retarded-effect-f'-to-c'}),
(\ref{eq:xarm-retarded-effect-c-to-f-mod}), and
(\ref{eq:yarm-retarded-effect-c-to-f-mod}) are determined by
\begin{eqnarray}
  \label{eq:retarded-effect-cavity-XITM}
  \tau_{x}' + \frac{1}{c}\hat{X}_{XITM}(t)
  &=&
      \frac{l_{x} + \hat{X}_{XITM}(t)}{c}
      ,
  \\
  \label{eq:retarded-effect-cavity-YITM}
  \tau_{y}' + \frac{1}{c}\hat{X}_{YITM}(t)
  &=&
      \frac{l_{y} + \hat{X}_{YITM}(t)}{c}
      .
\end{eqnarray}
Furthermore, we applied the tuning conditions
(\ref{eq:omega0-lxly-tune-cond}) to $(l_{x}+l_{y})/(2c)$.
However, similarly to the cases of
Eqs.~(\ref{eq:retarded-effect-cavity-X}) and
(\ref{eq:retarded-effect-cavity-Y}), $l_{x}$ and $l_{y}$ are also
modified by the constant term in $\hat{X}_{XITM}(t)$ and
$\hat{X}_{YITM}(t)$.
As shown in
Eqs.~(\ref{eq:Eq-of-motion-XITM-separation-classical-quantm}) and
(\ref{eq:Eq-of-motion-YITM-separation-classical-quantm}), we can
separate the solutions $\hat{X}_{XITM}$ and $\hat{X}_{YITM}$ to
Eqs.~(\ref{eq:Eq-of-motion-XITM-separation-classical-quantm}) and
(\ref{eq:Eq-of-motion-YITM-separation-classical-quantm}) into the
constant terms $\scrD_{ITM}$ and the fluctuation terms
$\hat{X}_{XITM}^{(fluc)}$ and $\hat{X}_{YITM}^{(fluc)}$.
Then, as in the cases of
Eqs.~(\ref{eq:retarded-effect-cavity-X-renorm}) and
(\ref{eq:retarded-effect-cavity-Y-renom}), we may replace
Eqs.~(\ref{eq:retarded-effect-cavity-XITM}) and
(\ref{eq:retarded-effect-cavity-YITM}) as
\begin{eqnarray}
  \tau_{x}' + \frac{1}{c}\hat{X}_{XITM}(t)
  &=&
      \frac{l_{x} + \scrD_{ITM}}{c}
      +
      \frac{1}{c} \hat{X}_{XITM}^{(fluc)}(t)
      \nonumber\\
  &=:&
      \tau_{x}^{(o)'}
      +
      \frac{1}{c} \hat{X}_{XITM}^{(fluc)}(t)
      ,
       \label{eq:retarded-effect-cavity-XITM-renorm}
  \\
  \tau_{y}' + \frac{1}{c}\hat{X}_{YITM}(t)
  &=&
      \frac{l_{y} + \scrD_{ITM}}{c}
      +
      \frac{1}{c} \hat{X}_{YITM}^{(fluc)}(t)
      \nonumber\\
  &=:&
      \tau_{y}^{(o)'}
      +
      \frac{1}{c} \hat{X}_{YITM}^{(fluc)}(t)
      .
       \label{eq:retarded-effect-cavity-YITM-renorm}
\end{eqnarray}
As in the cases of Eqs.~(\ref{eq:retarded-effect-cavity-X-renorm}) and
(\ref{eq:retarded-effect-cavity-Y-renom}), $\tau_{x}^{(o)'}$ and
$\tau_{y}^{(o)'}$ is regarded as the operation points of the length
between BS and ITMs in the Fabry-P\'erot interferometer.
Equation (\ref{eq:retarded-effect-cavity-XITM-renorm}) and
(\ref{eq:retarded-effect-cavity-YITM-renorm}) indicate that we may replace
\begin{eqnarray}
  \label{eq:operation-point-replacements-BS-XITM-length-tauxprime}
  && \tau_{x}' \rightarrow \tau_{x}^{(o)'}, \\
  \label{eq:operation-point-replacements-BS-YITM-length-tauyprime}
  && \tau_{y}' \rightarrow \tau_{y}^{(o)'}, \\
  \label{eq:operation-point-replacements-XITM}
  && \hat{X}_{XITM}(t) \rightarrow \hat{X}_{XITM}^{(fluc)}(t), \\
  \label{eq:operation-point-replacements-YITM}
  && \hat{X}_{YITM}(t) \rightarrow \hat{X}_{YITM}^{(fluc)}(t)
\end{eqnarray}
in Eqs.~(\ref{eq:xarm-retarded-effect-f'-to-c'}),
(\ref{eq:yarm-retarded-effect-f'-to-c'}),
(\ref{eq:xarm-retarded-effect-c-to-f-mod}),
(\ref{eq:yarm-retarded-effect-c-to-f-mod}), and any equations derived
from them.


We note that Eqs.(\ref{eq:operation-point-replacements-cavity-tau}),
(\ref{eq:operation-point-replacements-BS-XITM-length-tauxprime}), and
(\ref{eq:operation-point-replacements-BS-YITM-length-tauyprime}) are
regarded as the ``renormalizations'' of $L$, $l_{x}$, and $l_{y}$,
respectively.
After these renormalizations of $L$, $l_{x}$, and $l_{y}$, we apply
the tuning conditions
\begin{eqnarray}
  \label{eq:omega0-L-tune-cond-renorm}
  && \omega_{0} \tau^{(o)} = 2n\pi, \quad n\in\NF, \\
  \label{eq:omega0-lxly-tune-cond-renorm}
  && \omega_{0} \frac{\tau_{x}^{(o)'}+\tau_{y}^{(o)'}}{2} = 2m\pi,
     \quad m\in\NF,
\end{eqnarray}
instead of the tuning conditions (\ref{eq:omega0-L-tune-cond})
and (\ref{eq:omega0-lxly-tune-cond}).
After these renormalizations, the phase offset $\theta$ is defined by
(\ref{eq:offset-def}) is replaced by
\begin{eqnarray}
  \label{eq:offset-def-renorm}
  \theta := \omega_{0}(\tau_{y}^{(o)'}-\tau_{x}^{(o)'}).
\end{eqnarray}
Furthermore, through the renormalization
(\ref{eq:operation-point-replacements-cavity-tau}) and the tuning
condition (\ref{eq:omega0-L-tune-cond-renorm}), $\scrD_{EM}$ and
$\scrD_{ITM}$ defined by Eqs.~(\ref{eq:classical-terms-of-RPF-EM}) and
(\ref{eq:classical-terms-of-RPF-ITM}) are given by
\begin{eqnarray}
  \scrD_{EM}
  &=&
      T^{2} \frac{I_{0}}{m_{EM}c\omega_{p}^{2}}
      \left[ 1 - \sqrt{1-T} \right]^{-2}
      \label{eq:classical-terms-of-RPF-EM-renorm-retune}
      \\
  &\sim&
         + \frac{4I_{0}}{m_{EM}c\omega_{p}^{2}} \frac{1}{T}
      \label{eq:classical-terms-of-RPF-EM-renorm-retune-leading}
  \\
  &\sim&
         + 3.0\times 10^{-7}\;\mbox{m}
         \left(
         \frac{I_{0}}{10^{2}\;\mbox{W}}
         \right)
         \left(
         \frac{40\;\mbox{kg}}{m_{EM}}
         \right)
         \nonumber\\
  && \quad\quad\quad
     \times
         \left(
         \frac{2\pi\times 1\;\mbox{Hz}}{\omega_{p}}
         \right)
         \left(
         \frac{10^{-2}}{T}
         \right)
         .
      \label{eq:classical-terms-of-RPF-EM-renorm-retune-leading-order}
\end{eqnarray}
and
\begin{eqnarray}
  \scrD_{ITM}
  &=&
      - 2 \sqrt{1-T} \frac{I_{0}}{m_{ITM}c\omega_{p}^{2}}
      \left[ 1 - \sqrt{1-T} \right]^{-1}
  \label{eq:classical-terms-of-RPF-ITM-renorm-retune}
  \\
  &\sim&
         - \frac{4I_{0}}{m_{ITM}c\omega_{p}^{2}} \frac{1}{T}
  \label{eq:classical-terms-of-RPF-ITM-renorm-retune-leading}
  \\
  &\sim&
         - 3.0\times 10^{-7}\;\mbox{m}
         \left(
         \frac{I_{0}}{10^{2}\;\mbox{W}}
         \right)
         \left(
         \frac{40\;\mbox{kg}}{m_{ITM}}
         \right)
         \nonumber\\
  && \quad\quad\quad
     \times
         \left(
         \frac{2\pi\times 1\;\mbox{Hz}}{\omega_{p}}
         \right)
         \left(
         \frac{10^{-2}}{T}
         \right)
         .
  \label{eq:classical-terms-of-RPF-ITM-renorm-retune-leading-order}
\end{eqnarray}


Since all $\hat{X}_{x,y}$, $\hat{X}_{XITM}$, and $\hat{X}_{YITM}$ are
replaced by $\hat{X}_{x,y}^{(fluc)}$, $\hat{X}_{XITM}^{(fluc)}$, and
$\hat{X}_{YITM}^{(fluc)}$, respectively, their Fourier transformations
$\hat{Z}_{x,y}$, $\hat{Z}_{XITM}$, and $\hat{Z}_{YITM}$ are also
replaced to $\hat{Z}_{x,y}^{(fluc)}$, $\hat{Z}_{XITM}^{(fluc)}$, and
$\hat{Z}_{YITM}^{(fluc)}$.
These $\hat{Z}_{x,y}^{(fluc)}$, $\hat{Z}_{XITM}^{(fluc)}$, and
$\hat{Z}_{YITM}^{(fluc)}$ does not includes the classical term
which proportional to $2\pi\delta(\Omega)$.
Therefore, the resulting input-output relation corresponding to
Eq.~(\ref{eq:input-output-rel-explicit-quad-I0-kappabetahSQL}) is
given by the expression of
Eq.~(\ref{eq:input-output-rel-explicit-quad-I0-kappabetahSQL}) with
$\FrakR=0$.
In this expression of the input-output relation, there is no
additional term in the classical carrier fields $\FrakR$, we obtain
the noise spectral density $S_{H}(\Omega)= S_{(K)}(\Omega)/2$ instead
of Eq.~(\ref{eq:strain-referred-noise-spectrum-density}).
This recovers the conventional noise-spectral density except for the
fact that we have to use the definitions (\ref{eq:kappa-def}) and
(\ref{eq:hSQL-def}) of the variables $h_{SQL}$ and $\kappa$ with the
replacement $\tau\rightarrow\tau^{(o)}$, respectively.
The difference between the resulting noise spectral density
$S_{H}(\Omega)=S_{(K)}(\Omega)/2$ and Kimble's noise spectral density
$S_{(K)}^{(K)}$~\cite{H.J.Kimble-Y.Levin-A.B.Matsko-K.S.Thorne-S.P.Vyatchanin-2001}
are depicted in Fig.~\ref{fig:SKEQ-SKK-ratio-1-20k} except for the
factor $1/2$ in $S_{H}(\Omega)$ as discussed in
Sec.~\ref{sec:DC_readout_scheme_for_FP_GW_Detector}.


Since $\hat{X}_{XITM}=\scrD_{ITM}$, $\hat{X}_{YITM}=\scrD_{ITM}$,
$\hat{X}_{XEM} = \scrD_{EM}$, and $\hat{X}_{YEM} = \scrD_{EM}$ are
equilibrium points of the pendulum for mirrors', we can completely
exclude the classical signals due to the classical radiation pressure
forces $\FrakR$ in the feedback electric current.
For this reason, we call the set of the tuning conditions
(\ref{eq:omega0-L-tune-cond-renorm}) and
(\ref{eq:omega0-lxly-tune-cond-renorm}) as ``complete equilibrium
tuning.''
This complete equilibrium tuning is one of the justifications of the
ignorance of the effects of the classical part $\FrakR$, which comes 
from the radiation pressure forces.


\subsection{Incomplete equilibrium tuning}
\label{sec:Incomplete_equilibrium_tuning}


If we achieve the above complete equilibrium tuning in
experiments, we can realize Kimble's noise spectral
density~\cite{H.J.Kimble-Y.Levin-A.B.Matsko-K.S.Thorne-S.P.Vyatchanin-2001}
with the modifications of $\kappa$ and $h_{SQL}$ as an idealized
case.
However, in some experiments, we might not be able to achieve complete
equilibrium tuning for various reasons.
In this subsection, we consider the case where we cannot realize the
above complete equilibrium tuning and how this incompleteness
appears in the noise spectral densities through our consideration
within this paper.


To consider the incomplete equilibrium tuning, we introduce
``operation points'' so that
\begin{eqnarray}
  \frac{l_{x} + \hat{X}_{XITM}(t)}{c}
  &=:&
       \frac{l_{x} + (1-\epsilon) \scrD_{ITM}}{c} + \frac{\hat{X}_{XITM}^{\epsilon(o)}(t)}{c}
       \nonumber\\
  &=:&
       \frac{l_{x}^{\epsilon(o)}}{c} + \frac{1}{c} \hat{X}_{XITM}^{\epsilon(o)}(t)
       \nonumber\\
  &=:&
       \tau_{x}^{\epsilon(o)'} + \frac{1}{c} \hat{X}_{XITM}^{\epsilon(o)}(t)
       ,
       \label{eq:lx-incomplete-operation-point}
  \\
  \hat{X}_{XITM}^{\epsilon(o)}
  &:=&
       \epsilon \scrD_{ITM} + \hat{X}_{XITM}^{(fluc)}(t)
       ,
       \label{eq:XITM-incomplete-operation-point}
  \\
  \frac{l_{y} + \hat{X}_{XITM}(t)}{c}
  &=:&
       \frac{l_{y} + (1-\epsilon) \scrD_{ITM}}{c} + \frac{1}{c}\hat{X}_{YITM}^{\epsilon(o)}(t)
       \nonumber\\
  &=:&
       \frac{l_{y}^{\epsilon(o)}}{c} + \frac{1}{c}\hat{X}_{YITM}^{\epsilon(o)}(t)
       \nonumber\\
  &=:&
       \tau_{y}^{\epsilon(o)'} + \frac{1}{c}\hat{X}_{YITM}^{\epsilon(o)}(t)
       ,
       \label{eq:ly-incomplete-operation-point}
  \\
  \hat{X}_{YITM}^{\epsilon(o)}
  &:=&
       \epsilon \scrD_{ITM} + \hat{X}_{YITM}^{(fluc)}(t)
       ,
       \label{eq:YITM-incomplete-operation-point}
\end{eqnarray}
through $\scrD_{ITM}$, $\hat{X}_{XITM}^{(fluc)}(t)$, and
$\hat{X}_{YITM}^{(fluc)}(t)$ defined by
Eqs.~(\ref{eq:classical-terms-of-RPF-ITM}),
(\ref{eq:EOM-XITM-separation-classical-quantm-sol}), and
(\ref{eq:EOM-YITM-separation-classical-quantm-sol}), respectively.
Here, $\epsilon$ is a parameter of incompleteness of the tuning within
$[0,1]$.
We choose $\epsilon$ so that $\epsilon=0$ corresponds to the tuning
point, which is the equilibrium point of the pendulum, and
$\epsilon=1$ corresponds to the tuning without taking into account the
displacement of the equilibrium point of the pendulum due to the
classical radiation pressure from the laser.


As in the case of the complete equilibrium tuning, we regard
$\tau_{x}^{\epsilon(o)'}$ and $\tau_{y}^{\epsilon(o)'}$ as the
operation points of the length between BS and ITMs in the
Fabry-P\'erot interferometer.
Equations~(\ref{eq:lx-incomplete-operation-point}) and
(\ref{eq:ly-incomplete-operation-point}) indicate that we may replace
\begin{eqnarray}
  \label{eq:operation-point-replacements-BS-XITM-length-tauxprime-incomp}
  && \tau_{x}' \rightarrow \tau_{x}^{\epsilon(o)'}, \\
  \label{eq:operation-point-replacements-BS-YITM-length-tauyprime-incomp}
  && \tau_{y}' \rightarrow \tau_{y}^{\epsilon(o)'}, \\
  \label{eq:operation-point-replacements-XITM-incomp}
  && \hat{X}_{XITM}(t) \rightarrow \hat{X}_{XITM}^{\epsilon(o)}(t), \\
  \label{eq:operation-point-replacements-YITM-incomp}
  && \hat{X}_{YITM}(t) \rightarrow \hat{X}_{YITM}^{\epsilon(o)}(t)
\end{eqnarray}
in Eqs.~(\ref{eq:xarm-retarded-effect-f'-to-c'}),
(\ref{eq:yarm-retarded-effect-f'-to-c'}),
(\ref{eq:xarm-retarded-effect-c-to-f-mod}),
(\ref{eq:yarm-retarded-effect-c-to-f-mod}), and any equations derived
from them.


Similarly, we define the incomplete equilibrium operation point for
the retarded effect $\tau+\hat{X}_{x}(t-\tau)$ and
$\tau+\hat{X}_{y}(t-\tau)$ so that
\begin{eqnarray}
  \tau + \frac{1}{c} \hat{X}_{x}(t-\tau)
  &=&
      \frac{1}{c} \left( L + \hat{X}_{XEM} - \hat{X}_{XITM} \right)
      \nonumber\\
  &=:&
       \frac{1}{c} \left( L + (1-\epsilon)(\scrD_{EM}-\scrD_{ITM})
       \right.
       \nonumber\\
  && \quad
       \left.
       + \hat{X}_{XEM}^{\epsilon(o)} - \hat{X}^{\epsilon(o)}_{XITM}
       \right)
       \nonumber\\
  &=:&
       \tau^{\epsilon(o)} + \frac{1}{c}
       \hat{X}_{x}^{\epsilon(o)}(t-\tau^{\epsilon(o)}) + O(X_{x}^{2})
       ,
       \nonumber\\
       \label{eq:L-incomplete-operation-point-x}
  \\
  \hat{X}_{XEM}^{\epsilon(o)}
  &:=&
       \epsilon \scrD_{EM} + \hat{X}_{XEM}^{(fluc)},
     \label{eq:XEM-incomplete-operation-point-x}
  \\
  \hat{X}_{x}^{\epsilon(o)}
  &:=&
       \hat{X}^{\epsilon(o)}_{XEM} - \hat{X}_{XITM}^{\epsilon(o)},
     \label{eq:Xx-incomplete-operation-point-x}
\end{eqnarray}
\begin{eqnarray}
  \tau + \frac{1}{c} \hat{X}_{y}(t-\tau)
  &=&
      \frac{1}{c} \left( L + \hat{X}_{YEM} - \hat{X}_{YITM} \right)
      \nonumber\\
  &=:&
       \frac{1}{c} \left( L + (1-\epsilon)(\scrD_{EM}-\scrD_{ITM})
       \right.
       \nonumber\\
  && \quad
       \left.
       + \hat{X}_{YEM}^{\epsilon(o)} - \hat{X}^{\epsilon(o)}_{YITM}
       \right)
       \nonumber\\
  &=:&
       \tau^{\epsilon(o)} + \frac{1}{c}
       \hat{X}_{y}^{\epsilon(o)}(t-\tau^{\epsilon(o)}) + O(X_{y}^{2})
       ,
       \nonumber\\
       \label{eq:L-incomplete-operation-point-y}
  \\
  \hat{X}_{YEM}^{\epsilon(o)}
  &:=&
       \epsilon \scrD_{EM} + \hat{X}_{YEM}^{(fluc)},
     \label{eq:XEM-incomplete-operation-point-y}
  \\
  \hat{X}_{y}^{\epsilon(o)}
  &:=&
       \hat{X}^{\epsilon(o)}_{YEM} - \hat{X}_{YITM}^{\epsilon(o)}.
     \label{eq:Xx-incomplete-operation-point-y}
\end{eqnarray}


Here, we emphasize that $\tau^{\epsilon(o)}$ includes the relative
constant displacements between XEM and XITM (or equivalently, the
relative constant displacements between YEM and YITM) due to the
constant displacements $\scrD_{EM}$ and $\scrD_{ITM}$ from the
radiation pressure forces as
\begin{eqnarray}
  \label{eq:tauepsilono-incomp-tuning-point}
  \tau^{\epsilon(o)} = \frac{1}{c} \left( L + (1-\epsilon) (\scrD_{EM}
  - \scrD_{ITM})\right),
\end{eqnarray}
while the operator $\hat{X}_{x}^{\epsilon(o)}$ and
$\hat{X}_{y}^{\epsilon(o)}$ includes constant displacements
$\scrD_{EM}$ and $\scrD_{ITM}$ as
\begin{eqnarray}
  \label{eq:tauepsilono-incomp-tuning-hatXx-hatXy}
  \hat{X}_{x}^{\epsilon(o)}
  :=
  \epsilon \left( \scrD_{EM} - \scrD_{ITM} \right)
  + \hat{X}_{x}^{(fluc)}.
\end{eqnarray}


As in the case of the complete equilibrium tuning, we regard
$\tau^{\epsilon(o)}$ as the operation point of the cavity length
between ITMs and EMs in the Fabry-P\'erot interferometer.
Equations~(\ref{eq:L-incomplete-operation-point-x}) and
(\ref{eq:L-incomplete-operation-point-y}) indicate that we may replace
\begin{eqnarray}
  \label{eq:operation-point-replacements-cavity-tau-incomp}
  && \tau \rightarrow \tau^{\epsilon(o)}, \\
  \label{eq:operation-point-replacements-cavity-Xxy-incomp}
  && \hat{X}_{x,y}(t-\tau) \rightarrow \hat{X}_{x,y}^{\epsilon(o)}(t-\tau^{(o)})
\end{eqnarray}
in Eqs.~(\ref{eq:xarm-retarded-effect-g-to-gprime}),
(\ref{eq:yarm-retarded-effect-g-to-gprime}),
(\ref{eq:xyarm-prop-fprime-and-f-g})--(\ref{eq:yarm-prop-g-and-f-g}),
(\ref{eq:radiation-pressure-force-XEM}),
(\ref{eq:radiation-pressure-force-YEM}), and any equations derived
from them within the accuracy up to $O(X_{x,y}^{2})$.


We note that
Eqs.~(\ref{eq:operation-point-replacements-BS-XITM-length-tauxprime-incomp}),
(\ref{eq:operation-point-replacements-BS-YITM-length-tauyprime-incomp}),
and (\ref{eq:operation-point-replacements-cavity-tau-incomp}) are
regarded as the ``renormalizations'' of $l_{x}$, $l_{y}$, and $L$
which are different from that in
Sec.~\ref{sec:Complete_equilibrium_tuning}.
After these renormalizations of $l_{x}$, $l_{y}$, and $L$, we apply
the tuning condition
\begin{eqnarray}
  \label{eq:omega0-L-tune-cond-renorm-incomp}
  && \omega_{0} \tau^{\epsilon(o)} = 2n\pi, \quad n\in\NF, \\
  \label{eq:omega0-lxly-tune-cond-renorm-incomp}
  && \omega_{0} \frac{\tau_{x}^{\epsilon(o)'}+\tau_{y}^{\epsilon(o)'}}{2} = 2m\pi,
     \quad m\in\NF,
\end{eqnarray}
instead of the tuning conditions (\ref{eq:omega0-L-tune-cond})
and (\ref{eq:omega0-lxly-tune-cond}).
After these renormalizations, the phase offset $\theta$ is defined by
(\ref{eq:offset-def}) is replaced by
\begin{eqnarray}
  \label{eq:offset-def-renorm-incomp}
  \theta := \omega_{0}(\tau_{y}^{\epsilon(o)'}-\tau_{x}^{\epsilon(o)'}).
\end{eqnarray}


\begin{figure}
  \begin{center}
    \includegraphics[width=0.5\textwidth]{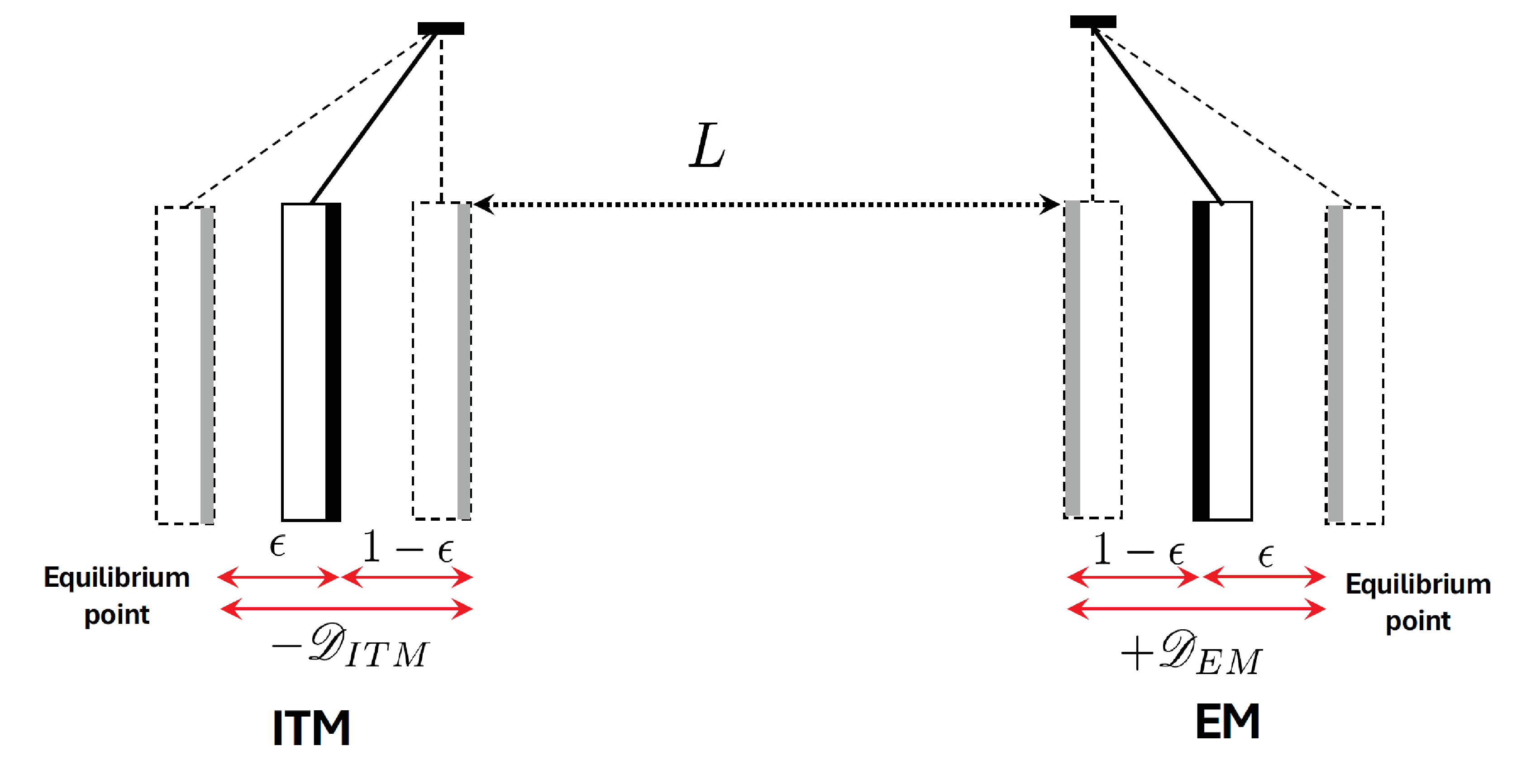}
  \end{center}
  \caption{
    The schematic picture of the incomplete equilibrium tuning.
    The incomplete parameter $\epsilon$ represents the
    deviations of the tuning point from the equilibrium point of the
    pendulum for the mirrors.
    $\epsilon=0$ corresponds to the complete equilibrium tuning.
    Even if we keep the mirrors at the operation point $\epsilon\neq
    0$, the mirrors still feel acceleration and try to move.
    These accelerations for each mirror are included in the feedback
    electric current.
  }
  \label{fig:Incomplete_Equilibrium_Tuning_Image}
\end{figure}


As in the case of the above complete equilibrium tuning, all
$\hat{X}_{x,y}$, $\hat{X}_{XITM}$, and $\hat{X}_{YITM}$ are replaced
by $\hat{X}^{\epsilon(o)}_{x,y}$, $\hat{X}_{XITM}^{\epsilon(o)}$, and
$\hat{X}_{YITM}^{\epsilon(o)}$, respectively.
Furthermore, their Fourier transformations $\hat{Z}_{x,y}$,
$\hat{Z}_{XITM}$, and $\hat{Z}_{YITM}$ are also replaced by
$\hat{Z}_{x,y}^{\epsilon(o)}$, $\hat{Z}_{XITM}^{\epsilon(o)}$, and
$\hat{Z}_{YITM}^{\epsilon(o)}$.
These $\hat{Z}_{x,y}^{\epsilon(o)}$, $\hat{Z}_{XITM}^{\epsilon(o)}$,
and $\hat{Z}_{YITM}^{\epsilon(o)}$ includes the classical term
proportional to $2\pi\delta(\Omega)$ in the order of $O(\epsilon)$.
Therefore, the resulting input-output relation corresponding to
Eq.~(\ref{eq:input-output-rel-explicit-quad-I0-kappabetahSQL}) is
given by the expression of
Eq.~(\ref{eq:input-output-rel-explicit-quad-I0-kappabetahSQL}) with
the replacement $\FrakR$ $\rightarrow$ $\epsilon\FrakR$.
In this new input-output relation, $\tau$ in the expressions of
$e^{\pm 2i\beta}$ given by Eq.~(\ref{eq:beta-def}), $\kappa$ given by
Eq.~(\ref{eq:kappa-def}), and $h_{SQL}$ given by
Eq.~(\ref{eq:hSQL-def}) must be replaced as
Eq.~(\ref{eq:operation-point-replacements-cavity-tau-incomp}).
We note that $\epsilon=0$ corresponds to the complete equilibrium
tuning discussed in Sec.~\ref{sec:Complete_equilibrium_tuning}.


Note that $\hat{X}_{XITM}=\scrD_{ITM}$,  $\hat{X}_{YITM}=\scrD_{ITM}$,
$\hat{X}_{XEM} = \scrD_{EM}$, and $\hat{X}_{YEM} = \scrD_{EM}$ are
equilibrium points of the pendulum for mirrors, as mentioned in
Sec.~\ref{sec:Complete_equilibrium_tuning}.
Even if we make the mirrors stay at the operation point so that
$\epsilon\neq 0$, the mirrors feel their accelerations due to the
classical radiation pressure and try to move with the frequency
$\omega_{p}$ of the pendulum.
In this case, these accelerations are measured by the feedback control
system, and their signals are included in the feedback electric
current.
In this case, the noise in the feedback electric current may differ
from that at the photodetection.
For this reason, we call the tuning conditions $\epsilon\neq 0$ as
``incomplete equilibrium tuning.''
The schematic picture of the incomplete equilibrium tuning is depicted
in Fig.~\ref{fig:Incomplete_Equilibrium_Tuning_Image}.


\begin{turnpage}
\begin{figure*}
  \begin{center}
    \includegraphics[width=1.2\textwidth]{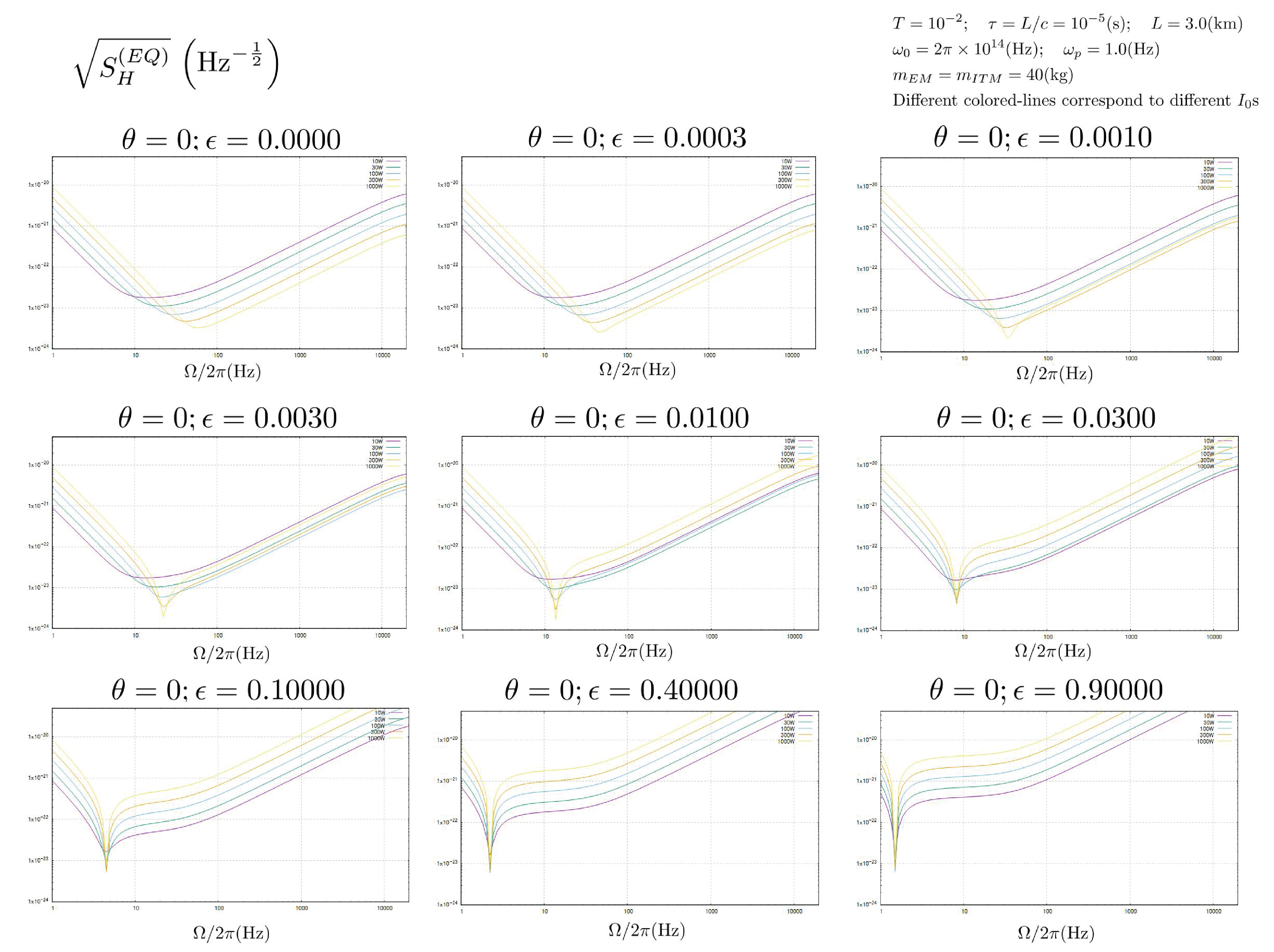}
  \end{center}
  \caption{
    The square root of the explicit signal referred noise spectral
    density $\sqrt{S_{(H)}^{EQ}(\Omega,\epsilon,\theta=0)}$ for each
    $\epsilon$ is depicted in the range 1Hz to 20 kHz with different
    values of the laser power $I_{0}$.
    $S_{(H)}^{(EQ)}$ is given by
    Eq.~(\ref{eq:strain-referred-noise-spectrum-density-incomplete-theta=0})
    with $m=m_{EM}=m_{ITM}$.
  }
  \label{fig:CompSHEQFULL-dark-1-20k-mod}
\end{figure*}
\end{turnpage}


We also note that the classical part $\FrakR$ from the radiation
pressure forces are given by the displacements $\scrD_{EM}$ and
$\scrD_{ITM}$ as
\begin{eqnarray}
  \FrakR
  &=&
      \frac{2\omega_{0}T}{c} \left[1 - \sqrt{1-T}\right]^{-2}
      \nonumber\\
  && \quad
     \times
     \left[
     \scrD_{EM} - \left(1 - \frac{[1 - \sqrt{1-T}]^{2}}{T}\right) \scrD_{ITM}
     \right]
     .
     \nonumber\\
     \label{eq:FrakR-scrDEM-scrDITM-relation}
\end{eqnarray}
These displacements $\scrD_{EM}$ and $\scrD_{ITM}$ are given by
Eqs.~(\ref{eq:classical-terms-of-RPF-EM-renorm-retune}) and
(\ref{eq:classical-terms-of-RPF-ITM-renorm-retune}), respectively.
Equation~(\ref{eq:FrakR-scrDEM-scrDITM-relation}) shows that the
origin of the additional classical carrier $\FrakR$ from the radiation
pressure forces is determined by the deviation from the complete
equilibrium tuning.


Even in this incomplete equilibrium tuning, the expectation value of
the signal operator $\hat{s}_{\calN_{b}}(\Omega)$ defined by
Eq.~(\ref{eq:signal-operator-Fourier-def}) is also given by
Eq.~(\ref{eq:GW-signal-operator-exp-value}).
This is because the expectation value
$\langle\hat{s}_{\calN_{b}}(\Omega)\rangle$ does not depend on
$\FrakR$ due to the reality condition of the gravitational-wave signal
$H(\Omega)=H^{*}(-\Omega)$.


On the other hand, the strain-referred noise spectral density
(\ref{eq:signal-referred-noise-spectral-density-def}) defined by the
signal-to-noise ratio at the photodetector is given by
Eq.~(\ref{eq:strain-referred-noise-spectrum-density}) with the
replacement $\FrakR\rightarrow\epsilon\FrakR$, i.e.,
\begin{eqnarray}
  &&
     \!\!\!\!\!\!\!
     S_{H}(\Omega)
     \nonumber\\
  &=&
      \frac{h_{SQL}^{2}}{4\cos^{2}\theta}
      \left[
      \left( \frac{1}{\kappa} + \frac{\kappa}{2} \left( 1 + \cos(2\theta) \right) \right)
      - 2 \epsilon\FrakR \cos(2\theta)
      \right.
      \nonumber\\
  && \quad\quad\quad\quad
      \left.
      +   \left( \frac{1}{\kappa} + \frac{\kappa}{2} \left( 1 - \cos(2\theta) \right) \right) \epsilon^{2}\FrakR^{2}
      \right]
      .
      \label{eq:strain-referred-noise-spectrum-density-incomplete}
\end{eqnarray}


Here, we consider the case of the complete dark port limit of the
expectation value (\ref{eq:GW-signal-operator-exp-value}) and the
signal-referred stationary noise spectrum density
(\ref{eq:strain-referred-noise-spectrum-density-incomplete}).
In this limit, the expectation value
(\ref{eq:GW-signal-operator-exp-value}) vanishes as expected.
This is the property of the DC-readout.
In the DC-readout scheme, we cannot measure the signal, i.e.,
gravitational waves in the complete dark port $\theta=0$.
However, if we consider the situation where $\theta\ll 1$ but
$\theta\neq 0$, we can measure the small expectation value
(\ref{eq:GW-signal-operator-exp-value}) and we can estimate the
gravitational-wave signal from this small expectation value.
In this case, the noise spectral density
(\ref{eq:strain-referred-noise-spectrum-density-incomplete}) has its
meaning and it is given by
\begin{eqnarray}
  S_{H}(\Omega)
  =
  \frac{h_{SQL}^{2}}{4}
  \left[
  \frac{1}{\kappa} + \kappa
  - 2 \left(\epsilon\FrakR\right)
  +   \frac{1}{\kappa} \left(\epsilon\FrakR\right)^{2}
  \right]
  .
  \label{eq:strain-referred-noise-spectrum-density-incomplete-theta=0}
\end{eqnarray}
In the case of the complete equilibrium tuning $\epsilon=0$, the noise
spectral density
(\ref{eq:strain-referred-noise-spectrum-density-incomplete-theta=0})
yields Kimble's noise spectral
density
$S_{(K)}/2$~\cite{H.J.Kimble-Y.Levin-A.B.Matsko-K.S.Thorne-S.P.Vyatchanin-2001}
apart from the overall factor $1/2$.
Here, we consider the case where $\epsilon\neq 0$.
In this case, the noise spectral density
(\ref{eq:strain-referred-noise-spectrum-density-incomplete-theta=0})
is also expressed as
\begin{eqnarray}
  S_{H}(\Omega)
  =
  \frac{h_{SQL}^{2}}{4\kappa}
  \left[
  1 + \left(\kappa - \epsilon\FrakR\right)^{2}
  \right]
  \geq
  \frac{h_{SQL}^{2}}{4\kappa}
  .
  \label{eq:strain-referred-noise-spectrum-density-incomplete-theta=0-2}
\end{eqnarray}
This indicates that the noise spectral density
(\ref{eq:strain-referred-noise-spectrum-density-incomplete-theta=0-2})
have the minimum value $h_{SQL}^{2}/(4\kappa)$.
The behavior of the noise spectral density
(\ref{eq:strain-referred-noise-spectrum-density-incomplete-theta=0-2})
depicted in Fig.~\ref{fig:CompSHEQFULL-dark-1-20k-mod}.


Equality in the last inequality in
Eq.~(\ref{eq:strain-referred-noise-spectrum-density-incomplete-theta=0-2})
is achieved when
\begin{eqnarray}
  \label{eq:theta=0-equality}
  \kappa = \epsilon \FrakR.
\end{eqnarray}
We evaluate Eq.~(\ref{eq:theta=0-equality}) through the definition
(\ref{eq:kappa-def}) of $\kappa$ and the definition
(\ref{eq:FrakR-def}) of $\FrakR$.
Here, we consider the situation where $\Omega\tau^{\epsilon(o)}\ll 1$
and use the approximation forms
(\ref{eq:kappa-def-ITM-inclusion-explicit-approx-eq-mass}) with
Eq.~(\ref{eq:ISQL-def}) and (\ref{eq:FrakR-defq-2-approx}) of $\kappa$
and $\FrakR$ with the equal mass condition $m:=m_{EM}=m_{ITM}$.
Then, we obtain the estimation
\begin{eqnarray}
  \label{eq:kappaoverepsilonFrakR-estimation}
  1 = \frac{\kappa}{\epsilon\FrakR} \sim
  \frac{2\omega_{p}^{2}\gamma^{2}}{\Omega^{2}(\Omega^{2}+\gamma^{2})}
  \sim
  \frac{2\omega_{p}^{2}}{\epsilon\Omega^{2}}.
\end{eqnarray}
From this evaluation the minimum of $S_{H}(\Omega,\theta=0)$ is
achieved at
\begin{eqnarray}
  \label{eq:SHtheta=0-minimum-achievement}
  \Omega \sim \frac{\sqrt{2}}{\epsilon} \omega_{p}.
\end{eqnarray}
This estimation is supported by the profiles in
Fig.~\ref{fig:CompSHEQFULL-dark-1-20k-mod}.


In Fig.~\ref{fig:CompSHEQFULL-dark-1-20k-mod}, the left-top panel
shows the conventional Kimble's noise spectral density with the
modification of $\kappa$ and $h_{SQL}$.
However, if the incompleteness $\epsilon$ increases, the deviation
from Kimble's noise spectral density can be seen.
After all, in the high-frequency region, the noise spectral density
shows that the shot-noise contribution in the signal-referred noise
spectral density does not decrease due to the increase of the incident
laser power $I_{0}$.
This is merely due to the contribution of the additional classical
carrier field $\epsilon\FrakR$, which may be dominant in the incomplete
equilibrium tuning.


Furthermore, the minimum at
Eq.~(\ref{eq:SHtheta=0-minimum-achievement}) of the noise spectral
density can be seen as a dip around the frequency
(\ref{eq:SHtheta=0-minimum-achievement}) in
Fig.~\ref{fig:CompSHEQFULL-dark-1-20k-mod}.
This dip shows a violation of the so-called ``standard quantum
limit'' which is the envelope of the left-top figure in
Fig.~\ref{fig:CompSHEQFULL-dark-1-20k-mod}.
However, as discussed in
Sec.~\ref{sec:Quantum_Mechanical_model_for_mirror_motions_and_its_sol},
this effect is not the violation of the conventional arguments of the
Heisenberg uncertainty principle which arise from the non-commutation
of the position operator $\hat{X}$ and the momentum operator
$\hat{P}$, i.e., $[\hat{X},\hat{P}]=i\hbar$.
This is because the quantum mirrors' initial conditions, which have
the information $[\hat{X},\hat{P}]=i\hbar$, do not affect our
consideration as discussed in
Sec.~\ref{sec:Quantum_Mechanical_model_for_mirror_motions_and_its_sol}.
Therefore, we have to say that this dip has nothing to do with the
arguments of the violation of the conventional arguments of the
Heisenberg uncertainty principle in Ref.~\cite{M.Ozawa-2004}.
In the gravitational-wave community, it is well known that the
standard quantum limit can be surpassed using squeezed laser
states. This indicates that violating the ``standard quantum limit''
does not imply violating a physical principle.
Therefore, we may regard the dip in
Fig.~\ref{fig:CompSHEQFULL-dark-1-20k-mod} as not a violation of the
physical principle.


\begin{turnpage}
\begin{figure*}
  \begin{center}
    \includegraphics[width=1.2\textwidth]{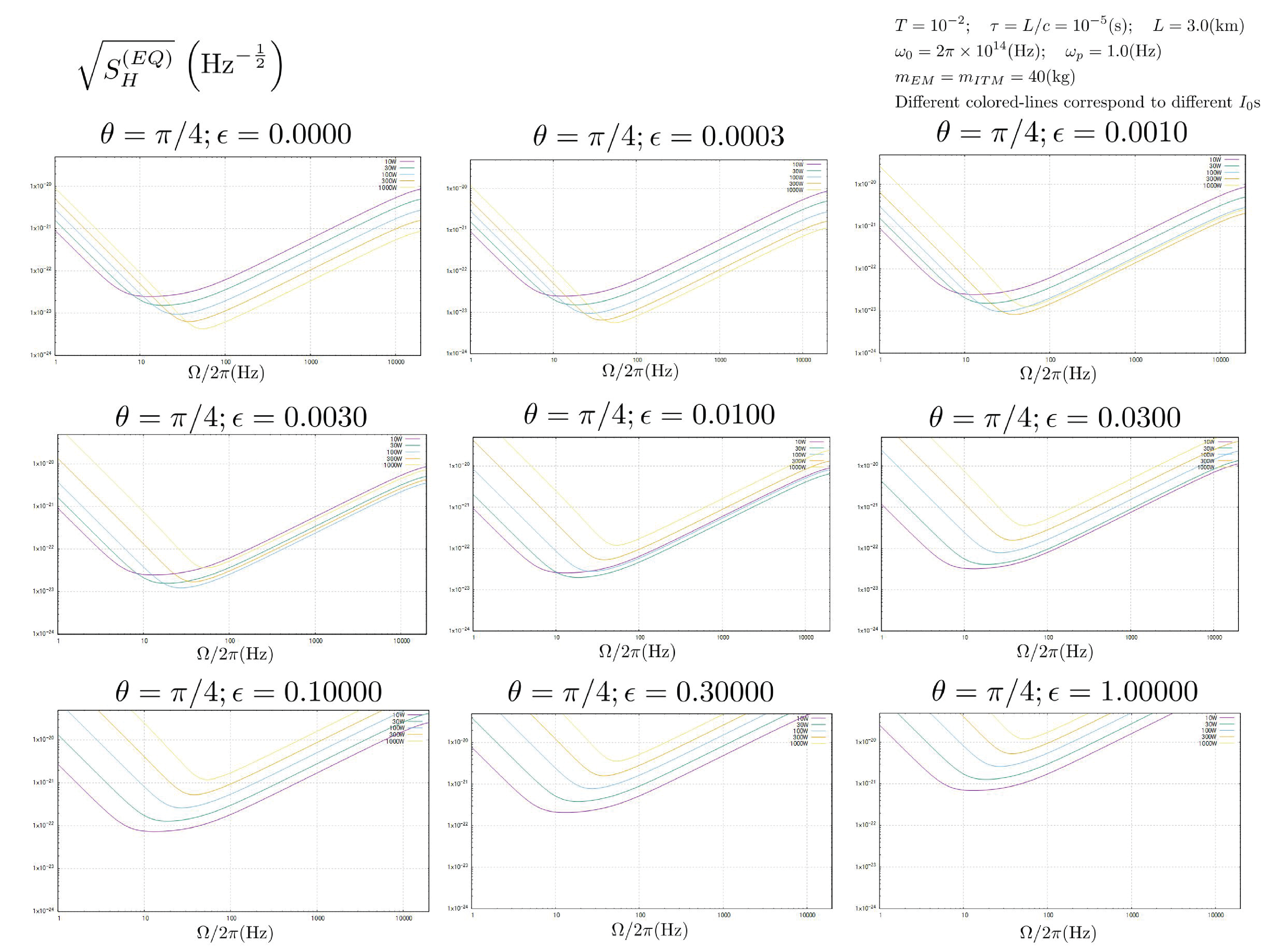}
  \end{center}
  \caption{
    The square root of the explicit signal-referred noise spectral
    density $\sqrt{S_{(H)}^{EQ}(\Omega,\epsilon,\theta=\pi/4)}$ for each
    $\epsilon$ is depicted in the range 1Hz to 20 kHz with different
    values of the laser power $I_{0}$.
    $S_{(H)}^{(EQ)}$ is given by
    Eq.~(\ref{eq:strain-referred-noise-spectrum-density-opt-thetapiover4}).
  }
  \label{fig:CompSHEQFULL-quaterPI-1-20k-mod}
\end{figure*}
\end{turnpage}


Next, we examine the situation where
$\langle\hat{s}_{\calN}(\Omega)\rangle$ is the maximum for the fixed
gravitational-wave signal $H(\Omega,L)$, i.e., $\theta=\pi/4$.
In this case, the noise spectral density $S_{H}(\Omega)$ in
Eq.~(\ref{eq:strain-referred-noise-spectrum-density-incomplete}) is
given by
\begin{eqnarray}
  S_{H}(\Omega)
  &=&
      \frac{h_{SQL}^{2}}{2}
      \left[
      \left( \frac{1}{\kappa} + \frac{\kappa}{2} \right)
      +   \left( \frac{1}{\kappa} + \frac{\kappa}{2} \right) \epsilon^{2}\FrakR^{2}
      \right]
      \nonumber\\
  &=&
      \frac{h_{SQL}^{2}}{2}
      \left( \frac{1}{\kappa} + \frac{\kappa}{2} \right)
      \left[
      1
      +  \epsilon^{2}\FrakR^{2}
      \right]
      .
      \label{eq:strain-referred-noise-spectrum-density-opt-thetapiover4}
\end{eqnarray}
The behavior of the noise-spectral density
$S_{H}(\Omega,\theta=\pi/4)$ in
Eq.~(\ref{eq:strain-referred-noise-spectrum-density-opt-thetapiover4})
is depicted in Fig.~\ref{fig:CompSHEQFULL-quaterPI-1-20k-mod}.
Since the effects of the incomplete equilibrium tuning are factorized
$[1+\epsilon^{2}\FrakR^{2}]$, the shape of the noise spectral
densities does not show any drastic changes.
However, we can see that if the effect of the additional classical
carrier $\epsilon\FrakR$ is considered, the power dependence of the
noise spectral density is changed.
As a result, the bottom figures in
Fig.~\ref{fig:CompSHEQFULL-quaterPI-1-20k-mod} show the property that
the shot-noise contribution is not reduced even if the input power
$I_{0}$ is increased.
Incidentally, the bottom right panel in
Fig.~\ref{fig:CompSHEQFULL-quaterPI-1-20k-mod}
is the same figure as
Fig.~\ref{fig:SHEQFULL-1-20k}.


Finally, we discuss the realization of the ideal noise spectral
density $S_{H}(\Omega)$ with $\epsilon\FrakR=0$.
As shown in both Fig.~\ref{fig:CompSHEQFULL-dark-1-20k-mod} and
Fig.~\ref{fig:CompSHEQFULL-quaterPI-1-20k-mod}, if we have
sufficiently small $\epsilon$, we can realize the ideal noise spectral
density $S_{H}(\Omega)$ with $\epsilon\FrakR\sim 0$.
From the input-output relation
(\ref{eq:input-output-rel-explicit-quad-I0-kappabetahSQL}), if the
term $\FrakR$ in the classical carrier field, which is proportional to
$2\pi\delta(\Omega)$ is negligible compared with $1$ in the same term,
we can realize the ideal input-output relation.
In the case of the incomplete equilibrium tuning, the incomplete
parameter $\epsilon$, which we introduced, appears as $\epsilon\FrakR$
in this classical carrier field of the input-output relation as
mentioned above.
Therefore, if we achieve the $\epsilon$ so that
\begin{eqnarray}
  \label{eq:ideallization-epsilon}
  \epsilon\FrakR \lesssim O(10^{-2}),
\end{eqnarray}
we can realize the idealized input-output relation which is shown in Ref.~\cite{H.J.Kimble-Y.Levin-A.B.Matsko-K.S.Thorne-S.P.Vyatchanin-2001}.
From the order of magnitude (\ref{eq:FrakR-eval-smallT-leading}) of
$\FrakR$, $\epsilon$ can be estimated as
\begin{eqnarray}
  \label{eq:ideallization-epsilon-estimation}
  \epsilon &\lesssim& O(10^{-2}) \FrakR^{-1} \\
  &\sim&
         O(10^{-2})
         \frac{ m c^{2} \omega_{p}^{2} T^{2}}{ 64 I_{0} \omega_{0} }
         \nonumber\\
  &\sim&
      3 \times 10^{-5}
      \times
      \left(
      \frac{10^{2}W}{I_{0}}
      \right)
      \left(
      \frac{2\pi\times 10^{14}\mbox{Hz}}{\omega_{0}}
      \right)
      \left(
      \frac{m}{40 \mbox{kg}}
      \right)
      \nonumber\\
  && \quad\quad\quad\quad
     \times
     \left(
     \frac{\omega_{p}}{2\pi\times 1 \mbox{Hz}}
     \right)^{2}
     \left(
     \frac{T}{10^{-2}}
     \right)^{2}
\end{eqnarray}
for the equal mass case $m=m_{EM}=m_{ITM}$.
In terms of the mirror displacements $\epsilon\scrD_{EM}$ and
$\epsilon\scrD_{ITM}$, which are estimated in
Eqs.~(\ref{eq:classical-terms-of-RPF-EM-renorm-retune})--(\ref{eq:classical-terms-of-RPF-ITM-renorm-retune-leading-order})
as
\begin{eqnarray}
  \epsilon \scrD_{EM}
  &\lesssim&
             +
             O(10^{-2})
             \frac{cT}{16\omega_{0}}
      \label{eq:classical-terms-of-RPF-EM-renorm-retune-require}
  \\
  &\sim&
         +
         3
         \times
         10^{-12}
         \; \mbox{m}
         \times
         \left(
         \frac{T}{10^{-2}}
         \right)
         \left(
         \frac{2\pi\times 10^{14}\;\mbox{Hz}}{\omega_{0}}
         \right)
         \nonumber\\
      \label{eq:classical-terms-of-RPF-EM-renorm-retune-require-leading-order}
\end{eqnarray}
and
\begin{eqnarray}
  \left|\epsilon\scrD_{ITM}\right|
  &\lesssim&
             O(10^{-2})
             \frac{cT}{16\omega_{0}}
             \label{eq:classical-terms-of-RPF-ITM-renorm-retune-require}
  \\
  &\sim&
         3
         \times
         10^{-12}
         \; \mbox{m}
         \times
         \left(
         \frac{T}{10^{-2}}
         \right)
         \left(
         \frac{2\pi\times 10^{14}\;\mbox{Hz}}{\omega_{0}}
         \right)
         .
         \nonumber\\
  \label{eq:classical-terms-of-RPF-ITM-renorm-retune-require-leading-order}
\end{eqnarray}
Thus, if we can control the mirrors' positions whose deviations from
the equilibrium points of the pendulum and the radiation pressure
forces are less than $\epsilon\scrD_{EM}$ and $\epsilon\scrD_{ITM}$,
we can realize the ideal Kimble noise spectral density
with the modification of $\kappa$ and $h_{SQL}$.
These $\epsilon\scrD_{EM}$ and $\epsilon\scrD_{ITM}$ are given by
Eqs.~(\ref{eq:classical-terms-of-RPF-EM-renorm-retune-require}) and
(\ref{eq:classical-terms-of-RPF-ITM-renorm-retune-require}),
respectively.


\section{Summary and Discussion}
\label{sec:Summary_and_Discussion}


The primary purpose of this paper was to develop a detailed
theoretical description of quantum noise in gravitational-wave
detectors, motivated by advances in measurement theory in quantum
field theory.
In the mathematically rigorous measurement theory in quantum
mechanics~\cite{M.Ozawa-2004}, the specification of the final observed
quantum operator is an essential issue in the theory.
We must extend this mathematical theory to quantum field theories to
apply it to gravitational-wave detectors.
Besides this extension of quantum measurement theory to quantum field
theories, the above specification of the final observed quantum
operator is crucial.
Actually, in quantum measurement theories, there is a famous problem
of the von Neumann chain~\cite{J.vonNeumann-2018}, i.e., where we
should regard the measurement outcomes as the classical information in
the sequence of the quantum measurements as emphasized in
Sec.~\ref{sec:Introduction}.
In this paper, we adopt the view that the von Neumann chain should be
cut at the point where photodetectors convert photon signals into
electric currents.
Note that actual gravitational-wave detectors employ a feedback system
to control the stability of the detector.
However, from our standpoint, we regard the feedback current as a
classical electric current.
Therefore, we do not need to consider the quantum properties of the
feedback control system.
If this premise is wrong, we have to discuss quantum feedback control
systems~\cite{H.M.Wiseman-G.J.Milburn-2010} as emphasized in
Sec.~\ref{sec:Introduction}, which is beyond the current scope of this
paper.


After the introduction of the notions that were used within
this paper in Sec.~\ref{sec:Preliminary}, we first developed the
description of the DC readout scheme in terms of quantum
electrodynamics.
Although the analysis in these arguments of the DC readout scheme is
just an extension of the balanced homodyne detection, which was
discussed in Ref.~\cite{K.Nakamura-2021}, the essence of the DC
readout scheme is the leakage of the large classical carrier field
from the output port.
Due to this large classical carrier field, we can separate the output
electric field into orders of the large classical carrier field.
After excluding the first dominant term, we can measure the output
field, which includes gravitational signals as the second dominant
term.
Since the leaked large classical carrier field from the output port
serves as the reference for the output signal in the DC readout
scheme, the leaked large classical carrier field is essential in the
DC readout scheme.


After the general arguments of the DC readout scheme in
Sec.~\ref{sec:General_arguments_for_the_DC-readout_scheme},
we carefully discussed the input-output relation of the Fabry-P\'erot
gravitational-wave detectors, because the details of the input-output
relation are necessary to examine the properties of the DC readout
scheme.
Therefore, in
Secs.~\ref{sec:Input-Output-relation-for-Fabry-Perot_interferometer},
\ref{sec:Eq_for_mirrors'_motions_and_their_solutions}, and
\ref{sec:Final_input-output-relation_for_Fabry-Perot_GW_Detector},
we described the detailed theoretical description of the derivation of
the input-output relation of the Fabry-P\'erot interferometer.


In
Sec.~\ref{sec:Input-Output-relation-for-Fabry-Perot_interferometer},
we derived the input-output relation through the propagation effects
and the reflection effects due to the mirrors in the Fabry-P\'erot
interferometer.
In the previous works of gravitational-wave detectors, it has been
considered only the end mirrors' motions.
However, in this paper, we also account for the motions of the input
test masses.
We obtained the input-output relation without specifying the mirror
displacements.
The equations of motion of mirrors determine these mirror
displacements.
Since we discussed the quantum motion of mirrors, we used the
Heisenberg equations of motion in quantum mechanics as the equations
of motion for mirrors.
Furthermore, we introduce the fundamental frequency $\omega_{p}$ of
the mirrors' pendulum, where $\omega_{p}$ is estimated as $\sim 1$Hz.


In Sec.~\ref{sec:Eq_for_mirrors'_motions_and_their_solutions}, we
briefly reviewed a quantum mechanical forced harmonic oscillator,
which is essential to our arguments on the motions of mirrors in
gravitational-wave detectors.
Based on the understanding of this quantum forced harmonic oscillator,
we explicitly derived the equations of motion for the mirrors in the
Fabry-P\"erot interferometer in
Sec.~\ref{sec:Heisenberg_Equations_for_mirrors'_motions}.
Then, we noted that evaluating the effects of radiation pressure is
essential.
Although the details are described in
Appendix~\ref{sec:Explicit_evaluation_of_the_radiation_pressure_forces},
we review the outline of the derivation of the radiation pressure
forces to the input test masses and to the
end test masses in Fabry-P\'erot gravitational-wave detectors in
Sec.~\ref{sec:Evaluation_of_Radiation_pressure_forces}.
From the derived radiation pressure forces, we obtained the solutions
to the Heisenberg equations in
Sec.~\ref{sec:Sol_to_Heisenberg_Eq_for_mirrors}.


From the derived solutions to the Heisenberg equations, we pointed out
that the initial conditions of the position $\hat{X}(-\infty)$ and the
momentum $\hat{P}(-\infty)$ for the harmonic oscillator with
$[\hat{X}(-\infty),\hat{P}(-\infty)]=i\hbar$ are ignored in the
solutions in Sec.~\ref{sec:Sol_to_Heisenberg_Eq_for_mirrors}, which
contribute to the commutation relation
$[\hat{X}(t),\hat{P}(t)]=i\hbar$ of the usual position and the
momentum.
This commutation relation is usually regarded as a realization of the
uncertainty relations in quantum mechanics.
However, these initial conditions of the position and the momentum are
excluded from our consideration due to their concentration on the
pendulum frequency $\omega=\pm\omega_{p}$, which is out of the
frequency range of our interest.
Therefore, we may say that the solutions in
Sec.~\ref{sec:Sol_to_Heisenberg_Eq_for_mirrors} do not include the
quantum uncertainties which arise from the commutation relations
$[\hat{X}(-\infty),\hat{P}(-\infty)]=i\hbar$.
This was clarified by the introduction of the pendulum frequency
$\omega_{p}$.


Together with the input-output relation derived in
Sec.~\ref{sec:Input-Output-relation-for-Fabry-Perot_interferometer}
and the solutions to the Heisenberg equation derived in
Sec.~\ref{sec:Sol_to_Heisenberg_Eq_for_mirrors}, we obtain the final
input-output relation of the Fabry-P\'erot interferometer in Sec.~\ref{sec:Final_input-output-relation_for_Fabry-Perot_GW_Detector}.
Then, we reach the input-output relation for the Fabry-P\'erot
interferometer (\ref{eq:input-output-rel-explicit-quad-I0-kappabetahSQL}).
Although the parameters $\beta$, $\kappa$, and $h_{SQL}$ in the
input-output relation are different from these parameters in
Ref.~\cite{H.J.Kimble-Y.Levin-A.B.Matsko-K.S.Thorne-S.P.Vyatchanin-2001},
the resulting input-output relation is almost the same as that derived
in Ref.~\cite{H.J.Kimble-Y.Levin-A.B.Matsko-K.S.Thorne-S.P.Vyatchanin-2001}.
The difference between $\kappa$ in this paper and the $\kappa$ in
Ref.~\cite{H.J.Kimble-Y.Levin-A.B.Matsko-K.S.Thorne-S.P.Vyatchanin-2001},
which denote $\kappa^{(K)}$ in this paper are depicted in
Fig.~\ref{fig:KappaEQKappaK-ratio}.
The difference between $h_{SQL}^{(EQ)}$ ($m=m_{EM}=m_{ITM}$) and
$h_{SQL}$ in
Ref.~\cite{H.J.Kimble-Y.Levin-A.B.Matsko-K.S.Thorne-S.P.Vyatchanin-2001}
which denote $h_{SQL}^{(K)}$ within this paper are depicted in
Fig.~\ref{fig:hSQLEQhSQLK-ratio}.
The frequency dependence of these differences arises from the phase
difference between the motion of EMs and that of ITMs.
Due to these differences, we reach the difference of the
$S_{(K)}^{(EQ)}$ and $S_{(K)}^{(K)}$ as depicted in
Fig.~\ref{fig:SKEQ-SKK-ratio-1-20k}, where $S_{(K)}^{(EQ)}$ is the
equal mass version of
Eq.~(\ref{eq:Kimble-noise-spectral-density-explicit}) and
$S_{(K)}^{(K)}$ is defined by Eq.~(\ref{eq:SKK-def}).


In addition to the above difference between the input-output relation
(\ref{eq:input-output-rel-explicit-quad-I0-kappabetahSQL}), the
difference also exists in the classical carrier part as the parameter
$\FrakR$.
This classical carrier field, modified by $\FrakR$, is used as the
reference field to measure the  gravitational-wave signal in the DC
readout scheme.
Therefore, the classical carrier field is essential in the DC readout
scheme.
As mentioned in Sec.~\ref{sec:DC_readout_scheme_for_FP_GW_Detector},
the expectation value of the signal is not affected by the additional
parameter $\FrakR$ in the classical carrier field due to the reality
condition of the gravitational-wave signal
$H(\Omega)=H^{*}(-\Omega)$.
The contribution of $\FrakR$ in the classical carrier field affects
the signal-referred noise spectral density
(\ref{eq:strain-referred-noise-spectrum-density}).
For this reason, we discussed the difference from the Kimble noise
spectral density in
Ref.~\cite{H.J.Kimble-Y.Levin-A.B.Matsko-K.S.Thorne-S.P.Vyatchanin-2001}.
As a result of the additional classical carrier field $\FrakR$, we
have to conclude that the shot noise contribution to the
high-frequency range of the signal-referred noise spectral density
does not decrease even if the incident laser power is increased.
This contradicts the description in
Ref.~\cite{H.J.Kimble-Y.Levin-A.B.Matsko-K.S.Thorne-S.P.Vyatchanin-2001},
which is the common knowledge in the community of gravitational-wave
experiments.


In this situation, we reconsidered the tuning points
(\ref{eq:omega0-L-tune-cond}) and (\ref{eq:omega0-lxly-tune-cond}) in
Sec.~\ref{sec:Changing_Tuning-Point}.
The origin of the additional modification $\FrakR$ of the classical
carrier in the input-output relation
(\ref{eq:input-output-rel-explicit-quad-I0-kappabetahSQL}) is the
leakage of the classical carrier from the radiation pressure forces,
which are proportional to $2\pi\delta(\Omega)$ in the Fourier
transformation of the radiation pressure forces.
In the time domain, these terms, proportional to $2\pi\delta(\Omega)$,
represent the constant forces in the radiation pressure forces.
Here, we note that the equations of motion for mirrors are expressed
as a forced harmonic oscillator.
In simple classical mechanics, or even in quantum mechanics, a
constant force in a harmonic oscillator yields the deviation of the
equilibrium point.
If there exists a constant force in the Heisenberg equation of
motion for a harmonic oscillator, we have to take into account the
change of the equilibrium points of the mirror displacements.
For this reason, we reconsidered the Heisenberg equation of motion for
the mirrors and the tuning points (\ref{eq:omega0-L-tune-cond}) and
(\ref{eq:omega0-lxly-tune-cond}) in
Sec.~\ref{sec:Changing_Tuning-Point}.


First, we considered the complete elimination of the deviations of the
mirror displacements from their equilibrium points in
Sec.~\ref{sec:Changing_Tuning-Point}.
Then, we showed that the deviation of the equilibrium point due to the
laser radiation pressure forces is eliminated through the
``renormalization'' of the cavity arm length $L$, the distances
$l_{x}$ and $l_{y}$ between BS and ITMs.
Then, we change the tuning point as
Eqs.~(\ref{eq:omega0-L-tune-cond-renorm}) and
(\ref{eq:omega0-lxly-tune-cond-renorm}).
We call this elimination of the classical forces from the radiation
pressure forces the ``complete equilibrium tuning.''
We estimate the deviations $\scrD_{EM}$ and $\scrD_{ITM}$ of the
equilibrium points as
Eqs.~(\ref{eq:classical-terms-of-RPF-EM-renorm-retune-leading-order})
and (\ref{eq:classical-terms-of-RPF-ITM-renorm-retune-leading-order}).


We also considered incomplete equilibrium tuning, in which deviations
in the mirror displacements from their equilibrium points are
partially eliminated through the introduction of the incomplete
parameter $\epsilon$.
We call this elimination the ``incomplete equilibrium tuning.''
If $\epsilon=0$, we can realize the complete equilibrium tuning
discussed above.
Instead of the complete equilibrium tuning
(\ref{eq:omega0-L-tune-cond-renorm}) and
(\ref{eq:omega0-lxly-tune-cond-renorm}), we apply the tuning condition
(\ref{eq:omega0-L-tune-cond-renorm-incomp}) and
(\ref{eq:omega0-lxly-tune-cond-renorm-incomp}).
We also showed that the additional modification $\FrakR$ of the
classical carrier in the input-output relation is explicitly expressed
by the displacement of the equilibrium points $\scrD_{EM}$ and
$\scrD_{ITM}$.
In this incomplete equilibrium tuning, the stationary noise spectral
density is given by
(\ref{eq:strain-referred-noise-spectrum-density-incomplete}).
To clarify the properties of the noise spectral density
(\ref{eq:strain-referred-noise-spectrum-density-incomplete}), we
consider the typical two cases.
The first case is $\theta\ll 1$ but $\theta\neq 0$ and the second case
is $\theta=\pi/4$.
These are depicted in Figs.~\ref{fig:CompSHEQFULL-dark-1-20k-mod}
and~\ref{fig:CompSHEQFULL-quaterPI-1-20k-mod}.


In the first case, the noise spectral density is given by
Eq.~(\ref{eq:strain-referred-noise-spectrum-density-incomplete-theta=0-2}).
If $\epsilon$ is sufficiently large,
Fig.~\ref{fig:CompSHEQFULL-dark-1-20k-mod} indicates the dip in the
low-frequency region.
Although this dip violates the ``standard quantum limit'' which is the
envelope of the curves with different powers $I_{0}$ in $\epsilon=0$
case, this is not surprising fact that there is no contradiction with
the commutation relation $[\hat{X}(t),\hat{P}(t)]=i\hbar$ as discussed
in Sec.~\ref{sec:Eq_for_mirrors'_motions_and_their_solutions}.
In the second case, the effect of $\FrakR$ appears as the factor of
the noise spectral density as
Eq.~(\ref{eq:strain-referred-noise-spectrum-density-opt-thetapiover4}).
In both cases, if the incompleteness $\epsilon$ is sufficiently large,
there are cases where the shot-noise contribution to the
high-frequency region in the signal-referred noise spectral density
does not decrease even if the incident power is increased.
We also evaluate the incompleteness $\epsilon$ when we can realize the
behavior near the complete equilibrium tuning in
Eqs.~(\ref{eq:classical-terms-of-RPF-EM-renorm-retune-require-leading-order})
and
(\ref{eq:classical-terms-of-RPF-ITM-renorm-retune-require-leading-order}).
This is the main result of this paper.


Here, we note that we cannot discuss the recent experimental
results of the violation of the ``standard quantum
limit''~\cite{H.Yu-et-al-2020,W.Jia-et-al-2024} through the
ingredients of this paper, because we did not discuss the power
recycling, the signal recycling, or the squeezed input techniques.
We cannot say any relation between the dips depicted in
Fig.~\ref{fig:CompSHEQFULL-dark-1-20k-mod} and works in
Refs.~\cite{H.Yu-et-al-2020,W.Jia-et-al-2024} due to the same reason.
In other words, there is ample room for development through an
accurate understanding of the ingredients in this paper.
We leave these rooms for future work.


Going back to the mathematically rigorous quantum measurement theory
in quantum mechanics, the developed mathematical measurement
theory~\cite{M.Ozawa-2004} was correct.
This is supported by the experimental realizations~\cite{J.Erhart-et-al-Nature-Phys-2012,L.A.Rozema-et-al-2012,F.Kaneda-S.Baek-M.Ozawa-K.Edamatsu-2014}
of the derived error-disturbance relations, which are different from
the Heisenberg uncertainty principle~\cite{Heisenberg-1927}.
However, due to the arguments within this paper, the uncertainty
relation of the position and the momentum for a point mass has nothing
to do with the so-called ``standard quantum limit'' in the
gravitational-wave detection community.
We may emphasize that the developed mathematical quantum
measurement theory~\cite{M.Ozawa-2004} is physically correct.
However, their motivation in the gravitational-wave detectors, as the
precise measurement of the mirrors' positions, was pointless.


Of course, it is well-known that this ``standard quantum limit'' can
be estimated based on the position and the momentum uncertainty in
quantum mechanics~\cite{H.Miao-PhDthesis-2010}.
The arguments presented in this paper do not provide any answer or
insight into this point.
At this moment, we have to say that this point requires the delicate
discussions on the quantum measurement theory in the quantum fields,
because it is true that the ``standard quantum limit'' in
gravitational-wave detectors arises from the noncommutativity
(\ref{eq:K.Nakamura-M.-K.Fujimoto-2018-15}) and
(\ref{eq:K.Nakamura-M.-K.Fujimoto-2018-16}) in the quantum
electrodynamics for the optical field, namely the uncertainty relation
of the electromagnetic field.
In this sense, we have to say that the application of the mathematical
rigorous quantum measurement theory in quantum
mechanics~\cite{M.Ozawa-2004} requires their extension to the quantum
measurement theory for quantum field theories.
Although the answer to this question is beyond the current scope of
this paper, we expect that the mathematically rigorous quantum
measurement theory in quantum field theory will exist.
We hope that our consideration within this paper will motivate us to
develop the measurement theory in quantum field theories.


Finally, we emphasize that the purpose of this paper is not to propose
new techniques nor to develop new ideas for gravitational-wave
detectors themselves.
However, in this paper, we aim to explicitly show the derivation of
the common knowledge in the community of gravitational-wave
detections, thereby clarifying their concepts from a more
theoretically accurate perspective.
Due to the more precise theoretical arguments on the noise spectral
density, we could discuss the incomplete equilibrium tuning of the
Frabry-P\'erot interferometer as an imperfection of the
interferometric gravitational-wave detectors.
We hope the ingredients of this paper and this kind of research will
be helpful for the further development of detector science in
gravitational-wave detectors.


\appendix

\section{Power counting photodetection}
\label{sec:Power_counting_photodetection}


In this appendix, we consider the photodetector model in which the
photocurrent is proportional to the power operator $\hat{P}_{b}(t)$ of
the output optical field $\hat{E}_{b}(t)$,
\begin{eqnarray}
  \label{eq:power-operator-def}
  \hat{P}_{b}(t)
  :=
  \frac{\kappa_{p}c}{4\pi\hbar} {\cal A} \left(\hat{E}_{b}(t)\right)^{2},
\end{eqnarray}
while we discussed the model in which the photocurrent is proportional
to the Glauber photon number~(\ref{eq:Glauber-photon-number}).
In Eq.~(\ref{eq:power-operator-def}), $\kappa_{p}$ is a
phenomenological constant whose dimension is [time].
This coefficient $\kappa_{p}$ includes so-called
``quantum efficiency.''
However, $\kappa_{p}$ is not important within our discussion, though
quantum efficiency is crucial for the actual experiment.
We use the notations in Sec.~\ref{sec:Electric_field_notation} for the
output electric field $\hat{E}_{b}(t)$ for the laser.


In terms of the quadrature $\hat{B}(\omega)$ defined by
Eq.~(\ref{eq:hatB-def}), the power operator
(\ref{eq:power-operator-def}) is given by
\begin{eqnarray}
  \label{eq:power-operator-quadrature}
  \hat{P}_{b}(t)
  &=&
      \frac{\kappa_{p}}{2}
      \int_{-\infty}^{+\infty} \frac{d\omega_{1}}{2\pi}
      \int_{-\infty}^{+\infty} \frac{d\omega_{2}}{2\pi}
      \sqrt{|\omega_{1}\omega_{2}|}
      \nonumber\\
  && \quad
     \times
     \hat{B}(\omega_{1})
     \hat{B}(\omega_{2})
     e^{-i(\omega_{1}+\omega_{2})t}
     ,
\end{eqnarray}
and its Fourier transformation is given by
\begin{eqnarray}
  &&
     \hat{\calP}_{b}(\Omega)
     \nonumber\\
  &:=&
       \int_{-\infty}^{+\infty} dt \hat{P}_{b}(t) e^{+i\Omega t}
       \nonumber\\
  &=&
      \frac{\kappa_{p}}{2}
      \int_{-\infty}^{+\infty} \frac{d\omega_{1}}{2\pi}
      \sqrt{|\omega_{1}(\Omega-\omega_{1})|}
      \hat{B}(\omega_{1})
      \hat{B}(\Omega-\omega_{1})
      .
      \nonumber\\
      \label{eq:power-operator-quadrature-Fourier}
\end{eqnarray}
Here, we assume the output quadrature $\hat{b}(\Omega)$ is given in
the form (\ref{eq:output-quadrature-general}) with the expectation
value (\ref{eq:exp-valu-hatb-def}).
Substituting Eqs.~(\ref{eq:output-quadrature-general}) and
(\ref{eq:exp-valu-hatb-def}) through Eq.~(\ref{eq:hatB-def}), and
considering the situation $\Omega\ll\omega_{0}$, we obtain the
expectation value of the power operator $\hat{\calP}_{b}(\Omega)$ as
\begin{eqnarray}
  \langle\hat{\calP}_{b}(\Omega)\rangle
  &=&
      \frac{\kappa_{p}}{2}
      \omega_{0}
      \left(
      \FrakB^{2}
      2\pi \delta(\Omega-2\omega_{0})
      +
      |\FrakB|^{2}
      2\pi \delta(\Omega)
      \right.
      \nonumber\\
  && \quad\quad\quad
      \left.
      +
      |\FrakB|^{2}
      2\pi \delta(\Omega)
      +
      (\FrakB^{*})^{2}
      2\pi \delta(\Omega+2\omega_{0})
      \right)
     \nonumber\\
  && \quad
     +
     \kappa_{p}
     \omega_{0}
     \left[
     + \FrakB \FrakA^{*}(\omega_{0}-\Omega)
     + \FrakB^{*} \FrakA(\Omega+\omega_{0})
     \right]
     \nonumber\\
  && \quad
     +
     O\left(
     |\FrakB|^{0}
     \right)
     .
     \label{eq:output-power-Fourier-interfero-operator-exp-eval}
\end{eqnarray}
Since gravitational-wave signals are included in $\FrakA(\omega)$, we
may define the signal operator $\hat{s}_{\calP_{b}}(\Omega)$ in the
situation $\Omega\ll\omega_{0}$ by
\begin{eqnarray}
  &&
     \hat{s}_{\calP_{b}}(\Omega)
     \nonumber\\
  &:=&
       \frac{1}{\kappa_{p}}
       \left[
       \hat{\calP}_{b}(\Omega)
       \right.
       \nonumber\\
  && \quad\quad
     \left.
     -
     \frac{\kappa_{p}}{2}
     \omega_{0}
     \left(
     \FrakB^{2}
     2\pi \delta(\Omega-2\omega_{0})
     +
     |\FrakB|^{2}
     2\pi \delta(\Omega)
     \right.
     \right.
     \nonumber\\
  && \quad\quad\quad\quad\quad
     \left.
     \left.
     +
     |\FrakB|^{2}
     2\pi \delta(\Omega)
     +
     (\FrakB^{*})^{2}
     2\pi \delta(\Omega+2\omega_{0})
     \right)
     \right]
     .
     \nonumber\\
     \label{eq:output-power-signal-operator-def-freq-domain}
\end{eqnarray}
For the frequency range $\Omega\ll\omega_{0}$, we obtain
\begin{eqnarray}
  \hat{s}_{\calP_{b}}(\Omega)
  &\sim&
         \omega_{0}
         \left[
         \FrakB \FrakA^{*}(\omega_{0}-\Omega)
         + \FrakB^{*} \FrakA(\Omega+\omega_{0})
         \right.
         \nonumber\\
  && \quad\quad
     \left.
     + \FrakB \hat{b}_{n}^{\dagger}(\omega_{0}-\Omega)
     + \FrakB^{*} \hat{b}_{n}(\omega+\Omega_{0})
     \right]
     \nonumber\\
  &&
     +
     O\left(|\FrakB|^{0}\right)
     .
     \label{eq:output-power-signal-operator-def-freq-domain-eval}
\end{eqnarray}
In this situation, the expectation value of $\hat{s}_{\calP_{b}}(\Omega)$ is
given by
\begin{eqnarray}
  \langle\hat{s}_{\calP_{b}}(\Omega)\rangle
  &\sim&
      \omega_{0}
      \left[
      \FrakB \FrakA^{*}(\omega_{0}-\Omega)
      + \FrakB^{*} \FrakA(\omega+\Omega_{0})
      \right]
     \nonumber\\
  && \quad
     +
     O\left(|\FrakB|^{0}\right)
     .
     \label{eq:output-power-signal-operator-freq-domain-eval-exp}
\end{eqnarray}
The dominant term in
Eq.~(\ref{eq:output-power-signal-operator-freq-domain-eval-exp}) is
same as
Eqs.~(\ref{eq:signal-operator-Fourier-exp}) in the case of the Glauber
photon number $\calN_{b}(\Omega)$ with the situation
$\Omega\ll\omega_{0}$.
The signal operator in the time domain $\hat{s}_{P_{b}}(t)$ is given by
the inverse Fourier transformation of $\hat{s}_{\calP_{b}}(\Omega)$ as
\begin{eqnarray}
  \hat{s}_{P_{b}}(t)
  &=&
      \int_{-\infty}^{+\infty} \frac{d\Omega}{2\pi}
      e^{-i\Omega t}
      \hat{s}_{\calP_{b}}(\Omega)
      .
      \label{eq:InvFourier-hatsPomega}
\end{eqnarray}


From the definition of the signal operator
(\ref{eq:output-power-signal-operator-def-freq-domain}), we can also
define the noise operator $\hat{s}_{Pn}(t)$ for this signal operator
$\hat{s}_{P_{b}}(t)$ by
\begin{eqnarray}
  &&
     \hat{s}_{Pn}(t)
     \nonumber\\
  &:=&
       \hat{s}_{P_{b}}(t) - \langle\hat{s}_{P_{b}}(t)\rangle
       \nonumber\\
  &=&
      \sqrt{\omega_{0}}
      \left(
      \FrakB e^{ - i \omega_{0} t}
      +
      \FrakB^{*} e^{ + i \omega_{0} t}
      \right)
      \nonumber\\
  &&
     \times
     \int_{-\infty}^{+\infty} \frac{d\omega_{1}}{2\pi}
     \sqrt{|\omega_{1}|}
     \left[
     \hat{b}_{n}(\omega_{1}) \Theta(\omega_{1})
     \right.
     \nonumber\\
  && \quad\quad\quad\quad\quad\quad\quad\quad\quad
     \left.
     + \hat{b}_{n}^{\dagger}(-\omega_{1}) \Theta(-\omega_{1})
     \right]
     e^{ -  i \omega_{1} t}
     \nonumber\\
  &&
     +
     O\left(|\FrakB|^{0}\right)
     .
     \label{eq:output-power-noise-operator-def-time-domain-2}
\end{eqnarray}
Through this definition of the noise operator $\hat{s}_{Pn}(t)$, we
can evaluate the time-averaged noise correlation function as in
Sec.~\ref{sec:General_arguments_for_the_DC-readout_scheme} by
\begin{eqnarray}
  &&
     C_{({\rm av})s_{Pn}}(\tau)
     \nonumber\\
  &:=&
       \lim_{T\rightarrow\infty} \frac{1}{T} \int_{-T/2}^{T/2} dt
       \frac{1}{2} \langle\mbox{in}|
       \hat{s}_{Pn}(t+\tau)
       \hat{s}_{Pn}(t)
       \nonumber\\
  && \quad\quad\quad\quad\quad\quad\quad\quad\quad\quad
  +
  \hat{s}_{Pn}(t)
  \hat{s}_{Pn}(t+\tau)
  |\mbox{in}\rangle
  .
     \nonumber\\
  \label{eq:noise-correlation-function-def-pwer}
\end{eqnarray}
Similarly to the arguments in
Sec.~\ref{sec:General_arguments_for_the_DC-readout_scheme}, the noise
spectral density $S_{s_{P}}(\omega)$ is also defined as
\begin{eqnarray}
  \label{eq:signal-spectral-density-def-power}
  &&
     S_{s_{P_{b}}}(\Omega)
     \nonumber\\
  &:=&
       \int_{-\infty}^{+\infty}
       d\tau
       C_{({\rm av})s_{N}}(\tau)
       e^{+i\Omega \tau}
       \\
  &=&
      \int_{-\infty}^{+\infty}
      d\tau
      e^{+i\omega \tau}
      \lim_{T\rightarrow\infty} \frac{1}{T} \int_{-T/2}^{T/2} dt
      \nonumber\\
  &&
     \times
      \frac{1}{2}
      \left[
      \langle
      \hat{s}_{Pn}(t+\tau)
      \hat{s}_{Pn}(t)
      \rangle
      +
      \langle
      \hat{s}_{Pn}(t)
      \hat{s}_{Pn}(t+\tau)
      \rangle
      \right]
     .
     \nonumber\\
  \label{eq:signal-spectral-density-def-2-power}
\end{eqnarray}
Through the original definition
(\ref{eq:power-operator-quadrature-Fourier}) of the power operator is
different from the definition of Glauber's photon number
(\ref{eq:Fourier-Galuber-photon-number}),
in the situation $\omega_{0}\gg\Omega$, the tedious but
straightforward calculations lead to the expression of the stationary
noise spectral density including the one-point support function
defined by Eq.~(\ref{eq:one-point-support-func-def-1}).
As in the case of Glauber's photon number, through the final
input-output relation
(\ref{eq:input-output-rel-explicit-quad-I0-kappabetahSQL}) which was
derived in
Sec.~\ref{sec:Final_input-output-relation_for_Fabry-Perot_GW_Detector},
we obtain the formulae
\begin{eqnarray}
  \langle
  \hat{b}_{n}^{\dagger}(\omega_{0}-\omega)
  \hat{b}_{n}^{\dagger}(\omega_{1})
  \rangle
  &\propto&
  2 \pi \delta(\omega_{0}+\omega-\omega_{1})
  .
  \label{eq:bndaggeromega0-omegabndagger-omega1-exp-delta-contribution}
  \\
  \langle
  \hat{b}_{n}(\omega_{0}+\omega)
  \hat{b}_{n}^{\dagger}(-\omega_{1})
  \rangle
  &\propto&
  2 \pi \delta(\omega_{0}+\omega+\omega_{1})
  .
  \label{eq:bnomega0+omegabndagger-omega1-exp-delta-contribution}
     \\
  \langle
  \hat{b}_{n}^{\dagger}(\omega_{0}-\omega)
  \hat{b}_{n}(\omega_{1})
  \rangle
  &\propto&
  2 \pi \delta(\omega-\omega_{0}+\omega_{1})
  .
  \label{eq:bndaggeromega0-omegabnomega1-exp-delta-contribution}
      \\
  \langle
  \hat{b}_{n}(\omega_{0}+\omega)
  \hat{b}_{n}(\omega_{1})
  \rangle
  &\propto&
  2 \pi \delta(\omega+\omega_{1}-\omega_{0})
  ,
  \label{eq:bnomega0+omegabnomega1-exp-delta-contribution}
\end{eqnarray}
in the same way to obtain Eqs.~(\ref{eq:bndaggeromega0-omegabndaggeromega3-exp-delta-contribution})--(\ref{eq:bnomega0+omegabnomega3-exp-delta-contribution})
  in Sec.~\ref{sec:Multi-mode_number_and_power_operators}.
Using these formulae,
(\ref{eq:bndaggeromega0-omegabndagger-omega1-exp-delta-contribution})--(\ref{eq:bnomega0+omegabnomega1-exp-delta-contribution}) we
reach the expression
\begin{eqnarray}
  &&
     2\pi \delta(\Omega-\Omega')
     S_{s_{P}}(\Omega)
     \nonumber\\
  &=&
      \frac{1}{2}
      \omega_{0}^{2}
      |\FrakB|^{2}
      \left\langle
      \left[
      e^{+2i\Theta}
      \hat{b}_{n-}^{\dagger}(\Omega)
      \hat{b}_{n+}^{\dagger}(\Omega')
      +
      \hat{b}_{n+}(\Omega)
      \hat{b}_{n+}^{\dagger}(\Omega')
      \right.
      \right.
      \nonumber\\
  && \quad\quad\quad\quad
     \left.
     \left.
      +
      \hat{b}_{n-}^{\dagger}(\Omega)
      \hat{b}_{n-}(\Omega')
      +
      e^{-2i\Theta}
      \hat{b}_{n+}(\Omega)
      \hat{b}_{n-}(\Omega')
      \right.
      \right.
      \nonumber\\
  && \quad\quad\quad\quad
     \left.
     \left.
     +
     e^{+2i\Theta}
     \hat{b}_{n+}^{\dagger}(\Omega')
     \hat{b}_{n-}^{\dagger}(\Omega)
     +
     \hat{b}_{n-}(\Omega')
     \hat{b}_{n-}^{\dagger}(\Omega)
      \right.
      \right.
      \nonumber\\
  && \quad\quad\quad\quad
     \left.
     \left.
     +
     \hat{b}_{n+}^{\dagger}(\Omega')
     \hat{b}_{n+}(\Omega)
     +
     e^{-2i\Theta}
     \hat{b}_{n-}(\Omega')
     \hat{b}_{n+}(\Omega)
     \right]
     \right\rangle
     \nonumber\\
  &&
     +
     O\left(|\FrakB|^{1},|\FrakB|^{0}\right)
     ,
     \label{eq:signal-spectral-density-explicit-tmp-2}
\end{eqnarray}
where we used the sideband picture
(\ref{eq:upper-lower-sideband-quadrature-def}) and the situation
$\Omega\ll\omega_{0}$.


Furthermore, we introduce the amplitude- and the phase-quadratures
$\hat{b}_{n1}(\Omega)$ and $\hat{b}_{n2}(\Omega)$ by
Eqs.~(\ref{eq:amplitude-phase-quadrature-def-1}) and
(\ref{eq:amplitude-phase-quadrature-def-2}).
Moreover, we define the operator $\hat{b}_{n\Theta}(\Omega)$ defined
by Eq.~(\ref{eq:bTheta-noise-quadrature-def}).
Then, we obtain
\begin{eqnarray}
  &&
     2\pi \delta(\Omega-\Omega')
     S_{s_{P}}(\Omega)
     \nonumber\\
  &=&
      \omega_{0}^{2}
      |\FrakB|^{2}
      \left\langle
      \left[
      \hat{b}_{n\Theta}(\Omega)
      \hat{b}_{n\Theta}^{\dagger}(\Omega)
      +
      \hat{b}_{n\Theta}^{\dagger}(\Omega)
      \hat{b}_{n\Theta}(\Omega)
      \right]
      \right\rangle
      \nonumber\\
  &&
     +
     O\left(|\FrakB|^{1},|\FrakB|^{0}\right)
     ,
     \label{eq:signal-spectral-density-explicit-power-final}
\end{eqnarray}
with the commutation relation (\ref{eq:commutation-of--hat-bTheta}).
This coincides with the expression of the noise spectral density
(\ref{eq:Measured-noise-spectral-density-DCreadout-bTheta}) in the
case of Glauber's photon number model of the photodetection.


\section{Explicit evaluation of the radiation pressure forces to the mirrors}
\label{sec:Explicit_evaluation_of_the_radiation_pressure_forces}


In this appendix, we show the evaluation of the radiation pressure
forces on the mirrors in
Sec.~\ref{sec:Heisenberg_Equations_for_mirrors'_motions}.
The crucial premise of this evaluation is that the radiation pressure
forces on the mirrors are determined by the laser's power, which
corresponds to the laser's pointing flux and affects the mirrors.
As described in
Ref.~\cite{H.J.Kimble-Y.Levin-A.B.Matsko-K.S.Thorne-S.P.Vyatchanin-2001},
the power operator of the laser is given by ${\cal A}
\hat{E}^{2}(t)/(4\pi)$, where $\hat{E}(t)$ is the electric field
operator which touch to mirrors and ${\cal A}$ is the cross-sectional
area of the optical beam which introduced in
Eq.~(\ref{eq:K.Nakamura-M.-K.Fujimoto-2018-14}).


As depicted in
Fig.~\ref{fig:arm-propagation-Fabry-Perot-setup-notation},
the electric field operator of the laser at EMs is determined by
$\hat{E}_{j_{x,y}}(t)$.
As depicted in
Fig.~\ref{fig:arm-propagation-Fabry-Perot-setup-notation}, the electric
field $\hat{E}_{j_{x,y}}(t)$ at EMs are the propagated fields of
$\hat{E}_{g_{x,y}}(t)$ at ITMs as
\begin{eqnarray}
  \label{eq:hatEjxy-is-propagated-from-hatEgxy}
  \hat{E}_{j_{x,y}}(t)
  =
  \hat{E}_{g_{x,y}}\left[t - \left(\tau + \frac{1}{c} \hat{X}_{x,y}(t-\tau)\right)\right]
  .
\end{eqnarray}
In this paper, we assumed the perfect reflection at EMs as shown in
Eq.~(\ref{eq:End-Mirror-Perfect-reflection-field}).
If we take into account the imperfection of the mirrors, we have to
change the condition (\ref{eq:End-Mirror-Perfect-reflection-field}) as
in Ref.~\cite{H.J.Kimble-Y.Levin-A.B.Matsko-K.S.Thorne-S.P.Vyatchanin-2001}.


For example, the radiation pressure force to XEM is given by
Eq.~(\ref{eq:radiation-pressure-force-XEM}).
Through the notation introduced in
Sec.~\ref{sec:Electric_field_notation} and
Eq.~(\ref{eq:hatXx-Fouriler-def}), the Fourier transformation of
the radiation pressure force $\hat{F}_{rpXEM}(t)$ to XEM per the XEM
mass $m_{EM}$ is given by
\begin{widetext}
\begin{eqnarray}
     \frac{1}{m_{EM}}
     \int_{-\infty}^{+\infty} dt e^{+i\omega t}
     \hat{F}_{rpXEM}(t)
  &=&
      \frac{\hbar}{m_{EM}c}
      e^{ + i \omega \tau }
      \int_{-\infty}^{+\infty} \frac{d\omega_{1}}{2\pi}
      \sqrt{|\omega_{1}(\omega-\omega_{1})|}
      \hat{G}_{x}(\omega_{1})
      \hat{G}_{x}(\omega - \omega_{1})
      \nonumber\\
  &&
     +
     i
     \frac{2\hbar}{m_{EM}c^{2}}
     e^{ + i \omega \tau }
     \int_{-\infty}^{+\infty} \frac{d\omega_{1}}{2\pi}
     \int_{-\infty}^{+\infty} \frac{d\omega_{2}}{2\pi}
     \omega_{1}
     \sqrt{|\omega_{1}\omega_{2}|}
     \hat{G}_{x}(\omega_{1})
     \hat{G}_{x}(\omega_{2})
     \hat{Z}_{x}(\omega - \omega_{1} - \omega_{2})
     \nonumber\\
  &&
     +
     O\left(\hat{X}_{x}^{2}\right)
     .
     \label{eq:XEM-radiation-pressure-force-Gx}
\end{eqnarray}
Substituting
Eq.~(\ref{eq:xarm-input-field-junction-hatCx}),
(\ref{eq:xarm-retarded-effect-f'-to-c'-mod-Fourier-result}), and
(\ref{eq:xarm-prop-gx-is-f-z-Fourier}) into
Eq.~(\ref{eq:XEM-radiation-pressure-force-Gx}), we obtain
\begin{eqnarray}
  &&
     \frac{1}{m_{EM}}
     \int_{-\infty}^{+\infty} dt e^{+i\omega t}
     \hat{F}_{rpXEM}(t)
     \nonumber\\
  &=&
      T
      \frac{\hbar}{2m_{EM}c}
      e^{ + i \omega \tau }
      e^{ + i \omega \tau_{x}' }
      \int_{-\infty}^{+\infty} \frac{d\omega_{1}}{2\pi}
      \sqrt{|\omega_{1}(\omega-\omega_{1})|}
      \left[ 1 - \sqrt{1-T} e^{+ 2 i \omega_{1} \tau } \right]^{-1}
     \left[ 1 - \sqrt{1-T} e^{+ 2 i (\omega-\omega_{1}) \tau } \right]^{-1}
      \nonumber\\
  && \quad\quad\quad\quad\quad\quad\quad\quad\quad\quad\quad\quad
     \times
     \left(
     \hat{D}(\omega_{1}) - \hat{A}(\omega_{1})
     \right)
     \left(
     \hat{D}(\omega-\omega_{1}) - \hat{A}(\omega-\omega_{1})
     \right)
     \nonumber\\
  &&
     +
     i T
     \frac{\hbar}{m_{EM}c}
     e^{ + i \omega \tau }
     \int_{-\infty}^{+\infty} \frac{d\omega_{1}}{2\pi}
     \int_{-\infty}^{+\infty} \frac{d\omega_{2}}{2\pi}
     e^{ + i (\omega_{1}+\omega_{2}) \tau_{x}' }
     \frac{\omega_{2}}{c}
     \sqrt{|\omega_{2}\omega_{1}|}
     \left[ 1 - \sqrt{1-T} e^{+ 2 i \omega_{1} \tau } \right]^{-1}
     \left[ 1 - \sqrt{1-T} e^{+ 2 i (\omega-\omega_{1}) \tau } \right]^{-1}
     \nonumber\\
  && \quad\quad\quad\quad\quad\quad\quad\quad\quad\quad\quad\quad
     \times
     \left(
     \hat{D}(\omega_{1}) - \hat{A}(\omega_{1})
     \right)
     \left(
     \hat{D}(\omega_{2}) - \hat{A}(\omega_{2})
     \right)
     \hat{Z}_{XITM}(\omega-\omega_{1}-\omega_{2})
     \nonumber\\
  &&
     +
     i T
     \frac{\hbar}{m_{EM}c^{2}}
     e^{ + i \omega \tau }
     \int_{-\infty}^{+\infty} \frac{d\omega_{1}}{2\pi}
     \int_{-\infty}^{+\infty} \frac{d\omega_{2}}{2\pi}
     e^{ + i (\omega_{1}+\omega_{2}) \tau_{x}' }
     \omega_{2}
     \sqrt{|\omega_{1}\omega_{2}|}
     \left[ 1 - \sqrt{1-T} e^{+ 2 i \omega_{2} \tau } \right]^{-1}
     \nonumber\\
  && \quad\quad\quad\quad\quad\quad\quad\quad\quad\quad\quad\quad\quad\quad\quad
     \times
     \left[ 1 - \sqrt{1-T} e^{+ 2 i \omega_{1} \tau } \right]^{-1}
     \left[ 1 - \sqrt{1-T} e^{+ 2 i (\omega-\omega_{1}) \tau } \right]^{-1}
     \nonumber\\
  && \quad\quad\quad\quad\quad\quad\quad\quad\quad\quad\quad\quad\quad\quad\quad
     \times
     \left[
     1
     + 2 \sqrt{1-T} e^{+ i ( \omega_{2} + \omega - \omega_{1} ) \tau }
     -     \sqrt{1-T} e^{+ 2 i (\omega-\omega_{1}) \tau }
     \right]
     \nonumber\\
  && \quad\quad\quad\quad\quad\quad\quad\quad\quad\quad\quad\quad\quad\quad\quad
     \times
     \left(
     \hat{D}(\omega_{1}) - \hat{A}(\omega_{1})
     \right)
     \left(
     \hat{D}(\omega_{2}) - \hat{A}(\omega_{2})
     \right)
     \hat{Z}_{x}(\omega - \omega_{1} - \omega_{2})
     \nonumber\\
  &&
     +
     O\left(\hat{X}^{2}_{x},\hat{X}_{XITM}^{2}\right)
     .
     \label{eq:radiation-pressure-Fourier-X-first-term-in-DAZx-sum}
\end{eqnarray}
\end{widetext}
Similarly, the radiation pressure force on YEM is given by
(\ref{eq:radiation-pressure-force-YEM}).
Substituting Eqs.~(\ref{eq:yarm-input-field-junction-hatCy}),
(\ref{eq:yarm-retarded-effect-f'-to-c'-mod-Fourier-result}), and
(\ref{eq:xarm-prop-gx-is-f-z-Fourier}) with the replacement
$x\rightarrow y$, we obtain
\begin{widetext}
\begin{eqnarray}
  &&
     \frac{1}{m_{EM}}
     \int_{-\infty}^{+\infty} dt e^{+i\omega t}
     \hat{F}_{rpYEM}(t)
     \nonumber\\
  &=&
      T
      \frac{\hbar}{2m_{EM}c}
      e^{ + i \omega \tau }
      e^{ + i \omega \tau_{y}' }
      \int_{-\infty}^{+\infty} \frac{d\omega_{1}}{2\pi}
      \sqrt{|\omega_{1}(\omega-\omega_{1})|}
      \left[ 1 - \sqrt{1-T} e^{+ 2 i \omega_{1} \tau } \right]^{-1}
      \left[ 1 - \sqrt{1-T} e^{+ 2 i (\omega-\omega_{1}) \tau } \right]^{-1}
     \nonumber\\
  && \quad\quad\quad\quad\quad\quad\quad\quad\quad\quad\quad\quad
     \times
     \left(
     \hat{D}(\omega_{1}) + \hat{A}(\omega_{1})
     \right)
     \left(
     \hat{D}(\omega-\omega_{1}) + \hat{A}(\omega-\omega_{1})
     \right)
     \nonumber\\
  &&
     +
     i T
     \frac{\hbar}{m_{EM}c}
     e^{ + i \omega \tau }
     \int_{-\infty}^{+\infty} \frac{d\omega_{1}}{2\pi}
     \int_{-\infty}^{+\infty} \frac{d\omega_{2}}{2\pi}
     e^{ + i ( \omega_{1} + \omega_{2} ) \tau_{y}' }
     \frac{\omega_{2}}{c}
     \sqrt{|\omega_{1}\omega_{2}|}
     \left[ 1 - \sqrt{1-T} e^{+ 2 i \omega_{1} \tau } \right]^{-1}
     \left[ 1 - \sqrt{1-T} e^{+ 2 i (\omega-\omega_{1}) \tau } \right]^{-1}
     \nonumber\\
  && \quad\quad\quad\quad\quad\quad\quad\quad\quad\quad\quad\quad
     \times
     \left(
     \hat{D}(\omega_{1}) + \hat{A}(\omega_{1})
     \right)
     \left(
     \hat{D}(\omega_{2}) + \hat{A}(\omega_{2})
     \right)
     \hat{Z}_{YITM}(\omega-\omega_{1}-\omega_{2})
     \nonumber\\
  &&
     +
     i T
     \frac{\hbar}{m_{EM}c^{2}}
     e^{ + i \omega \tau }
     \int_{-\infty}^{+\infty} \frac{d\omega_{1}}{2\pi}
     \int_{-\infty}^{+\infty} \frac{d\omega_{2}}{2\pi}
     e^{ + i ( \omega_{1} + \omega_{2} ) \tau_{y}' }
     \omega_{2}
     \sqrt{|\omega_{1}\omega_{2}|}
     \left[ 1 - \sqrt{1-T} e^{+ 2 i \omega_{2} \tau } \right]^{-1}
     \nonumber\\
  && \quad\quad\quad\quad\quad\quad\quad\quad\quad\quad\quad\quad\quad\quad\quad
     \times
     \left[ 1 - \sqrt{1-T} e^{+ 2 i \omega_{1} \tau } \right]^{-1}
     \left[ 1 - \sqrt{1-T} e^{+ 2 i (\omega-\omega_{1}) \tau } \right]^{-1}
     \nonumber\\
  && \quad\quad\quad\quad\quad\quad\quad\quad\quad\quad\quad\quad\quad\quad\quad
     \times
     \left[
     1
     + 2 \sqrt{1-T} e^{+ i ( \omega_{2} + \omega - \omega_{1} ) \tau }
     -     \sqrt{1-T} e^{+ 2 i (\omega-\omega_{1}) \tau }
     \right]
     \nonumber\\
  && \quad\quad\quad\quad\quad\quad\quad\quad\quad\quad\quad\quad\quad\quad\quad
     \times
     \left(
     \hat{D}(\omega_{1}) + \hat{A}(\omega_{1})
     \right)
     \left(
     \hat{D}(\omega_{2}) + \hat{A}(\omega_{2})
     \right)
     \hat{Z}_{y}(\omega-\omega_{1}-\omega_{2})
     \nonumber\\
  &&
     +
     O\left(\hat{X}_{y}^{2},\hat{X}_{YITM}^{2}\right)
     .
     \label{eq:radiation-pressure-Fourier-Y-first-term-in-DAZy-sum}
\end{eqnarray}
\end{widetext}


Next, we consider the radiation pressure force on XITM.
As depicted in
Fig.~\ref{fig:arm-propagation-Fabry-Perot-setup-notation}, the
radiation pressure force to XITM is given by
Eq.~(\ref{eq:total-radiation-pressure-force-XITM}).
Through the junction condition (\ref{eq:Kimble-B8-correspoind-2}) for
the electric field operator at the XITM and the propagation effects
(\ref{eq:xarm-retarded-effect-g-to-gprime}), the radiation pressure
force (\ref{eq:total-radiation-pressure-force-XITM}) on XITM yields
\begin{widetext}
\begin{eqnarray}
     \hat{F}_{rpXITM}(t)
  &=&
      \frac{{\cal A}}{4\pi}\left(\hat{E}_{f_{x}}(t)\right)^{2}
      -
      \frac{{\cal A}}{4\pi}\left(\hat{E}_{g_{x}}(t)\right)^{2}
      +
      \frac{{\cal A}}{4\pi}\left(
      - \sqrt{1-T} \hat{E}_{f_{x}}(t)
      +
      \sqrt{T} \hat{E}_{g_{x}}\left[
      t - 2 \left(\tau + \frac{1}{c} \hat{X}_{x}(t-\tau) \right)
      \right]
      \right)^{2}
      \nonumber\\
  &&
      -
      \frac{{\cal A}}{4\pi}
      \left(\hat{E}_{g_{x}}\left[t-2\left(\tau+\frac{1}{c}\hat{X}_{x}(t-\tau)\right)\right]\right)^{2}
      .
      \label{eq:total-radiation-pressure-force-XITM-append}
\end{eqnarray}
\end{widetext}
Through the notation same as
Eq.~(\ref{eq:K.Nakamura-M.-K.Fujimoto-2018-18}), the Fourier
transformation (\ref{eq:hatXx-Fouriler-def}),
Eqs.~(\ref{eq:xarm-input-field-junction-hatCx}),
(\ref{eq:XITM-Fourier-def}),
(\ref{eq:xarm-retarded-effect-f'-to-c'-mod-Fourier-result}), and
(\ref{eq:xarm-prop-gx-is-f-z-Fourier}), the tedious calculations
lead to the Fourier transformation of the radiation pressure force
$\hat{F}_{rpXITM}(t)$ as
\begin{widetext}
\begin{eqnarray}
  &&
     \int_{-\infty}^{+\infty} dt e^{+i\omega t}
     \frac{1}{m_{ITM}}
     \hat{F}_{rpXITM}(t)
     \nonumber\\
  &=&
      -
      \sqrt{1-T}
      \frac{\hbar}{2m_{ITM}c}
      e^{ + i \omega \tau_{x}' }
      \int_{-\infty}^{+\infty} \frac{d\omega_{1}}{2\pi}
      \sqrt{|\omega_{1}(\omega-\omega_{1})|}
      \left[ 1 - \sqrt{1-T} e^{+ 2 i \omega_{1} \tau } \right]^{-1}
     \nonumber\\
  && \quad\quad\quad\quad\quad\quad\quad\quad\quad\quad\quad\quad\quad\quad
     \times
     \left[ 1 - \sqrt{1-T} e^{+ 2 i (\omega - \omega_{1}) \tau } \right]^{-1}
     \left[
     2 e^{+ 2 i \omega_{1} \tau }
     - \sqrt{1-T} ( 1 + e^{+ 2 i \omega  \tau } )
     \right]
     \nonumber\\
  && \quad\quad\quad\quad\quad\quad\quad\quad\quad\quad\quad\quad\quad\quad
     \times
     \left(
     \hat{D}(\omega_{1}) - \hat{A}(\omega_{1})
     \right)
     \left(
     \hat{D}(\omega-\omega_{1}) - \hat{A}(\omega-\omega_{1})
     \right)
      \nonumber\\
  &&
     -
     i \sqrt{1-T}
     \frac{\hbar}{m_{ITM}c}
     \int_{-\infty}^{+\infty} \frac{d\omega_{1}}{2\pi}
     \int_{-\infty}^{+\infty} \frac{d\omega_{2}}{2\pi}
     e^{ + i ( \omega_{1} + \omega_{2} ) \tau_{x}' }
     \frac{\omega_{2}}{c}
     \sqrt{|\omega_{1}\omega_{2}|}
     \left[ 1 - \sqrt{1-T} e^{+ 2 i \omega_{1} \tau } \right]^{-1}
     \left[ 1 - \sqrt{1-T} e^{+ 2 i (\omega - \omega_{1}) \tau } \right]^{-1}
     \nonumber\\
  && \quad\quad\quad\quad\quad\quad\quad\quad\quad\quad\quad\quad\quad\quad\quad\quad
     \times
     \left[
     e^{+ 2 i (\omega-\omega_{1}) \tau }
     + e^{+ 2 i \omega_{1} \tau }
     -  \sqrt{1-T} ( 1 + e^{+ 2 i \omega  \tau } )
     \right]
     \nonumber\\
  && \quad\quad\quad\quad\quad\quad\quad\quad\quad\quad\quad\quad\quad\quad\quad\quad
     \times
     \left(
     \hat{D}(\omega_{1}) - \hat{A}(\omega_{1})
     \right)
     \left(
     \hat{D}(\omega_{2}) - \hat{A}(\omega_{2})
     \right)
     \hat{Z}_{XITM}(\omega-\omega_{1}-\omega_{2})
     \nonumber\\
  &&
     -
     i T \sqrt{1-T}
     \frac{2\hbar}{m_{ITM}c^{2}}
     \int_{-\infty}^{+\infty} \frac{d\omega_{1}}{2\pi}
     \int_{-\infty}^{+\infty} \frac{d\omega_{2}}{2\pi}
     \sqrt{|\omega_{1}\omega_{2}|}
     \omega_{2}
     e^{ +  i ( \omega_{2} + \omega - \omega_{1} ) \tau }
     e^{ + i ( \omega_{1} + \omega_{2} ) \tau_{x}' }
     \left[ 1 - \sqrt{1-T} e^{+ 2 i \omega_{1} \tau } \right]^{-1}
     \nonumber\\
  && \quad\quad\quad\quad\quad\quad\quad\quad\quad\quad\quad\quad\quad\quad\quad\quad
     \times
     \left[ 1 - \sqrt{1-T} e^{+ 2 i ( \omega - \omega_{1} ) \tau } \right]^{-1}
     \left[ 1 - \sqrt{1-T} e^{+ 2 i \omega_{2} \tau } \right]^{-1}
     \nonumber\\
  && \quad\quad\quad\quad\quad\quad\quad\quad\quad\quad\quad\quad\quad\quad\quad\quad
     \times
     \left(
     \hat{D}(\omega_{1}) - \hat{A}(\omega_{1})
     \right)
     \left(
     \hat{D}(\omega_{2}) - \hat{A}(\omega_{2})
     \right)
     \hat{Z}_{x}(\omega-\omega_{1}-\omega_{2})
     \nonumber\\
  &&
     +
     O\left(\hat{X}_{x}^{2},\hat{X}_{XITM}^{2}\right)
     .
     \label{eq:total-radiation-pressure-force-to-XITM-DAZx-sum}
\end{eqnarray}


Similarly, the radiation pressure force on the YITM is given by
Eq.~(\ref{eq:total-radiation-pressure-force-YITM}), which yields
\begin{eqnarray}
     \hat{F}_{rpYITM}(t)
  &=&
      \frac{{\cal A}}{4\pi}\left(\hat{E}_{f_{y}}(t)\right)^{2}
      -
      \frac{{\cal A}}{4\pi}\left(\hat{E}_{g_{y}}(t)\right)^{2}
      +
      \frac{{\cal A}}{4\pi}\left(
      - \sqrt{1-T} \hat{E}_{f_{y}}(t)
      + \sqrt{T} \hat{E}_{g_{y}}\left[t - 2 \left( \tau + \frac{1}{c} \hat{X}_{y}(t-\tau)\right)\right]
      \right)^{2}
      \nonumber\\
  &&
      -
      \frac{{\cal A}}{4\pi}\left(
      \hat{E}_{g_{y}}\left[t - 2 \left(\tau + \frac{1}{c} \hat{X}_{y}(t-\tau)\right)\right]
      \right)^{2}
      .
      \label{eq:total-radiation-pressure-force-YITM-append}
\end{eqnarray}
\end{widetext}
As in the case of $\hat{F}_{rpXITM}(t)$, we used
Eqs.~(\ref{eq:yarm-retarded-effect-g-to-gprime}) and
(\ref{eq:yarm-prop-fprime-and-f-g}).
Furthermore, through the notation same as
Eq.~(\ref{eq:K.Nakamura-M.-K.Fujimoto-2018-18}), the Fourier
transformation (\ref{eq:hatXx-Fouriler-def}),
Eqs.~(\ref{eq:yarm-input-field-junction-hatCy}),
(\ref{eq:YITM-Fourier-def}),
(\ref{eq:yarm-retarded-effect-f'-to-c'-mod-Fourier-result}), and
(\ref{eq:xarm-prop-gx-is-f-z-Fourier}) with the replacement
$x\rightarrow y$, the tedious calculations lead to the Fourier
transformation of the radiation pressure force $\hat{F}_{rpYITM}(t)$
as
\begin{widetext}
\begin{eqnarray}
  &&
     \int_{-\infty}^{+\infty} dt e^{+i\omega t}
     \frac{1}{m_{ITM}}
     \hat{F}_{rpYITM}(t)
     \nonumber\\
  &=&
      -
      \sqrt{1-T}
      \frac{\hbar}{2m_{ITM}c}
      e^{ + i \omega \tau_{y}' }
      \int_{-\infty}^{+\infty} \frac{d\omega_{1}}{2\pi}
      \sqrt{|\omega_{1}(\omega-\omega_{1})|}
      \left[ 1 - \sqrt{1-T} e^{+ 2 i \omega_{1} \tau } \right]^{-1}
      \nonumber\\
  && \quad\quad\quad\quad\quad\quad\quad\quad\quad\quad\quad\quad\quad\quad
     \times
      \left[ 1 - \sqrt{1-T} e^{+ 2 i (\omega - \omega_{1}) \tau } \right]^{-1}
     \left[
     2 e^{+ 2 i \omega_{1} \tau }
     - \sqrt{1-T} ( 1 + e^{+ 2 i \omega  \tau } )
     \right]
      \nonumber\\
  && \quad\quad\quad\quad\quad\quad\quad\quad\quad\quad\quad\quad\quad\quad
     \times
     \left(
     \hat{D}(\omega_{1}) + \hat{A}(\omega_{1})
     \right)
     \left(
     \hat{D}(\omega-\omega_{1}) + \hat{A}(\omega-\omega_{1})
     \right)
     \nonumber\\
  &&
     -
     i \sqrt{1-T}
     \frac{\hbar}{m_{ITM}c}
     \int_{-\infty}^{+\infty} \frac{d\omega_{1}}{2\pi}
     \int_{-\infty}^{+\infty} \frac{d\omega_{2}}{2\pi}
     e^{ + i ( \omega_{1} + \omega_{2} ) \tau_{y}' }
     \frac{\omega_{2}}{c}
     \sqrt{|\omega_{1}\omega_{2}|}
     \left[ 1 - \sqrt{1-T} e^{+ 2 i \omega_{1} \tau } \right]^{-1}
     \left[ 1 - \sqrt{1-T} e^{+ 2 i (\omega - \omega_{1}) \tau } \right]^{-1}
     \nonumber\\
  && \quad\quad\quad\quad\quad\quad\quad\quad\quad\quad\quad\quad\quad\quad\quad\quad
     \times
     \left[
     e^{+ 2 i (\omega - \omega_{1}) \tau }
     + e^{+ 2 i \omega_{1} \tau }
     -  \sqrt{1-T} ( 1 + e^{+ 2 i \omega  \tau } )
     \right]
     \nonumber\\
  && \quad\quad\quad\quad\quad\quad\quad\quad\quad\quad\quad\quad\quad\quad\quad\quad
     \times
     \left(
     \hat{D}(\omega_{1}) + \hat{A}(\omega_{1})
     \right)
     \left(
     \hat{D}(\omega_{2}) + \hat{A}(\omega_{2})
     \right)
     \hat{Z}_{YITM}(\omega-\omega_{1}-\omega_{2})
     \nonumber\\
  &&
     -
     i T \sqrt{1-T}
     \frac{2\hbar}{m_{ITM}c^{2}}
     \int_{-\infty}^{+\infty} \frac{d\omega_{1}}{2\pi}
     \int_{-\infty}^{+\infty} \frac{d\omega_{2}}{2\pi}
     \sqrt{|\omega_{1}\omega_{2}|}
     \omega_{2}
     e^{ +  i ( \omega_{2} + \omega - \omega_{1} ) \tau }
     e^{ + i ( \omega_{1} + \omega_{2} ) \tau_{y}' }
     \left[ 1 - \sqrt{1-T} e^{+ 2 i \omega_{1} \tau } \right]^{-1}
     \nonumber\\
  && \quad\quad\quad\quad\quad\quad\quad\quad\quad\quad\quad\quad\quad\quad\quad\quad
     \times
     \left[ 1 - \sqrt{1-T} e^{+ 2 i ( \omega - \omega_{1} ) \tau } \right]^{-1}
     \left[ 1 - \sqrt{1-T} e^{+ 2 i \omega_{2} \tau } \right]^{-1}
     \nonumber\\
  && \quad\quad\quad\quad\quad\quad\quad\quad\quad\quad\quad\quad\quad\quad\quad\quad
     \times
     \left(
     \hat{D}(\omega_{1}) + \hat{A}(\omega_{1})
     \right)
     \left(
     \hat{D}(\omega_{2}) + \hat{A}(\omega_{2})
     \right)
     \hat{Z}_{y}(\omega-\omega_{1}-\omega_{2})
     \nonumber\\
  &&
     +
     O\left(\hat{X}_{y}^{2},\hat{X}_{YITM}^{2}\right)
     .
     \label{eq:total-radiation-pressure-force-to-YITM-DA-sum}
\end{eqnarray}
\end{widetext}
Here, we note that the radiation pressure forces
(\ref{eq:radiation-pressure-Fourier-X-first-term-in-DAZx-sum}),
(\ref{eq:radiation-pressure-Fourier-Y-first-term-in-DAZy-sum}),
(\ref{eq:total-radiation-pressure-force-to-XITM-DAZx-sum}), and
(\ref{eq:total-radiation-pressure-force-to-YITM-DA-sum}) are given by
the input quadrature $\hat{D}(\omega)$ and $\hat{A}(\omega)$ with the
Fourier transformations of the displacements $\hat{Z}_{x}$,
$\hat{Z}_{y}$, $\hat{Z}_{XITM}$, and $\hat{Z}_{YITM}$.
Within this paper, we do not consider the power recycling technique,
nor the signal recycling techniques~\cite{H.J.Kimble-Y.Levin-A.B.Matsko-K.S.Thorne-S.P.Vyatchanin-2001,H.Miao-PhDthesis-2010,A.Buonanno-Y.Chen-2001,A.Buonanno-Y.Chen-2002,A.Buonanno-Y.Chen-2003}.
If we take into account the power recycling technique, we have to use
the reflected optical field quadrature by the power recycling mirror
as the input quadrature $\hat{D}(\omega)$.
On the other hand, if we take into account the signal recycling
technique, we have to use the reflected optical field quadrature by
the signal recycling mirror as the input quadrature $\hat{A}(\omega)$.
In any case, we have to note that even when we consider the power
recycling technique and the signal recycling technique, the
expressions of the radiation pressure forces
(\ref{eq:radiation-pressure-Fourier-X-first-term-in-DAZx-sum}),
(\ref{eq:radiation-pressure-Fourier-Y-first-term-in-DAZy-sum}),
(\ref{eq:total-radiation-pressure-force-to-XITM-DAZx-sum}), and
(\ref{eq:total-radiation-pressure-force-to-YITM-DA-sum}) should be
used.


We also note that we have to evaluate the Fourier transformations of
the mirror displacement operators
$\FrakD_{d}^{\dagger}\hat{Z}_{x}\FrakD_{d}$,
$\FrakD_{d}^{\dagger}\hat{Z}_{y}\FrakD_{d}$,
$\FrakD_{d}^{\dagger}\hat{Z}_{XITM}\FrakD_{d}$, and
$\FrakD_{d}^{\dagger}\hat{Z}_{YITM}\FrakD_{d}$ instead of operators
$\hat{Z}_{x}$, $\hat{Z}_{y}$, $\hat{Z}_{XITM}$, and $\hat{Z}_{YITM}$
when we evaluate the input-output relation
(\ref{eq:input-output-relation-general-pert-sideband-approx}).
Here, the operator $\FrakD_{d}$ is the displacement operator
associated with the coherent state for the optical quadrature
$\hat{D}(\omega):=\hat{d}(\omega)\Theta(\omega)+\hat{d}^{\dagger}(-\omega)\Theta(-\omega)$
which is defined by Eq.~(\ref{eq:displacement-operator-for-hatd}).
For the quadrature $\hat{D}(\omega)$,
$\FrakD_{d}^{\dagger}\hat{D}(\omega)\FrakD_{d}$ are given by
Eq.~(\ref{eq:coherent-state-hatD-Heisenberg}).
Trivially, the quadrature $\hat{A}(\omega)$ commutes with the
displacement operator $\FrakD_{d}$.
Keep in our mind the properties of the displacement operator
$\FrakD_{d}$, we evaluate the radiation pressure forces
(\ref{eq:radiation-pressure-Fourier-X-first-term-in-DAZx-sum}),
(\ref{eq:radiation-pressure-Fourier-Y-first-term-in-DAZy-sum}),
(\ref{eq:total-radiation-pressure-force-to-XITM-DAZx-sum}), and
(\ref{eq:total-radiation-pressure-force-to-YITM-DA-sum}) by the
operations $\FrakD_{d}^{\dagger}$ from the left and $\FrakD_{d}$ from
the right.
Furthermore, we ignore the terms include
$\hat{D}_{v}(\omega)\hat{D}_{v}(\omega')$,
$\hat{D}_{v}(\omega)\hat{A}(\omega')$, and
$\hat{A}(\omega)\hat{A}(\omega')$.
We symbolically denote these terms $\hat{D}_{v}^{2}$,
$\hat{D}_{v}\hat{A}$, and $\hat{A}^{2}$.
Moreover, we also ignored the terms that include
$\hat{D}_{v}(\omega)\hat{Z}_{x}(\omega')$,
$\hat{D}_{v}(\omega)\hat{Z}_{XITM}(\omega')$,
$\hat{D}_{v}(\omega)\hat{Z}_{y}(\omega')$,
$\hat{D}_{v}(\omega)\hat{Z}_{YITM}(\omega')$,
$\hat{A}(\omega)\hat{Z}_{x}(\omega')$,
$\hat{A}(\omega)\hat{Z}_{XITM}(\omega')$,
$\hat{A}(\omega)\hat{Z}_{y}(\omega')$,
$\hat{A}(\omega)\hat{Z}_{YITM}(\omega')$.
We symbolically denote these terms as $\hat{D}_{v}\hat{X}$ and
$\hat{A}\hat{X}$.
Through these evaluations, the radiation pressure forces
(\ref{eq:radiation-pressure-Fourier-X-first-term-in-DAZx-sum}),
(\ref{eq:radiation-pressure-Fourier-Y-first-term-in-DAZy-sum}),
(\ref{eq:total-radiation-pressure-force-to-XITM-DAZx-sum}), and
(\ref{eq:total-radiation-pressure-force-to-YITM-DA-sum}) are given by
\begin{widetext}
\begin{eqnarray}
  &&
     \frac{1}{m_{EM}}
     \int_{-\infty}^{+\infty} dt e^{+i\omega t}
     \FrakD_{d}^{\dagger}\hat{F}_{rpXEM}(t)\FrakD_{d}
     \nonumber\\
  &=&
      \frac{\hbar T}{2m_{EM}c}
      e^{ + i \omega (\tau+\tau_{x}' ) }
      \int_{-\infty}^{+\infty} \frac{d\omega_{1}}{2\pi}
      \sqrt{|\omega_{1}(\omega-\omega_{1})|}
      \left[ 1 - \sqrt{1-T} e^{+ 2 i \omega_{1} \tau } \right]^{-1}
      \left[ 1 - \sqrt{1-T} e^{+ 2 i (\omega-\omega_{1}) \tau } \right]^{-1}
     \nonumber\\
  && \quad\quad\quad\quad\quad\quad\quad\quad\quad\quad\quad\quad
     \times
     \hat{D}_{c}(\omega_{1})
     \hat{D}_{c}(\omega-\omega_{1})
     \nonumber\\
  &&
     +
     \frac{\hbar T}{m_{EM}c}
     e^{ + i \omega (\tau+\tau_{x}' ) }
     \int_{-\infty}^{+\infty} \frac{d\omega_{1}}{2\pi}
     \sqrt{|\omega_{1}(\omega-\omega_{1})|}
     \left[ 1 - \sqrt{1-T} e^{+ 2 i \omega_{1} \tau } \right]^{-1}
     \left[ 1 - \sqrt{1-T} e^{+ 2 i (\omega-\omega_{1}) \tau } \right]^{-1}
     \nonumber\\
  && \quad\quad\quad\quad\quad\quad\quad\quad\quad\quad\quad\quad
     \times
     \hat{D}_{c}(\omega_{1})
     \left(
     \hat{D}_{v}(\omega-\omega_{1}) - \hat{A}(\omega-\omega_{1})
     \right)
     \nonumber\\
  &&
     +
     i
     \frac{\hbar T}{m_{EM}c}
     e^{ + i \omega \tau }
     \int_{-\infty}^{+\infty} \frac{d\omega_{1}}{2\pi}
     \int_{-\infty}^{+\infty} \frac{d\omega_{2}}{2\pi}
     e^{ + i (\omega_{1}+\omega_{2}) \tau_{x}' }
     \frac{\omega_{2}}{c}
     \sqrt{|\omega_{2}\omega_{1}|}
     \left[ 1 - \sqrt{1-T} e^{+ 2 i \omega_{1} \tau } \right]^{-1}
     \nonumber\\
  && \quad\quad\quad\quad\quad\quad\quad\quad\quad\quad\quad\quad\quad\quad\quad
     \times
     \left[ 1 - \sqrt{1-T} e^{+ 2 i (\omega-\omega_{1}) \tau } \right]^{-1}
     \hat{D}_{c}(\omega_{1})
     \hat{D}_{c}(\omega_{2})
     \FrakD_{d}^{\dagger}\hat{Z}_{XITM}(\omega-\omega_{1}-\omega_{2})\FrakD_{d}
     \nonumber\\
  &&
     +
     i
     \frac{\hbar T}{m_{EM}c^{2}}
     e^{ + i \omega \tau }
     \int_{-\infty}^{+\infty} \frac{d\omega_{1}}{2\pi}
     \int_{-\infty}^{+\infty} \frac{d\omega_{2}}{2\pi}
     e^{ + i (\omega_{1}+\omega_{2}) \tau_{x}' }
     \omega_{2}
     \sqrt{|\omega_{1}\omega_{2}|}
     \left[ 1 - \sqrt{1-T} e^{+ 2 i \omega_{2} \tau } \right]^{-1}
     \nonumber\\
  && \quad\quad\quad\quad\quad\quad\quad\quad\quad\quad\quad\quad\quad\quad\quad
     \times
     \left[
     1
     + 2 \sqrt{1-T} e^{+ i ( \omega_{2} + \omega - \omega_{1} ) \tau }
     -     \sqrt{1-T} e^{+ 2 i (\omega-\omega_{1}) \tau }
     \right]
     \nonumber\\
  && \quad\quad\quad\quad\quad\quad\quad\quad\quad\quad\quad\quad\quad\quad\quad
     \times
     \left[ 1 - \sqrt{1-T} e^{+ 2 i \omega_{1} \tau } \right]^{-1}
     \left[ 1 - \sqrt{1-T} e^{+ 2 i (\omega-\omega_{1}) \tau } \right]^{-1}
     \nonumber\\
  && \quad\quad\quad\quad\quad\quad\quad\quad\quad\quad\quad\quad\quad\quad\quad
     \times
     \hat{D}_{c}(\omega_{1})
     \hat{D}_{c}(\omega_{2})
     \FrakD_{d}^{\dagger}\hat{Z}_{x}(\omega - \omega_{1} - \omega_{2})\FrakD_{d}
     \nonumber\\
  &&
      +
     O\left(
     \left(\hat{X}\right)^{2},
     \hat{D}_{v}\hat{A},
     \left(\hat{A}\right)^{2},
     \left(\hat{D}_{v}\right)^{2},
     \hat{D}_{v}\hat{X},
     \hat{A}\hat{X}
     \right)
     ,
     \label{eq:radiation-pressure-Fourier-X-first-term-in-DAZx-cohe-sum}
\end{eqnarray}
\begin{eqnarray}
  &&
     \frac{1}{m_{EM}}
     \int_{-\infty}^{+\infty} dt e^{+i\omega t}
     \FrakD_{d}^{\dagger}\hat{F}_{rpYEM}(t)\FrakD_{d}
     \nonumber\\
  &=&
      \frac{\hbar T}{2m_{EM}c}
      e^{ + i \omega (\tau+\tau_{y}') }
      \int_{-\infty}^{+\infty} \frac{d\omega_{1}}{2\pi}
      \sqrt{|\omega_{1}(\omega-\omega_{1})|}
      \left[ 1 - \sqrt{1-T} e^{+ 2 i \omega_{1} \tau } \right]^{-1}
      \left[ 1 - \sqrt{1-T} e^{+ 2 i (\omega-\omega_{1}) \tau } \right]^{-1}
     \nonumber\\
  && \quad\quad\quad\quad\quad\quad\quad\quad\quad\quad\quad\quad
     \times
     \hat{D}_{c}(\omega_{1})
     \hat{D}_{c}(\omega-\omega_{1})
     \nonumber\\
  &&
     +
     \frac{\hbar T}{m_{EM}c}
     e^{ + i \omega (\tau+\tau_{y}' ) }
     \int_{-\infty}^{+\infty} \frac{d\omega_{1}}{2\pi}
     \sqrt{|\omega_{1}(\omega-\omega_{1})|}
     \left[ 1 - \sqrt{1-T} e^{+ 2 i \omega_{1} \tau } \right]^{-1}
     \left[ 1 - \sqrt{1-T} e^{+ 2 i (\omega-\omega_{1}) \tau } \right]^{-1}
     \nonumber\\
  && \quad\quad\quad\quad\quad\quad\quad\quad\quad\quad\quad\quad
     \times
     \hat{D}_{c}(\omega-\omega_{1})
     \left(
     \hat{D}_{v}(\omega_{1}) + \hat{A}(\omega_{1})
     \right)
     \nonumber\\
  &&
     +
     i
     \frac{\hbar T}{m_{EM}c}
     e^{ + i \omega \tau }
     \int_{-\infty}^{+\infty} \frac{d\omega_{1}}{2\pi}
     \int_{-\infty}^{+\infty} \frac{d\omega_{2}}{2\pi}
     e^{ + i ( \omega_{1} + \omega_{2} ) \tau_{y}' }
     \frac{\omega_{2}}{c}
     \sqrt{|\omega_{1}\omega_{2}|}
     \left[ 1 - \sqrt{1-T} e^{+ 2 i \omega_{1} \tau } \right]^{-1}
     \nonumber\\
  && \quad\quad\quad\quad\quad\quad\quad\quad\quad\quad\quad\quad\quad\quad\quad
     \times
     \left[ 1 - \sqrt{1-T} e^{+ 2 i (\omega-\omega_{1}) \tau } \right]^{-1}
     \hat{D}_{c}(\omega_{1})
     \hat{D}_{c}(\omega_{2})
     \FrakD_{d}^{\dagger}\hat{Z}_{YITM}(\omega-\omega_{1}-\omega_{2})\FrakD_{d}
     \nonumber\\
  &&
     +
     i
     \frac{\hbar T}{m_{EM}c^{2}}
     e^{ + i \omega \tau }
     \int_{-\infty}^{+\infty} \frac{d\omega_{1}}{2\pi}
     \int_{-\infty}^{+\infty} \frac{d\omega_{2}}{2\pi}
     e^{ + i ( \omega_{1} + \omega_{2} ) \tau_{y}' }
     \omega_{2}
     \sqrt{|\omega_{1}\omega_{2}|}
     \left[ 1 - \sqrt{1-T} e^{+ 2 i \omega_{2} \tau } \right]^{-1}
     \nonumber\\
  && \quad\quad\quad\quad\quad\quad\quad\quad\quad\quad\quad\quad\quad\quad\quad
     \times
     \left[
     1
     + 2 \sqrt{1-T} e^{+ i ( \omega_{2} + \omega - \omega_{1} ) \tau }
     -     \sqrt{1-T} e^{+ 2 i (\omega-\omega_{1}) \tau }
     \right]
     \nonumber\\
  && \quad\quad\quad\quad\quad\quad\quad\quad\quad\quad\quad\quad\quad\quad\quad
     \times
     \left[ 1 - \sqrt{1-T} e^{+ 2 i \omega_{1} \tau } \right]^{-1}
     \left[ 1 - \sqrt{1-T} e^{+ 2 i (\omega-\omega_{1}) \tau } \right]^{-1}
     \nonumber\\
  && \quad\quad\quad\quad\quad\quad\quad\quad\quad\quad\quad\quad\quad\quad\quad
     \times
     \hat{D}_{c}(\omega_{1})
     \hat{D}_{c}(\omega_{2})
     \FrakD_{d}^{\dagger}\hat{Z}_{y}(\omega-\omega_{1}-\omega_{2})\FrakD_{d}
     \nonumber\\
  &&
     +
     O\left(
     \left(\hat{X}\right)^{2},
     \hat{D}_{v}\hat{A},
     \left(\hat{A}\right)^{2},
     \left(\hat{D}_{v}\right)^{2},
     \hat{D}_{v}\hat{X},
     \hat{A}\hat{X}
     \right)
     ,
     \label{eq:radiation-pressure-Fourier-Y-first-term-in-DAZy-coh-sum}
\end{eqnarray}
\begin{eqnarray}
  &&
     \int_{-\infty}^{+\infty} dt e^{+i\omega t}
     \frac{1}{m_{ITM}}
     \FrakD_{d}^{\dagger}\hat{F}_{rpXITM}(t)\FrakD_{d}
     \nonumber\\
  &=&
      -
      \sqrt{1-T}
      \frac{\hbar}{2m_{ITM}c}
      e^{ + i \omega \tau_{x}' }
      \int_{-\infty}^{+\infty} \frac{d\omega_{1}}{2\pi}
      \sqrt{|\omega_{1}(\omega-\omega_{1})|}
      \left[ 1 - \sqrt{1-T} e^{+ 2 i \omega_{1} \tau } \right]^{-1}
      \left[ 1 - \sqrt{1-T} e^{+ 2 i (\omega - \omega_{1}) \tau } \right]^{-1}
     \nonumber\\
  && \quad\quad\quad\quad\quad\quad\quad\quad\quad\quad\quad\quad\quad\quad
     \times
     \left[
     2 e^{+ 2 i \omega_{1} \tau }
     - \sqrt{1-T} ( 1 + e^{+ 2 i \omega  \tau } )
     \right]
     \hat{D}_{c}(\omega_{1})
     \hat{D}_{c}(\omega-\omega_{1})
     \nonumber\\
  &&
     -
     2 \sqrt{1-T}
     \frac{\hbar}{2m_{ITM}c}
     e^{ + i \omega \tau_{x}' }
     \int_{-\infty}^{+\infty} \frac{d\omega_{1}}{2\pi}
     \sqrt{|\omega_{1}(\omega-\omega_{1})|}
     \left[ 1 - \sqrt{1-T} e^{+ 2 i \omega_{1} \tau } \right]^{-1}
     \left[ 1 - \sqrt{1-T} e^{+ 2 i (\omega - \omega_{1}) \tau } \right]^{-1}
     \nonumber\\
  && \quad\quad\quad\quad\quad\quad\quad\quad\quad\quad\quad\quad\quad\quad
     \times
     \left[
     e^{+ 2 i \omega_{1} \tau }
     + e^{+ 2 i (\omega-\omega_{1}) \tau }
     -  \sqrt{1-T} ( 1 + e^{+ 2 i \omega  \tau } )
     \right]
     \nonumber\\
  && \quad\quad\quad\quad\quad\quad\quad\quad\quad\quad\quad\quad\quad\quad
     \times
     \hat{D}_{c}(\omega_{1})
     \left(
     \hat{D}_{v}(\omega-\omega_{1}) - \hat{A}(\omega-\omega_{1})
     \right)
     \nonumber\\
  &&
     -
     i \sqrt{1-T}
     \frac{\hbar}{m_{ITM}c}
     \int_{-\infty}^{+\infty} \frac{d\omega_{1}}{2\pi}
     \int_{-\infty}^{+\infty} \frac{d\omega_{2}}{2\pi}
     e^{ + i ( \omega_{1} + \omega_{2} ) \tau_{x}' }
     \frac{\omega_{2}}{c}
     \sqrt{|\omega_{1}\omega_{2}|}
     \left[ 1 - \sqrt{1-T} e^{+ 2 i \omega_{1} \tau } \right]^{-1}
     \nonumber\\
  && \quad\quad\quad\quad\quad\quad\quad\quad\quad\quad\quad\quad\quad\quad\quad\quad
     \times
     \left[ 1 - \sqrt{1-T} e^{+ 2 i (\omega - \omega_{1}) \tau } \right]^{-1}
     \left[
     e^{+ 2 i (\omega-\omega_{1}) \tau }
     + e^{+ 2 i \omega_{1} \tau }
     -  \sqrt{1-T} ( 1 + e^{+ 2 i \omega  \tau } )
     \right]
     \nonumber\\
  && \quad\quad\quad\quad\quad\quad\quad\quad\quad\quad\quad\quad\quad\quad\quad\quad
     \times
     \hat{D}_{c}(\omega_{1})
     \hat{D}_{c}(\omega_{2})
     \FrakD_{d}^{\dagger}\hat{Z}_{XITM}(\omega-\omega_{1}-\omega_{2})\FrakD_{d}
     \nonumber\\
  &&
     -
     i T \sqrt{1-T}
     \frac{2\hbar}{m_{ITM}c^{2}}
     \int_{-\infty}^{+\infty} \frac{d\omega_{1}}{2\pi}
     \int_{-\infty}^{+\infty} \frac{d\omega_{2}}{2\pi}
     \sqrt{|\omega_{1}\omega_{2}|}
     \omega_{2}
     e^{ +  i ( \omega_{2} + \omega - \omega_{1} ) \tau }
     e^{ + i ( \omega_{1} + \omega_{2} ) \tau_{x}' }
     \left[ 1 - \sqrt{1-T} e^{+ 2 i \omega_{1} \tau } \right]^{-1}
     \nonumber\\
  && \quad\quad\quad\quad\quad\quad\quad\quad\quad\quad\quad\quad\quad\quad\quad\quad
     \times
     \left[ 1 - \sqrt{1-T} e^{+ 2 i ( \omega - \omega_{1} ) \tau } \right]^{-1}
     \left[ 1 - \sqrt{1-T} e^{+ 2 i \omega_{2} \tau } \right]^{-1}
     \nonumber\\
  && \quad\quad\quad\quad\quad\quad\quad\quad\quad\quad\quad\quad\quad\quad\quad\quad
     \times
     \hat{D}_{c}(\omega_{1})
     \hat{D}_{c}(\omega_{2})
     \FrakD_{d}^{\dagger}\hat{Z}_{x}(\omega-\omega_{1}-\omega_{2})\FrakD_{d}
     \nonumber\\
  &&
     +
     O\left(
     \left(\hat{X}\right)^{2},
     \hat{D}_{v}\hat{A},
     \left(\hat{A}\right)^{2},
     \left(\hat{D}_{v}\right)^{2},
     \hat{D}_{v}\hat{X},
     \hat{A}\hat{X}
     \right)
     ,
     \label{eq:total-radiation-pressure-force-to-XITM-DAZx-coh-sum}
\end{eqnarray}
\begin{eqnarray}
  &&
     \int_{-\infty}^{+\infty} dt e^{+i\omega t}
     \frac{1}{m_{ITM}}
     \FrakD_{d}^{\dagger}\hat{F}_{rpYITM}(t)\FrakD_{d}
     \nonumber\\
  &=&
      -
      \sqrt{1-T}
      \frac{\hbar}{2m_{ITM}c}
      e^{ + i \omega \tau_{y}' }
      \int_{-\infty}^{+\infty} \frac{d\omega_{1}}{2\pi}
      \sqrt{|\omega_{1}(\omega-\omega_{1})|}
      \left[ 1 - \sqrt{1-T} e^{+ 2 i \omega_{1} \tau } \right]^{-1}
      \left[ 1 - \sqrt{1-T} e^{+ 2 i (\omega - \omega_{1}) \tau } \right]^{-1}
     \nonumber\\
  && \quad\quad\quad\quad\quad\quad\quad\quad\quad\quad\quad\quad\quad\quad
     \times
     \left[
     2 e^{+ 2 i \omega_{1} \tau }
     - \sqrt{1-T} ( 1 + e^{+ 2 i \omega  \tau } )
     \right]
     \hat{D}_{c}(\omega_{1})
     \hat{D}_{c}(\omega-\omega_{1})
     \nonumber\\
  &&
     -
     2 \sqrt{1-T}
     \frac{\hbar}{2m_{ITM}c}
     e^{ + i \omega \tau_{y}' }
     \int_{-\infty}^{+\infty} \frac{d\omega_{1}}{2\pi}
     \sqrt{|\omega_{1}(\omega-\omega_{1})|}
     \left[ 1 - \sqrt{1-T} e^{+ 2 i \omega_{1} \tau } \right]^{-1}
     \left[ 1 - \sqrt{1-T} e^{+ 2 i (\omega - \omega_{1}) \tau } \right]^{-1}
     \nonumber\\
  && \quad\quad\quad\quad\quad\quad\quad\quad\quad\quad\quad\quad\quad\quad
     \times
     \left[
     e^{+ 2 i \omega_{1} \tau }
     + e^{+ 2 i (\omega-\omega_{1}) \tau }
     -  \sqrt{1-T} ( 1 + e^{+ 2 i \omega  \tau } )
     \right]
     \nonumber\\
  && \quad\quad\quad\quad\quad\quad\quad\quad\quad\quad\quad\quad\quad\quad
     \times
     \hat{D}_{c}(\omega_{1})
     \left(
     \hat{D}_{v}(\omega-\omega_{1}) + \hat{A}(\omega-\omega_{1})
     \right)
     \nonumber\\
  &&
     -
     i \sqrt{1-T}
     \frac{\hbar}{m_{ITM}c}
     \int_{-\infty}^{+\infty} \frac{d\omega_{1}}{2\pi}
     \int_{-\infty}^{+\infty} \frac{d\omega_{2}}{2\pi}
     e^{ + i ( \omega_{1} + \omega_{2} ) \tau_{y}' }
     \frac{\omega_{2}}{c}
     \sqrt{|\omega_{1}\omega_{2}|}
     \left[ 1 - \sqrt{1-T} e^{+ 2 i \omega_{1} \tau } \right]^{-1}
     \nonumber\\
  && \quad\quad\quad\quad\quad\quad\quad\quad\quad\quad\quad\quad\quad\quad\quad\quad
     \times
     \left[ 1 - \sqrt{1-T} e^{+ 2 i (\omega - \omega_{1}) \tau } \right]^{-1}
     \left[
     e^{+ 2 i (\omega - \omega_{1}) \tau }
     + e^{+ 2 i \omega_{1} \tau }
     -  \sqrt{1-T} ( 1 + e^{+ 2 i \omega  \tau } )
     \right]
     \nonumber\\
  && \quad\quad\quad\quad\quad\quad\quad\quad\quad\quad\quad\quad\quad\quad\quad\quad
     \times
     \hat{D}_{c}(\omega_{1})
     \hat{D}_{c}(\omega_{2})
     \FrakD_{d}^{\dagger}\hat{Z}_{YITM}(\omega-\omega_{1}-\omega_{2})\FrakD_{d}
     \nonumber\\
  &&
     -
     i T \sqrt{1-T}
     \frac{2\hbar}{m_{ITM}c^{2}}
     \int_{-\infty}^{+\infty} \frac{d\omega_{1}}{2\pi}
     \int_{-\infty}^{+\infty} \frac{d\omega_{2}}{2\pi}
     \sqrt{|\omega_{1}\omega_{2}|}
     \omega_{2}
     e^{ +  i ( \omega_{2} + \omega - \omega_{1} ) \tau }
     e^{ + i ( \omega_{1} + \omega_{2} ) \tau_{y}' }
     \left[ 1 - \sqrt{1-T} e^{+ 2 i \omega_{1} \tau } \right]^{-1}
     \nonumber\\
  && \quad\quad\quad\quad\quad\quad\quad\quad\quad\quad\quad\quad\quad\quad\quad\quad
     \times
     \left[ 1 - \sqrt{1-T} e^{+ 2 i ( \omega - \omega_{1} ) \tau } \right]^{-1}
     \left[ 1 - \sqrt{1-T} e^{+ 2 i \omega_{2} \tau } \right]^{-1}
     \nonumber\\
  && \quad\quad\quad\quad\quad\quad\quad\quad\quad\quad\quad\quad\quad\quad\quad\quad
     \times
     \hat{D}_{c}(\omega_{1})
     \hat{D}_{c}(\omega_{2})
     \FrakD_{d}^{\dagger}\hat{Z}_{y}(\omega-\omega_{1}-\omega_{2})\FrakD_{d}
     \nonumber\\
  &&
     +
     O\left(
     \left(\hat{X}\right)^{2},
     \hat{D}_{v}\hat{A},
     \left(\hat{A}\right)^{2},
     \left(\hat{D}_{v}\right)^{2},
     \hat{D}_{v}\hat{X},
     \hat{A}\hat{X}
     \right)
     .
     \label{eq:total-radiation-pressure-force-to-YITM-DA-coh-sum}
\end{eqnarray}
\end{widetext}


Next, we consider the case of the monochromatic incident laser.
In this case, the classical part $\hat{D}_{c}(\omega)$ of the incident
quadrature $\hat{D}(\omega)$ is given by
Eq.~(\ref{eq:hatDc-def-monochromatic}) with
Eq.~(\ref{eq:N-classical-power-relation}).
Substituting Eq.~(\ref{eq:hatDc-def-monochromatic}) into
Eqs.~(\ref{eq:radiation-pressure-Fourier-X-first-term-in-DAZx-cohe-sum})--(\ref{eq:total-radiation-pressure-force-to-YITM-DA-coh-sum}),
we can evaluate the linearized radiation pressure forces in the case
of the monochromatic incident laser with the central frequency
$\omega_{0}$.
The resulting expressions are naturally expressed in the sideband
picture for the frequency of $\omega_{0}+\Omega$ with the sideband
frequency $\Omega$.
As discussed in
Sec.~\ref{sec:Final_Input-output_relation_with_mirrors_motion}, we
ignore the rapidly oscillating terms in the frequencies
$2\omega_{0}\pm\Omega$.
Then, we obtain
\begin{widetext}
\begin{eqnarray}
  &&
     \frac{1}{m_{EM}}
     \int_{-\infty}^{+\infty} dt e^{+i\Omega t}
     \FrakD_{d}^{\dagger}\hat{F}_{rpXEM}(t)\FrakD_{d}
     \nonumber\\
  &=&
      +
      \frac{ N^{2} T\hbar\omega_{0}}{m_{EM}c}
      \left[ 1 - \sqrt{1-T} e^{+ 2 i \omega_{0} \tau } \right]^{-1}
      \left[ 1 - \sqrt{1-T} e^{ - 2 i \omega_{0} \tau } \right]^{-1}
      2 \pi \delta(\Omega)
      \nonumber\\
  &&
     +
     \frac{ N T\hbar}{m_{EM}c}
     e^{ + i \Omega (\tau+\tau_{x}' ) }
     \sqrt{|\omega_{0}(\Omega-\omega_{0})|}
     \left[ 1 - \sqrt{1-T} e^{+ 2 i \omega_{0} \tau } \right]^{-1}
     \left[ 1 - \sqrt{1-T} e^{+ 2 i (\Omega-\omega_{0}) \tau } \right]^{-1}
     \left(
     \hat{D}_{v}(\Omega-\omega_{0}) - \hat{A}(\Omega-\omega_{0})
     \right)
     \nonumber\\
  &&
     +
     \frac{ N T\hbar}{m_{EM}c}
     e^{ + i \Omega (\tau+\tau_{x}') }
     \sqrt{|\omega_{0}(\Omega+\omega_{0})|}
     \left[ 1 - \sqrt{1-T} e^{+ 2 i (\Omega+\omega_{0}) \tau } \right]^{-1}
     \left[ 1 - \sqrt{1-T} e^{ -  2 i \omega_{0} \tau } \right]^{-1}
     \left(
     \hat{D}_{v}(\Omega+\omega_{0}) - \hat{A}(\Omega+\omega_{0})
     \right)
     \nonumber\\
  &&
     +
     2 N^{2} T \sqrt{1-T}
     \sin(2 \omega_{0} \tau )
     \frac{\hbar\omega_{0}^{2}}{m_{EM}c^{2}}
     e^{ + i \Omega \tau }
     \left[ 1 - e^{+ 2 i \Omega \tau } \right]
     \left[ 1 - \sqrt{1-T} e^{+ 2 i \omega_{0} \tau } \right]^{-1}
     \left[ 1 - \sqrt{1-T} e^{ - 2 i \omega_{0} \tau } \right]^{-1}
     \nonumber\\
  && \quad\quad
     \times
     \left[ 1 - \sqrt{1-T} e^{+ 2 i (\Omega+\omega_{0}) \tau } \right]^{-1}
     \left[ 1 - \sqrt{1-T} e^{+ 2 i (\Omega-\omega_{0}) \tau } \right]^{-1}
     \FrakD_{d}^{\dagger}\hat{Z}_{XITM}(\Omega)\FrakD_{d}
     \nonumber\\
  &&
     -
     4 N^{2} T \sqrt{1-T}
     \sin( 2 \omega_{0} \tau )
     \frac{\hbar\omega_{0}^{2}}{m_{EM}c^{2}}
     e^{ + 2 i \Omega \tau }
     \left[ 1 - \sqrt{1-T} e^{ - 2 i \omega_{0} \tau } \right]^{-1}
     \left[ 1 - \sqrt{1-T} e^{+ 2 i \omega_{0} \tau } \right]^{-1}
     \nonumber\\
  && \quad\quad
     \times
     \left[ 1 - \sqrt{1-T} e^{+ 2 i (\Omega-\omega_{0}) \tau } \right]^{-1}
     \left[ 1 - \sqrt{1-T} e^{+ 2 i (\Omega+\omega_{0}) \tau } \right]^{-1}
     \FrakD_{d}^{\dagger}\hat{Z}_{x}(\Omega)\FrakD_{d}
     \nonumber\\
  &&
     +
     O\left(
     \left(\hat{X}\right)^{2},
     \hat{D}_{v}\hat{A},
     \left(\hat{A}\right)^{2},
     \left(\hat{D}_{v}\right)^{2},
     \hat{D}_{v}\hat{X},
     \hat{A}\hat{X}
     \right)
     \nonumber\\
  &&
     +
     \mbox{``rapid oscillation terms with the frequency $2\omega_{0}\pm\omega$"}
     ,
     \label{eq:DdaggerD-rad-pres-F-XEM-in-DA-Dmono-hineg}
\end{eqnarray}
\begin{eqnarray}
  &&
     \frac{1}{m_{EM}}
     \int_{-\infty}^{+\infty} dt e^{+i\Omega t}
     \FrakD_{d}^{\dagger}\hat{F}_{rpYEM}(t)\FrakD_{d}
     \nonumber\\
  &=&
      +
      \frac{ N^{2} T\hbar\omega_{0}}{m_{EM}c}
      \left[ 1 - \sqrt{1-T} e^{ - 2 i \omega_{0} \tau } \right]^{-1}
      \left[ 1 - \sqrt{1-T} e^{+ 2 i \omega_{0} \tau } \right]^{-1}
      2 \pi \delta(\Omega)
      \nonumber\\
  &&
     +
     \frac{ N T\hbar}{m_{EM}c}
     e^{ + i \Omega (\tau+\tau_{y}') }
     \sqrt{|(\Omega-\omega_{0})\omega_{0}|}
     \left[ 1 - \sqrt{1-T} e^{+ 2 i \omega_{0} \tau } \right]^{-1}
     \left[ 1 - \sqrt{1-T} e^{+ 2 i (\Omega-\omega_{0}) \tau } \right]^{-1}
     \left(
     \hat{D}_{v}(\Omega-\omega_{0}) + \hat{A}(\Omega-\omega_{0})
     \right)
     \nonumber\\
  &&
     +
     \frac{ N T\hbar}{m_{EM}c}
     e^{ + i \Omega (\tau+\tau_{y}') }
     \sqrt{|(\Omega+\omega_{0})\omega_{0}|}
     \left[ 1 - \sqrt{1-T} e^{ - 2 i \omega_{0} \tau } \right]^{-1}
     \left[ 1 - \sqrt{1-T} e^{+ 2 i (\Omega+\omega_{0}) \tau } \right]^{-1}
     \left(
     \hat{D}_{v}(\Omega+\omega_{0}) + \hat{A}(\Omega+\omega_{0})
     \right)
     \nonumber\\
  &&
     +
     2 N^{2} T \sqrt{1-T}
     \sin( 2 \omega_{0} \tau )
     \frac{\hbar\omega_{0}^{2}}{m_{EM}c^{2}}
     e^{ + i \Omega \tau }
     \left[ 1 - e^{+ 2 i \Omega \tau } \right]
     \left[ 1 - \sqrt{1-T} e^{+ 2 i \omega_{0} \tau } \right]^{-1}
     \left[ 1 - \sqrt{1-T} e^{ - 2 i \omega_{0} \tau } \right]^{-1}
     \nonumber\\
  && \quad\quad
     \times
     \left[ 1 - \sqrt{1-T} e^{+ 2 i (\Omega+\omega_{0}) \tau } \right]^{-1}
     \left[ 1 - \sqrt{1-T} e^{+ 2 i (\Omega-\omega_{0}) \tau } \right]^{-1}
     \FrakD_{d}^{\dagger}\hat{Z}_{YITM}(\Omega)\FrakD_{d}
     \nonumber\\
  &&
     -
     4 N^{2} T \sqrt{1-T}
     e^{+ 2 i \Omega \tau }
     \sin( 2 \omega_{0} \tau )
     \frac{\hbar\omega_{0}^{2}}{m_{EM}c^{2}}
     \left[ 1 - \sqrt{1-T} e^{ - 2 i \omega_{0} \tau } \right]^{-1}
     \left[ 1 - \sqrt{1-T} e^{+ 2 i \omega_{0} \tau } \right]^{-1}
     \nonumber\\
  && \quad\quad
     \times
     \left[ 1 - \sqrt{1-T} e^{+ 2 i (\Omega-\omega_{0}) \tau } \right]^{-1}
     \left[ 1 - \sqrt{1-T} e^{+ 2 i (\Omega+\omega_{0}) \tau } \right]^{-1}
     \FrakD_{d}^{\dagger}\hat{Z}_{y}(\Omega)\FrakD_{d}
     \nonumber\\
  &&
     +
     O\left(
     \left(\hat{X}\right)^{2},
     \hat{D}_{v}\hat{A},
     \left(\hat{A}\right)^{2},
     \left(\hat{D}_{v}\right)^{2},
     \hat{D}_{v}\hat{X},
     \hat{A}\hat{X}
     \right)
     \nonumber\\
  &&
     +
     \mbox{``rapid oscillation terms with the frequency $2\omega_{0}\pm\omega$"}
     ,
     \label{eq:DdaggerD-rad-pres-F-YEM-in-DA-Dmono-hineg}
\end{eqnarray}
\begin{eqnarray}
  &&
     \FrakD_{d}^{\dagger}\scrF_{rpXITM}(\Omega)\FrakD_{d}
     \nonumber\\
  &=&
     \int_{-\infty}^{+\infty} dt e^{+i\Omega t}
     \frac{1}{m_{ITM}}
     \FrakD_{d}^{\dagger}\hat{F}_{rpXITM}(t)\FrakD_{d}
     \nonumber\\
  &=&
      -
      2 N^{2} \sqrt{1-T}
      \frac{\hbar\omega_{0}}{m_{ITM}c}
      \left[ 1 - \sqrt{1-T} e^{+ 2 i \omega_{0} \tau } \right]^{-1}
      \left[ 1 - \sqrt{1-T} e^{ - 2 i \omega_{0} \tau } \right]^{-1}
      \left[ \cos( 2 \omega_{0} \tau ) - \sqrt{1-T} \right]
      2 \pi \delta(\Omega)
      \nonumber\\
  &&
     -
     N \sqrt{1-T}
     \frac{\hbar}{m_{ITM}c}
     e^{ + i \Omega \tau_{x}' }
     \sqrt{|\omega_{0}(\Omega-\omega_{0})|}
     \left[ 1 - \sqrt{1-T} e^{+ 2 i \omega_{0} \tau } \right]^{-1}
     \left[ 1 - \sqrt{1-T} e^{+ 2 i (\Omega - \omega_{0}) \tau } \right]^{-1}
     \nonumber\\
  && \quad\quad
     \times
     \left[
     e^{+ 2 i \omega_{0} \tau }
     + e^{+ 2 i (\Omega-\omega_{0}) \tau }
     -  \sqrt{1-T} ( 1 + e^{+ 2 i \Omega  \tau } )
     \right]
     \left(
     \hat{D}_{v}(\Omega-\omega_{0}) - \hat{A}(\Omega-\omega_{0})
     \right)
     \nonumber\\
  &&
     -
     N \sqrt{1-T}
     \frac{\hbar}{m_{ITM}c}
     e^{ + i \Omega \tau_{x}' }
     \sqrt{|\omega_{0}(\Omega+\omega_{0})|}
     \left[ 1 - \sqrt{1-T} e^{ - 2 i \omega_{0} \tau } \right]^{-1}
     \left[ 1 - \sqrt{1-T} e^{+ 2 i (\Omega + \omega_{0}) \tau } \right]^{-1}
     \nonumber\\
  && \quad\quad
     \times
     \left[
     e^{ - 2 i \omega_{0} \tau }
     + e^{+ 2 i (\Omega+\omega_{0}) \tau }
     -  \sqrt{1-T} ( 1 + e^{+ 2 i \Omega  \tau } )
     \right]
     \left(
     \hat{D}_{v}(\Omega+\omega_{0}) - \hat{A}(\Omega+\omega_{0})
     \right)
     \nonumber\\
  &&
     -
     N^{2} T \sqrt{1-T}
     2 \sin(2 \omega_{0} \tau )
     \frac{\hbar\omega_{0}^{2}}{m_{ITM}c^{2}}
     \left[ 1 - \sqrt{1-T} e^{ - 2 i \omega_{0} \tau } \right]^{-1}
     \left[ 1 - \sqrt{1-T} e^{+ 2 i \omega_{0} \tau } \right]^{-1}
     \nonumber\\
  && \quad\quad
     \times
     \left[ 1 - \sqrt{1-T} e^{+ 2 i (\Omega - \omega_{0}) \tau } \right]^{-1}
     \left[ 1 - \sqrt{1-T} e^{+ 2 i (\Omega + \omega_{0}) \tau } \right]^{-1}
     \left[ 1 - e^{+ 2 i \Omega \tau } \right]
     \FrakD_{d}^{\dagger}\hat{Z}_{XITM}(\Omega)\FrakD_{d}
     \nonumber\\
  &&
     +
     4 N^{2} T \sqrt{1-T}
     e^{ +  i \Omega \tau }
     \sin( 2 \omega_{0} \tau )
     \frac{\hbar\omega_{0}^{2}}{m_{ITM}c^{2}}
     \left[ 1 - \sqrt{1-T} e^{ - 2 i \omega_{0} \tau } \right]^{-1}
     \left[ 1 - \sqrt{1-T} e^{+ 2 i \omega_{0} \tau } \right]^{-1}
     \nonumber\\
  && \quad\quad
     \times
     \left[ 1 - \sqrt{1-T} e^{+ 2 i ( \Omega - \omega_{0} ) \tau } \right]^{-1}
     \left[ 1 - \sqrt{1-T} e^{+ 2 i ( \Omega + \omega_{0} ) \tau } \right]^{-1}
     \FrakD_{d}^{\dagger}\hat{Z}_{x}(\Omega)\FrakD_{d}
     \nonumber\\
  &&
     +
     O\left(
     \left(\hat{X}\right)^{2},
     \hat{D}_{v}\hat{A},
     \left(\hat{A}\right)^{2},
     \left(\hat{D}_{v}\right)^{2},
     \hat{D}_{v}\hat{X},
     \hat{A}\hat{X}
     \right)
     \nonumber\\
  &&
     +
     \mbox{``rapid oscillation terms with the frequency $2\omega_{0}\pm\omega$"}
     ,
     \label{eq:DdaggerD-rad-pres-F-XITM-DAZ-Dmono-hineg}
\end{eqnarray}
\begin{eqnarray}
  &&
     \FrakD_{d}^{\dagger}\scrF_{rpYITM}(\Omega)\FrakD_{d}
     \nonumber\\
  &=&
     \int_{-\infty}^{+\infty} dt e^{+i\Omega t}
     \frac{1}{m_{ITM}}
     \FrakD_{d}^{\dagger}\hat{F}_{rpYITM}(t)\FrakD_{d}
     \nonumber\\
  &=&
      -
      2 N^{2} \sqrt{1-T}
      \frac{\hbar\omega_{0}}{m_{ITM}c}
      \left[ 1 - \sqrt{1-T} e^{+ 2 i \omega_{0} \tau } \right]^{-1}
      \left[ 1 - \sqrt{1-T} e^{ - 2 i \omega_{0} \tau } \right]^{-1}
      \left[ \cos( 2 \omega_{0} \tau ) -\sqrt{1-T} \right]
      2 \pi \delta(\Omega)
      \nonumber\\
  &&
     -
     N \sqrt{1-T}
     \frac{\hbar}{m_{ITM}c}
     e^{ + i \Omega \tau_{y}' }
     \sqrt{|\omega_{0}(\Omega-\omega_{0})|}
     \left[ 1 - \sqrt{1-T} e^{+ 2 i \omega_{0} \tau } \right]^{-1}
     \left[ 1 - \sqrt{1-T} e^{+ 2 i (\Omega - \omega_{0}) \tau } \right]^{-1}
     \nonumber\\
  && \quad\quad
     \times
     \left[
     e^{+ 2 i \omega_{0} \tau }
     + e^{+ 2 i (\Omega-\omega_{0}) \tau }
     -  \sqrt{1-T} ( 1 + e^{+ 2 i \Omega  \tau } )
     \right]
     \left(
     \hat{D}_{v}(\Omega-\omega_{0}) + \hat{A}(\Omega-\omega_{0})
     \right)
     \nonumber\\
  &&
     -
     2 N \sqrt{1-T}
     \frac{\hbar}{2m_{ITM}c}
     e^{ + i \Omega \tau_{y}' }
     \sqrt{|\omega_{0}(\Omega+\omega_{0})|}
     \left[ 1 - \sqrt{1-T} e^{ - 2 i \omega_{0} \tau } \right]^{-1}
     \left[ 1 - \sqrt{1-T} e^{+ 2 i (\Omega + \omega_{0}) \tau } \right]^{-1}
     \nonumber\\
  && \quad\quad
     \times
     \left[
     e^{ - 2 i \omega_{0} \tau }
     + e^{+ 2 i (\Omega+\omega_{0}) \tau }
     -  \sqrt{1-T} ( 1 + e^{+ 2 i \Omega  \tau } )
     \right]
     \left(
     \hat{D}_{v}(\Omega+\omega_{0}) + \hat{A}(\Omega+\omega_{0})
     \right)
     \nonumber\\
  &&
     -
     2 N^{2} T \sqrt{1-T}
     \sin( 2 \omega_{0} \tau )
     \frac{\hbar\omega_{0}^{2}}{m_{ITM}c^{2}}
     \left[ 1 - \sqrt{1-T} e^{ - 2 i \omega_{0} \tau } \right]^{-1}
     \left[ 1 - \sqrt{1-T} e^{+ 2 i \omega_{0} \tau } \right]^{-1}
     \nonumber\\
  && \quad\quad
     \times
     \left[ 1 - \sqrt{1-T} e^{+ 2 i (\Omega - \omega_{0}) \tau } \right]^{-1}
     \left[ 1 - \sqrt{1-T} e^{+ 2 i (\Omega + \omega_{0}) \tau } \right]^{-1}
     \left[ 1 - e^{+ 2 i \Omega \tau } \right]
     \FrakD_{d}^{\dagger}\hat{Z}_{YITM}(\Omega)\FrakD_{d}
     \nonumber\\
  &&
     +
     4 N^{2} T \sqrt{1-T}
     e^{ +  i  \Omega \tau }
     \sin( 2 \omega_{0} \tau )
     \frac{\hbar\omega_{0}^{2}}{m_{ITM}c^{2}}
     \left[ 1 - \sqrt{1-T} e^{ - 2 i \omega_{0} \tau } \right]^{-1}
     \left[ 1 - \sqrt{1-T} e^{+ 2 i \omega_{0} \tau } \right]^{-1}
     \nonumber\\
  && \quad\quad
     \times
     \left[ 1 - \sqrt{1-T} e^{+ 2 i ( \Omega - \omega_{0} ) \tau } \right]^{-1}
     \left[ 1 - \sqrt{1-T} e^{+ 2 i ( \Omega + \omega_{0} ) \tau } \right]^{-1}
     \FrakD_{d}^{\dagger}\hat{Z}_{y}(\Omega)\FrakD_{d}
     \nonumber\\
  &&
     +
     O\left(
     \left(\hat{X}\right)^{2},
     \hat{D}_{v}\hat{A},
     \left(\hat{A}\right)^{2},
     \left(\hat{D}_{v}\right)^{2},
     \hat{D}_{v}\hat{X},
     \hat{A}\hat{X}
     \right)
     \nonumber\\
  &&
     +
     \mbox{``rapid oscillation terms with the frequency $2\omega_{0}\pm\omega$"}
     .
     \label{eq:DdaggerD-rad-pres-F-YITM-in-DA-Dmono-hineg}
\end{eqnarray}
\end{widetext}
Here, $\scrF_{rpXITM}(\omega)$ and $\scrF_{rpYITM}(\omega)$ are
defined in Eq.~(\ref{eq:XITM-radi-press-Fourier}) and
(\ref{eq:YITM-radi-press-Fourier}), respectively.
Furthermore, from the definition
(\ref{eq:radiation-pressure-Fourier-X-def}) and
(\ref{eq:radiation-pressure-Fourier-Y-def}) of $\scrF_{rpx}(\omega)$
and $\scrF_{rpy}(\omega)$, we obtain
\begin{widetext}
\begin{eqnarray}
  &&
     \FrakD_{d}^{\dagger}\scrF_{rpx}(\Omega)\FrakD_{d}
     \nonumber\\
  &=&
      \frac{N^{2}\hbar\omega_{0}}{c}
      \left[ 1 - \sqrt{1-T} e^{+ 2 i \omega_{0} \tau } \right]^{-1}
      \left[ 1 - \sqrt{1-T} e^{ - 2 i \omega_{0} \tau } \right]^{-1}
     \left[
     \frac{T}{m_{EM}}
     +
     \frac{2\sqrt{1-T}}{m_{ITM}}
     \left[ \cos( 2 \omega_{0} \tau ) - \sqrt{1-T} \right]
     \right]
     2 \pi \delta(\Omega)
      \nonumber\\
  &&
     +
     \frac{N\hbar}{c}
     e^{ + i \Omega \tau_{x}' }
     e^{ + i \Omega \tau }
     \sqrt{|\omega_{0}(\Omega-\omega_{0})|}
     \left[ 1 - \sqrt{1-T} e^{+ 2 i \omega_{0} \tau } \right]^{-1}
     \left[ 1 - \sqrt{1-T} e^{+ 2 i (\Omega-\omega_{0}) \tau } \right]^{-1}
     \nonumber\\
  && \quad\quad
     \times
     \left[
     \frac{T}{m_{EM}}
     +
     \frac{2\sqrt{1-T}}{m_{ITM}}
     \left[ \cos( (2 \omega_{0}-\Omega) \tau ) - \sqrt{1-T} \cos(\Omega\tau) \right]
     \right]
     \left(
     \hat{D}_{v}(\Omega-\omega_{0}) - \hat{A}(\Omega-\omega_{0})
     \right)
     \nonumber\\
  &&
     +
     \frac{N\hbar}{c}
     e^{ + i \Omega \tau_{x}' }
     e^{ + i \Omega \tau }
     \sqrt{|\omega_{0}(\Omega+\omega_{0})|}
     \left[ 1 - \sqrt{1-T} e^{+ 2 i (\Omega+\omega_{0}) \tau } \right]^{-1}
     \left[ 1 - \sqrt{1-T} e^{ -  2 i \omega_{0} \tau } \right]^{-1}
     \nonumber\\
  && \quad\quad
     \times
     \left[
     \frac{T}{m_{EM}}
     +
     \frac{2\sqrt{1-T}}{m_{ITM}}
     \left[ \cos( ( 2 \omega_{0} + \Omega ) \tau ) - \sqrt{1-T} \cos( \Omega \tau ) \right]
     \right]
     \left(
     \hat{D}_{v}(\Omega+\omega_{0}) - \hat{A}(\Omega+\omega_{0})
     \right)
     \nonumber\\
  &&
     +
     \sin(2 \omega_{0} \tau )
     \frac{2 N^{2} T \sqrt{1-T} \hbar\omega_{0}^{2}}{c^{2}}
     \left[ \frac{1}{m_{EM}} e^{ + i \Omega \tau } + \frac{1}{m_{ITM}} \right]
     \left[ 1 - \sqrt{1-T} e^{+ 2 i \omega_{0} \tau } \right]^{-1}
     \left[ 1 - \sqrt{1-T} e^{ - 2 i \omega_{0} \tau } \right]^{-1}
     \nonumber\\
  && \quad\quad
     \times
     \left[ 1 - \sqrt{1-T} e^{+ 2 i (\Omega+\omega_{0}) \tau } \right]^{-1}
     \left[ 1 - \sqrt{1-T} e^{+ 2 i (\Omega-\omega_{0}) \tau } \right]^{-1}
     \left[ 1 - e^{+ 2 i \Omega \tau } \right]
     \FrakD_{d}^{\dagger}\hat{Z}_{XITM}(\Omega)\FrakD_{d}
     \nonumber\\
  &&
     -
     \sin( 2 \omega_{0} \tau )
     e^{ +  i \Omega \tau }
     \frac{4 N^{2} T \sqrt{1-T} \hbar\omega_{0}^{2}}{c^{2}}
     \left[ \frac{1}{m_{EM}} e^{ + i \Omega \tau } + \frac{1}{m_{ITM}} \right]
     \left[ 1 - \sqrt{1-T} e^{+ 2 i \omega_{0} \tau } \right]^{-1}
     \left[ 1 - \sqrt{1-T} e^{ - 2 i \omega_{0} \tau } \right]^{-1}
     \nonumber\\
  && \quad\quad
     \times
     \left[ 1 - \sqrt{1-T} e^{+ 2 i (\Omega-\omega_{0}) \tau } \right]^{-1}
     \left[ 1 - \sqrt{1-T} e^{+ 2 i (\Omega+\omega_{0}) \tau } \right]^{-1}
     \FrakD_{d}^{\dagger}\hat{Z}_{x}(\Omega)\FrakD_{d}
     \nonumber\\
  &&
     +
     O\left(
     \left(\hat{X}\right)^{2},
     \hat{D}_{v}\hat{A},
     \left(\hat{A}\right)^{2},
     \left(\hat{D}_{v}\right)^{2},
     \hat{D}_{v}\hat{X},
     \hat{A}\hat{X}
     \right)
     \nonumber\\
  &&
     +
     \mbox{``rapid oscillation terms with the frequency $2\omega_{0}\pm\omega$"}
     ,
     \label{eq:DdaggerD-rad-pres-Frpx-Dmono-hineg}
\end{eqnarray}
\begin{eqnarray}
  &&
     \FrakD_{d}^{\dagger}\scrF_{rpy}(\Omega)\FrakD_{d}
     \nonumber\\
  &=&
      N^{2}
      \frac{\hbar\omega_{0}}{c}
      \left[ 1 - \sqrt{1-T} e^{ - 2 i \omega_{0} \tau } \right]^{-1}
      \left[ 1 - \sqrt{1-T} e^{+ 2 i \omega_{0} \tau } \right]^{-1}
      \left[
      \frac{T}{m_{EM}}
      +
      \frac{2 \sqrt{1-T}}{m_{ITM}}
      \left[ \cos( 2 \omega_{0} \tau ) -\sqrt{1-T} \right]
      \right]
      2 \pi \delta(\Omega)
      \nonumber\\
  &&
     +
     N
     \frac{\hbar}{c}
     e^{ + i \Omega \tau }
     e^{ + i \Omega \tau_{y}' }
     \sqrt{|(\Omega-\omega_{0})\omega_{0}|}
     \left[ 1 - \sqrt{1-T} e^{+ 2 i \omega_{0} \tau } \right]^{-1}
     \left[ 1 - \sqrt{1-T} e^{+ 2 i (\Omega-\omega_{0}) \tau } \right]^{-1}
     \nonumber\\
  && \quad\quad
     \times
     \left[
     \frac{T}{m_{EM}}
     +
     \frac{2 \sqrt{1-T}}{m_{ITM}}
     \left[ \cos( ( 2 \omega_{0} - \Omega ) \tau ) -\sqrt{1-T} \cos( \Omega \tau ) \right]
     \right]
     \left(
     \hat{D}_{v}(\Omega-\omega_{0}) + \hat{A}(\Omega-\omega_{0})
     \right)
     \nonumber\\
  &&
     +
     N
     \frac{\hbar}{c}
     e^{ + i \Omega \tau }
     e^{ + i \Omega \tau_{y}' }
     \sqrt{|(\Omega+\omega_{0})\omega_{0}|}
     \left[ 1 - \sqrt{1-T} e^{ - 2 i \omega_{0} \tau } \right]^{-1}
     \left[ 1 - \sqrt{1-T} e^{+ 2 i (\Omega+\omega_{0}) \tau } \right]^{-1}
     \nonumber\\
  && \quad\quad
     \times
     \left[
     \frac{T}{m_{EM}}
     +
     \frac{2 \sqrt{1-T}}{m_{ITM}}
     \left[ \cos( ( 2 \omega_{0} + \omega ) \tau ) - \sqrt{1-T} \cos( \omega \tau )  \right]
     \right]
     \left(
     \hat{D}_{v}(\Omega+\omega_{0}) + \hat{A}(\Omega+\omega_{0})
     \right)
     \nonumber\\
  &&
     +
     2 N^{2} T \sqrt{1-T}
     \sin( 2 \omega_{0} \tau )
     \frac{\hbar\omega_{0}^{2}}{c^{2}}
     \left[ 1 - \sqrt{1-T} e^{ - 2 i \omega_{0} \tau } \right]^{-1}
     \left[ 1 - \sqrt{1-T} e^{+ 2 i \omega_{0} \tau } \right]^{-1}
     \left[ 1 - \sqrt{1-T} e^{+ 2 i (\Omega-\omega_{0}) \tau } \right]^{-1}
     \nonumber\\
  && \quad\quad
     \times
     \left[ 1 - \sqrt{1-T} e^{+ 2 i (\Omega+\omega_{0}) \tau } \right]^{-1}
     \left[ 1 - e^{+ 2 i \Omega \tau } \right]
     \left[ \frac{1}{m_{EM}} e^{ + i \Omega \tau } + \frac{1}{m_{ITM}} \right]
     \FrakD_{d}^{\dagger}\hat{Z}_{YITM}(\Omega)\FrakD_{d}
     \nonumber\\
  &&
     -
     4 N^{2} T \sqrt{1-T}
     \sin( 2 \omega_{0} \tau )
     e^{+ i \Omega \tau }
     \frac{\hbar\omega_{0}^{2}}{c^{2}}
     \left[ \frac{1}{m_{EM}} e^{+ i \Omega \tau } + \frac{1}{m_{ITM}} \right]
     \left[ 1 - \sqrt{1-T} e^{+ 2 i \omega_{0} \tau } \right]^{-1}
     \left[ 1 - \sqrt{1-T} e^{ - 2 i \omega_{0} \tau } \right]^{-1}
     \nonumber\\
  && \quad\quad
     \times
     \left[ 1 - \sqrt{1-T} e^{+ 2 i (\Omega+\omega_{0}) \tau } \right]^{-1}
     \left[ 1 - \sqrt{1-T} e^{+ 2 i (\Omega-\omega_{0}) \tau } \right]^{-1}
     \FrakD_{d}^{\dagger}\hat{Z}_{y}(\Omega)\FrakD_{d}
     \nonumber\\
  &&
     +
     O\left(
     \left(\hat{X}\right)^{2},
     \hat{D}_{v}\hat{A},
     \left(\hat{A}\right)^{2},
     \left(\hat{D}_{v}\right)^{2},
     \hat{D}_{v}\hat{X},
     \hat{A}\hat{X}
     \right)
     \nonumber\\
  &&
     +
     \mbox{``rapid oscillation terms with the frequency $2\omega_{0}\pm\omega$"}
     .
     \label{eq:DdaggerD-rad-pres-Frpy-Dmono-hineg}
\end{eqnarray}
\end{widetext}
In these expressions
(\ref{eq:DdaggerD-rad-pres-F-XITM-DAZ-Dmono-hineg}),
(\ref{eq:DdaggerD-rad-pres-F-YITM-in-DA-Dmono-hineg}),
(\ref{eq:DdaggerD-rad-pres-Frpx-Dmono-hineg}), and
(\ref{eq:DdaggerD-rad-pres-Frpy-Dmono-hineg}), we have the terms proportional to
$\FrakD_{d}^{\dagger}\hat{Z}_{XITM}(\Omega)\FrakD_{d}$,
$\FrakD_{d}^{\dagger}\hat{Z}_{x}(\Omega)\FrakD_{d}$,
$\FrakD_{d}^{\dagger}\hat{Z}_{YITM}(\Omega)\FrakD_{d}$,
$\FrakD_{d}^{\dagger}\hat{Z_{y}}(\Omega)\FrakD_{d}$ which includes the
factor $\sin(2\omega_{0}\tau)$.
These terms correspond to the optical spring effects of the
Fabry-P\'erot
interferometer~\cite{A.DiVirgillio-et-al-2006,A.Rai-G.S.Agarwai-2008}.
However, these optical spring effects can be ignored through the tuning
condition (\ref{eq:omega0-L-tune-cond}) due to the factor
$\sin(2\omega_{0}\tau)$.
In this paper, we only consider the situation of this tuning, in which
there is no optical spring effect.


\section{Consistency relation
  (\ref{eq:calF-different-omega-commutation-relation-Fourier}) of the
  operator $\hat{\calF}(\omega)$}
\label{sec:Consistency_relation_of_calF}


In this appendix, we consider the consistency relation
(\ref{eq:calF-different-omega-commutation-relation-Fourier}) through
the radiation pressure forces~(\ref{eq:DdaggerD-rad-pres-F-XITM-DAZ-Dmono-hineg-toXITM-tautune})--(\ref{eq:DdaggerD-rad-pres-Frpy-Dmono-hineg-tautune}).
However, the confirmation of the consistency relation
(\ref{eq:calF-different-omega-commutation-relation-Fourier}) for these
four radiation pressure forces are essentially identical.
Therefore, in this appendix, we evaluate the consistency relation
(\ref{eq:calF-different-omega-commutation-relation-Fourier}) only for
the radiation pressure force
(\ref{eq:DdaggerD-rad-pres-F-XITM-DAZ-Dmono-hineg-toXITM-tautune}).


The radiation pressure force
(\ref{eq:DdaggerD-rad-pres-F-XITM-DAZ-Dmono-hineg-toXITM-tautune})
includes a classical part.
We consider the commutation relation of the radiation pressure force,
and this classical part does not contribute to the commutation
relation.
Therefore, we ignore the classical part in
Eq.~(\ref{eq:DdaggerD-rad-pres-F-XITM-DAZ-Dmono-hineg-toXITM-tautune}).
Furthermore, we only consider the situation $\Omega\ll\omega_{0}$.
In this case, the radiation pressure
force~(\ref{eq:DdaggerD-rad-pres-F-XITM-DAZ-Dmono-hineg-toXITM-tautune})
is symbolically represented by
\begin{eqnarray}
  &&
     \!\!\!\!\!\!\!\!\!\!\!\!\!
  - \frac{1}{\mu} \hat{\cal F}(\Omega)
     \nonumber\\
  &=&
      \frac{\omega_{0}}{\mu_{XITM}(\Omega)}
      \left(
      \hat{D}_{v}(\Omega-\omega_{0}) - \hat{A}(\Omega-\omega_{0})
      \right.
      \nonumber\\
  && \quad\quad\quad\quad\quad\quad
      \left.
      +
      \hat{D}_{v}(\Omega+\omega_{0}) - \hat{A}(\Omega+\omega_{0})
      \right)
     \nonumber\\
  &=&
      \frac{\omega_{0}}{\mu_{XITM}(\Omega)}
      \left(
      \hat{d}^{\dagger}(\omega_{0}-\Omega) - \hat{a}^{\dagger}(\omega_{0}-\Omega)
      \right.
      \nonumber\\
  && \quad\quad\quad\quad\quad\quad
      \left.
      +
      \hat{d}(\omega_{0}+\Omega) - \hat{a}(\omega_{0}+\Omega)
      \right)
      ,
      \label{eq:DdaggerD-rad-pres-F-XITM-symbolic}
\end{eqnarray}
where
\begin{eqnarray}
  \frac{1}{\mu_{XITM}(\Omega)}
  &:=&
       \frac{2N\hbar\sqrt{1-T}}{m_{ITM}c}
       e^{ + i \Omega (\tau+\tau_{x}') }
       \nonumber\\
  && \quad
     \times
     \left[ 1 - \sqrt{1-T} e^{+ 2 i \Omega \tau } \right]^{-1}
     \cos(\Omega\tau)
     .
     \nonumber\\
  \label{eq:1overmuXITM}
\end{eqnarray}


Now, we evaluate the consistency relation
(\ref{eq:calF-different-omega-commutation-relation-Fourier}) for the
radiation pressure force (\ref{eq:DdaggerD-rad-pres-F-XITM-symbolic}) as
\begin{eqnarray}
  &&
     \int_{-\infty}^{+\infty} \frac{d\omega_{1}}{2\pi}
     \frac{
     \omega-\omega_{1}
     }{
     \left(\omega_{1}^{2} - \omega_{p}^{2}\right)
     \left((\omega-\omega_{1})^{2} - \omega_{p}^{2}\right)
     }
     \nonumber\\
  && \quad\quad
     \times
     \left[
     \frac{1}{\mu}
     \hat{\calF}(\omega_{1})
     ,
     \frac{1}{\mu}
     \hat{\calF}(\omega-\omega_{1})
     \right]
     \nonumber\\
  &=&
      \int_{-\infty}^{+\infty} \frac{d\omega_{1}}{2\pi}
      \frac{
      \omega-\omega_{1}
      }{
      \left(\omega_{1}^{2} - \omega_{p}^{2}\right)
      \left((\omega-\omega_{1})^{2} - \omega_{p}^{2}\right)
      }
     \nonumber\\
  && \quad\quad
     \times
      \frac{
      \omega_{0}^{2}
      }{
      \mu_{XITM}(\omega_{1})
      \mu_{XITM}(\omega-\omega_{1})
      }
     \nonumber\\
  && \quad\quad
     \times
     \left[
      \hat{d}^{\dagger}(\omega_{0}-\omega_{1}) - \hat{a}^{\dagger}(\omega_{0}-\omega_{1})
     \right.
     \nonumber\\
  && \quad\quad\quad\quad
     \left.
      +
      \hat{d}(\omega_{0}+\omega_{1}) - \hat{a}(\omega_{0}+\omega_{1})
     ,
     \right.
     \nonumber\\
  && \quad\quad\quad\quad
     \left.
      \hat{d}^{\dagger}(\omega_{0}-(\omega-\omega_{1})) - \hat{a}^{\dagger}(\omega_{0}-(\omega-\omega_{1}))
     \right.
     \nonumber\\
  && \quad\quad\quad\quad
     \left.
      +
      \hat{d}(\omega_{0}+(\omega-\omega_{1})) - \hat{a}(\omega_{0}+(\omega-\omega_{1}))
     \right]
     \nonumber\\
  &=&
      \int_{-\infty}^{+\infty} \frac{d\omega_{1}}{2\pi}
      \frac{
      \omega-\omega_{1}
      }{
      \left(\omega_{1}^{2} - \omega_{p}^{2}\right)
      \left((\omega-\omega_{1})^{2} - \omega_{p}^{2}\right)
      }
     \nonumber\\
  && \quad\quad
     \times
      \frac{
      \omega_{0}^{2}
      }{
      \mu_{XITM}(\omega_{1})
      \mu_{XITM}(\omega-\omega_{1})
      }
     \nonumber\\
  && \quad\quad
     \times
     \left\{
     -
     2\pi \delta(\omega)
     -
     2\pi \delta(\omega)
     +
     2\pi \delta(\omega)
     +
     2 \pi \delta(\omega)
     \right\}
     \nonumber\\
  &=&
      0
      .
      \label{eq:calF-commutation-relation-Fourier-XITM-symbolic}
\end{eqnarray}
Then, we have confirmed the consistency relation
(\ref{eq:calF-different-omega-commutation-relation-Fourier}) for the
radiation pressure force (\ref{eq:DdaggerD-rad-pres-F-XITM-symbolic}),
i.e., the radiation pressure force
(\ref{eq:DdaggerD-rad-pres-F-XITM-DAZ-Dmono-hineg-toXITM-tautune}).


For the other radiation pressure forces
(\ref{eq:DdaggerD-rad-pres-F-YITM-in-DA-Dmono-hineg-toYITM-tautune})--(\ref{eq:DdaggerD-rad-pres-Frpy-Dmono-hineg-tautune}),
the evaluations of the consistency relation
(\ref{eq:calF-different-omega-commutation-relation-Fourier}) are
similar to that for the radiation pressure force
(\ref{eq:DdaggerD-rad-pres-F-XITM-DAZ-Dmono-hineg-toXITM-tautune})
shown above.


\begin{acknowledgments}
  The author acknowledges Masa-Katsu Fujimoto for valuable discussions
  and his continuous encouragement.
  He also thanks Tomotada Akutsu, Takayuki Tomaru, and colleagues in
  the Gravitational Science Project at the National Astronomical
  Observatory of Japan for their continuous encouragement and
  conversations.
\end{acknowledgments}


\begin{thebibliography}{999}
\bibitem{K.Nakamura-2025-footnote1}
  As shown in
  Sec.~\ref{sec:General_arguments_for_the_DC-readout_scheme}, we use
  the monochromatic classical carrier field as the reference for the
  measurement of the small signal.
  This classical carrier field is also described by the
  interferometer's input-output relation.
\bibitem{K.Nakamura-2025-footnote2}
  In this paper, we consider a closed quantum system comprising the
  laser and test masses, in the presence of a classical external
  gravitational-wave signal.
  From the viewpoint of test masses, the radiation pressure from the
  laser and the gravitational-wave signal are the quantum and
  classical external forces that have nothing to do with the degree of
  freedom of the motion of the test masses.
  Furthermore, to regularize the effects of the classical forces in
  the solutions discussed in
  Sec.~\ref{sec:Sol_to_Heisenberg_Eq_for_mirrors}, we consider the
  model of a forced harmonic oscillator with the fundamental frequency
  $\omega_{p}$.
  This model is a minimal model for the case where we ignore any
  optical spring effects discussed in
  Appendix~\ref{sec:Explicit_evaluation_of_the_radiation_pressure_forces}.
  Instead of a harmonic oscillator, the reader might consider the
  friction term in the equation of motion for test masses to
  regularize the effects of the classical forces in
  Sec.~\ref{sec:Sol_to_Heisenberg_Eq_for_mirrors}.
  However, the quantum mechanical treatment of the friction term
  implies the existence of an external system to which the energy of
  the test masses escapes due to the friction, and we have to care
  about this external system as a quantum system.
  To avoid such problems, in this paper, the author did not employ
  this friction term but instead used a forced harmonic oscillator as
  a quantum-mechanical model of the test-mass motion.
\bibitem{K.Nakamura-2025-footnote3}
  More precisely, from the Heisenberg equation of mirrors' motion
  which is discussed in Sec.~\ref{sec:Eq_for_mirrors'_motions_and_their_solutions},
  $\FrakD_{d}^{\dagger}\hat{Z}_{diff}(2\omega_{0}\pm\Omega)\FrakD_{d}$
  and
  $\FrakD_{d}^{\dagger}\hat{Z}_{com}(2\omega_{0}\pm\Omega)\FrakD_{d}$
  includes $\FrakD_{d}^{\dagger}\hat{Z}_{diff}(\Omega)\FrakD_{d}$ and
  $\FrakD_{d}^{\dagger}\hat{Z}_{com}(\Omega)\FrakD_{d}$,
  respectively.
  However, these terms include the factor $1/\omega_{0}^{2}$ in their
  coefficients, reflecting the mirror's inertia.
  Since we consider the situation $\omega_{0}$ is sufficiently large,
  the terms $\FrakD_{d}^{\dagger}\hat{Z}_{diff}(\Omega)\FrakD_{d}$ or
  $\FrakD_{d}^{\dagger}\hat{Z}_{com}(\Omega)\FrakD_{d}$ come from
  $\FrakD_{d}^{\dagger}\hat{Z}_{diff}(2\omega_{0}\pm\Omega)\FrakD_{d}$
  and
  $\FrakD_{d}^{\dagger}\hat{Z}_{com}(2\omega_{0}\pm\Omega)\FrakD_{d}$
  through the Heisenberg equations of motion are negligibly small.
\end{thebibliography}

\end{document}